\documentclass[longauth]{aa} 

\bibpunct{(}{)}{;}{a}{}{,} 
\usepackage[normal]{threeparttable}
\usepackage{txfonts}
\usepackage{graphicx}           
\usepackage{natbib}  
\usepackage{hyperref}
\hypersetup{colorlinks,linkcolor={blue},citecolor={blue},urlcolor={red}} 
\usepackage[capitalise]{cleveref}
\usepackage{subfig}
\usepackage[hang,flushmargin]{footmisc}

\newcommand{\vlsr}{\mbox{$V_{\rm LSR}$\,}}

\newcommand{\pc}{\mbox{pc}}

\newcommand{\nthp}{\mbox{N$_2$H$^+$}}

\newcommand{\cleanest}{\texttt{cleanest}\xspace}

\newcommand\statcont{\texttt{statcont}\xspace}
\newcommand\tclean{\texttt{tclean}\xspace}

\newcommand{\msun}{\ensuremath{M_{\odot}}\xspace}                       

\newcommand{\kms}{\textrm{km~s}\ensuremath{^{-1}}\xspace}       

\usepackage{afterpage}

\begin{document}

\title{ALMA-IMF VII - First release of the full spectral line cubes:\\Core kinematics traced by DCN J=(3-2)}
\author{N. Cunningham\inst{1}\and 
A. Ginsburg\inst{2}\and 
R. Galv\'an-Madrid\inst{3}\and 
F. Motte\inst{1}\and 
T. Csengeri\inst{4}\and 
A. M.\ Stutz\inst{5}\and
M. Fern\'andez-L\'opez\inst{6}\and 
R. H. Álvarez-Gutiérrez\inst{5}\and
M. Armante\inst{7,8}\and 
T. Baug\inst{9}\and 
M. Bonfand\inst{4,21}\and 
S. Bontemps\inst{4}\and 
J. Braine\inst{4}\and 
N. Brouillet\inst{4}\and 
G. Busquet\inst{10,11,12}\and 
D. J. D\'iaz-Gonz\'alez\inst{3}\and 
J. Di Francesco\inst{18}\and
A. Gusdorf\inst{7,8}\and 
F. Herpin\inst{4}\and 
H. Liu\inst{19}\and 
A. L\'opez-Sepulcre\inst{1,13}\and 
F. Louvet\inst{1,22}\and 
X. Lu\inst{20}\and 
L. Maud\inst{14}\and 
T. Nony\inst{3}\and 
F. A. Olguin\inst{15}\and 
Y. Pouteau\inst{1}\and 
R. Rivera-Soto\inst{3}\and 
N. A. Sandoval-Garrido\inst{5}\and
P. Sanhueza\inst{16,17}\and 
K. Tatematsu\inst{16,17}\and 
A. P. M. Towner\inst{2}\and
M. Valeille-Manet\inst{4}}

\institute{Universit\'e Grenoble Alpes, CNRS, Institut de Plan\'etologie et d'astrophysique de Grenoble, F-38000, Grenoble, France\and
Department of Astronomy, University of Florida, PO Box 112055, USA \and
Instituto de Radioastronom\'ia y Astrof\'isica, Universidad Nacional Aut\'onoma de M\'exico, Morelia, Michoac\'an 58089, M\'exico\and
Laboratoire d'astrophysique de Bordeaux, Univ. Bordeaux, CNRS, B18N, allée Geoffroy Saint-Hilaire, 33615 Pessac, France\and
Departamento de Astronom\'{i}a, Universidad de Concepci\'{o}n, Casilla 160-C, 4030000 Concepci\'{o}n, Chile\and
Instituto Argentino de Radioastronom\'{\i} a (CCT-La Plata, CONICET; CICPBA), C.C. No. 5, 1894, Villa Elisa, Buenos Aires, Argentina\and
Laboratoire de Physique de l\'{E}cole Normale Sup\'{e}rieure, ENS, Universit\'{e} PSL, CNRS, Sorbonne Universit\'{e}, Universit\'{e} Paris Cit\'{e}, F-75005, Paris, France\and
Observatoire de Paris, PSL University, Sorbonne Universit\'{e}, LERMA, 75014, Paris, France\and
S. N. Bose National Centre for Basic Sciences, Block JD, Sector III, Salt Lake, Kolkata 700106, India\and
Department de Frisca Quantica i Astrofrisca, Universitat de Barcelona (UB), c. Mart\'i i Franqu\`es, 1, 08028 Barcelona, Catalonia, Spain \and
Instituto de Ci\`encias del Cosmos (ICCUB), Universitat de Barcelona (UB), c. Mart\'i i Franqu\`es, 1, 08028 Barcelona, Catalonia, Spain \and
Institut d'Estudis Espacials de Catalunya (IEEC), Gran Capita, 2-4, 08340, Barcelona, Catalonia, Spain\and
Institut de Radioastronomie Millim\'etrique (IRAM), 300 rue de la Piscine, 38406 Saint-Martin-D'H\`eres, France\and
ESO Headquarters, Karl-Schwarzchild-Str 2 D-85748 Garching\and
Institute of Astronomy, National Tsing Hua University, Hsinchu 30013, Taiwan\and
National Astronomical Observatory of Japan, National Institutes of Natural Sciences, 2-21-1 Osawa, Mitaka, Tokyo 181-8588, Japan\and
Astronomical Science Program,
Graduate Institute for Advanced Studies, SOKENDAI,
2-21-1 Osawa, Mitaka, Tokyo 181-8588, Japan\and
Herzberg Astronomy and Astrophysics Research Centre, National Research Council of Canada, 5071 West Saanich Road, Victoria, BC V9E 2E7, Canada\and
Department of Astronomy, Yunnan University, Kunming, 650091, PR China\and
Shanghai Astronomical Observatory, Chinese Academy of Sciences, 80 Nandan Road, Shanghai 200030, People’s Republic of China\and 
Department of Astronomy, University of Virginia, Charlottesville, VA 22904, USA\and
DAS, Universidad de Chile, 1515 camino el observatorio, Las Condes, Santiago, Chile
}

\date{Received November 10, 2022; accepted May 26, 2023}

\authorrunning{Cunningham et al}

\abstract{ALMA-IMF is an Atacama Large Millimeter/submillimeter Array (ALMA) Large Program designed to measure the core mass function (CMF) of 15 protoclusters chosen to span their early evolutionary stages.  It further aims to understand their kinematics, chemistry, and the impact of gas inflow, accretion, and dynamics on the CMF. We present here the first release of the ALMA-IMF line data cubes (DR1), produced from the combination of two ALMA 12m-array configurations. The data include 12 spectral windows, with eight at 1.3~mm and four at 3~mm. The broad spectral coverage of ALMA-IMF ($\sim$6.7\,GHz bandwidth coverage per field) hosts a wealth of simple atomic, molecular, ionised, and complex organic molecular lines. We describe the line cube calibration done by ALMA and the subsequent calibration and imaging we performed. We discuss our choice of calibration parameters and optimisation of the cleaning parameters, and we demonstrate the utility and necessity of additional processing compared to the ALMA archive pipeline. As a demonstration of the scientific potential of these data, we present a first analysis of the DCN~(3-2) line. We find that DCN~(3-2) traces a diversity of morphologies and complex velocity structures, which tend to be more filamentary and widespread in evolved regions and are more compact in the young and intermediate-stage protoclusters. Furthermore, we used the DCN~(3-2) emission as a tracer of the gas associated with 595 continuum cores across the 15 protoclusters, providing the first estimates of the core systemic velocities and linewidths within the sample. We find that DCN (3-2) is detected towards a higher percentage of cores in evolved regions than the young and intermediate-stage protoclusters and is likely a more complete tracer of the core population in more evolved protoclusters. The full ALMA 12m-array cubes for the ALMA-IMF Large Program are provided with this DR1 release. 
}

\maketitle

\section{Introduction}

The relative number of stars born with different masses between 0.01~M$_{\odot}$ and $>$100~M$_{\odot}$, described by the initial mass function (IMF), is thought to be universal in studies of the cosmic history of star formation, including within our own Galaxy (e.g. \citealt{bastian10}; \citealt{kroupa13}). Early studies of the core mass distribution in local star-forming regions (SFRs) found distributions with shapes similar to the stellar IMF, suggesting a direct mapping of the core mass function (CMF) to the IMF with a constant efficiency factor  (e.g. \citealt{Motte1998}, \citealt{motte01}; \citealt{TeSa98}; \citealt{alves07}; \citealt{Enoch2008}; \citealt{konyves15}). These studies, however, were based on nearby star-forming clouds, where the core mass was limited to $<$5~M$_{\odot}$. More recent studies on a larger selection of clouds (e.g. \citealt{zhang15}; \citealt{ohashi16}; \citealt{csengeri17a}; \citealt{lu20}) and the pilot studies of the ALMA-IMF Large Program (e.g. \citealt{ginsburg17}; \citealt{Motte2018}, and \citealt{sanhueza19}) identified varying CMF shapes, with evidence for top-heavy CMFs towards regions forming massive stars. Moreover, the dependence of the IMF on the environment remains the subject of debate (see reviews by \citealt{offner14}; \citealt{krumholz15}; \citealt{ballesteros20}; \citealt{lee20}). These circumstances highlighted the need for a larger, more statistically robust sample of protoclusters in different environments to test the universality of the CMF and determine if the cloud characteristics impact its shape. 

ALMA-IMF is an ALMA Large Program (\#2017.1.01355.L, PIs: Motte, Ginsburg, Louvet, Sanhueza) to survey 15 nearby high-mass SFRs in the Galactic plane with the goal of characterising the CMF and its evolution. To understand the CMF fully, it is imperative to also investigate the distribution, dynamics, and kinematics of the gas from clumps to clusters down to core scales and determine if inflows, outflows, or the formation of filaments may be correlated with the CMF shape or impact its shape over time. One of the main objectives of the ALMA-IMF Large Program is to discriminate between the quasi-static and dynamic scenarios of cloud-scale star formation by quantifying the role of cloud and core kinematics in defining core mass and core mass growth over time. 

An overview of ALMA-IMF is presented in Paper I by \cite{Motte2022}, who describes the sample selection, classification of the evolutionary nature of individual protoclusters in the three subgroups (i.e. young, intermediate, and evolved), and early results, highlighting the complex velocity and filamentary structures in the protoclusters. In Paper II, \cite{Ginsburg2022} describes the first data release of the continuum images and presents an analysis of the spectral indices on the continuum data. Paper III, \cite{Pouteau22}, and \cite{louvet22} utilise the ALMA-IMF continuum data to characterise the continuum cores, providing the first core catalogues and building the first individual region and global CMFs for the 15 protoclusters. The next step in understanding the origin of the CMF is to measure the internal kinematics with evolution and determine their importance on the shape and evolution of the CMF over time. The ALMA-IMF spectral setup hosts a wealth of molecular emission well suited for this purpose, such as the $^{12}$CO (2-1), SiO (5-4), and SO (6-5), which can be used to explore the outflow population and provide a means to characterise further the nature of the cores as being pre- or proto-stellar (see \citealt{Nony2023}). The core population, gas inflow, and filamentary structure towards the 15 protoclusters will be probed using dense gas tracers in the ALMA-IMF spectral coverage, such as $^{13}$CS (5-4), DCN (3-2), \nthp (1-0), and N$_2$D$^+$ (3-2) to understand further inflow onto the cores and characterise core mass growth with time and its implications on the CMF. In addition, the multitude of emission lines, from species tracing ionised gas to the interstellar complex organic molecules (iCOMs) present in the ALMA-IMF spectral coverage, will provide the community with an unprecedented database with high legacy value for cores, hot cores, shocks, and outflows.
We present the data reduction steps and imaging strategies implemented to obtain the ALMA-IMF spectral line cubes. This DR1 data release provides the position-position-velocity image cubes of the 12 spectral windows (spws), $\sim$6.7~GHz of bandwidth coverage per protocluster, resulting in 180 image cubes for the whole ALMA-IMF sample. 
As a demonstration of the scientific potential of these data, we also present
a first analysis of the DCN (3-2) emission towards the 15 protoclusters. As DCN (3-2) is expected to be an optically thin, dense gas tracer, we further utilised the DCN emission to explore the core kinematics of the $\sim$600 thermal dust cores identified and described in \citet{louvet22}. 

This paper is organised as follows: in Section 2, we report a summary of the observations taken by ALMA and an overview of the target regions. Section 3 describes the data processing and imaging strategy performed to produce the final deconvolved line cubes, the data products, and the post-processing steps. In sections 4, and 5 we provide an analysis and discussion of DCN line emission towards the 15 protoclusters. In Section 6, we give a summary of the DR1 release and the results obtained.

\section{Observations}
\begin{table*}[htp]
\centering
\begin{threeparttable}[c]
\caption{Overview of the ALMA-IMF protocluster clouds and their evolutionary stage.}
\label{tab:sample}
 \begin{tabular}{lcccccllll}
\hline \noalign {\smallskip}
Protocluster  & RA\tnote{1}    & Dec\tnote{1}    & \vlsr\tnote{1} & $d$\tnote{1}   & Evolutionary & \multicolumn{2}{c}{Imaged areas\tnote{3}} & \multicolumn{2}{c}{f$_{BW}$ lines\tnote{4}}\\  
cloud name\tnote{1}    & \multicolumn{2}{c}{[ICRS (J2000)]}   & [$\kms$]      & [kpc]  &  stage\tnote{2}  & \multicolumn{2}{c}{[$\pc \times\pc$]} &B3 & B6\\
&&&&&& $A_{\rm 1.3\,mm}$ & $A_{\rm 3\,mm}$ &\\
\hline \noalign {\smallskip}

W43-MM1     
    & 18:47:47.00 &  $-$01:54:26.0 & $+97$    & 5.5$\pm 0.4$  
    & Y &  $3.1\times2.3$       &$5.1 \times 4.0$  & 0.64 & 0.55\\
    
W43-MM2     
    & 18:47:36.61 & $-$02:00:51.1 & $+97$   & 5.5$\pm 0.4$  
    & Y & $2.6 \times 2.4$& $5.1 \times 4.0$ & 0.46 & 0.52\\
    
G338.93     
    & 16:40:34.42 & $-$45:41:40.6 & $-62$   & 3.9$\pm 1.0$ 
    & Y & $1.6 \times 1.6$      & $2.9 \times 2.8$ & 0.49 & 0.25\\      

G328.25     
    & 15:57:59.68 & $-$53:58:00.2 & $-43$   & 2.5$\pm 0.5$  
    & Y & $1.4 \times 1.4$      & $2.2 \times 1.9$ & 0.2 & 0.53\\

G337.92         
    & 16:41:10.62 & $-$47:08:02.9 & $-40$   & 2.7$\pm 0.7$  
    & Y & $1.2 \times 1.1$      & $2.1 \times 2.0$ & 0.37 & 0.53\\ 
    
G327.29     
    & 15:53:08.13 & $-$54:37:08.6 & $-45$   & 2.5$\pm 0.5$  
    & Y & $1.3 \times 1.3$      & $1.9 \times 1.8$ & 0.65 & 0.48\\ 
       
\hline

G351.77    
    & 17:26:42.62 & $-$36:09:20.5 & $-3$   & 2.0$\pm 0.7$  
    &  I & $1.3 \times 1.3$    & $1.8 \times 1.7$ & 0.64 & 0.56\\
    
G008.67     
    & 18:06:21.12 & $-$21:37:16.7 & $+37.6$   & 3.4$\pm 0.3$  
    & I &  $2.2 \times 1.4$     & $3.1 \times 2.1$  &0.22 &0.29  \\

W43-MM3     
    & 18:47:41.46 & $-$02:00:27.6 & $+97$   & 5.5$\pm 0.4$ 
    & I &  $2.7 \times 2.4$     & $5.1 \times 4.0$ &0.13 & 0.08  \\
    
W51-E       
    & 19:23:44.18 & $+$14:30:29.5 & $+55$   & 5.4$\pm 0.3$  
    & I & $2.6 \times 2.4$      & $4.2 \times 3.9$ & 0.3 & 0.37 \\ 
    
G353.41     
    & 17:30:26.28 & $-$34:41:49.7 & $-17$   & 2.0$\pm 0.7$  
    &  I & $1.3 \times 1.3$ &  $1.8 \times 1.7$ & 0.14 & 0.08  \\

\hline
G010.62    & 18:10:28.84 & $-$19:55:48.3 & $-2$ & 4.95$\pm 0.5$  & E & $2.3 \times 2.2$     & $3.8 \times 3.6$ &0.11 & 0.28  \\
W51-IRS2  & 19:23:39.81 & $+$14:31:03.5 & $+55$   & 5.4$\pm 0.3$  & E & $2.6 \times 2.4$     & $4.2 \times 3.9$ & 0.26 & 0.42 \\   
G012.80    & 18:14:13.37 & $-$17:55:45.2 & $+37$   & 2.4$\pm 0.2$ & E & $1.5 \times 1.5$    & $2.2 \times 2.1$ &0.12 & 0.16\\         
G333.60  & 16:22:09.36 & $-$50:05:58.9 & $-47$   & 4.2$\pm 0.7$ &  E & $2.9 \times 2.9$    & $3.9 \times 3.7$ & 0.16 & 0.2  \\
\hline \noalign {\smallskip}
\end{tabular}
\begin{tablenotes}
\item[1] Protocluster name, central position, and velocity relative to the local standard of rest used for the ALMA-IMF observations.
$\vlsr$ values are taken from the high-density gas studies by \cite{wienen15}, \cite{ginsburg15} for W51, \cite{nguyen13} for W43, and \cite{immer14} for G012.80. The phase centre of W43-MM1 in the pilot study is R.A.  $=$ 18:47:46.50, Dec $= -01$:54:29.5. Distances are taken from \cite{Motte2022} and references therein.
\item[2] Classification of the ALMA-IMF protocluster clouds: Young (Y), Intermediate (I), and Evolved (E). See Section 4.1 of  \citealt{Motte2022}.  
\item[3] Physical areas encompassing the combined primary beams of the 1.3~mm and 3~mm mosaics. The released maps cover an area down to the full width at $10\%$ of the maximum of the primary beam response.
\item[4] This is the fraction of the bandwidth found to contain bright line emission, which was excluded to make the \cleanest continuum cubes, 
i.e. $1-f_\mathrm{BW,cleanest}$ from Table 3 of \citet{Ginsburg2022}. To first order, this fraction highlights the percentage of bandwidth containing bright line emission and illustrates the variation of molecular emission between the regions.
\end{tablenotes}
\end{threeparttable}
\end{table*}

\label{sec:observations}
ALMA-IMF\footnote{\url{https://www.almaimf.com}} (the ALMA Large Program \#2017.1.01355.L, PIs: Motte, Ginsburg, Louvet, Sanhueza), targets 15 protocluster clouds in our Galaxy using ALMA with its 12m, 7m and total power (TP) arrays and in two frequency bands; Band-3 (B3; $\sim$91-106~GHz) and Band-6 (B6; $\sim$216-234~GHz).
 We combine Tables 1 and 2 from \citet{Motte2022} to provide here a global overview of the 15 observed protoclusters in \cref{tab:sample}, listing their central positions, \vlsr estimates, distances from the Sun, evolutionary nature, and imaged areas of the mosaics in ALMA's B3 (3~mm) and B6 (1~mm). In addition, \cref{tab:sample} lists the fraction of the bandwidth containing bright line emission (f$_{BW}$ line). This is the fraction of bandwidth that was excluded from making the \texttt{cleanest} continuum images (taken from Table 3 of \citealt{Ginsburg2022}) and which highlights, to first order, the relative amount of molecular line emission present in the brightest regions of the protoclusters. A detailed description of the observing details and target selection for ALMA-IMF is provided in \citet{Ginsburg2022} and \citet{Motte2022}, respectively. ALMA-IMF was designed to observe homogeneously 15 of the most extreme and massive Galactic clouds within a distance range of 2-5.5~kpc. This distance range allows coverage of $\gtrsim$1$\times$1~pc$^2$ of the highest column density regions towards each protocluster determined from ATLASGAL imaging \citep{csengeri17b}, while also allowing for a reasonable observing time towards the more distant protoclusters in the sample. The observing setup and array configurations were chosen to achieve a spatial resolution of $\sim$2000~au for all protoclusters, regardless of distance. In B3, all targets were observed with two 12m array configurations. In B6, the more distant regions were observed in two configurations (see \citealt{Ginsburg2022} for the full description of the 12m array configurations per protocluster). The released cubes are produced from the combination of the two ALMA 12m-array configurations. The resulting angular resolutions of the continuum data are between 0.3\arcsec\, and 1.5\arcsec\, using a robust weighting of 0 for the Briggs weighting. The major and minor axis of the synthesised beam for each spectral window towards each protocluster is provided in \cref{tab:spwsbeams}.
  
\begin{figure*}
    \begin{center}
    \includegraphics[width=0.49\textwidth]{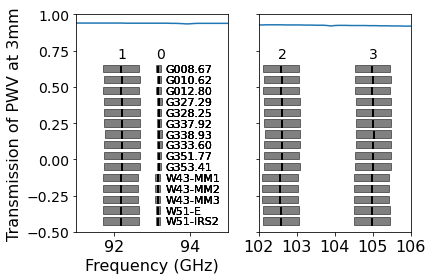} 
    \includegraphics[width=0.49\textwidth]{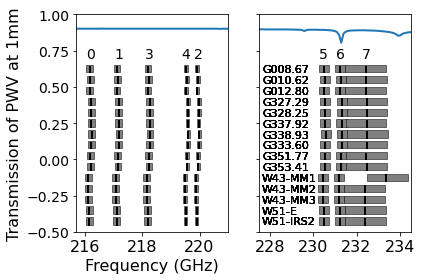} 
        \caption[]{Overview of the bandwidth covered by the spectral windows (spws, grey boxes), four in B3 (left panels) and eight in B6 (right panels) towards the 15 ALMA-IMF fields. The black line represents the central frequency for each spectral window. The numbers at the top denote the spectral windows as described in \cref{tab:spws_fred}. The exact frequency coverage is provided in \cref{tab:spwsfreq}. The blue line is the atmospheric transmission at a precipitable water vapour (PWV){\footnotemark} of 1.796~mm in B6 and 5.18~mm in B3 (the typical values used in the ALMA sensitivity calculator), for a source at an elevation of 45 degrees.} The coverage of the ALMA-IMF pilot program (B6 spectral window 7 for the W43-MM1 field) is offset by 1~GHz.
    \label{fig:spws}
    \end{center}
  \end{figure*}

An overview of the spectral setup of the 12 ALMA spectral windows is displayed in \cref{fig:spws}. The corresponding bandwidth, spectral resolution, along with main spectral lines in their respective spectral windows are also provided in \cref{tab:spws_fred}\footnote{This is the same as Table 3 of \citet{Motte2022} but with slightly updated values for the achieved spectral resolutions.\label{note:newres}}. For W43-MM1, the B6 data were taken as part of the pilot program, 2013.1.01365.S \citep{Motte2018}. The exact frequency coverage for each spectral window and field are given in \cref{tab:spwsfreq}.

\begin{table*}[htbp!]
\centering
\small
\begin{threeparttable}[c]
\caption{Spectral setups of the ALMA-IMF Large Program.}
\label{tab:spws_fred}
 \begin{tabular}{lcccccl}
\hline \noalign {\smallskip}
ALMA  & Spectral  & Central & Bandwidth & \multicolumn{2}{c}{Resolution\tnote{1}}  & Main spectral lines \\
Band  & Window & Frequency &  &  &  &  \\
      &  (SPW)      & [GHz] & [MHz] & [kHz] & [\kms] &  \\

\hline \noalign {\smallskip}
Band 6  & SPW0 & 216.200  & 234 & 122 & 0.17 & DCO$^+$~(3-2), CH$_{3}$OCHO, OC$^{33}$S~(18-17), HCOOH \\ 
        & SPW1 & 217.150  & 234 & 244 & 0.34  &
        SiO~(5-4), DCN~(3-2), $^{13}$CH$_{3}$OH, CH$_{3}$OCH$_{3}$ \\ 
        & SPW2 & 219.945  & 117 & 244 & 0.33  & SO~(6-5), H$_{2}^{13}$CO~(3$_{1,2}$-2$_{1, 1}$), CH$_{3}$OH \\ 
        & SPW3 & 218.230  & 234 & 122 & 0.17  & H$_{2}$CO~(3-2), O$^{13}$CS~(18-17), HC$_{3}$N~(24-23), CH$_{3}$OCHO \\
        & SPW4 & 219.560  & 117 & 122 & 0.17   & C$^{18}$O~(2-1), C$_{2}$H$_{5}$CN \\ 
        & SPW5 & 230.530  & 468 & 974 & 1.27  & CO~(2-1), CH$_{3}$CHO, CH$_{3}$OH, C$_{2}$H$_{3}$CN, C$_{2}$H$_{5}$OH \\
        & SPW6 & 231.280  & 468 & 244 & 0.32 & $^{13}$CS~(5-4), N$_{2}$D$^{+}$~(3-2), OCS~(19-18), CH$_{3}$CHO, CH$_{3}$OH,\\
        & & & & & &     CH$_{3}^{18}$OH, C$_{2}$H$_{5}$CN \\ 
        & SPW7 & 232.450  & 1875   & 976 & 1.26 & H30$\alpha$, CH$_{3}$CHO, CH$_{3}$OH, CH$_{3}$OCHO, C$_{2}$H$_{5}$OH, C$_{2}$H$_{5}$CN, \\
        & & & & & & CH$_{3}$OCH$_{3}$, CH$_{3}$COCH$_{3}$, $^{13}$CH$_{3}$CN~(13-12), H$_{2}$C$^{34}$S~(7$_{1,7}$-6$_{1,6}$)  \\      
        \hline
Band 3  & SPW0 & 93.1734 & 117 & 61 & 0.2 & N$_{2}$H$^{+}$~(1-0), CH$_{3}$OH \\ 
        & SPW1 & 92.2000 & 938 & 488 & 1.6 &  CH$_3$CN~(5-4), H41$\alpha$, CH$_{3}$$^{13}$CN, $^{13}$CS~(2-1), $^{13}$CH$_{3}$OH, CH$_{3}$OCHO\\
        & SPW2 & 102.600 & 938 & 488 & 1.4 &  CH$_3$CCH~(6-5), CH$_{3}$OH, H$_{2}$CS,C$_{2}$H$_{5}$CN, C$_{2}$H$_{5}$OH, CH$_{3}$NCO \\
        & SPW3 & 105.000 & 938 & 488 & 1.4 & H$_{2}$CS, CH$_{3}$OH, C$_{2}$H$_{3}$CN, C$_{2}$H$_{5}$OH, CH$_{3}$OCH$_{3}$ \\
\hline \noalign {\smallskip}

\end{tabular}
\begin{tablenotes}
\item[1]This is the same as Table 3 of \citet{Motte2022}, but with slightly updated values for the achieved spectral resolutions.
\end{tablenotes}
\end{threeparttable}
\end{table*}

\afterpage{}
  \footnotetext{The values over the observed frequency range are taken from \url{https://www.apex-telescope.org/sites/chajnantor/atmosphere/transpwv/index\_ns.php}
}

\section{Data products}
\label{sec:data}
Together with this paper, we provide access to the DR1 line cube release\footnote{The ALMA-IMF project page is at \url{https://www.almaimf.com/data.html}. This Data Release is hosted at Harvard Dataverse \url{https://dataverse.harvard.edu/dataverse/alma-imf}}, which includes the calibrated and imaged full spectral windows (both in B3 and B6) of the 12m array configurations towards the 15 protoclusters. 
ALMA-IMF also includes 7m array and TP observations using the same spectral setup. The combination of these data, which is particularly important for lines with significant extended emission, is ongoing, and those cubes will be added to the repository as they are created for future planned studies. For spectral window 0 in B3, which includes the \nthp\, line, the 12~m only data still contain artefacts due to the extended emission and missing short spacings, we thus exclude this window from this DR1 release. The combined \nthp\, data for this spectral window will be added to the repository as part of future works (e.g. \citealt{Stutz22} \'Alvarez-Guti\'errez et al., in prep, and Sandoval et al., in prep). We include the properties of spectral window 0 in B3 (e.g. beam size, and noise estimates) in this paper for completeness.

The data were restored to measurement sets using the \texttt{scriptForPI.py} files provided by the ALMA archive, and further batch processed with the custom scripts and imaging parameters of the ALMA-IMF data pipeline \citep{Ginsburg2022}. All measurement sets underwent QA3 reprocessing: the FAUST Large Program  \citep[Project code: 2018.1.01205.L,][]{Codella2021} reported that the calibration of the system temperature adopted by ALMA could result in the artificial suppression of bright lines\footnote{\url{https://help.almascience.org/kb/articles/607}}. 
The ALMA-IMF data were also affected by these issues, particularly spectral window 5 in B6 and spectral window 0 in B3, containing the brightest emission lines of CO~(2-1) and \nthp~(1-0), respectively. The measurement sets were returned to the Joint ALMA Observatory for further QA3 processing in November 2020, and the reprocessing was completed in March 2021. The W43-MM1 B6 data from the pilot program 2013.1.01365.S \citep[][]{Motte2018} were also reprocessed following the same QA3 procedure as for the 2017.1.01355.L data. 
\begin{figure}
    \centering
    \includegraphics[width=0.45\textwidth]{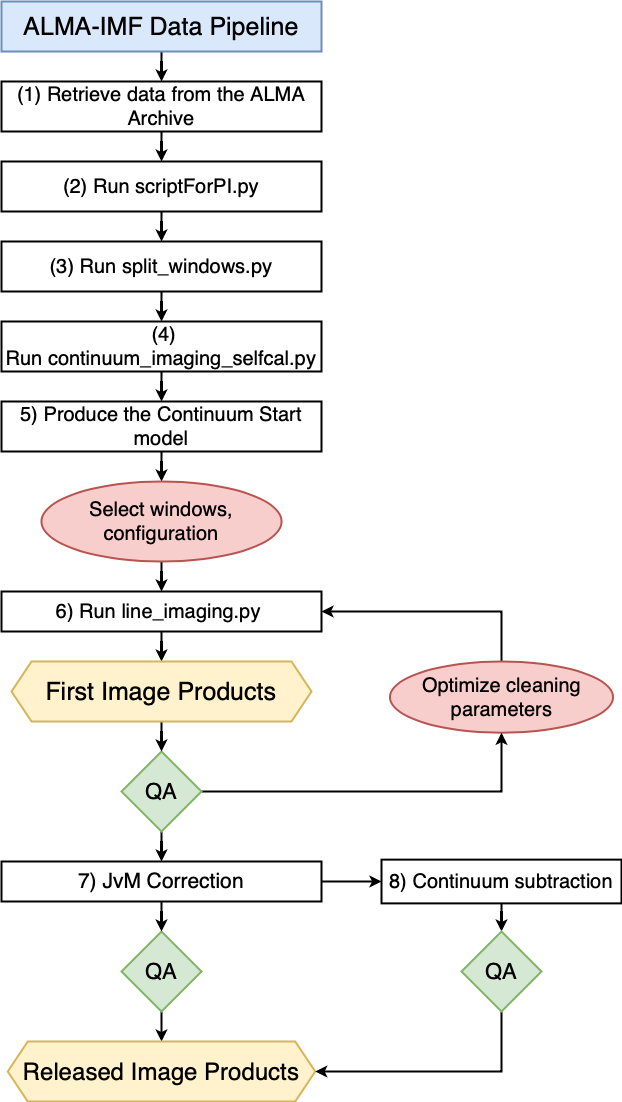}
    \caption{Flow chart providing an overview of the framework of the ALMA-IMF data pipeline employed to produce the line cubes provided in this release. The white boxes labelled from 1-8 describe the steps in the pipeline that are defined by running scripts or procedures (as described in more detail in \cref{sec:almaimfpipeline}). Red ellipses highlight points where manual input was required to either select which line, spectral window, or configuration on which to perform \tclean, or to optimise the \tclean parameters after internal quality assessments (green diamonds). }
    \label{fig:pipeline}
\end{figure}

\subsection{ALMA-IMF data pipeline\label{sec:almaimfpipeline}}
The custom ALMA-IMF data pipeline developed to produce the calibrated and imaged continuum data (as described in \citealt{Ginsburg2022}) was subsequently adapted to process the full spectral line windows. It runs in the CASA \citep{McMullin2007} environment and is described in the following eight steps. The full data pipeline and custom python scripts can be found on the ALMA-IMF GitHub repository\footnote{\url{https://github.com/ALMA-IMF/reduction}\label{github}}.
\begin{enumerate}
    \item Retrieve and extract the data from the ALMA archive using astroquery \citep{Ginsburg2019b}.
    \item Run \texttt{scriptForPI.py} to restore the measurement sets.
    \item Separate the continuum and line measurement sets with \texttt{split\_windows.py}\footref{github} and combine the different 12m array configurations.
    \item Run the \texttt{continuum\_imaging\_selfcal.py} script\footref{github} to perform the continuum imaging and self calibration\footnote{We note that the self-calibration solutions were not applied to the line data}.
    \item Produce the continuum start model from the continuum data.
    \item Run the \texttt{line\_imaging.py} script\footref{github} to perform the line imaging.
    \item Apply the `JvM' correction \citep{Jorsater1995} to the cleaned cubes (see Section \ref{sec:jvm} for more details).
    \item Run \statcont \citep{statcont2017} on the imaged line cubes to produce the continuum subtracted cubes (this step is optional).
\end{enumerate}
For illustration, \cref{fig:pipeline} displays a flow diagram of the ALMA-IMF imaging pipeline and image processing steps taken to obtain the final line cubes provided in this release. The initial four steps are identical to the continuum data processing as detailed in Section 3.1.1 of \citet{Ginsburg2022}. Step 5) onward details the processing of the line cubes. In Step 5) the input start model for the line cleaning is derived from the output model of the continuum cleaning (see Section \ref{sec:startmodel}). To produce the optimal \tclean parameters in CASA used for these data, we performed several internal quality assessments (QAs) of the \tclean products and several iterations to test a range of \tclean parameters (see \ref{sec:cubes} for more details). As with the continuum cleaning, the most important input file is \texttt{imaging\_parameters.py}\footref{github}, which includes the user-specified \tclean parameters for both the continuum and now the line cubes for all fields and spectral windows. Finally, the last two steps, 7 and 8, refer to post-processing actions after the final CASA cubes are produced. The optional continuum subtraction on the line cubes described in Step 8 is performed in the image plane using the \statcont procedure, as described in \citet{statcont2017}. The \statcont continuum-subtracted cubes are included in the data release. We also provide the continuum estimates and the primary beam responses for each field and spectral window so that the unsubtracted and non-primary beam-corrected cubes can be reproduced if required. 
\subsection{Optimisation of cleaning and imaging parameters\label{sec:cubes}}
 The ALMA-IMF dataset covers a multitude of molecular lines with varying dynamic ranges and morphology across the 15 protoclusters. Therefore, the \tclean parameters are optimised to be as homogeneous as possible, allowing the pipeline to run in an automated way. In the following, we discuss the selection of \tclean parameters implemented to reach the imaged line cubes of the released ALMA-IMF dataset. 

\subsubsection{Continuum start-model}
\label{sec:startmodel}
Given the large extent of the mosaic coverage, the complexity of the data, and the varying dynamic range across an individual field, we chose not to perform a simple continuum subtraction in the uv-plane before running \tclean. This choice led to difficulties, however, with \tclean diverging for some fields. Thus, we use a continuum model cube as the input \texttt{startmodel} parameter in \tclean. This model cube is constructed from the .tt0 and .tt1 products of the mostly line-free \texttt{cleanest} continuum imaging (see \citealt{Ginsburg2022}), \tclean then stores this as an initial model and subtracts it from the visibilities at the beginning of the deconvolution process.

\subsubsection{Cleaning optimisation and masking}
We performed several iterations of \tclean using the Hogbom and multi-scale deconvolvers and tested methods to mask the line emission: 1) an internally developed auto-masking, 2) the CASA auto multi-threshold masking, and 3) a primary beam mask. Multi-scale deconvolution with a primary beam mask produced the most stable and optimal results over the full set of spectral windows. In spectral windows with bright and extended emission, the routines that mask in an automated way could not adequately cover  the full extent of the emission. Furthermore, auto multi-threshold had additional convergence problems when working with multi-scale deconvolution. For these reasons, we opted to use the primary beam mask (pbmask) as our \tclean mask, setting its limit to 10\% of the primary beam response. For the deconvolver, we found that multi-scale clean performed better at recovering larger-scale emission than Hogbom deconvolution. This difference was expected since the latter uses only point sources as a model. We defined the scales used in multi-scale \tclean using a geometric series starting at 0 and increasing by $\sim \times 2$ the ratio of the minor beam axis to the cell size until a scale of $\sim$5.5\arcsec\ is reached. This largest scale is selected to be a factor of 2-3 smaller than the typical largest recoverable scale in the data. The scales used for each field and spectral window are provided in the \texttt{imaging\_parameters.py} file\footref{github}.
\begin{figure}
    \centering
    \includegraphics[width=0.44\textwidth]{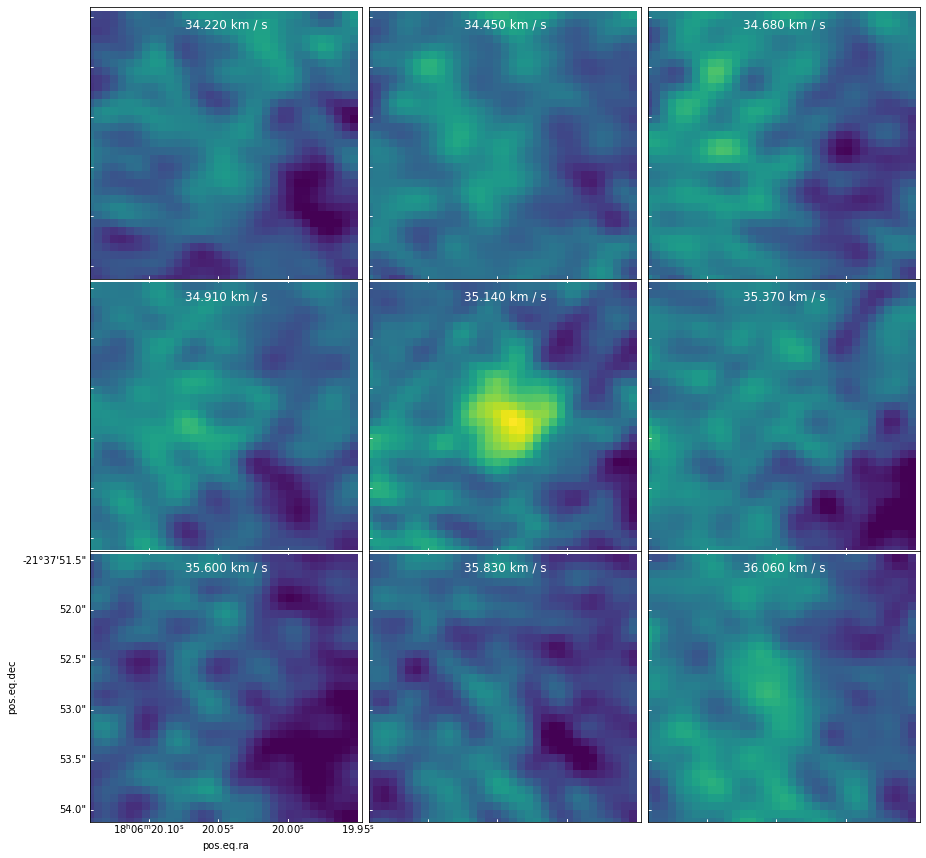} 
    \includegraphics[width=0.46\textwidth]{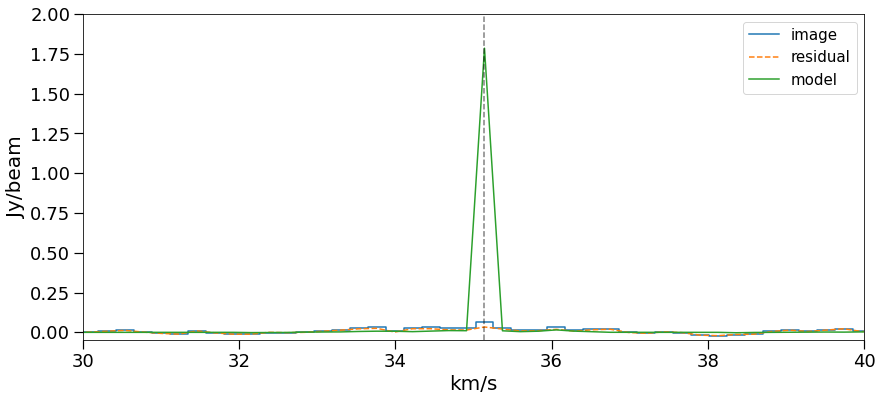}
    \caption{Example of the `stippling' effect as previously described in \citet{Czekala2021}. The top panel shows nine channels of a zoomed-in region in a restored image cube, cleaned using a 3$\sigma$ threshold.  The central channel shows the resulting spurious emission peak. This bright emission, of order the beam size, is only present in a single channel, while the neighbouring pixels show no equivalent emission. The bottom panel shows the model extracted at the central pixel, where the bright model component (green line) is only found in a single pixel and channel. This model component is then convolved with the Gaussian clean beam during the \tclean run to produce the restored image, resulting in spurious compact emission in a single channel}.
    \label{fig:stipling}
\end{figure}

\begin{figure}
    \centering
    \includegraphics[width=0.49\textwidth]{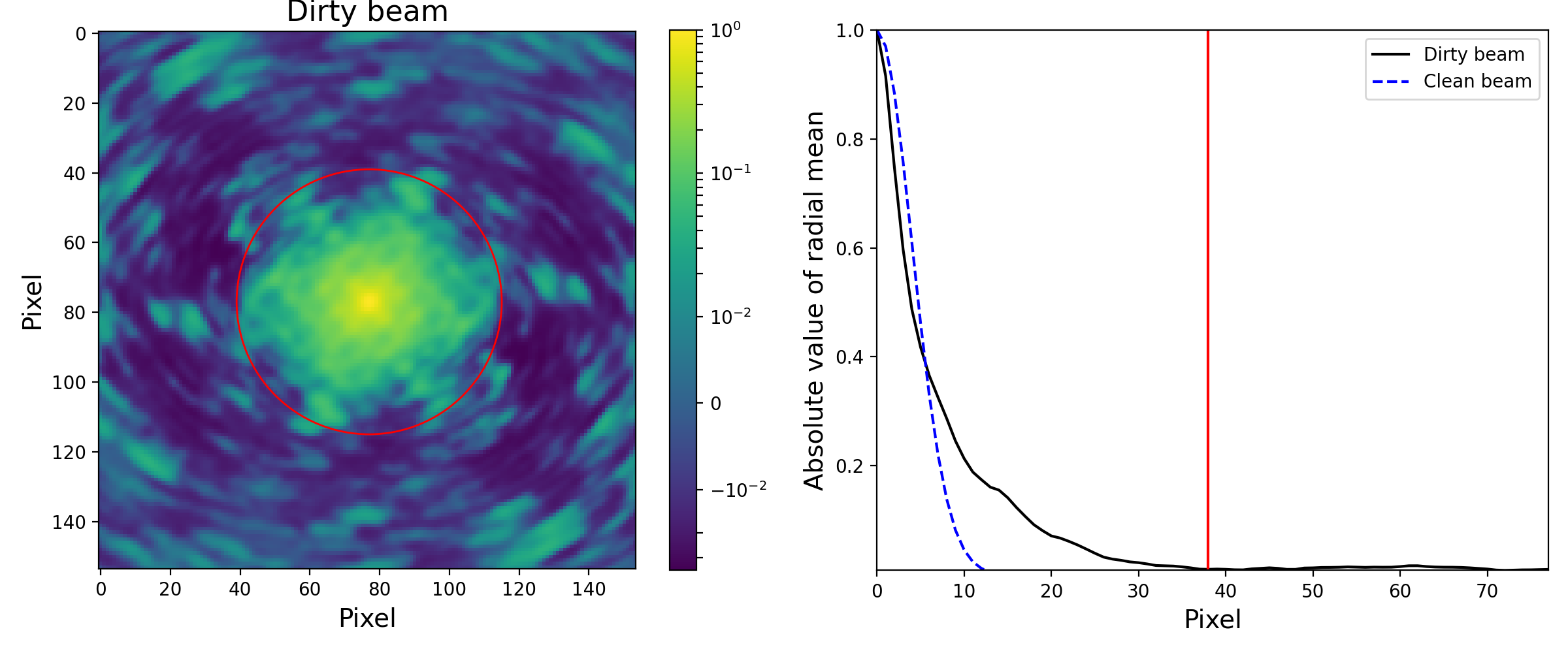} 
    \includegraphics[width=0.49\textwidth]{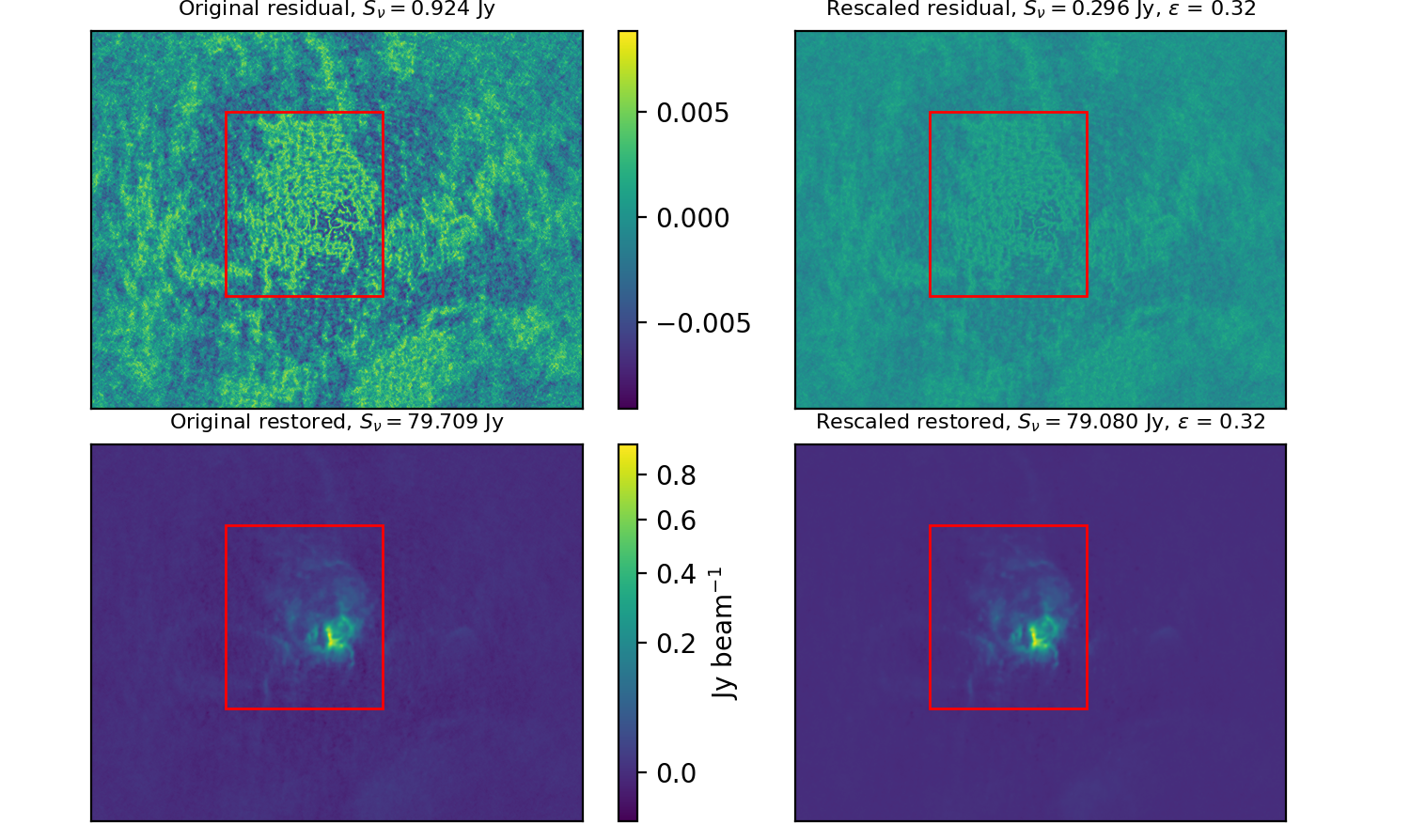} 
    \caption{Example of the JvM correction for the peak channel of the H$41\alpha$ line in spectral window 1 of Band 3 for protocluster G333.60. The top left panel shows the original PSF (i.e. the dirty beam), where the red circle marks the first null. The top right panel shows the absolute value of the radial profile of the beam (black solid line) and the corresponding approximation to a Gaussian clean beam (blue dashed line). The $\epsilon$ factor, defined as the ratio of the clean-to-dirty beam volumes, is 0.32 in this example. The residuals are shown for the original image (middle left panel) and after (middle right), the JvM correction is applied. The red rectangle shows the aperture where the quoted fluxes are measured. 
    The restored images without (bottom left) and with JvM (bottom right) correction are shown in the bottom panels.}
    \label{fig:jvmcorrection}
\end{figure}
\subsubsection{Cleaning threshold}

During the QA of the line cubes, we iterated over several cleaning thresholds. We found that using a relatively shallow threshold of 5$\sigma$ resulted in the most stable outcomes, that is, mitigating the effects of divergence and `stippling' over the full dataset. As described in \citet{Czekala2021} for the ALMA MAPS Large Program, a stippling pattern can be present in deeply cleaned cubes ($\leq3\sigma$), which then results in artificial clean components appearing as spurious sources -- of order the beam size -- in the deconvolved image cubes. In \cref{fig:stipling}, we show an example of this  stippling effect in one of the ALMA-IMF line cubes. When cleaning down to 3$\sigma$, a bright model component was found in only one channel and pixel, resulting in a spurious emission peak in the cleaned cube, present in only a single channel. A Jupyter notebook describing this effect can be found in the ALMA-IMF pipeline repository\footnote{\url{https://github.com/ALMA-IMF/notebooks/blob/master/ConfettiQA.ipynb}}. We estimate that this  issue only affects $<$0.1\% of the pixels in the cubes of this data release. In addition to the stippling, we found that cleaning to levels of 4$\sigma$ or deeper also resulted in divergence across multiple spectral windows, compared with setting the noise threshold to $\geq$5$\sigma$. For the latter thresholds, however, we still found divergence in a few channels towards spectral windows containing bright and extended emission. These mainly were identified around the \vlsr of the $^{12}$CO (2-1) and C$^{18}$O (2-1) transitions. In these cases, we chose to mask the channels that showed divergence and re-run the \tclean, leaving those channels uncleaned in the final cubes (the masked channel ranges can be obtained in the \texttt{imaging\_parameters.py} file\footref{github}). Towards several spectral windows for the protocluster G351.77, however, we still identified divergence. In the end, for this source, we ran \tclean only down to 10$\sigma$ level for all spectral windows. The full set of cleaning parameters and masked channels can be found in the \texttt{imaging\_parameters.py} file\footref{github}.

\begin{figure}
    \centering
    \includegraphics[width=0.49\textwidth]{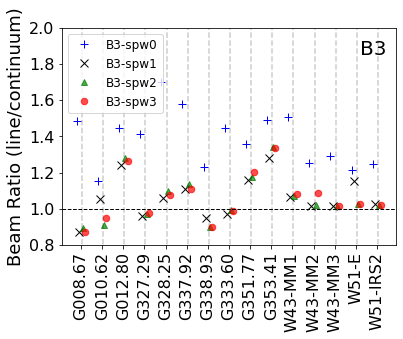}
    \includegraphics[width=0.49\textwidth]{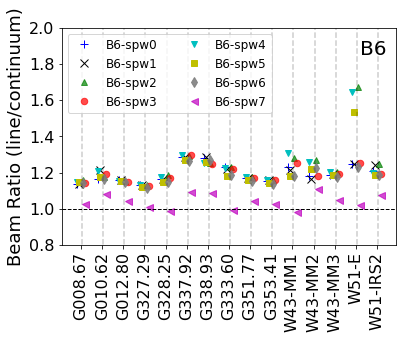} 
    \caption{Ratio of the average line to continuum beam for each protocluster and spectral window (B3 top; B6 bottom panels). Where the average beam is defined as $\theta_{\rm{ave}}=\sqrt{\theta_{\rm{maj}}\,\times\,\theta_{\rm{min}}}$. We scale the ratio linearly to account for the frequency difference between the central frequency of the spectral window and the frequency of the continuum beam. The native beam major and minor axis for each spectral window is listed in \cref{tab:spwsbeams}. The proposed beam and the average recovered continuum beam for all fields are given in \cref{tab:spwsbeams_contprop}. The central frequency of the continuum images for B3 and B6 for each protocluster is given in \citet{Ginsburg2022}, and the spectral coverage for each of the line cubes are shown in \cref{tab:spwsfreq}. The dashed horizontal line shows a one-to-one ratio between the line and continuum beams.
    } 
    \label{fig:beams}
\end{figure}

\begin{figure*}
    \centering
    \includegraphics[width=0.49\textwidth]{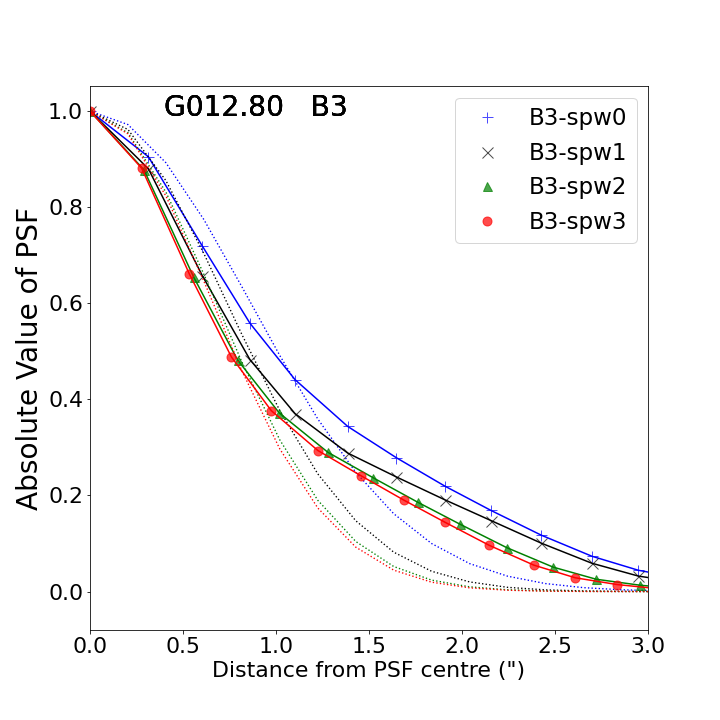} 
    \includegraphics[width=0.49\textwidth]{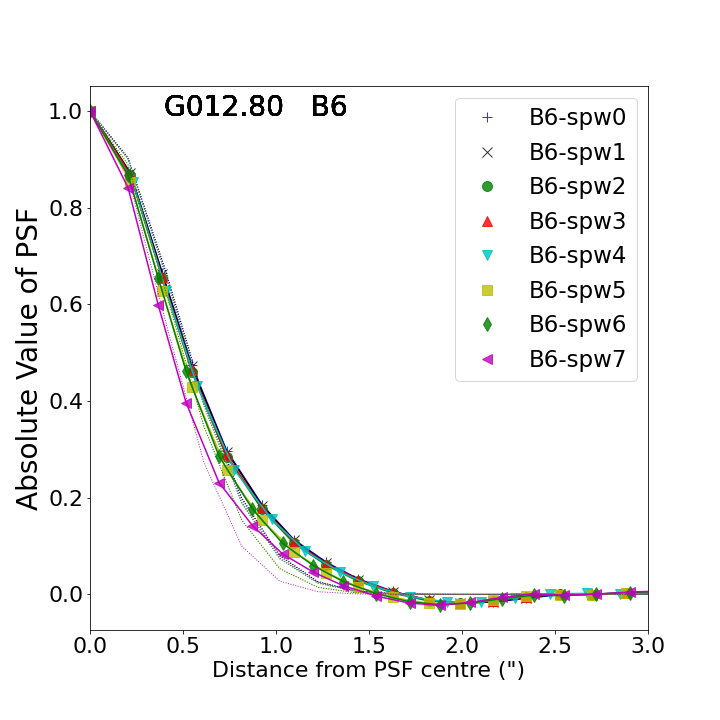}

    \caption{Absolute value of the radial mean of the PSF (dirty beam) for the protocluster G012.80 as a function of the distance from the PSF centre (in arcseconds). B3 spectral windows are shown on the left and B6 spectral windows are on the right. The solid lines use the \texttt{.psf} outputs from \tclean and the dashed lines are the expected Gaussian clean beams (the FWHM is taken from the geometric mean of the beam major and minor axes). For the B3 spectral windows, the Gaussian clean beam is a good approximation to the PSF only within $\sim$50\% of its FWHM. B3 spectral window 0 shows the largest deviation, resulting in a larger estimation of the respective Gaussian clean beam in \tclean. The B6 spectral windows typically show a better correspondence between the PSFs and the Gaussian clean beam. 
    }
    \label{fig:psfs}
\end{figure*}

\begin{figure*}
    \centering
    \includegraphics[width=0.47\textwidth]{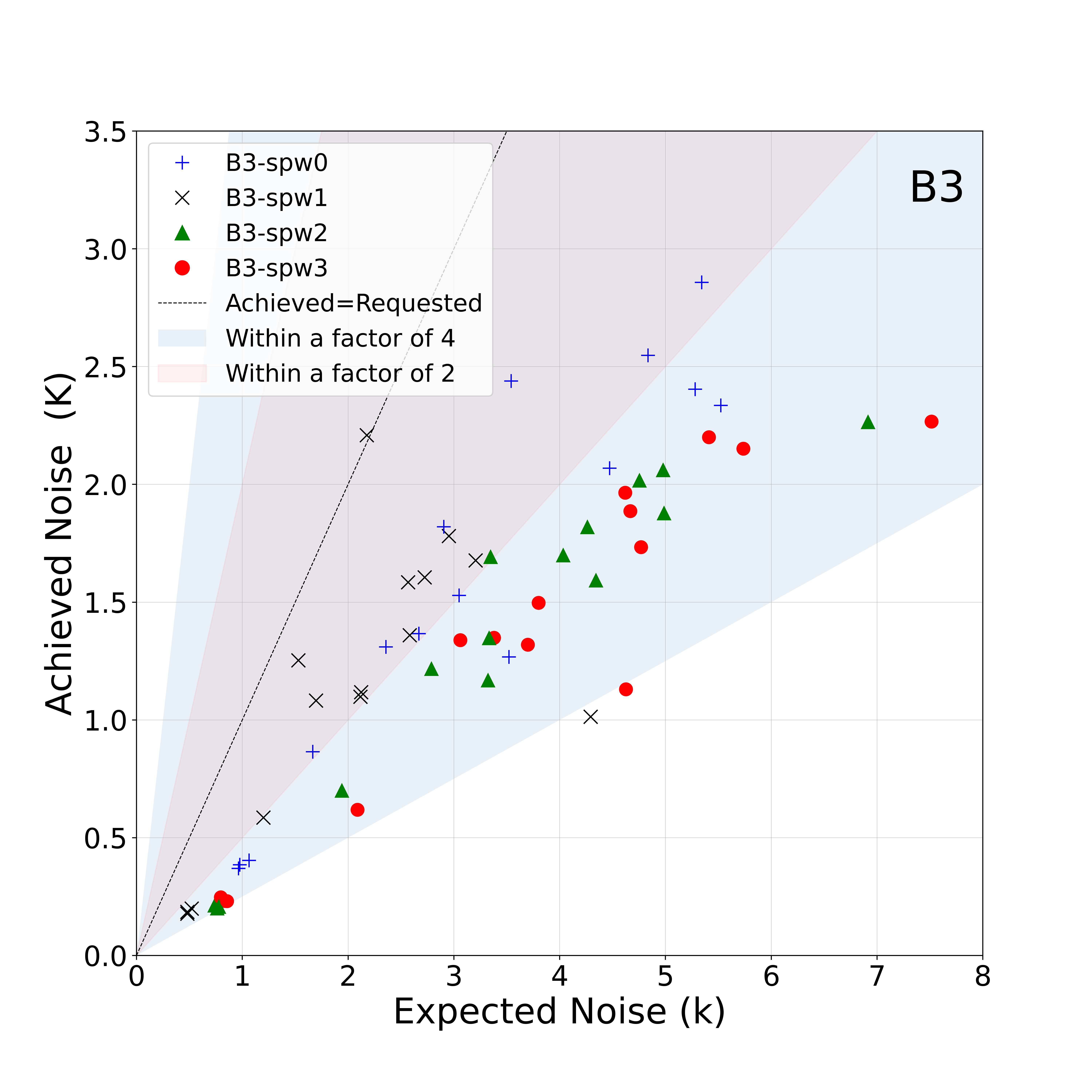} 
    \includegraphics[width=0.47\textwidth]{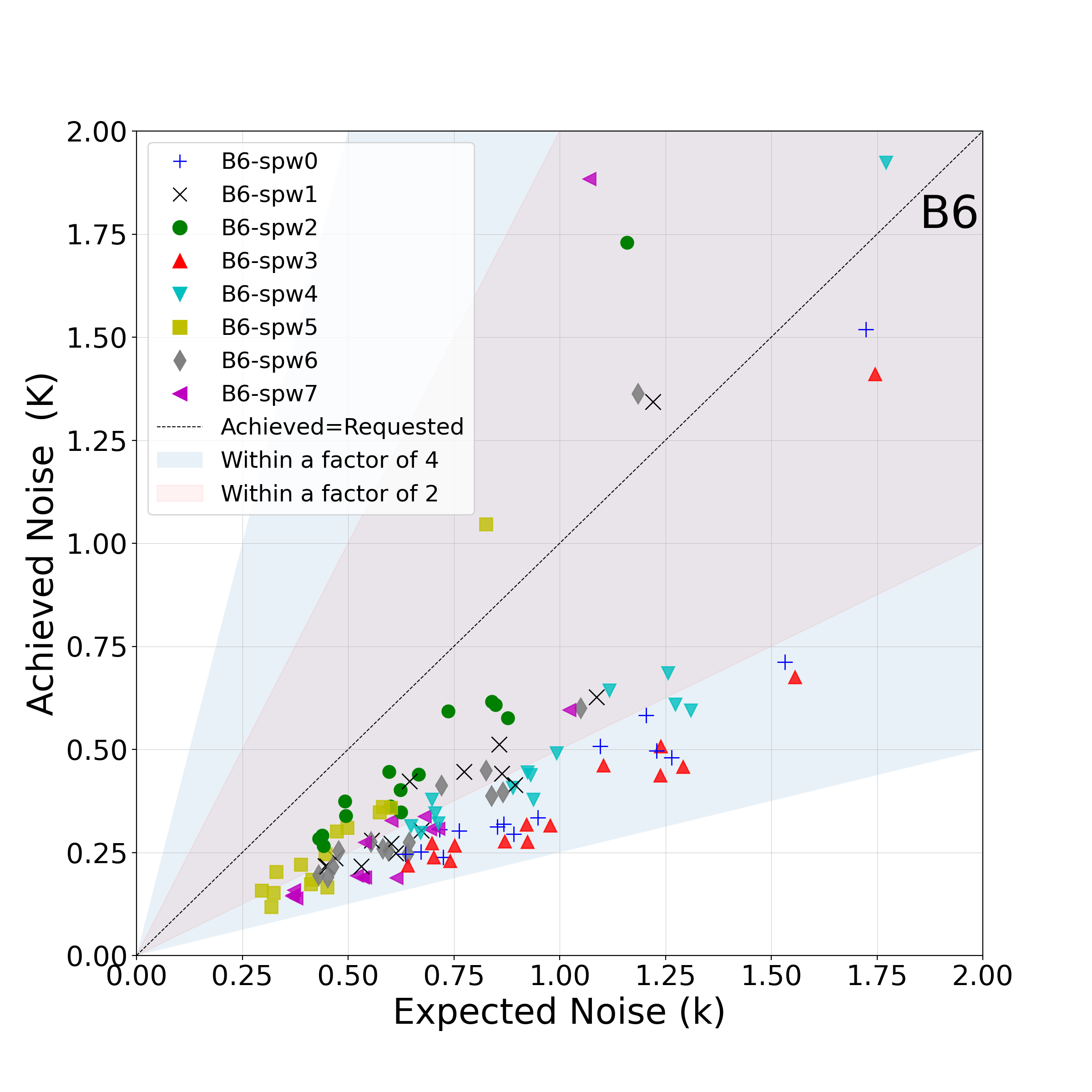} 
    \caption{Expected thermal noise from the observations vs the achieved noise in the cubes for each spectral window in Kelvin (B3 left panel and B6 right panel). 
    The Y-axis shows the measured noise in the released cubes in Jy per dirty beam converted to brightness temperature as described in Section \ref{sec:beamsizes}.
    The X-axis shows the expected brightness temperature sensitivity for the proposed beam sizes (see \cref{tab:spwsbeams_contprop}). Furthermore, to directly compare the achieved noise in the data to the expected noise from the proposed observations, we scale the expected noise by the frequency difference between the requested representative frequency and the central frequency of each spectral window. We also scale the expected noise by the difference in the representative channel width and the achieved channel width for each spectral window. 
    As shown in Figure \ref{fig:beams}, the achieved beam area is typically larger than the continuum beam by $\sim$1-1.5$\times$. Thus, the expected noise is also scaled to account for the difference in the beam areas between the achieved and proposed beams (see \cref{tab:spwsbeams} for the exact proposed beams and achieved beams per field). 
    The shaded regions show a noise within a factor of two (pink) and four (blue) from the expected noise, and the dashed line represents a 1 to 1 ratio. We typically achieve a noise on average $\sim$2-2.5 times lower than expected.
    }
    \label{fig:noise}
\end{figure*}

\subsection{JvM correction}
\label{sec:jvm}
The JvM correction \citep{Jorsater1995} is applied to the CASA-generated cubes to correct the flux scale of the residuals, and thus of the restored image, since the volumes of the clean and dirty beams are not the same \citep[see also][]{walter2008,Czekala2021}. \cref{fig:jvmcorrection} illustrates this correction for the G333.60 region's peak channel of the H$41\alpha$ line (situated in spectral window 1 of Band 3). The top panels show that a Gaussian clean beam is a reasonable approximation for the central part of the original PSF (dirty beam) but not beyond $\sim$50\% of its FWHM. A factor $\epsilon$ is defined as the ratio of the clean-to-dirty beam volumes. 
The dirty beam is only considered up to its first null since this is the part approximated to a Gaussian during deconvolution and because the volume of a dirty beam without zero spacing integrated over its full domain is formally zero.
The output CASA cubes are the channel-by-channel addition of a $\texttt{.model}$ image convolved with the Gaussian clean beam (resulting in units of Jy per clean beam), the residuals contained in the \texttt{.residual} file, which are still in units of Jy per dirty beam, are then added. The JvM correction rescales the residuals to the same units (Jy per clean beam) by multiplying them by the $\epsilon$ factor before the final image restoration\footnote{A public tool to apply the JvM correction to interferometric images will be released in \url{https://github.com/radio-astro-tools/beam-volume-tools}}.

\cref{fig:jvmcorrection} shows a case where $\epsilon$ is as low as 0.32 (typically $\epsilon > 0.5$). The mean, median and standard deviation for $\epsilon$ over all spectral windows and fields are 0.62, 0.60, and 0.17, respectively. The effect of the JvM correction on the flux in a restored-image is minimal when the line is relatively bright and  deeply cleaned. The JvM correction becomes important as the flux in the residuals becomes larger than the flux in the model and as $\epsilon$ decreases. The first condition can occur in cubes with very bright and extended line emission that cannot be cleaned too deeply or for faint lines that reside in the same cube of a much brighter line. A small $\epsilon$ appears when the dirty beam has a substantial plateau beyond its core, which might occur due to the combination of different array configurations. The $\epsilon$ factor for each protocluster and spectral window can be found in its FITS header and in  \cref{appendix:tab:epsilon}.   
The \texttt{CASA}-generated cubes generally have a different beam for every channel. 
The common beam found by \texttt{radio-beam} is the minimum beam that contains all of the per-channel beams, excluding outliers (which sometimes occur at the edge of spectral windows). Although the channel-to-channel beam should vary smoothly with frequency, sometimes significant jumps in beam size can occur due to software instabilities. 
To make the cubes more readily usable, we convolve the model in every channel to a common Gaussian beam per spectral window using the \texttt{radio-beam} tool\footnote{\url{https://github.com/radio-astro-tools/radio-beam}}.
Using a common beam has the advantage of eliminating channel-to-channel variations, resulting in a cube with the same spatial resolution across all channels and allowing direct comparisons of full-bandwidth spectra in units of brightness temperature. 
  
The JvM cubes are in a common unit (Jy per clean beam), meaning flux measurements from these cubes can be interpreted. As discussed previously, these cubes are not deeply cleaned due to divergence and stippling issues over the full bandwidth, which can result in significant real emission remaining in the residuals. Thus, the JvM correction provides a more accurate estimation of the flux in our data. An important consequence, however, of performing the JvM correction is that the noise in the residual is also scaled by the $\epsilon$ factor. The most conservative approach to obtaining the noise, which we adopt here, is to estimate the noise from emission and line-free regions in the JvM cubes (in units of Jy per clean beam) and scale by 1/$\epsilon$ factor (see \cref{sec:noise}), thus, taking the higher noise estimates in units of Jy per dirty beam. In other words, we assume the noise estimated from a smaller clean beam should be larger by the ratio of the dirty and clean beam areas. This conservative approach means we adopt a higher noise level (in units of Jy per dirty beam), reducing the number of sources we consider significant. Still, we suggest further investigation of noise behaviour with non-Gaussian beams would be helpful. We note that \citet{walter2008} and \citet{Czekala2021} adopt the noise in Jy per dirty beam and Jy per clean beam as their noise levels, respectively. The latter is more representative of the telescope sensitivity for point-source detection prior to deconvolution.

\subsection{Beam sizes}
\label{sec:beamsizes}
In \cref{tab:spwsbeams_contprop}, we present the proposed beam and the average recovered continuum beam for all fields. In \cref{tab:spwsbeams}, we provide the major and minor axes along with the beam position angle for all spectral windows towards all protoclusters. Plots of the continuum PSFs and central average frequencies are presented in \citet{Ginsburg2022}. In \cref{fig:beams}, we show the ratio of the average recovered beam for each spectral window and protocluster compared to the average continuum beam. We further scale the beam ratio by the frequency difference between the central frequency in the given spectral window to the frequency used to determine the continuum beam to account for any beam size differences due to frequency. The average ratio of the line beams and continuum beam is $\sim$1.2, except for B3, spectral window 0 (which includes the \nthp\, transition), where the ratio is on average $\sim$1.4. The larger beams recovered in this spectral window are due to a plateau in the PSF, which may have occurred due to the combination of two considerably different array configurations, which resulted in a broader Gaussian fit during the determination of the synthesised beam in \tclean. An example of the PSF profiles and synthesised beams for all spectral windows towards G012.80 is shown in \cref{fig:psfs}. Additional broadening can be seen in B3, particularly in spectral window 0.

\subsection{Noise estimation}
\label{sec:noise}
Estimating the noise homogeneously over the full sample of protoclusters and spectral windows is non-trivial, given that several lines are present in the same spectral window, with varying morphology and intensity across a given protocluster. To estimate noise levels, we use the median-absolute-deviation (MAD) estimator\footnote{Implemented in \texttt{astropy.stats.median\_abs\_deviation} and we scale this by 1.4826 such that the reported value is equivalent to the standard deviation if the underlying data are normally distributed.\label{note:mad}} and take a threshold cut in intensity so that the noise is estimated using only the 25\% of the channels with the lowest intensity across the full spectral window, and using the manually defined regions\footnote{\url{https://github.com/ALMA-IMF/reduction/tree/master/reduction/noise\_estimation\_regions}} which were created for the continuum estimates of the noise in \citet{Ginsburg2022}. 
The noise estimates for each field and spectral window are taken on the non-continuum subtracted line cubes with the JvM correction applied but uncorrected by the primary beam response.
As described in \cref{sec:jvm}, we adopt the noise in units of Jy per dirty beam and provide the noise estimates for all fields and spectral windows (in units of mJy per dirty beam) in \cref{tab:spwsrms}. We also provide the requested (theoretical) noise in \cref{tab:spwsrms} from the original ALMA-IMF proposal scaled to the obtained channel width and central frequency. In \cref{fig:noise}, we compare the requested to achieved noise in brightness temperature units. To provide a direct comparison between the achieved noise (taken from \ref{tab:spwsrms}) and the requested noise, we scaled the requested noise to the same spectral resolution, beam size, and central frequency as those obtained in the observed data cubes for all spectral windows.  
We show in \cref{fig:noise} that the achieved noise is better than expected (in all but one case) and, on average, $\sim$2.3 times lower than anticipated, accounting for the larger beam sizes achieved. We note that the data were observed in better conditions for most fields and with lower system temperatures (typically 20-30\% better, but in some cases up to $\sim$50\%) than those used for the sensitivity calculator's noise estimates. The continuum noise is typically within a factor of 2 \citep{Ginsburg2022} of the theoretical value.

\subsection{Released line cubes and post-processed image products}
The DR1 line cube release discussed here is made available with this work\footnote{see the ALMA-IMF website; \url{https://www.almaimf.com/data.html} and is hosted by Harvard Dataverse; \url{https:
//dataverse.harvard.edu/dataverse/alma-imf}}. We provide the continuum-subtracted cubes corrected for their respective primary beam response and the JvM correction. The \texttt{STATCONT} procedure employed simultaneously over the full spectral window can satisfactorily remove the continuum for most of the bandwidth and fields, however, towards, the positions of hot core candidates or bright outflows, the continuum subtraction could be improved. 
For completeness, we also include the continuum estimates from the cubes, the models, residuals and primary beam files in this release. 
The naming structure of the cubes is given as the protocluster name, the ALMA band (i.e. B3 or B6), array configuration (we release here only the 12m array data), and spectral window (see \cref{tab:spws_fred} for reference). Meanwhile \texttt{.JvM} refers to the JvM correction (we have applied the correction to all released cubes; the $\epsilon$ factors used can be found in \cref{tab:spwsrms}), \texttt{.image.pbcor} refers to the primary beam correction, and \texttt{.statcont.contsub} refers to the continuum subtraction applied using the \texttt{STATCONT} procedure. The cubes range from the smallest at several MBs to the largest, which are $\sim$50~GB. The full DR1 dataset is $\sim$5~TB.

\section{Analysis}
\begin{figure*}
    \centering
    \includegraphics[width=1\textwidth]{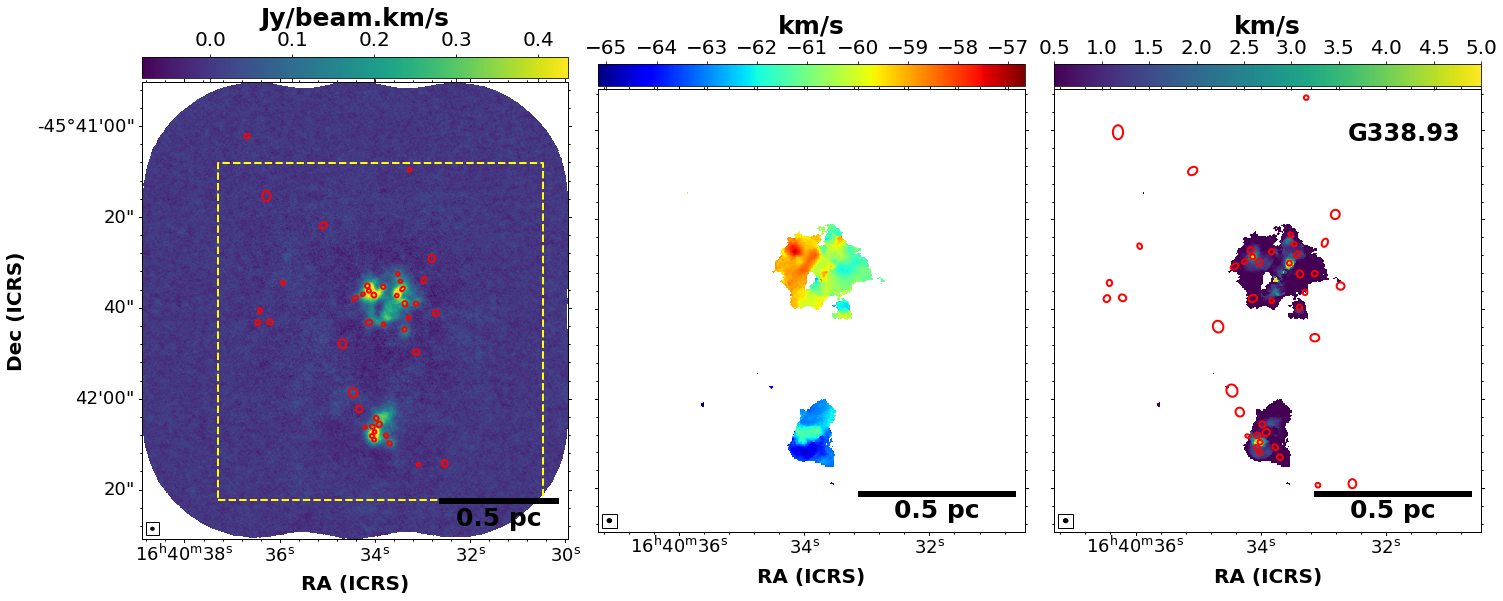}
    \includegraphics[width=1\textwidth]{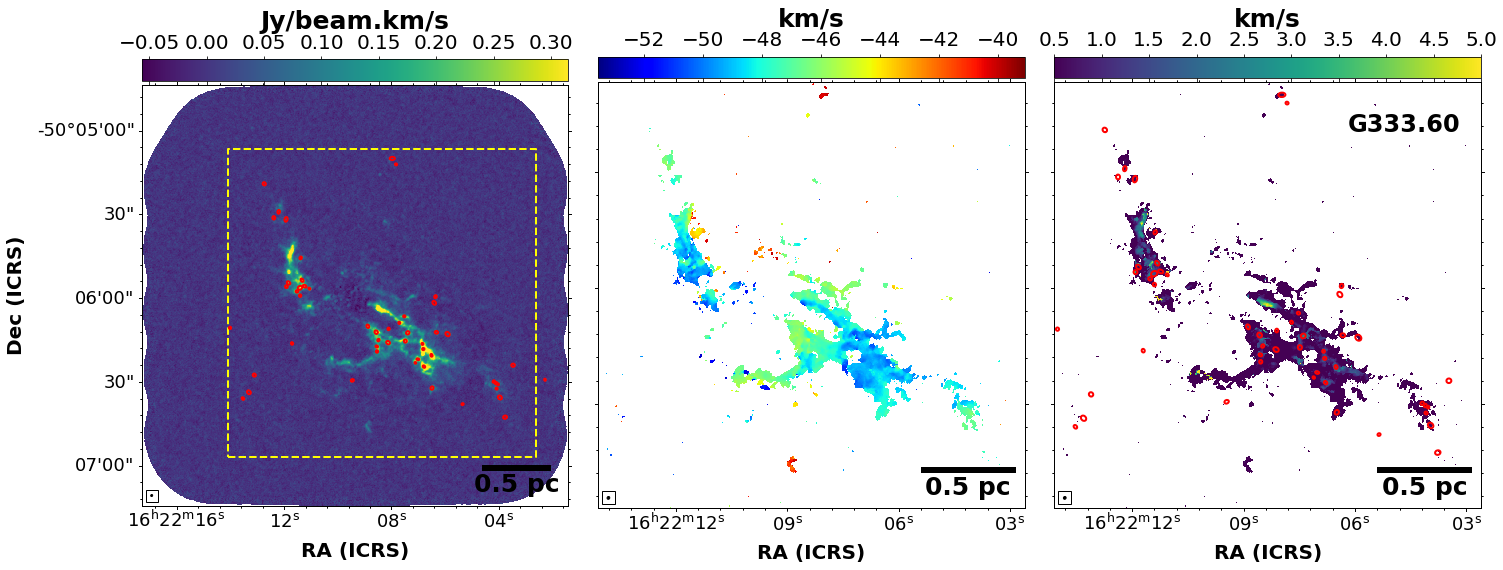} 
    \caption{Moment maps (moment 0, 1,  and 2 in the left, centre, and right panels, respectively) of DCN (3-2) emission towards two example protoclusters with different evolutionary classifications, the young protocluster G338.93 (top), and the evolved G333.60 (bottom). All three moments have been determined over a velocity range of $-53.4$ to $-68.6$~\kms and $-40$ to $-52.7$~\kms for G338.93 and G333.60, respectively. For the moment 0 and moment 2 maps, we overlay the core positions and sizes (red ellipses) from the continuum core catalogue described in \cite{louvet22}. For the moment 1 and moment 2 maps, we show a zoom-in of the area highlighted by the yellow dashed box overlaid on the moment 0 map. The moment 1 and moment 2 maps have an additional threshold cut per channel of 4~$\sigma$. The synthesised beam is shown in the bottom left corner of each image. These maps highlight differences in the morphology traced by the DCN (3-2) emission, where the emission is more widespread and filamentary towards the evolved region G333.60 than in the young region G338.93. This trend in the DCN (3-2) emission morphology with the evolutionary stage is consistent across the full ALMA-IMF sample.} 
    \label{fig:G333G338_momentmaps}
\end{figure*}
The full set of ALMA-IMF line cubes contains a wealth of emission from a variety of molecular line species (see \cref{tab:spws_fred}) and will provide the community with an unprecedented database with high legacy value for cores, hot cores (e.g. \citealt{Bonfand23}; \citealt{Brouillet2022}), outflows, and inflows. To illustrate the richness of the ALMA-IMF survey data, we present a first look at the DCN (3-2) emission towards the 15 protoclusters, along with an analysis of the DCN (3-2) emission extracted from the compact continuum core population. We utilise the molecular transition DCN (3-2) as a proxy for the gas associated with the core as it is typically an optically thin tracer with a critical density of $\sim$10$^7$~cm$^{-3}$. Furthermore, towards several low, intermediate, and massive star-forming regions DCN (3-2) emission has been previously observed to coincide well with the thermal dust emission associated with star-forming cores (see, e.g. \citealt{Cunningham2016}; \citealt{Tatematsu2020}; \citealt{Sakai2022}) and is not typically observed as an outflow tracer (however, it has been associated with shocks in, e.g. L1157-B1: \citealt{Busquet2017}). 

\subsection{Global scale DCN (3-2) morphology}
\label{sec:momentmaps}

DCN (3-2) has a rest frequency of 217.23854~GHz \citep{muller2001,muller2005} and is located in B6, spectral window 1, with a spectral resolution of 0.34~\kms. We use the continuum subtracted (\texttt{STATCONT}) line cubes provided in this release and do not perform any additional baseline subtraction. We then extract a 50~\kms wide subset of channels around the \vlsr reported in \citet{Motte2022} and create the first three moment maps towards the 15 protoclusters, using the non-primary beam corrected cubes. The DCN (3-2) emission is integrated over a manually determined velocity range for each protocluster, using a 4$\sigma$ threshold to identify channels where DCN (3-2) emission is present over an area larger than the beam size (the velocity ranges used for each protocluster are provided in the figure captions). The $\sigma$ level is the noise estimated for spectral window 1 as described in \cref{sec:noise} (i.e. in units of mJy per dirty beam) and listed for each protocluster in \cref{tab:spwsrms}. Furthermore, for the first and second-moment maps, we take an additional threshold cut per channel and include only pixels that contain emission $>$4$\sigma$, preventing noisy pixels from being added into them. To highlight the contrast in the DCN (3-2) emission across the sample, we selected two protoclusters with different evolutionary classifications: young G338.93 and evolved G333.60, and display their first three moments in \cref{fig:G333G338_momentmaps}. Towards G338.93, there are two prominent $\sim$0.2~pc clumps of emission with a spread of $\sim$10~\kms in velocity. On the other hand, towards G333.60, the emission is tracing a more filamentary structure spread over a larger extent, that is, over $\sim$2~pc. This morphological difference appears to be part of a general trend (see \cref{fig:appendixmomments}), where dense gas is less widespread in the young regions. Still, the trend needs to be confirmed using various gas tracers \citep{Cunningham23}. The full set of moment maps towards all protoclusters is presented in \cref{fig:appendixmomments}. We note that in a few regions (e.g. G327.29 and G351.77), the moment 0 maps display negative bowls, resulting from the missing short spacings in this 12m-array only data.

\begin{figure*}
    \centering
\includegraphics[width=0.32\textwidth]{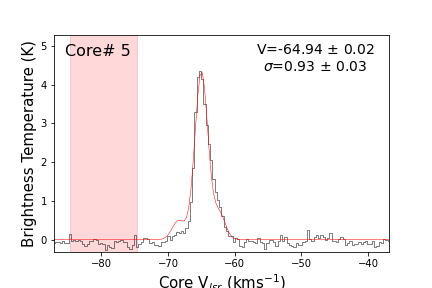} 
\includegraphics[width=0.32\textwidth]{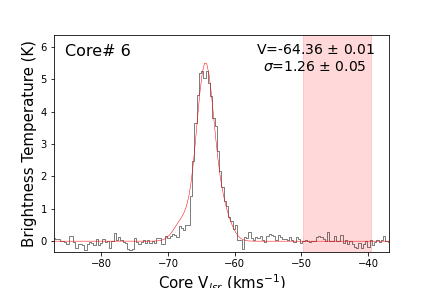} 
\includegraphics[width=0.32\textwidth]{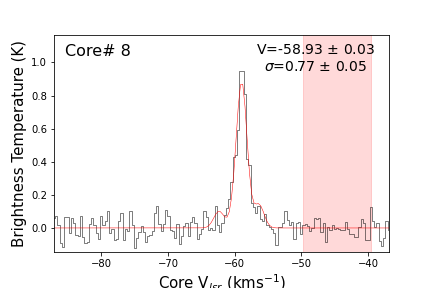}    
\includegraphics[width=0.32\textwidth]{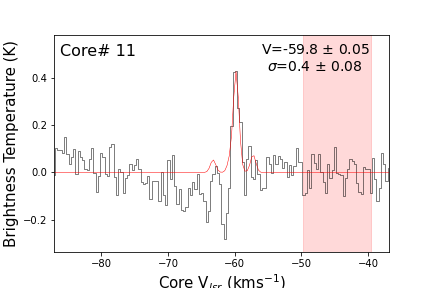} 
\includegraphics[width=0.32\textwidth]{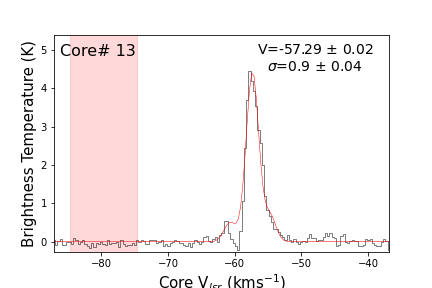} 
\includegraphics[width=0.32\textwidth]{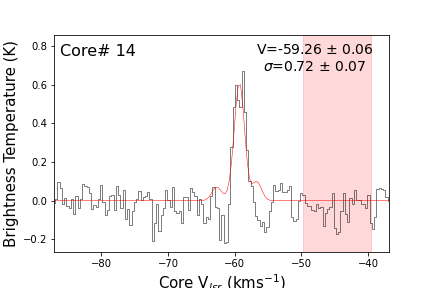}   

    \caption{Example of the core-averaged, background-subtracted DCN (3-2) Single-type spectra for six cores in the young protocluster G338.93. The associated continuum core number is given in the top left of each panel (the core numbering is taken from \citet{louvet22}). The core \vlsr (V) and velocity dispersion ($\sigma$) in units of \kms from the HSF fit are given in the top right. The line fit parameters for each core are also presented in \cref{tab:coretables_g338}. The pink-shaded region represents the part of the spectrum used to estimate the MAD noise.}
    \label{fig:dcnspectra_split_G338_s}
\end{figure*}

\begin{figure*}
    \centering
\includegraphics[width=0.32\textwidth]{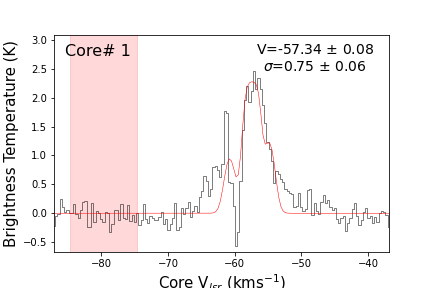} 
\includegraphics[width=0.32\textwidth]{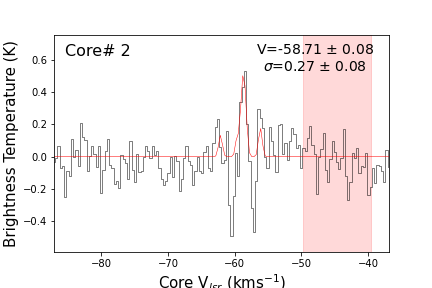} 
\includegraphics[width=0.32\textwidth]{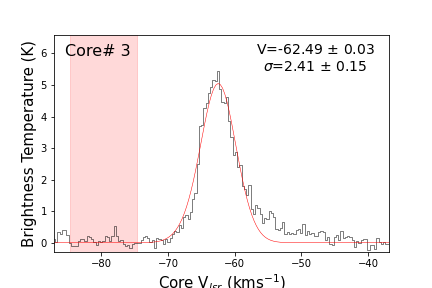} 
\includegraphics[width=0.32\textwidth]{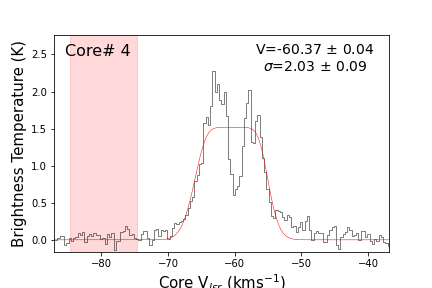} 
\includegraphics[width=0.32\textwidth]{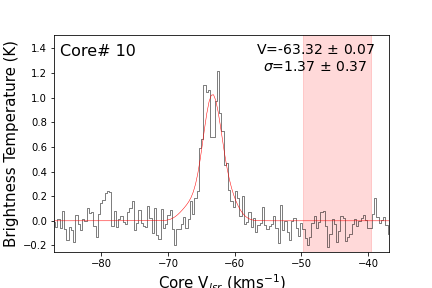} 
\includegraphics[width=0.32\textwidth]{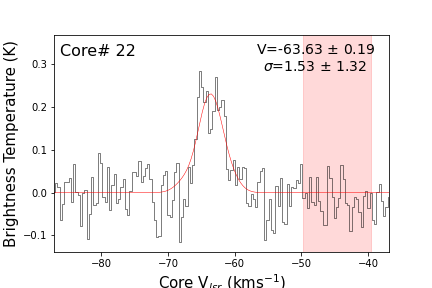} 
    \caption{Example of the core-averaged, background-subtracted DCN (3-2) Complex-type spectra for six cores in the young protocluster G338.93. The associated continuum core number is given in the top left of each panel (the core numbering is taken from \citet{louvet22}). The core \vlsr (V) and velocity dispersion ($\sigma$) in units of \kms from the HSF fit are given in the top right. Given the complex structure in their spectra, the resulting fits are excluded from the analysis and labelled as Complex-type spectra. The pink-shaded region represents the part of the spectrum used to estimate the MAD noise.}
    \label{fig:dcnspectra_split_G338_c}
\end{figure*}

\begin{table*}[htbp!]
\centering
\small
\begin{threeparttable}[c]
\caption{Example of the DCN fits towards the first 15 cores of the young protocluster G338.93.}
\label{tab:coretables_g338}
\begin{tabular}{llllllllccc}
\hline \noalign {\smallskip}
 n\tnote{1}  & Core\tnote{1} &RA\tnote{1}  & DEC\tnote{1} & F$_{A}$\tnote{1}   & F$_{B}$\tnote{1}    & PA\tnote{1} & T & \vlsr\tnote{2}  & Linewidth\tnote{2}   & Spectral\tnote{3} \\
 & Name  & [ICRS]  &  [ICRS]   &  [\arcsec] &[\arcsec] &  [deg] & [K] & [\kms]  & [\kms] & Type \\
\hline \noalign {\smallskip}
1  &  250.1422011-45.6934028  &  16:40:34.13  &  -45:41:36.25  &  0.89  &  0.79  &  84  &  100 $\pm$ 50  &  -57.34 $\pm$ 0.08  &  --  &  Complex \\
2  &  250.1427268-45.6936217  &  16:40:34.25  &  -45:41:37.04  &  0.97  &  0.73  &  124  &  100 $\pm$ 50  &  -58.71 $\pm$ 0.08  &  --  &  Complex \\
3  &  250.1417044-45.7020244  &  16:40:34.01  &  -45:42:7.290  &  0.86  &  0.79  &  92  &  100 $\pm$ 50  &  -62.49 $\pm$ 0.03  &  --  &  Complex \\
4  &  250.1397543-45.6936984  &  16:40:33.54  &  -45:41:37.31  &  0.86  &  0.82  &  7  &  100 $\pm$ 50  &  -60.37 $\pm$ 0.04  &  --  &  Complex \\
5  &  250.1417392-45.7025004  &  16:40:34.02  &  -45:42:9.000  &  0.86  &  0.79  &  89  &  36 $\pm$ 7  &  -64.94 $\pm$ 0.02  &  2.19 $\pm$ 0.07  &  Single \\
6  &  250.1419183-45.7022799  &  16:40:34.06  &  -45:42:8.210  &  1.0  &  0.86  &  65  &  35 $\pm$ 10  &  -64.36 $\pm$ 0.01  &  2.98 $\pm$ 0.12  &  Single \\
7  &  250.1378537-45.7040233  &  16:40:33.08  &  -45:42:14.48  &  0.81  &  0.76  &  56  &  30 $\pm$ 6  &  --  &  --  &  -- \\
8  &  250.1433977-45.6938542  &  16:40:34.42  &  -45:41:37.88  &  1.42  &  0.88  &  119  &  27 $\pm$ 6  &  -58.93 $\pm$ 0.03  &  1.81 $\pm$ 0.12  &  Single \\
9  &  250.1497218-45.6929092  &  16:40:35.93  &  -45:41:34.47  &  0.98  &  0.74  &  29  &  24 $\pm$ 5  &  --  &  --  &  -- \\
10  &  250.1403862-45.7027264  &  16:40:33.69  &  -45:42:9.820  &  1.02  &  0.88  &  63  &  100 $\pm$ 50  &  -63.32 $\pm$ 0.07  &  --  &  Complex \\
11  &  250.1409277-45.6954519  &  16:40:33.82  &  -45:41:43.63  &  0.87  &  0.74  &  15  &  27 $\pm$ 6  &  -59.8 $\pm$ 0.05  &  0.95 $\pm$ 0.2  &  Single \\
12  &  250.1444989-45.69665  &  16:40:34.68  &  -45:41:47.94  &  1.99  &  1.75  &  16  &  24 $\pm$ 5  &  --  &  --  &  -- \\
13  &  250.1423231-45.6931039  &  16:40:34.16  &  -45:41:35.17  &  1.06  &  1.04  &  129  &  29 $\pm$ 6  &  -57.29 $\pm$ 0.02  &  2.12 $\pm$ 0.09  &  Single \\
14  &  250.1394346-45.6928199  &  16:40:33.46  &  -45:41:34.15  &  0.87  &  0.7  &  72  &  28 $\pm$ 6  &  -59.26 $\pm$ 0.06  &  1.69 $\pm$ 0.17  &  Single \\
15  &  250.1461922-45.6894135  &  16:40:35.09  &  -45:41:21.89  &  1.68  &  1.26  &  124  &  24 $\pm$ 5  &  --  &  --  &  -- \\

\hline \noalign {\smallskip}
\end{tabular}
\begin{tablenotes}
\item[1] The core numbering (n), core name, RA, DEC, position angle (PA), core major (F$_{A}$) and core minor (F$_{B}$) FWHM and dust temperatures (T) are taken from the smoothed core catalogues in \citet{louvet22}. The dust temperature estimates provided here are also used to determine the core masses and are discussed in more detail in \citet{louvet22}. 
\item[2] The \vlsr and linewidth are taken from the HFS fits. For reference, we include the \vlsr for the Complex-type spectra, however, as the spectra are complex, often containing multiple components, we do not use them in the analysis or provide the linewidths. 
\item[3] Dashes are given for those cores with no DCN (3-2) detection.
\end{tablenotes}
\end{threeparttable}
\end{table*}

\begin{table*}[htbp!]
\centering
\begin{threeparttable}[c]
\caption{Characteristic parameters of the DCN (3-2) hyperfine fits}
\label{tab:dcnfits_average}
 \begin{tabular}{lllllllllllll}
\hline \noalign {\smallskip}
Protocluster         & Number\tnote{1}  & Number\tnote{2}   & Detection\tnote{2} & Velocity\tnote{3} & \vlsr\tnote{3}   & \vlsr\tnote{3}    & Linewidth\tnote{3}  & Linewidth\tnote{3}   &  Linewidth\tnote{3}  \\ 
cloud name & of Cores & detected    & rate [\%]   & range    & mean   & std     & mean  & median &  std \\
&  &(complex)        &  (complex)      & [km~s$^{-1}$]   & [km~s$^{-1}$] & [km~s$^{-1}$]  & [km~s$^{-1}$]& [km~s$^{-1}$] & [km~s$^{-1}$] \\

\hline \noalign {\smallskip}
\multicolumn{10}{c}{Young} \\
G327.29 & 32 & 13 (17)&  41 (53)\% & 5.7 & -45.1 & 1.9  & 1.3 & 1.3 & 0.3 \\
G328.25 & 11 & 3  (4)&  27 (36)\% & 2.5 & -42.7 & 1.0  & 2.0 & 1.2 & 1.2 \\
G338.93 & 42 & 18 (26)&  43 (62)\% & 7.7 & -61.0 & 2.0  & 1.6 & 1.6 & 0.7 \\
G337.92 & 22 & 8  (15)&  36 (68)\% & 3.9 & -39.2 & 1.3  & 2.3 & 2.0 & 1.2 \\
W43-MM1 & 70 & 15 (29)&  21 (41)\% & 7.0 & 98.5 & 2.1  & 1.1 & 0.9 & 0.5 \\
W43-MM2 & 40 & 18 (25)&  45 (62)\% & 4.7 & 91.1 & 1.4  & 1.4 & 1.3 & 0.5 \\
Average &  36 & 12 (19) & 35 (53)\% &  5.2 &  & 1.9  & 1.5 & 1.3 & 0.76 \\
Total &  217 & 75 (116)  \\
\hline
\multicolumn{10}{c}{Intermediate} \\
G351.77 & 18 & 4  (13)&  22 (72)\% & 6.0 & -3.9 & 2.4  & 1.4 & 1.5 & 0.3 \\
G353.41 & 45 & 14 (15)&  31 (33)\% & 7.6 & -17.4 & 1.9  & 1.4 & 1.1 & 0.7 \\
G008.67 & 19 & 9  (13)&  47 (68)\% & 7.3 & 35.9 & 2.5  & 1.5 & 1.4 & 0.5 \\
W43-MM3 & 36 & 9  (13)&  25 (36)\% & 4.6 & 92.8 & 1.4  & 1.4 & 1.2 & 0.7 \\
W51-E & 23 & 7  (10)&  30 (43)\% & 11.7 & 56.2 & 4.5  & 1.8 & 2.0 & 0.7 \\
Average &  28 & 9  (13) & 30 (45)\% &  7.4 &  & 2.9  & 1.5 & 1.3 & 0.64 \\
Total &  141 & 43 (64)  \\
\hline
\multicolumn{10}{c}{Evolved} \\
G333.60 &  52 & 28 (38) &  54 (73)\% & 10.4 & -47.8 & 2.2  & 1.3 & 1.2 & 0.4 \\
G010.62 &  42 & 28 (34) &  67 (81)\% & 10.1 & -2.7 & 2.4  & 1.1 & 1.1 & 0.3 \\
G012.80 &  46 & 37 (38) &  80 (83)\% & 7.4 & 36.0 & 1.5  & 1.3 & 1.2 & 0.6 \\
W51-IRS2 &  97 & 55 (67) &  57 (69)\% & 13.7 & 61.8 & 2.4  & 1.2 & 1.1 & 0.5 \\
Average &  59 & 37 (44) & 62 (75)\% &  10.4 &  & 2.3  & 1.2 & 1.1 & 0.51 \\
Total &  237 & 148 (177)  \\

\hline \noalign {\smallskip}
\end{tabular}
\begin{tablenotes}
\item[1] The number of continuum cores is taken from the smoothed core catalogue from \citet{louvet22}. We note that four cores in the sample overlap: two cores from W51-E (cores 20 and 30) overlap with two cores in W51-IRS2 (cores 22 and 38), and two cores in W43-MM2 (cores 10 and 46) overlap with two cores in W43-MM3 (cores 2 and 37). However, as they are either non-detections or Complex-type spectra, we do not explicitly assign them to a given field as done in \citet{louvet22}. 
\item[2] The total count/percentages in the brackets include both Single- and Complex-type DCN (3-2) detections. Spectra are described as Complex-type spectra and are not well fit by a single component. We provide the percentages and number of detections, but cores with a Complex-type spectrum are not included in the estimates of the Velocity range, \vlsr, or linewidth.
\item[3] Only Single-type DCN spectra (fit with a single component) and listed as Single in the Tables of the DCN fits (e.g. \cref{tab:coretables_g338}) are included in the estimates of the Velocity ranges, \vlsr and linewidth for each protocluster. 

\end{tablenotes}
\end{threeparttable}
\end{table*}

\subsection{DCN (3-2) line extraction and fitting}

 The DCN (3-2) line emission is extracted from each core using \texttt{spectral-cube}\footnote{\url{https://spectral-cube.readthedocs.io/en/latest/}} \citep{Ginsburg2019b}. We use an elliptical aperture with a major (minor) axis length twice that of the continuum source major (minor) FWHM to extract the spectrum, where the continuum core sizes are taken from \cite{louvet22}\footnote{We use here the smoothed core catalogue, where the continuum maps are smoothed to the same angular resolution of 2700~au.}. As with the continuum cores, where the background is filtered to estimate the core properties more accurately, we chose to perform background subtraction on the DCN (3-2) emission to limit contamination from background and foreground DCN (3-2) within the dense regions of the protocluster. 
 We extract the background emission from an elliptical annulus with an inner major (minor) axis size equal to the size of the spectral aperture (i.e. twice the continuum source major (minor) FWHM) and outer radius size 1.5x larger (i.e. three times the continuum source major (minor) FWHM). In these complex regions, particularly towards the densest parts of the protoclusters, nearby cores can overlap with the background annulus. We excluded pixels from neighbouring cores that spatially overlap with the background annulus in these instances. An example of the core spectral aperture and resulting background annulus is given in \cref{appendix:dcn_fitting}. A core-averaged spectrum is then extracted for each of the 595 cores across the 15 ALMA-IMF clouds, and the average spectrum from its background annulus is subtracted from it. The resulting core-average, background-subtracted spectrum for each core is then fitted using a single component hyperfine structure model (HFS) adapted for the DCN (3-2) transition in \texttt{PySpecKit} \citep{Ginsburg2011}. The methodology of the line extraction, background subtraction, and HSF fitting are discussed further in \cref{appendix:dcn_fitting}.

\subsection{DCN (3-2) detection of cores}
\label{line_fitting_complex}
A core is determined to have a DCN (3-2) detection if a velocity dispersion of $>$0.2\,\kms and a signal-to-noise ratio (S/N) $>$4$\sigma$ are found in the core averaged, background-subtracted spectrum. The noise in the spectrum is estimated from the MAD\footref{note:mad} in 30 continuous channels ($\sim$10\,\kms) from either the lower or upper part of the spectrum that were identified by eye to be in a line free part of the spectrum. Furthermore, as described in \cref{sec:jvm}, we use the higher noise estimates in units of mJy per dirty beam. 

Of the 595 continuum cores, 357 (60\%) have a DCN (3-2) detection. Within each protocluster, however, several detected cores display complex spectra (91 over the full sample), thus not well fit by the single component HFS. We classify spectra as Complex-type if multiple components or clear signs of structure affecting the single component fits are present in the spectrum. This is done manually by visually inspecting all fitted DCN (3-2) spectra. 
While the \vlsr from Complex-type spectral fits are provided for reference, the linewidths are not given, and these cores are excluded from the following analysis. Only cores determined to be well fit by a single component (e.g. listed as Single in \cref{tab:coretables_g338}) are considered further. The Complex-type spectra may harbour multiple velocity components or arise from line self-absorption due to optical depth effects. We note that a few of the DCN (3-2) spectra (i.e. core 3, G338.93) show broad wings, which may indicate that it is tracing shocked gas (e.g. \citealt{Busquet2017}). Additional work is required, however, to understand the nature of the Complex-type spectra fully and extract accurate fits, which is beyond this paper's scope. After filtering out the cores with a Complex DCN (3-2) spectrum, the sample is reduced to 266 continuum cores, that is, $\sim$45\% of the continuum cores have a single component HFS fit. 

We show examples of the Single- and Complex-type DCN (3-2) spectra in Figs \ref{fig:dcnspectra_split_G338_s} and \ref{fig:dcnspectra_split_G338_c}, respectively, extracted towards a sample of cores in the protocluster G338.93. In \cref{tab:coretables_g338}, we give an example of the resulting linewidth and \vlsr from the hyperfine fitting for G338.93. The linewidths are taken as $\sigma_{obs}\times\sqrt{8ln2}$, where $\sigma_{obs}$ is the observed velocity dispersion from single component HFS fits to the core-averaged, background-subtracted spectra. The full set of Single- and Complex-type spectra and tables for G338.93 and all protoclusters can be found in \cref{appendix:spectral}. Towards G338.93, 26 of its 42 cores have DCN (3-2) detections ($\sim$62\%), with 8 of these cores displaying complex spectra, thus after filtering, $\sim$43\% of the cores in G338.93 have a Single-type DCN (3-2) detection. In the evolved region G333.60, 38 of its 52 cores have DCN (3-2) detections ($\sim$73\%), with 10 of these cores displaying complex spectra, filtering out the Complex-type spectra, $\sim$54\% of the cores have Single-type detection. In \cref{tab:dcnfits_average}, we show the detection rate of the DCN (3-2) fits for both the Single- and Complex-type spectra extracted for all protoclusters. We find that DCN (3-2) is detected more often in cores situated in evolved regions with an average detection rate of 62\% for cores with Single-type spectra, compared to a lower average detection rate ($<$35\%) in the young and intermediate regions. This difference is seen whether the Complex-type spectra are included or not, suggesting DCN detects a higher fraction of the continuum cores in the evolved protoclusters. This trend is in line with the more widespread detection of the DCN (3-2) emission in evolved protoclusters, as described in \cref{sec:momentmaps}.

To assess if the background subtraction could bias the results, we also perform an extraction on the non-background subtracted DCN (3-2) spectra towards each core in \cref{appendix:dcn_fitting}. While more cores are detected overall with Single-type spectra ($\sim$56\%), DCN still detects a higher fraction of the continuum cores in  evolved protoclusters ($\sim$72\%) compared with young and intermediate regions ($<$50\%). We also note that the absorption features in several spectra may come from the background subtraction (e.g. core 2, G338.93). Absorption features are, however, still present in the non-background subtracted spectra and are likely a consequence of the missing short spacings in these 12m~array only data.

\subsection{Protocluster \vlsr estimates from the DCN (3-2) fits}

\begin{figure}
    \centering
    \includegraphics[width=0.49\textwidth]{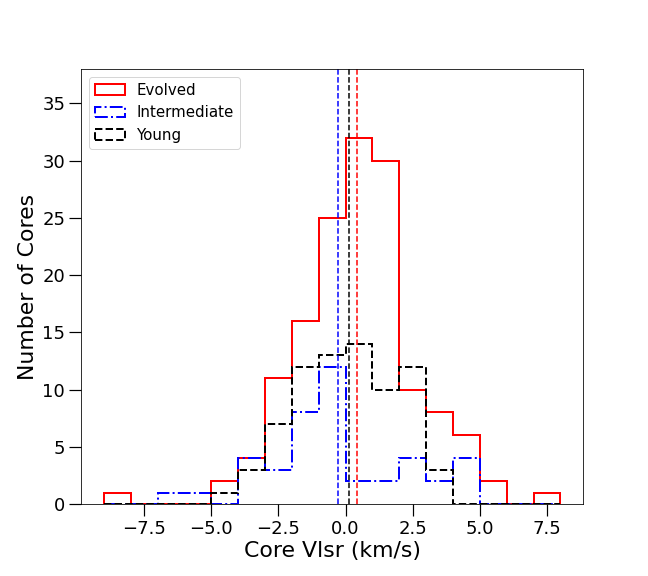}    
    \caption{Distribution of the core velocities from the DCN line fits  split by evolutionary stage. The velocity offset is determined as the central velocity of the fit and subtracting from it the average protocluster velocity (estimated from the average of the DCN (3-2) moment 1 maps). The vertical dashed lines are the average values for each subgroup (0.13\kms, -0.29\kms, and 0.44\kms for the young, intermediate and evolved regions, respectively). 
    We find no obvious dependence on the distribution of the core velocities over the sample.  }
    \label{fig:veloffset}
\end{figure}

In \cref{tab:dcnfits_average}, we also provide the average \vlsr and velocity ranges for the detected cores in each protocluster using only the fits from Single-type DCN (3-2) spectra. For most protoclusters, we find that the \vlsr estimates given in \cref{tab:sample} \citep{Motte2022} are consistent with the average centroid core \vlsr for each protocluster. For W51-IRS2, W43-MM2, and W43-MM3, however, there is a shift of 5-7~\kms between the average core \vlsr and that of the \vlsr estimates from \citet{Motte2022}. The shift in W51-IRS2 is likely due to the two distinct velocity distributions of its cores (which can be seen in the moment maps for W51-IRS2 in \cref{fig:appendixmomments}) and has been observed previously \citep{ginsburg15}. Furthermore, both W43-MM2 and W43-MM3 reside in the same complex (W43), along with W43-MM1, and the few \kms velocity offset in these regions is likely due to differences between the average bulk motions on larger scales compared to the internal motions of the embedded protoclusters. The \vlsr distribution for the cores situated within a given region span a range of values from $2.5<\Delta v_{LSR} < 13.7$ \kms; however, the higher values are produced by outliers (see \cref{fig:veloffset}).
In \cref{fig:veloffset}, we display the distribution of the core velocities per evolutionary stage. We estimate the velocity offset for each core by taking the central velocity of the fit and subtracting from it the average protocluster velocity (estimated from the average of the DCN (3-2) moment 1 maps). While the evolved protoclusters have cores with a larger velocity spread, that sample is dominated by the protocluster W51-IRS2, which likely has two distinct velocity distributions for its cores. Thus, we find no apparent difference in the spread of the core \vlsr with the evolutionary stage of the host region.

\begin{table*}[htbp!]
\centering
\begin{threeparttable}[c]
\caption{Comparison of molecular line detection towards the continuum core population of G338.93.}
\label{tab:linescomparison}
 \begin{tabular}{lcccccccc}
\hline \noalign {\smallskip}
Molecule      & Transition   & Frequency\tnote{1}   & {\it{{E}$_{u}$/$k_b$}}\tnote{1}& Critical Density\tnote{1} & \multicolumn{2}{c}{Cores Detected}\tnote{2}   &   \multicolumn{2}{c}{Percentage (\%)}  \\ 
\cline{6-7}\cline{8-9}
              &              &    [GHz]         &     [K]      & $n_{crit}$ [cm$^{-3}$] &  Single & Complex & Single & Complex  \\
\hline \noalign {\smallskip}
DCO$^+$   & 3-2      & 216.113 &  20.7    &   2.4x10$^6$    &  5    &  8  &  12  &  20   \\ 
OC$^{33}$S& 18-17 & 216.147 & 98.6 &  1.3x10$^{6}$ &   2   &  1  &  5  &   2  \\ 
DCN       & 3-2   & 217.238 & 20.9 &   1.8x10$^7$      &   18 & 8  & 43 & 19 \\ 
C$^{18}$O & 2-1   & 219.560 & 15.8 & 9.9x10$^3$  &   13 & 21 & 31 & 50 \\ 
OCS       & 19-18 & 231.061 & 110.9&  4.9x10$^5$ &   6   &  10 &  14  & 24    \\ 
$^{13}$CS & 5-4   & 231.220 & 33.3 &  4.3x10$^{6}$ &   3  & 7 & 7 &  17  \\ 
N$_2$D$^+$& 3-2   & 231.322 & 22.2  &  3.6x10$^6$      &  8    & 2   &  19  &  5   \\
\hline \noalign {\smallskip}
\end{tabular}
\begin{tablenotes}
\item[1] The rest frequencies were taken from the Cologne Database for Molecular Spectroscopy (CDMS). The temperatures and critical densities for OCS, OC$^{33}$S, C$^{18}$O and $^{13}$CS are taken from \citet{molet19}. For DCN (3-2), we take the values from \citet{Bocso2019} and for N$_2$D$^+$, DCO$^+$ we follow \citet{Bocso2019} and approximate the critical density assuming $n_{crit}\approx$A/$\Gamma$ \citep{Shirley2015}, for collisional rates at T=100~K where the Einstein A and $\Gamma$ coefficients are taken from the Leiden Atomic Molecular Database (LAMBDA) \citep{Schoier2005}. N$_2$D$^+$ is approximated by the values for N$_2$H$^+$.
\item[2] Number of continuum cores that have background subtracted spectrum which is well represented by either a single Gaussian component fit (Single) or has a spectrum that consists of multiple components or is too complex to fit with a single component Gaussian (Complex). The total number of continuum cores in G338.93 is 42. In the case of DCN (3-2), we use the adapted single-component HFS to fit the spectra.
\end{tablenotes}
\end{threeparttable}
\end{table*}

\begin{figure}
    \centering
\includegraphics[width=0.49\textwidth]{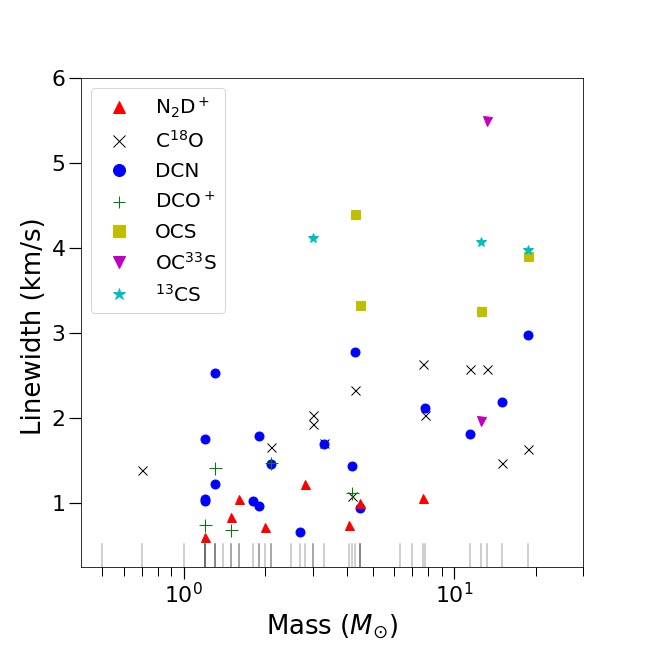} 
    \caption{Linewidths extracted towards the core population of G338.93 for the molecules listed in \cref{tab:linescomparison} as a function of the continuum core masses taken from \citet{louvet22}, assuming the core dust temperature estimates as given in \cref{tab:coretables_g338}. The linewidths shown are from fits classified as Single-type which can be well approximated by a single Gaussian (or single HFS fit for DCN) component fit to the extracted spectrum. The grey vertical lines represent the masses of all 42 cores in the G338.93 protocluster. DCN (3-2) well represents the full mass range of the cores in the protocluster.}
    \label{fig:linecomparison}
\end{figure}
\subsection{Comparison with other molecular lines}
\label{sec:line_comp}
We compare here the DCN (3-2) emission extracted towards the protocluster G338.93 to other molecular lines (situated within the ALMA-IMF spectral coverage) that have previously been used to study core properties and kinematics, such as DCO$^+$ (3-2), $^{13}$CS (5-4), OCS (19-18), C$^{18}$O (2-1), and N$_2$D$^+$ (3-2) (e.g. \citealt{Li2022}; \citealt{Sakai2022}; \citealt{Nony2020}; \citealt{Cunningham2016}; \citealt{Maud2015}). They are fitted and classified as Single- or Complex-type spectra in the same way as for DCN (3-2) (described in \cref{line_fitting_complex}), but using only a single component Gaussian fit in pyspeckit. To be consistent, for lines (e.g. DCO$^+$) situated in spectral windows with higher spectral resolution than B6 spectral window 1, we smooth these cubes to have the same spectral resolution as DCN. In \cref{tab:linescomparison}, we show the detection rate of several molecular transitions towards the core population of G338.93 along with their critical densities and the energies of their respective lower level above ground (in K).
DCN (3-2) exhibits Single-type spectra towards the highest percentile ($\sim$43\%) of continuum cores in G338.93, versus 5-31\% for the other species. While C$^{18}$O (2-1) is detected towards 1.3 times more cores overall ($\sim$81\%), a large fraction of those spectra (62\%) contains Complex-type spectra, which exhibit multiple peaks or non-Gaussian profiles (see \cref{line_fitting_complex}).
In \cref{fig:linecomparison}, we show the extracted linewidths for each molecular species as a function of the associated continuum core masses (taken from \citealt{louvet22}) for the protocluster G338.93. DCN (3-2) is extracted with a Single-type spectrum across the majority of the mass distribution of the cores. In comparison, $^{13}$CS (5-4) and OCS (19-18) are found predominantly towards the higher-mass continuum cores. Meanwhile, N$_2$D$^+$ (3-2) and DCO$^+$ (3-2) trace cores in the low/intermediate mass range with a Single-type spectrum. We note that several of the lowest mass cores in this region are not traced by DCN (3-2), but are traced by either C$^{18}$O (2-1), N$_2$D$^+$ (3-2) and DCO$^+$ (3-2), thus, DCN (3-2) may not be sensitive to the lowest mass cores (see \cref{sec:corenature} for further discussion). Furthermore, N$_2$D$^+$ (3-2) predominantly traces different cores than DCN (3-2) in G338.93 and could provide a complementary tracer of the low/intermediate mass core population. For the ALMA-IMF protoclusters, N$_2$D$^+$ may not be the optimal tracer of high-mass prestellar cores, contrary to the picture in infrared dark clouds (e.g. \citealt{Tan2013}, \citealt{Barnes23}) where N$_2$D$^+$ is observed towards massive prestellar cores. 
While \cref{fig:linecomparison} highlights the need for a multi-line analysis to recover the full core population in these protoclusters, such a task is beyond the scope of this paper. It will be performed on individual regions in the future (e.g. \citealt{Cunningham23}). DCN (3-2) recovers the highest percentage of continuum cores with Single-type spectra compared to other dense gas tracers. Thus, using only a single tracer, DCN (3-2) likely still provides the best proxy of the dense gas associated with the continuum cores in the ALMA-IMF bandwidth coverage. 
\section{Discussion}
\begin{figure*}
\centering
\includegraphics[width=0.9\textwidth]{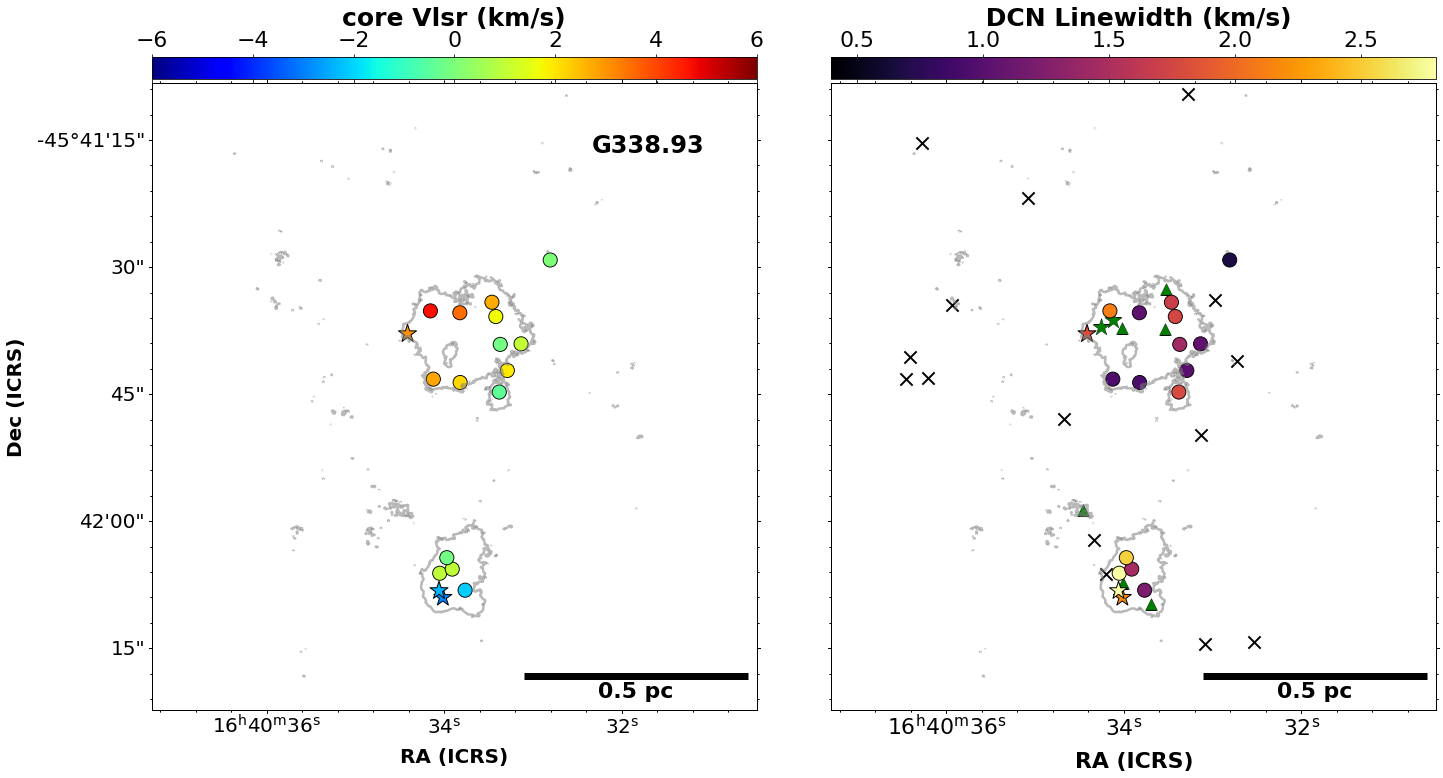}
\includegraphics[width=0.9\textwidth]{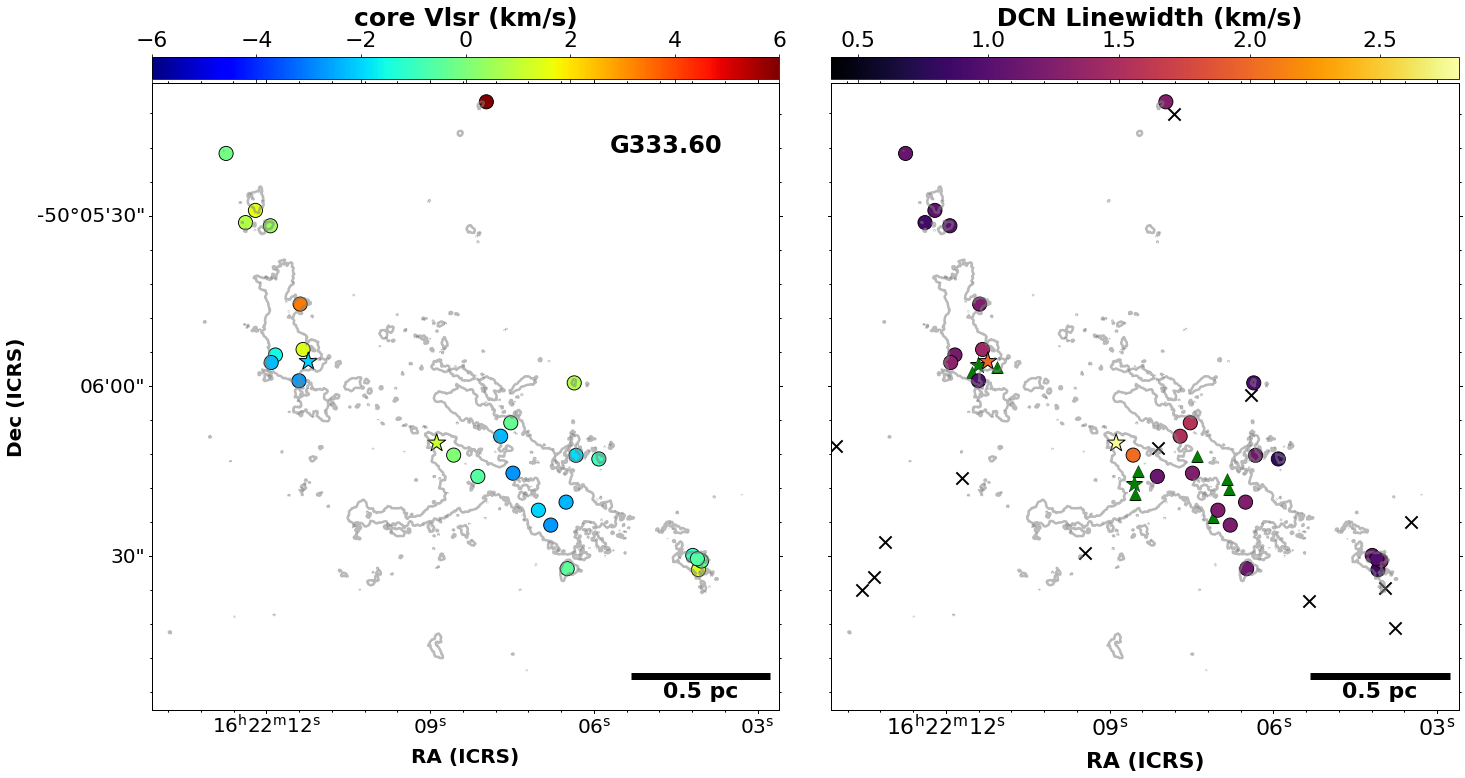} 
\caption{Core \vlsr (left) and DCN (3-2) linewidths (right) estimated from the DCN (3-2) fits to the continuum cores towards examples of a young and an evolved protocluster G338.93 (top) and G333.60 (bottom), respectively. The circles and stars represent the detected cores with mass estimates below and above 8\msun, respectively, with the colour scale displaying the fitted parameters from the DCN (3-2) fits (left: core \vlsr, right: linewidth). The core \vlsr is the centroid velocity of the DCN (3-2) fit minus the cloud \vlsr (taken as -62\kms, and -47\kms for G338.93, and G333.60, respectively). The grey contours are the 4 $\sigma$ level of the DCN (3-2) moment 0 map. In the right panels, the positions of cores without a DCN (3-2) detection are marked with a black cross and with a black star for cores with a mass estimate below and above 8\msun, respectively, and green triangles and stars represent cores with a Complex-type DCN (3-2) spectra, with a mass estimate below and above 8\msun, respectively. \label{fig:G333G338_coredisp}}
\end{figure*}

\begin{figure*}
    \centering
    \includegraphics[width=0.85\textwidth]{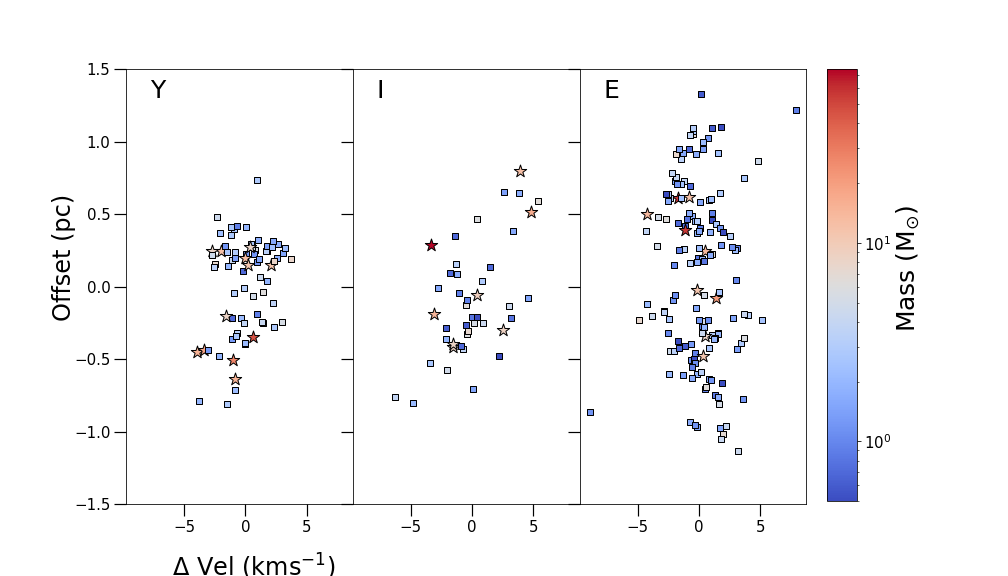} 
 
    \caption{Core \vlsr vs the core radial offset grouped by evolutionary classification of the host protocluster; young, intermediate and evolved (Y: left, I: centre, and E: right, respectively). The radial spatial offset for each core is determined by taking the position of the core and subtracting from it the mean position of all cores in its protocluster and assuming the distances given in \cref{tab:sample} for each region. The velocity offset for each core is determined by subtracting the centroid \vlsr found from all cores with Single-type spectra in a given region from it. The cores are colour-coded by their mass estimates taken from \citet{louvet22}, and star symbols represent those cores with mass estimates $>$8M$_{\odot}$. Cores with a declination smaller than the average position have a negative offset. For the young and intermediate regions, the spatial distribution of the cores is more compact compared to cores in evolved protoclusters, where the cores can be spatially distributed over larger areas ($>$2.0~pc.)}
    \label{fig:pv_ra_corevel}
\end{figure*}

\begin{figure*}
    \centering
    \includegraphics[width=0.49\textwidth]{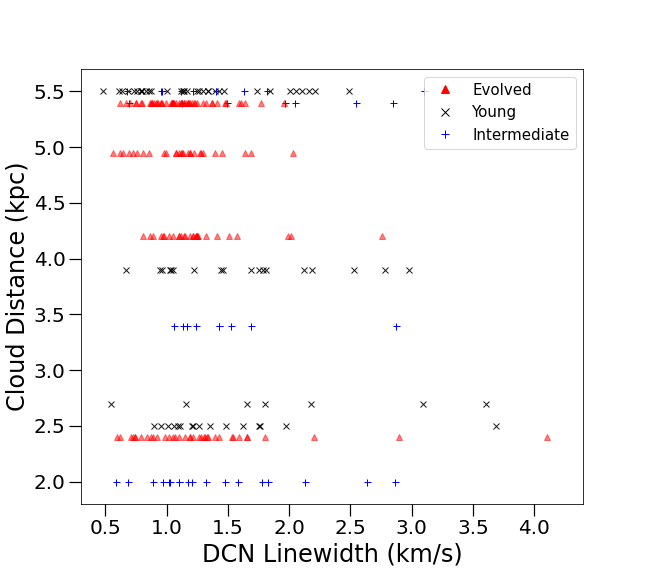} 
    \includegraphics[width=0.49\textwidth]{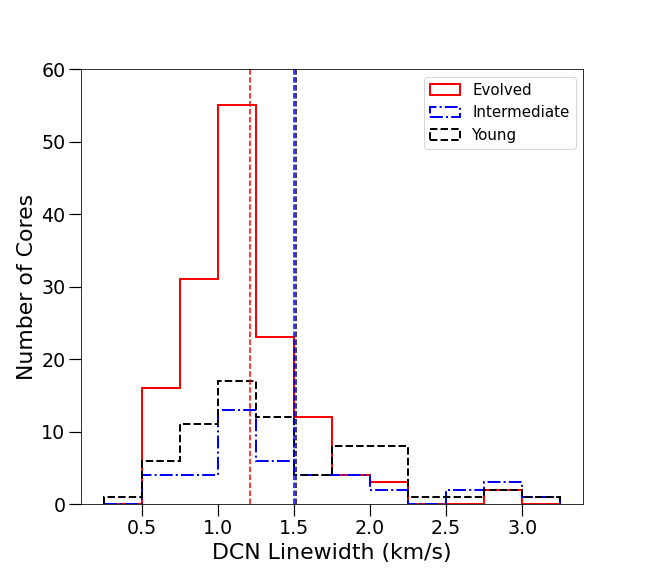} 
    \caption{Distribution of the core linewidths from DCN Single-type line fits as a function of protocluster distance (left). The red triangles, blue plusses, and black crosses represent the linewidths of core emission associated with evolved, intermediate, and young protoclusters, respectively. There is no obvious dependence on the protocluster distance and the linewidths. Right: Histogram of the linewidth of the DCN line fits grouped by evolutionary stage, where the solid red, blue dot-dashed, and black dashed lines represent the evolved, intermediate, and young protoclusters, respectively. The vertical lines display the mean linewidth for each evolutionary stage with the same colours described above. The evolved regions have a slightly smaller average linewidth (1.2\kms) than young and intermediate regions ($\sim$1.5\kms), significant within the standard errors.}
    \label{fig:distancefwhm}
\end{figure*}

\begin{figure}
    \centering
    \includegraphics[width=0.49\textwidth]{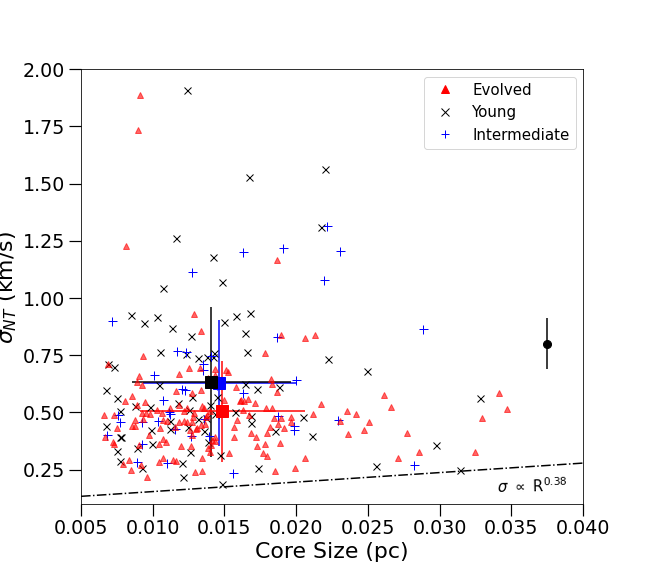} 
    \caption{Distribution of the core-size as a function of the respective non-thermal velocity dispersion from the DCN (3-2) spectra. The red triangles, blue plusses, and black crosses represent evolved, intermediate, and young protoclusters, respectively. The red, blue, and black squares are the average values (0.50\kms, 0.63\kms, and 0.63\kms) for the collective evolved, intermediate and young values, respectively, where the error bars represent the standard deviations (0.22\kms, 0.27\kms, and 0.33\kms, respectively). The black dashed line is the Larson relation \citep{Larson1981}, and nearly all cores are above it. The black circle and error bar provide an estimate of the typical errors on the linewidth estimates. We find no correlation between the non-thermal velocity dispersion with core-size in the ranges of values we sample here.}
    \label{fig:coresize_dispersion}
\end{figure}

\subsection{Global DCN (3-2) morphology and kinematics}
The DCN (3-2) emission displays a diversity of morphology and velocity structure across the ALMA-IMF fields. For the evolved regions, the emission traces more elongated, filamentary structures (e.g. W51-IRS2, G333.60, and G012.80) compared with more compact and less elongated emission for the young protoclusters (e.g. G328.25, and G338.93). 
The difference may be caused by a time evolution of density and temperature, where the evolved regions have had more time to concentrate gas, reaching temperatures and densities sufficient to produce DCN (3-2) emission, perhaps due to the development of expanding HII regions in the evolved protoclusters, which adds heating and compression and can shape their surrounding cloud (such as in G5.89-0.39, \citealt{Fernandez2021}).  
The DCN (3-2) moment 1 maps display a broad range of velocity dispersion across the protoclusters, typically around 12~\kms and up to 16~\kms in W51-IRS2. We use here the spread of core velocities and the global DCN (3-2) emission to identify two protoclusters in the sample, W51-IRS2 and W51-E, which likely consist of multiple clouds along the line of sight that may not be spatially coherent structures in position-velocity space (e.g. \citealt{ginsburg15}). 

\subsection{Core distribution and kinematics in individual protoclusters\label{sec:core_dist_individual}}
In \cref{fig:G333G338_coredisp}, we show the positions of the detected cores and their respective core \vlsr and linewidths overlaid on their DCN (3-2) moment 0 contour maps for the two example protoclusters, G338.93 and G333.60. The entire sample is shown in \cref{appendix:dcn_coredisperionplots}. We estimate their core-to-core velocity dispersions (i.e. the standard deviation of the core \vlsr) and find values of $\sim$2~\kms for both the young (G338.93) and evolved (G333.60) regions. In G338.93, however, the cores with DCN (3-2) detections are distributed in two separated $\leq$0.5~pc size hubs with different average bulk velocities, -60.2~\kms and -62.6~\kms for the north and south hub, respectively, compared to the global average of -61~\kms. If we consider these core populations separately, the estimated core-to-core velocity dispersion is smaller $\sim$1.5~\kms. In G333.60, the distribution of the detected cores is more widespread, over $\sim$3~pc. While it is less obvious to split the cores based on the DCN (3-2) emission, if we split the cores into three subgroups, the central $\sim$0.5pc and the two filaments in the north-east and south-west, we obtain core-to-core velocities of $\sim$1.2\kms, $\sim$1.9\kms, $\sim$1.0\kms, respectively. We find no striking difference between the two regions, however, depending on how the cores are grouped, the core-to-core velocities can change by up to a factor of 2. The core-to-core velocity dispersions for the sub-groups in G338.93 and G333.60 are still larger than nearby star-forming regions (e.g. $<$0.6\,\kms \citealt{Kirk2010}), but they are in line with those found by \citet{Cheng2020} in the massive star-forming region, G286, on similar $\sim$1~pc scales.

Furthermore, if we consider the DCN (3-2) linewidths of cores situated within the two $<$0.5~pc size hubs in G338.93, we find an average linewidth with a standard deviation of $\sim$1.6~\kms and $\sim$0.7~\kms, respectively, whereas the cores outside of these main hubs (which are low and intermediate-mass) are not detected. Towards the evolved protocluster G333.60, around the central $\sim$0.5~pc close to the positions of the known HII region (e.g. \citealt{Fujiyoshi2006}), the linewidths are similar to G338.93 (average and standard deviation of $\sim$1.5~\kms and $\sim$0.5~\kms, respectively) or associated with Complex-type spectra (green points). In contrast, cores with a DCN (3-2) detection outside of the central region are typically lower-mass and situated along the filamentary structures and have smaller linewidths (with an average and standard deviation of $\sim$1.2~\kms and $\sim$0.3~\kms, respectively).  

\subsection{Core kinematics and distribution with evolutionary classification}

\subsubsection{Core-to-core velocity distribution}

In \cref{fig:pv_ra_corevel}, we plot the core-to-core velocity dispersion vs the spatial spread (following, e.g. \citealt{Stutz2016}) for each core grouped by their host region's evolutionary stage. The radial spatial offset for each core is determined by taking the position of the core and subtracting from it the mean position of all cores in its protocluster and assuming the distances given in \cref{tab:sample} for each region. 
The velocity offset for each core is determined by subtracting the centroid \vlsr found from all cores with Single-type spectra in a given region from it. We find core-to-core velocity dispersions of 1.8$\pm$0.2~\kms, 2.6$\pm$0.4\kms, and 2.2$\pm$0.2\kms for the young, intermediate, and evolved protoclusters, respectively. This suggests that the young protoclusters have a slightly smaller core-to-core velocity dispersion compared with the intermediate and evolved protoclusters; however, a Kolmogorov-Smirnov (KS) test between the core-to-core velocities of the evolutionary stages does not give $P_{values}$<0.05. Furthermore, as discussed in \cref{sec:core_dist_individual} for G338.93, and G333.60, the core-to-core velocities can change by a factor of 2 depending on how cores are separated. Thus, we would need to separate the cores spatially and by velocity in each region to assess the core-to-core velocity and spatial distribution fully. 
In \cref{fig:pv_ra_corevel}, we also find that the most massive cores (red stars) are not necessarily at the central velocity or average central position of the protocluster, although we have a small number of statistics for the most massive cores. However, this also depends on how the cores are sub-clustered in a given protocluster. A complete analysis of the core-to-envelope dispersion will be done in future works, along with the separation of cores in individual protoclusters to estimate the core-to-core velocities more accurately.

\cref{fig:pv_ra_corevel} highlights again that the cores in evolved protoclusters have a more widespread spatial distribution than those in the young and intermediate regions. The standard deviation of the spatial spread for the young, intermediate, and evolved regions are 0.33~pc, 0.39~pc, and 0.60~pc, respectively, with standard errors $<$0.06~pc. A KS test on the core offsets in the young and intermediate regions suggests they are drawn from a different population than cores in evolved regions. Thus, the cores detected in DCN (3-2) are typically more concentrated in the young and intermediate regions. This is also the case if we consider the full core populations, not just those with DCN (3-2) detections. This suggests that the spatial distribution is inherent to the evolutionary stage and not biased by the cores with DCN (3-2) detections. 
Furthermore, we find that cores with large spatial offsets in evolved regions (i.e. $>$0.5~pc) are predominantly lower mass. As discussed above, this depends on where the spatial centre of the protocluster is defined and if the cores are further sub-clustered. However, if we take ten random cores in each protocluster as the central position, the evolved protoclusters always have larger standard deviations in their spatial spread, which is significant with a KS-test. Thus, the larger values are likely not due to a bias in the definition of the cluster centre. A full assessment of the sub-clustering or separation of core populations as done in (e.g. \citealt{Pouteau23}) should be performed to confirm this.

\subsubsection{DCN (3-2) linewidths}

Considering the DCN (3-2) linewidths grouped by evolutionary classification, we find slightly smaller average linewidths in the evolved regions of $\sim$1.2~\kms compared to the young and intermediate protoclusters of $\sim$1.5~\kms, and standard deviations of $\sim$0.5, and $\sim$0.7~\kms respectively. In \cref{fig:distancefwhm}, we show the histogram of the DCN (3-2) linewidths extracted from the continuum cores as a function of the evolutionary stage. 
Furthermore, a KS test on the DCN (3-2) linewidths of those cores situated in the evolved regions, when compared with the DCN (3-2) linewidths of cores from intermediate $+$ young regions, gives a $P_{values}$<0.001, suggesting that they are not drawn from the same population. We note, however, that we do not have linewidth estimates with DCN (3-2) for all cores, thus, this trend would need to be confirmed with additional lines.

While evolved protoclusters are typically at a slightly larger distance ($\sim$4.2~kpc) compared with the young and intermediate regions ($\sim$3.7~kpc), as our data are taken at a constant linear resolution, there is no reason to expect differences in the linewidths with protocluster distance, as shown in \cref{fig:distancefwhm}. 
Furthermore, if we consider the size-linewidth relation, we may expect an increase in the DCN linewidths with the core size. In \cref{fig:coresize_dispersion}, we show the non-thermal velocity dispersion-size relation for the 266 cores in our sample from all 15 protoclusters, where the non-thermal velocity dispersion ($\sigma_{NT}$) is estimated assuming:
\begin{equation}
 \label{eq:thermaldisp}
 \sigma_{NT} = \sqrt{\sigma^2 - \frac{kT_{k}}{\mu m_H}},
 \nonumber
\end{equation}
where  $k$ is the Boltzmann constant, T$_{k}$ is the gas kinetic temperature used for DCN (3-2), which is taken to be equal to the dust temperature assumed by \citet{louvet22} and has a range from 18~K to be 300~K for all cores (the assumed temperatures for individual cores are shown in, e.g. \cref{tab:coretables_g338}), $m_H$ is the hydrogen mass, and $\mu$ is the molecular weight of the DCN molecule. The core sizes are the geometric mean of the major and minor continuum core FWHMs from \citet{louvet22} deconvolved by the smoothed beam\footnote{Unless the deconvolved size is less than half of the non-deconvolved size, then the non-deconvolved source size is used.}.
We fit Pearson's correlation coefficient ($\rho$), which measures the linear correlation between two variables, and find a value of 0.18, indicating no correlation exists between the dispersion and core sizes in these data. A similar lack of correlation between the non-thermal velocity dispersion and radius was found by \citet{Traficante2018} towards a sample of Hi-Gal clumps, albeit on the larger size scales of 0.1-1~pc. If we consider the different evolutionary stages independently, there is similarly no correlation for any individual subgroup. However, we note that we are sampling a very small range of core sizes (i.e. a factor of $\sim$3).

\begin{figure}
    \centering
    \includegraphics[width=0.49\textwidth]{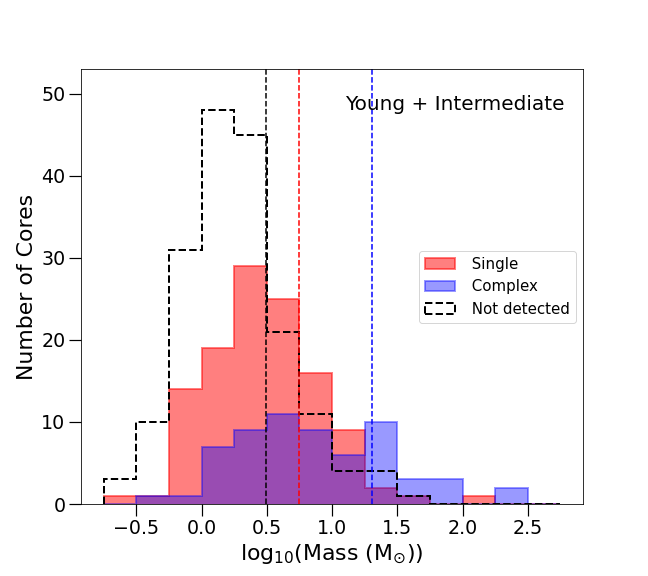} 
    \includegraphics[width=0.49\textwidth]{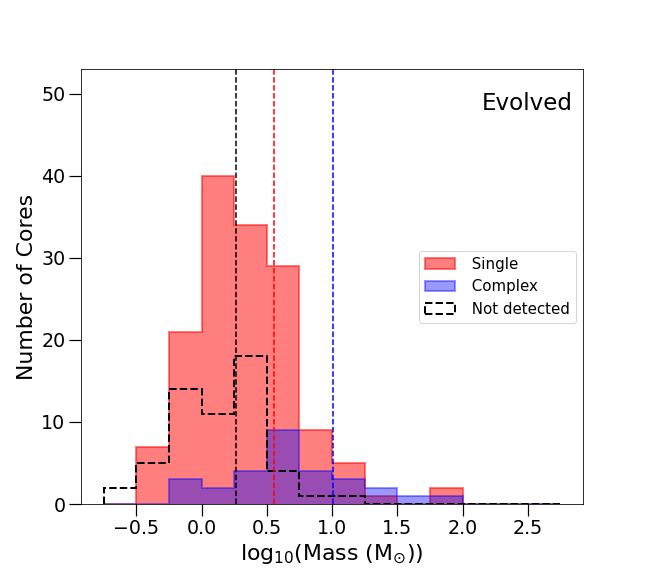} 
    \caption{Histograms of the mass distribution (in log scale) of the cores in the combined young $+$ intermediate regions (top panel) and evolved regions (bottom panel). The shaded red and blue areas show the cores with Single-type DCN (3-2) and Complex-type spectra, respectively. The dashed black line represents cores with a DCN (3-2) non-detection. The red, blue, and black dashed vertical lines represent the average mass values (top: 5.6\,M$_{\odot}$, 20.1\,M$_{\odot}$, and 3.1\,M$_{\odot}$; bottom: 3.6\,M$_{\odot}$, 10.2\,M$_{\odot}$, and 1.8\,M$_{\odot}$) for the Single-, Complex-type, and non-detections, respectively. We find that the Complex-type cores have an average mass nearly an order of magnitude higher than those without a DCN (3-2) detection and around three times higher than cores with a Single-type DCN detection, regardless of evolutionary class.}
    \label{fig:coremass_dispersion}
\end{figure}
\subsection{Nature of the core population traced by DCN (3-2)\label{sec:corenature}}

In \cref{fig:coremass_dispersion}, we show the distribution of core masses taken from \citet{louvet22} (assuming the core dust temperature estimates as given in e.g. \cref{tab:coretables_g338}) separated by their DCN classification (i.e. Single-,  Complex-type and non-detected). We find that the Complex-type spectra are typically extracted in cores with higher masses and densities ($\sim$17\,M$_{\odot}$, and $\sim$2.5$\times$10$^7$cm$^{-3}$, respectively) than those with Single-type DCN (3-2) spectra ($\sim$5\,M$_{\odot}$, and $\sim$1$\times$10$^7$cm$^{-3}$, respectively). Since complexity is associated with higher mass cores, we suggest the Complex-type spectra may be caused by high optical depths, either through self-absorption of the line or absorption of the background continuum. However, the presence of multiple velocity components cannot be ruled out. If we consider the non-detected cores, the average masses and densities are slightly lower, $\sim$3\,M$_{\odot}$, and $\sim$0.8$\times$10$^7$cm$^{-3}$, respectively. Moreover, there is a higher proportion of non-detected cores with masses less than a few M$_{\odot}$ in the young and intermediate regions than the evolved protoclusters. As shown previously, towards the young protocluster, G338.93 (see \cref{fig:linecomparison}), DCN (3-2) is not detected towards cores with masses below $\sim$1~M$_{\odot}$, which could indicate a sensitivity bias for the lowest mass cores.
We note that in G338.93, several of the lowest mass cores that are not detected by DCN (3-2) are, however, detected in other species (i.e. N$_2$D$^+$ (3-2) at $\sim$1M$_{\odot}$).
If we consider recent DCN (3-2) observations by \citet{Hsieh2023} towards SVS13A, a low mass protobinary system harbouring two subsolar protostars at 293pc, where they find peak brightness temperatures of around 15\,K, we would not be sensitive to those cores at the distance of G338.93.  We would expect, however, to detect them towards the ALMA-IMF protoclusters at distances $<$3kpc (e.g. G337.92), suggesting that a sensitivity bias does not account for all of the non-detected low mass cores. Furthermore, if we include all non-detected cores with a mass below 1M$_{\odot}$ into the detected core statistics, the evolved protoclusters would still have $\sim$27\% more DCN (3-2) detections. While the sensitivity limit has an impact on the DCN (3-2) detections and should be explored further, it does not fully explain the higher fraction of cores with a DCN detection in the evolved regions. 

The detection of a deuterated species is typically associated with cold temperatures and high densities (e.g. \citealt{RobertsandMillar2000}); however, DCN can form through both cold and high-temperature reaction channels, with reactions at higher temperatures being the dominant pathway (e.g. \citealt{Turner2001}; \citealt{Roueff2013}). The abundance of DCN is enhanced in warmer gas up to and even above 70K (e.g. \citealt{Millar1989}; \citealt{Turner2001}, \citealt{Roueff2007}; \citealt{Roueff2013}). Thus, unlike other deuterated species such as N$_2$D$^+$, where the abundance is enhanced in the cold ($<$25\,K) dense environments, where CO is frozen out and destroyed in higher temperatures, DCN (3-2) emission also traces warmer gas. The higher number of DCN (3-2) detected cores in evolved protoclusters may be due to temperature, with the evolved regions having higher temperatures over a larger fraction of the cloud than young and intermediate regions. Perhaps due to the presence of HII regions in the evolved protoclusters, which can add heating and compression in the surrounding cloud. We should also note the possibility that DCN enhancement may result from sputtering. In particular, for high-mass protostellar cores with powerful outflows and shocks, DCN emission frozen in the dust grains could be released back to the gas phase by the passage of strong shocks (e.g. \citealt{Busquet2017}). 

Cores with an N$_2$D$^+$ (3-2) detection but without a DCN detection are typically located outside the central hubs in G338.93 and may be in colder, more quiescent parts of the clouds. This has been observed in other works (e.g. \citealt{Tatematsu2020}; \citealt{Cheng2020}, \citealt{Li2022}; \citealt{Sakai2022}) where the spatial distribution between DCN (3-2) and N$_2$D$^+$ (3-2) is observed to be different. \citealt{Sakai2022} found DCN (3-2) traced regions associated with more active signs of ongoing star formation, compared to N$_2$D$^+$ (3-2), which they observed to be tracing the more quiescent, colder regions in their sample of infrared dark clouds. This spatial dichotomy between DCN (3-2) emission and N$_2$D$^+$/DCO$^+$ was similarly found in the young massive star-forming region G286 \citep{Cheng2020} where they identified the DCN emission to be again concentrated around the more active, likely warmer parts of the cloud compared with other deuterated species. To assess the dichotomy and nature of the cores fully across the ALMA-IMF regions requires a multi-line analysis, which will be performed in future works (e.g. \citealt{Cunningham23})

\section{Summary}

We have presented an overview of the calibration and imaging steps performed to produce the full set of ALMA-IMF 12m-only line cubes provided as part of the DR1 release with this paper. We have described the internal ALMA-IMF data pipeline available on the ALMA-IMF GitHub. We used the custom ALMA-IMF pipeline to produce a homogeneous and reproducible set of imaged line cubes for the 12 spectral windows and 6.7~GHz of total spectral coverage, containing a multitude of key molecular tracers that will provide a lasting legacy value to the community.   

In addition, we provide an overview of the DCN (3-2) emission across the ALMA-IMF protoclusters, where the global DCN (3-2) emission displays a diversity of morphology and kinematics across the sample, tracing more widespread, filamentary structures in the evolved regions, compared to young protoclusters where the DCN emission is found in more compact hub-type structures. We also extracted the DCN spectra from the 595 continuum cores, where DCN (3-2) emission is detected in $\sim$60\% of the continuum cores, of which $\sim$45\% have a Single-type DCN spectrum. The evolved protoclusters have a higher percentage of DCN-detected cores, suggesting that the DCN (3-2) transition is likely a more complete tracer of the core population in evolved protoclusters. Furthermore, the cores with DCN (3-2) detections in evolved regions tend to be more spatially extended over the region than in the younger protoclusters. This may be due to the time evolution of the gas temperature and density, where the evolved regions are warmer over a larger portion of the cloud. We provide first estimates of the core linewidths and core \vlsr from the DCN (3-2) fits and find an average core-to-core velocity and linewidth for cores with Single-type DCN spectra of $\sim$2.2~\kms and $\sim$1.3~\kms, respectively, across the 15 protoclusters. We find no obvious difference between the average core-to-core velocities and only a tentative difference in linewidths with the protocluster evolutionary stage. This may, however, be biased by the fact that we do not have estimates for all of the cores.


\bibliographystyle{aa}
\bibliography{almaimf_new}

\begin{acknowledgement}
We are grateful to the anonymous referee for their comments which improved this manuscript. This paper makes use of the following ALMA data: ADS/JAO.ALMA 2017.1.01355.L, 2013.1.01365.S, and 2015.1.01273.S. ALMA is a partnership of ESO (representing its member states), NSF (USA) and NINS (Japan), together with NRC (Canada), MOST and ASIAA (Taiwan), and KASI (Republic of Korea), in cooperation with the Republic of Chile. The Joint ALMA Observatory is operated by ESO, AUI/NRAO and NAOJ. 
This project has received funding from the European Research Council (ERC) via the ERC Synergy Grant ECOGAL (grant 855130), from the French Agence Nationale de la Recherche (ANR) through the project COSMHIC (ANR-20- CE31-0009).
The project leading to this publication has received support from ORP, which is funded by the European Union’s Horizon 2020 research and innovation programme under grant agreement No 101004719 [ORP].
A.G. acknowledges support from the National Science Foundation under grants AST-2008101 and CAREER 2142300.
T.Cs. has received financial support from the French State in the framework of the IdEx Universit\'e de Bordeaux Investments for the future Program. 
R.G.-M. acknowledges support from UNAM-PAPIIT project IN108822 and from CONACyT Ciencia de Frontera project ID: 86372.
AS gratefully acknowledges support by the Fondecyt Regular (projectcode 1220610), and ANID BASAL projects ACE210002 and FB210003.
R.A. gratefully acknowledges support from ANID Beca Doctorado Nacional 21200897.
T. B.  acknowledges the support from S. N. Bose National Centre for Basic Sciences under the Department of Science and Technology, Govt. of India. 
M.B. has received financial support from the French State in the framework of the IdEx Universit\'e de Bordeaux Investments for the future Program. 
S.B. acknowledges support by the French Agence Nationale de la Recherche (ANR) through the project \textit{GENESIS} (ANR-16-CE92-0035-01).
ALS and YP acknowledge funding from the European Research Council (ERC) under the European Union’s Horizon 2020 research and innovation programme, for the project 'the Dawn of Organic Chemistry' (DOC), grant agreement No 741002.
GB acknowledges support by the grant PID2020-117710GB-I00 (MCI-AEI-FEDER,UE).
\end{acknowledgement}

\begin{appendix}
\section{Properties of the line cubes and overview of spectral setups}
\label{appendix:spectral_setups}
In \cref{tab:spwsbeams_contprop} the geometric average of the expected continuum beam from the proposal ($\theta_{prop}$) and the obtained beams ($\theta_{cont}$) are given. In \cref{tab:spwsbeams}, we provide the beam major and minor axes, along with the position angle for all observed protoclusters and spectral windows. In \cref{tab:spwsfreq}, we provide the exact range of the frequency setups for each spectral window towards all 15 protoclusters. In \cref{tab:spwsrms}, we provide the rms noise extracted for each spectral window (in units of mJy per dirty beam) as well as the noise expected from the proposed observations, scaled to the same spectral resolution and central frequency. In \cref{appendix:tab:epsilon}, we provide the $\epsilon$ factors used in the JvM correction for each spectral window.
\begin{table}[htbp!]
\centering
\caption{Proposed and actual continuum beams for all spectral windows. \label{tab:spwsbeams_contprop}}
\begin{tabular}{@{\extracolsep{3pt}}lccc@{}}

\hline \noalign {\smallskip}

Protocluster & & B3  & B6   \\ 
cloud name &$\theta_{prop}$ &$\theta_{cont}$  &$\theta_{cont}$  \\
& [\arcsec]&[\arcsec]& [\arcsec] \\
\hline \noalign {\smallskip}

G008.67  & 0.67  & 0.52  & 0.66\\
G010.62  & 0.37   &0.36  & 0.47\\
G012.80  & 0.95  &1.3  & 0.88\\
G327.29  & 0.67 & 0.4   &  0.66\\
G328.25  & 0.67 & 0.51  & 0.54 \\
G337.92  & 0.51 & 0.37  & 0.54\\
G338.93  & 0.51  & 0.42   &0.53 \\
G333.60  & 0.51 & 0.46   &0.55 \\
G351.77  & 0.95 & 1.4   & 0.77\\
G353.41  & 0.95 & 1.2   & 0.79\\
W43-MM1  &0.37  & 0.41  & 0.43\\
W43-MM2  &0.37  & 0.27    &0.46 \\
W43-MM3  &0.37  & 0.35   &0.49\\
W51-E    &0.37  & 0.27   &0.3 \\
W51-IRS2 &0.37  & 0.28  & 0.47\\

\hline \noalign {\smallskip}
\end{tabular}

\end{table}

\begin{table*}[htbp!]
\centering
\caption{Obtained beams for B3 spectral windows (0-3).}
\label{tab:spwsbeams}
 \begin{tabular}{@{\extracolsep{3pt}}llllllllllllll@{}}

\hline \noalign {\smallskip}

Protocluster &  \multicolumn{3}{c}{spw0 }  & \multicolumn{3}{c}{spw1}  & \multicolumn{3}{c}{spw2 } & \multicolumn{3}{c}{spw3}  \\ 
\cline{2-4}\cline{5-7}\cline{8-10}\cline{11-13}
cloud name&  $\theta_{maj}$ & $\theta_{min}$ &PA&  $\theta_{maj}$ & $\theta_{min}$ &PA& $\theta_{maj}$ & $\theta_{min}$ & PA& $\theta_{maj}$ & $\theta_{min}$ &PA   \\
&(\arcsec) &(\arcsec)& &(\arcsec) &(\arcsec)& & (\arcsec) &(\arcsec)& & (\arcsec) &(\arcsec)&   \\
\hline \noalign {\smallskip}

G008.67  &    0.95&0.72&72&0.55&0.43&69&0.50&0.40&65&0.49&0.38&61 \\
G010.62  &    0.53&0.39&102&0.46&0.38&107&0.37&0.28&94&0.37&0.29&107 \\
G012.80  &   2.23&1.83&88&1.85&1.65&80&1.73&1.52&84&1.70&1.43&87 \\
G327.29  &   0.66&0.57&53&0.45&0.39&69&0.41&0.35&69&0.41&0.35&68 \\
G328.25  &   0.97&0.90&71&0.69&0.51&106&0.63&0.48&113&0.61&0.45&112 \\
G337.92  &   0.67&0.60&-72&0.49&0.42&73&0.44&0.39&73&0.43&0.36&75 \\
G338.93  &   0.59&0.54&-62&0.45&0.42&40&0.39&0.35&88&0.38&0.35&84 \\
G333.60  &   0.73&0.70&13&0.50&0.47&45&0.45&0.44&25&0.45&0.42&37 \\
G351.77  &   2.20&1.88&91&1.85&1.65&88&1.69&1.51&89&1.72&1.49&100 \\
G353.41  &   2.12&1.76&77&1.83&1.54&69&1.67&1.50&76&1.62&1.47&78 \\
W43-MM1  &    0.88&0.48&99&0.59&0.37&107&0.54&0.33&107&0.53&0.33&108 \\
W43-MM2  &   0.41&0.32&106&0.34&0.26&106&0.30&0.24&110&0.32&0.24&124 \\
W43-MM3  &  0.58&0.42&97&0.47&0.33&99&0.42&0.29&98&0.41&0.29&99 \\
W51-E  &   0.38&0.35&106&0.36&0.33&108&0.29&0.27&83&0.28&0.26&85 \\
W51-IRS2  &   0.39&0.36&111&0.33&0.30&131&0.29&0.27&130&0.28&0.27&128 \\

\hline \noalign {\smallskip}
\end{tabular}

%
\end{table*}
\begin{table*}[htbp!]
\ContinuedFloat
\centering
\caption{Continued: Obtained beams for B6 spectral windows (0-3).}
 \begin{tabular}{@{\extracolsep{3pt}}llllllllllllll@{}}

\hline \noalign {\smallskip}

Protocluster &  \multicolumn{3}{c}{spw0 }  & \multicolumn{3}{c}{spw1}  & \multicolumn{3}{c}{spw2 } & \multicolumn{3}{c}{spw3}  \\ 
\cline{2-4}\cline{5-7}\cline{8-10}\cline{11-13}
cloud name&  $\theta_{maj}$ & $\theta_{min}$ &PA&  $\theta_{maj}$ & $\theta_{min}$ &PA& $\theta_{maj}$ & $\theta_{min}$ & PA& $\theta_{maj}$ & $\theta_{min}$ &PA   \\
&(\arcsec) &(\arcsec)& &(\arcsec) &(\arcsec)& & (\arcsec) &(\arcsec)& & (\arcsec) &(\arcsec)&   \\
\hline \noalign {\smallskip}
G008.67  &  0.88&0.72&98&0.88&0.72&98&0.88&0.72&96&0.88&0.71&99 \\
G010.62  &  0.65&0.51&106&0.68&0.53&-72&0.64&0.50&106&0.66&0.52&-76 \\
G012.80  &  1.30&0.89&76&1.29&0.88&76&1.29&0.86&76&1.29&0.87&77 \\
G327.29  &  0.82&0.76&121&0.82&0.75&124&0.82&0.74&123&0.82&0.74&125 \\
G328.25  &  0.74&0.59&-13&0.74&0.58&-14&0.75&0.59&-14&0.74&0.58&-14 \\
G337.92  &  0.81&0.66&129&0.80&0.66&129&0.79&0.66&128&0.80&0.66&129 \\
G338.93  &  0.77&0.69&80&0.77&0.68&81&0.74&0.65&81&0.75&0.65&83 \\
G333.60  &  0.75&0.69&-35&0.75&0.68&-36&0.74&0.68&-37&0.74&0.68&-38 \\
G351.77  &  1.08&0.84&88&1.08&0.83&87&1.07&0.82&88&1.08&0.82&88 \\
G353.41  &  1.13&0.83&86&1.13&0.83&85&1.11&0.83&86&1.12&0.83&86 \\
W43-MM1  &  0.66&0.48&-81&0.65&0.47&100&0.67&0.49&103&0.65&0.49&105 \\
W43-MM2  &  0.63&0.52&100&0.62&0.51&-84&0.68&0.54&100&0.64&0.50&96 \\
W43-MM3  &  0.66&0.57&86&0.66&0.57&86&0.66&0.57&86&0.66&0.57&86 \\
W51-E  & 0.46&0.35&29&0.46&0.35&30&0.57&0.49&38&0.46&0.35&30 \\
W51-IRS2  &  0.64&0.57&-18&0.65&0.59&-23&0.65&0.58&-21&0.63&0.56&-24 \\
\hline \noalign {\smallskip}
\end{tabular}

\end{table*}

\begin{table*}
\ContinuedFloat
\centering
\caption{Continued: Obtained beams for B6 spectral windows (4-7).}
 \begin{tabular}{@{\extracolsep{3pt}}llllllllllllll@{}}

\hline \noalign {\smallskip}

Protocluster &  \multicolumn{3}{c}{spw4 }  & \multicolumn{3}{c}{spw5}  & \multicolumn{3}{c}{spw6 } & \multicolumn{3}{c}{spw7}  \\ 
\cline{2-4}\cline{5-7}\cline{8-10}\cline{11-13}
cloud name&  $\theta_{maj}$ & $\theta_{min}$ &PA&  $\theta_{maj}$ & $\theta_{min}$ &PA& $\theta_{maj}$ & $\theta_{min}$ & PA& $\theta_{maj}$ & $\theta_{min}$ &PA   \\
&(\arcsec) &(\arcsec)& &(\arcsec) &(\arcsec)& & (\arcsec) &(\arcsec)& & (\arcsec) &(\arcsec)&   \\
\hline \noalign {\smallskip}
G008.67  &   0.88&0.71&99&0.84&0.68&96&0.83&0.67&96&0.73&0.61&97 \\
G010.62  &   0.67&0.52&112&0.61&0.48&-73&0.61&0.47&106&0.58&0.43&110 \\
G012.80  &   1.29&0.87&77&1.21&0.83&76&1.21&0.82&76&1.13&0.72&76 \\
G327.29  &   0.82&0.74&125&0.77&0.70&128&0.77&0.69&127&0.69&0.62&-35 \\
G328.25  &   0.74&0.58&-14&0.69&0.54&-13&0.69&0.54&-13&0.61&0.45&-13 \\
G337.92  &   0.80&0.66&129&0.75&0.61&127&0.74&0.61&127&0.65&0.51&127 \\
G338.93  &   0.75&0.66&82&0.71&0.63&84&0.71&0.63&83&0.60&0.55&91 \\
G333.60  &   0.74&0.68&-36&0.69&0.62&-34&0.68&0.62&-34&0.57&0.52&-23 \\
G351.77  &   1.07&0.82&88&1.01&0.77&88&1.00&0.77&87&0.91&0.68&87 \\
G353.41  &   1.12&0.82&84&1.06&0.77&84&1.05&0.76&86&0.96&0.67&85 \\
W43-MM1  &   0.69&0.50&99&0.59&0.43&100&0.59&0.43&100&0.50&0.34&103 \\
W43-MM2  &   0.68&0.53&96&0.61&0.51&110&0.61&0.51&110&0.56&0.45&112 \\
W43-MM3  &   0.66&0.57&85&0.62&0.54&86&0.62&0.53&85&0.55&0.46&88 \\
W51-E  &   0.57&0.47&39&0.52&0.41&34&0.42&0.32&30&0.34&0.27&25 \\
W51-IRS2   &  0.63&0.56&-23&0.59&0.52&-23&0.59&0.52&-25&0.54&0.47&-26 \\
\hline \noalign {\smallskip}
\end{tabular}
\end{table*}

\begin{table*}[htbp!]
\centering
\caption{Minimum and maximum frequency ranges [GHz] for the spectral window setups for B3 spws (0-3).}
\label{tab:spwsfreq}

\begin{tabular}{@{\extracolsep{6pt}}lcccccccc@{}}
\hline \noalign {\smallskip}
Protocluster  & \multicolumn{2}{c}{spw0} & \multicolumn{2}{c}{spw1} & \multicolumn{2}{c}{spw2}& \multicolumn{2}{c}{spw3} \\ 
\cline{2-3}\cline{4-5}\cline{6-7}\cline{8-9}
cloud name\tnote{1}  & min & max  & min & max  & min & max  & min & max  \\
\hline \noalign {\smallskip}

G008.67 & 93.103&93.220&91.720&92.656&102.119&103.056&104.519&105.455 \\
G010.62 & 93.116&93.233&91.733&92.669&102.133&103.069&104.533&105.469 \\
G012.80 & 93.103&93.220&91.720&92.657&102.119&103.056&104.519&105.455 \\
G327.29 & 93.129&93.246&91.746&92.682&102.147&103.084&104.547&105.484 \\
G328.25 & 93.128&93.245&91.745&92.682&102.147&103.083&104.547&105.483 \\
G337.92 & 93.127&93.244&91.744&92.681&102.146&103.082&104.546&105.482 \\
G338.93 & 93.134&93.251&91.751&92.688&102.153&103.089&104.554&105.490 \\
G333.60 & 93.130&93.247&91.746&92.683&102.148&103.085&104.548&105.485 \\
G351.77 & 93.116&93.233&91.733&92.669&102.133&103.069&104.533&105.469 \\
G353.41 & 93.120&93.237&91.737&92.674&102.138&103.074&104.538&105.474 \\
W43-MM1 & 93.085&93.202&91.702&92.639&102.099&103.035&104.498&105.434 \\
W43-MM2 & 93.085&93.202&91.702&92.638&102.099&103.035&104.498&105.435 \\
W43-MM3 & 93.085&93.202&91.702&92.639&102.099&103.035&104.498&105.434 \\
W51-E & 93.098&93.215&91.715&92.651&102.113&103.049&104.513&105.449 \\
W51-IRS2 & 93.098&93.215&91.715&92.651&102.113&103.050&104.513&105.449 \\

\hline \noalign {\smallskip}
\end{tabular}

%
\end{table*}
\begin{table*}[htbp!]
\ContinuedFloat
\centering
\caption{Continued: Minimum and maximum frequency ranges [GHz] for the spectral window setups for for B6 spws (0-3).}
\begin{tabular}{@{\extracolsep{6pt}}lcccccccc@{}}
\hline \noalign {\smallskip}
Protocluster  & \multicolumn{2}{c}{spw0} & \multicolumn{2}{c}{spw1} & \multicolumn{2}{c}{spw2}& \multicolumn{2}{c}{spw3} \\ 
\cline{2-3}\cline{4-5}\cline{6-7}\cline{8-9}
cloud name\tnote{1}  & min & max  & min & max  & min & max  & min & max  \\
\hline \noalign {\smallskip}
G008.67 & 216.056&216.290&217.006&217.240&219.859&219.976&218.086&218.320 \\
G010.62 & 216.084&216.318&217.034&217.268&219.888&220.005&218.114&218.348 \\
G012.80 & 216.056&216.290&217.006&217.240&219.860&219.976&218.086&218.320 \\
G327.29 & 216.115&216.350&217.066&217.299&219.920&220.036&218.146&218.380 \\
G328.25 & 216.114&216.348&217.064&217.298&219.918&220.035&218.144&218.378 \\
G337.92 & 216.112&216.346&217.062&217.296&219.916&220.033&218.142&218.376 \\
G338.93 & 216.128&216.362&217.078&217.312&219.932&220.049&218.158&218.392 \\
G333.60 & 216.117&216.351&217.067&217.301&219.921&220.038&218.147&218.381 \\
G351.77 & 216.085&216.319&217.035&217.269&219.889&220.006&218.115&218.349 \\
G353.41 & 216.095&216.329&217.045&217.279&219.899&220.016&218.125&218.360 \\
W43-MM1 & 216.014&216.249&216.964&217.198&219.817&219.934&218.044&218.278 \\
W43-MM2 & 216.013&216.247&216.963&217.197&219.816&219.932&218.042&218.276 \\
W43-MM3 & 216.013&216.247&216.963&217.197&219.816&219.932&218.042&218.277 \\
W51-E & 216.043&216.277&216.993&217.227&219.846&219.963&218.073&218.307 \\
W51-IRS2 & 216.043&216.277&216.993&217.227&219.846&219.963&218.073&218.307 \\
\hline \noalign {\smallskip}
\end{tabular}

\end{table*}

\begin{table*}[htbp!]
\ContinuedFloat
\centering
\caption{Continued: Minimum and maximum frequency ranges [GHz] for the spectral window setups for B6 spws (4-7).}
\begin{tabular}{@{\extracolsep{6pt}}lcccccccc@{}}
\hline \noalign {\smallskip}
Protocluster  & \multicolumn{2}{c}{spw4} & \multicolumn{2}{c}{spw5} & \multicolumn{2}{c}{spw6}& \multicolumn{2}{c}{spw7} \\ 
\cline{2-3}\cline{4-5}\cline{6-7}\cline{8-9}
  
cloud name\tnote{1}  & min & max  & min & max  & min & max  & min & max  \\
\hline \noalign {\smallskip}
G008.67 & 219.474&219.591&230.268&230.735&231.017&231.485&231.480&233.353 \\
G010.62 & 219.503&219.620&230.298&230.765&231.047&231.516&231.508&233.381 \\
G012.80 & 219.474&219.591&230.269&230.735&231.017&231.486&231.480&233.353 \\
G327.29 & 219.534&219.651&230.331&230.798&231.081&231.549&231.541&233.414 \\
G328.25 & 219.533&219.650&230.330&230.797&231.079&231.547&231.540&233.413 \\
G337.92 & 219.531&219.648&230.328&230.794&231.077&231.545&231.538&233.361 \\
G338.93 & 219.547&219.664&230.345&230.811&231.094&231.562&231.554&233.426 \\
G333.60 & 219.536&219.653&230.333&230.800&231.082&231.550&231.543&233.415 \\
G351.77 & 219.504&219.621&230.299&230.766&231.048&231.517&231.510&233.383 \\
G353.41 & 219.514&219.631&230.310&230.777&231.059&231.527&231.521&233.393 \\
W43-MM1 & 219.432&219.549&230.224&230.690&230.973&231.441&232.490&234.362 \\
W43-MM2 & 219.431&219.548&230.222&230.689&230.971&231.439&231.436&233.308 \\
W43-MM3 & 219.431&219.547&230.222&230.688&230.971&231.439&231.435&233.308 \\
W51-E & 219.461&219.578&230.255&230.721&231.004&231.472&231.467&233.339 \\
W51-IRS2 & 219.461&219.578&230.255&230.721&231.003&231.472&231.467&233.339 \\
\hline \noalign {\smallskip}
\end{tabular}
\end{table*}

\begin{table*}[htbp!]
\centering
\caption{Spectral window noise estimates for B3 spws (0-3).}
\label{tab:spwsrms}
\begin{tabular}{@{\extracolsep{4pt}}lllllllllllll@{}}

\hline \noalign {\smallskip}
Protocluster  & \multicolumn{3}{c}{spw0}  & \multicolumn{3}{c}{spw1}  & \multicolumn{3}{c}{spw2}
    & \multicolumn{3}{c}{spw3}  \\ 
\cline{2-4}\cline{5-7}\cline{8-10}\cline{11-13}
cloud name\tnote{1}  & $\sigma_{MAD}$ & $\sigma_{req}$ & $\frac{\sigma_{MAD}}{\sigma_{req}}$  & $\sigma_{MAD}$ & $\sigma_{req}$ & $\frac{\sigma_{MAD}}{\sigma_{req}}$  & $\sigma_{MAD}$ & $\sigma_{req}$ & $\frac{\sigma_{MAD}}{\sigma_{req}}$  & $\sigma_{MAD}$ & $\sigma_{req}$ & $\frac{\sigma_{MAD}}{\sigma_{req}}$\\
& \multicolumn{2}{c}{[mJybeam$^{-1}$]}& & \multicolumn{2}{c}{[mJybeam$^{-1}$]}& &  \multicolumn{2}{c}{[mJybeam$^{-1}$]}& & \multicolumn{2}{c}{[mJybeam$^{-1}$]}\\
\hline \noalign {\smallskip}
G008.67  &  7.39 & 14.84 & 0.50 & 2.78 & 5.47 & 0.50 & 3.28 & 7.25 & 0.55 & 3.59 & 7.59 & 0.60 \\
G010.62  &  3.52 & 5.04 & 0.70 & 1.51 & 1.86 & 0.79 & 1.53 & 2.46 & 0.75 & 1.32 & 2.58 & 0.65 \\
G012.80  &  11.18 & 28.36 & 0.39 & 3.95 & 10.46 & 0.37 & 4.82 & 13.85 & 0.42 & 5.10 & 14.50 & 0.45 \\
G327.29  &  6.23 & 14.92 & 0.42 & 1.26 & 5.50 & 0.22 & 2.86 & 7.29 & 0.48 & 2.88 & 7.63 & 0.48 \\
G328.25  &  8.16 & 14.83 & 0.55 & 5.38 & 5.47 & 0.96 & 3.52 & 7.24 & 0.59 & 3.76 & 7.58 & 0.63 \\
G337.92  &  3.63 & 9.93 & 0.37 & 2.25 & 3.66 & 0.60 & 2.52 & 4.85 & 0.63 & 1.60 & 5.08 & 0.40 \\
G338.93  &  4.67 & 9.89 & 0.47 & 2.15 & 3.64 & 0.58 & 2.47 & 4.83 & 0.62 & 2.66 & 5.05 & 0.67 \\
G333.60  &  4.97 & 9.84 & 0.50 & 1.84 & 3.63 & 0.50 & 2.02 & 4.80 & 0.51 & 2.25 & 5.03 & 0.57 \\
G351.77  &  10.84 & 28.50 & 0.38 & 3.78 & 10.52 & 0.35 & 4.39 & 13.92 & 0.38 & 5.69 & 14.58 & 0.50 \\
G353.41  &  10.75 & 28.46 & 0.38 & 3.91 & 10.50 & 0.36 & 4.45 & 13.90 & 0.39 & 4.95 & 14.55 & 0.43 \\
W43-MM1  &  2.60 & 4.99 & 0.52 & 0.88 & 1.84 & 0.47 & 1.07 & 2.44 & 0.53 & 0.96 & 2.55 & 0.48 \\
W43-MM2  &  2.68 & 4.99 & 0.54 & 1.09 & 1.84 & 0.58 & 1.26 & 2.44 & 0.63 & 1.38 & 2.55 & 0.69 \\
W43-MM3  &  3.14 & 4.99 & 0.63 & 1.15 & 1.84 & 0.61 & 1.29 & 2.44 & 0.64 & 1.42 & 2.55 & 0.71 \\
W51-E  &  2.23 & 4.95 & 0.45 & 0.92 & 1.83 & 0.49 & 1.06 & 2.41 & 0.53 & 1.16 & 2.53 & 0.58 \\
W51-IRS2  &  2.58 & 4.95 & 0.52 & 0.93 & 1.83 & 0.50 & 1.24 & 2.41 & 0.62 & 1.29 & 2.53 & 0.65 \\
\hline \noalign {\smallskip}
\end{tabular}
\begin{tablenotes}
\item $\sigma_{MAD}$ is given in units of mJy per dirty beam as discussed in \cref{sec:jvm}, while $\sigma_{req}$ is the noise expected from the proposed ALMA observations.
\end{tablenotes}
\end{table*}
\begin{table*}[htbp!]
\ContinuedFloat
\centering
\caption{Spectral window noise estimates for B6 spws (0-3).}
\begin{tabular}{@{\extracolsep{4pt}}lllllllllllll@{}}

\hline \noalign {\smallskip}
Protocluster  & \multicolumn{3}{c}{spw0}  & \multicolumn{3}{c}{spw1}  & \multicolumn{3}{c}{spw2}
    & \multicolumn{3}{c}{spw3} \\ 
\cline{2-4}\cline{5-7}\cline{8-10}\cline{11-13}
cloud name\tnote{1}  & $\sigma_{MAD}$ & $\sigma_{req}$ & $\frac{\sigma_{MAD}}{\sigma_{req}}$  & $\sigma_{MAD}$ & $\sigma_{req}$ & $\frac{\sigma_{MAD}}{\sigma_{req}}$  & $\sigma_{MAD}$ & $\sigma_{req}$ & $\frac{\sigma_{MAD}}{\sigma_{req}}$  & $\sigma_{MAD}$ & $\sigma_{req}$ & $\frac{\sigma_{MAD}}{\sigma_{req}}$\\
& \multicolumn{2}{c}{[mJybeam$^{-1}$]}& & \multicolumn{2}{c}{[mJybeam$^{-1}$]}& &  \multicolumn{2}{c}{[mJybeam$^{-1}$]}& & \multicolumn{2}{c}{[mJybeam$^{-1}$]}\\
\hline \noalign {\smallskip}
G008.67  &  12.23 & 29.65 & 0.36 & 10.80 & 19.75 & 0.48 & 14.87 & 20.39 & 0.66 & 11.25 & 30.21 & 0.33 \\
G010.62  &  3.88 & 10.22 & 0.33 & 3.25 & 6.81 & 0.42 & 4.35 & 7.03 & 0.56 & 3.62 & 10.41 & 0.31 \\
G012.80  &  14.75 & 46.99 & 0.27 & 13.37 & 31.29 & 0.38 & 19.28 & 32.31 & 0.54 & 13.78 & 47.87 & 0.26 \\
G327.29  &  13.95 & 32.34 & 0.38 & 12.20 & 21.54 & 0.50 & 14.75 & 22.24 & 0.60 & 12.04 & 32.95 & 0.33 \\
G328.25  &  25.35 & 32.32 & 0.69 & 22.47 & 21.54 & 0.92 & 30.00 & 22.23 & 1.22 & 23.69 & 32.93 & 0.64 \\
G337.92  &  6.52 & 19.90 & 0.29 & 5.08 & 13.26 & 0.34 & 7.47 & 13.68 & 0.49 & 5.75 & 20.27 & 0.25 \\
G338.93  &  6.30 & 19.30 & 0.29 & 5.49 & 12.86 & 0.38 & 7.74 & 13.28 & 0.53 & 6.05 & 19.67 & 0.27 \\
G333.60  &  5.87 & 19.96 & 0.26 & 8.27 & 13.29 & 0.55 & 6.91 & 13.73 & 0.46 & 5.42 & 20.34 & 0.24 \\
G351.77  &  16.63 & 49.16 & 0.30 & 14.38 & 32.75 & 0.39 & 20.08 & 33.81 & 0.54 & 15.80 & 50.04 & 0.28 \\
G353.41  &  17.82 & 49.59 & 0.31 & 16.01 & 33.03 & 0.43 & 22.12 & 34.10 & 0.59 & 15.90 & 50.53 & 0.28 \\
W43-MM1  &  3.64 & 10.29 & 0.31 & 3.26 & 6.85 & 0.42 & 4.86 & 7.07 & 0.62 & 3.32 & 10.48 & 0.28 \\
W43-MM2  &  3.01 & 10.29 & 0.26 & 2.65 & 6.85 & 0.34 & 3.85 & 7.07 & 0.49 & 2.89 & 10.48 & 0.25 \\
W43-MM3  &  3.54 & 10.29 & 0.30 & 3.16 & 6.85 & 0.41 & 4.20 & 7.07 & 0.54 & 3.19 & 10.48 & 0.27 \\
W51-E  &  4.36 & 10.49 & 0.36 & 3.83 & 7.02 & 0.48 & 4.90 & 7.24 & 0.61 & 4.15 & 10.73 & 0.34 \\
W51-IRS2  &  3.52 & 10.53 & 0.29 & 3.19 & 7.02 & 0.40 & 4.35 & 7.24 & 0.54 & 3.25 & 10.73 & 0.27 \\

\hline \noalign {\smallskip}
\end{tabular}
\begin{tablenotes}
\item $\sigma_{MAD}$ is given in units of mJy per dirty beam as discussed in \cref{sec:jvm}, while $\sigma_{req}$ is the noise expected from the proposed ALMA observations.
\end{tablenotes}
\end{table*}

\begin{table*}[htbp!]
\ContinuedFloat
\centering
\caption{Spectral window noise estimates for B6 spws (4-7).}
\begin{tabular}{@{\extracolsep{4pt}}lllllllllllll@{}}

\hline \noalign {\smallskip}
Protocluster  & \multicolumn{3}{c}{spw4}  & \multicolumn{3}{c}{spw5}  & \multicolumn{3}{c}{spw6}
    & \multicolumn{3}{c}{spw7}  \\ 
\cline{2-4}\cline{5-7}\cline{8-10}\cline{11-13}
cloud name\tnote{1}  & $\sigma_{MAD}$ & $\sigma_{req}$ & $\frac{\sigma_{MAD}}{\sigma_{req}}$  & $\sigma_{MAD}$ & $\sigma_{req}$ & $\frac{\sigma_{MAD}}{\sigma_{req}}$  & $\sigma_{MAD}$ & $\sigma_{req}$ & $\frac{\sigma_{MAD}}{\sigma_{req}}$  & $\sigma_{MAD}$ & $\sigma_{req}$ & $\frac{\sigma_{MAD}}{\sigma_{req}}$\\
& \multicolumn{2}{c}{[mJybeam$^{-1}$]}& & \multicolumn{2}{c}{[mJybeam$^{-1}$]}& &  \multicolumn{2}{c}{[mJybeam$^{-1}$]}& & \multicolumn{2}{c}{[mJybeam$^{-1}$]}\\
\hline \noalign {\smallskip}
G008.67  &  15.89 & 30.58 & 0.47 & 7.67 & 12.38 & 0.62 & 10.09 & 24.73 & 0.41 & 6.43 & 11.71 & 0.56 \\
G010.62  &  5.15 & 10.54 & 0.44 & 2.61 & 4.26 & 0.61 & 3.21 & 8.52 & 0.38 & 1.75 & 4.03 & 0.44 \\
G012.80  &  21.63 & 48.46 & 0.40 & 10.76 & 19.61 & 0.55 & 11.90 & 39.19 & 0.30 & 7.01 & 18.56 & 0.38 \\
G327.29  &  16.42 & 33.35 & 0.44 & 8.14 & 13.50 & 0.60 & 10.45 & 26.98 & 0.39 & 6.41 & 12.78 & 0.51 \\
G328.25  &  32.64 & 33.34 & 0.88 & 16.99 & 13.49 & 1.25 & 22.07 & 26.97 & 0.82 & 22.71 & 12.77 & 1.80 \\
G337.92  &  8.48 & 20.52 & 0.37 & 3.67 & 8.31 & 0.44 & 5.05 & 16.60 & 0.30 & 2.80 & 7.86 & 0.36 \\
G338.93  &  8.45 & 19.91 & 0.38 & 3.38 & 8.06 & 0.42 & 5.11 & 16.10 & 0.32 & 2.74 & 7.63 & 0.36 \\
G333.60  &  7.50 & 20.58 & 0.33 & 3.04 & 8.34 & 0.36 & 4.57 & 16.65 & 0.27 & 2.45 & 7.89 & 0.31 \\
G351.77  &  20.76 & 50.70 & 0.37 & 12.15 & 20.53 & 0.59 & 13.36 & 41.01 & 0.33 & 8.49 & 19.43 & 0.44 \\
G353.41  &  22.05 & 51.10 & 0.39 & 12.74 & 20.71 & 0.61 & 13.58 & 41.37 & 0.33 & 8.77 & 19.59 & 0.45 \\
W43-MM1  &  4.69 & 10.61 & 0.40 & 2.43 & 4.29 & 0.56 & 3.05 & 8.58 & 0.35 & 2.09 & 4.06 & 0.52 \\
W43-MM2  &  4.25 & 10.61 & 0.36 & 1.58 & 4.29 & 0.37 & 2.56 & 8.58 & 0.30 & 1.63 & 4.06 & 0.40 \\
W43-MM3  &  4.64 & 10.61 & 0.39 & 2.27 & 4.29 & 0.53 & 2.77 & 8.58 & 0.32 & 1.62 & 4.06 & 0.40 \\
W51-E  &  4.71 & 10.86 & 0.39 & 2.78 & 4.40 & 0.63 & 3.57 & 8.79 & 0.41 & 2.45 & 4.16 & 0.59 \\
W51-IRS2  &  4.41 & 10.86 & 0.37 & 2.05 & 4.40 & 0.46 & 2.90 & 8.79 & 0.33 & 1.55 & 4.16 & 0.38 \\

\hline \noalign {\smallskip}
\end{tabular}
\begin{tablenotes}
\item $\sigma_{MAD}$ is given in units of mJy per dirty beam as discussed in \cref{sec:jvm}, while $\sigma_{req}$ is the noise expected from the proposed ALMA observations.

\end{tablenotes}
\end{table*}

\begin{table*}[htbp!]
\centering
\caption{\label{appendix:tab:epsilon} $\epsilon$ values used for the JvM correction for B3 spws (0-3) and B6 spws (0-7). }
\begin{tabular}{@{\extracolsep{6pt}}lllllllllllll@{}}
 \hline \noalign {\smallskip}
 Protocluster  & \multicolumn{4}{c}{B3 }& \multicolumn{8}{c}{B6} \\ 
 \cline{2-5} \cline{6-13}
cloud name & spw0& spw1  & spw2 & spw3 & spw0  & spw1  & spw2 & spw3 & spw4 & spw5 & spw6 & spw7 \\
\hline \noalign {\smallskip}
G008.67 & 0.37&0.49&0.47&0.47 & 0.67&0.67&0.67&0.67&0.66&0.67&0.67&0.64 \\
G010.62 & 0.32&0.47&0.94&0.46 & 0.61&0.61&0.61&0.61&0.61&0.6&0.6&0.87 \\
G012.80 & 0.66&0.56&0.58&0.59 & 0.84&0.84&0.85&0.84&0.84&0.84&0.84&0.81 \\
G327.29 & 0.23&0.88&0.38&0.93 & 0.7&0.7&0.69&0.7&0.7&0.68&0.69&0.67 \\
G328.25 & 0.32&0.33&0.52&0.52 & 0.6&0.6&0.59&0.6&0.6&0.6&0.6&0.61 \\
G337.92 & 0.4&0.31&0.32&0.48 & 0.59&0.59&0.58&0.59&0.59&0.58&0.58&0.52 \\
G338.93 & 0.31&0.38&0.25&0.25 & 0.58&0.58&0.58&0.57&0.58&0.58&0.57&0.5 \\
G333.60 & 0.31&0.48&0.5&0.46 & 0.54&0.54&0.54&0.54&0.54&0.52&0.52&0.49 \\
G351.77 & 0.67&0.58&0.58&0.6 & 0.79&0.81&0.79&0.74&0.79&0.74&0.76&0.7 \\
G353.41 & 0.67&0.56&0.6&0.58 & 0.78&0.79&0.74&0.79&0.75&0.76&0.79&0.71 \\
W43-MM1 & 0.47&0.95&0.37&0.4 & 0.6&0.6&0.62&0.62&0.62&0.61&0.61&0.61 \\
W43-MM2 & 0.26&0.94&0.97&0.95 & 0.65&0.62&0.66&0.65&0.65&0.64&0.64&0.65 \\
W43-MM3 & 0.32&0.85&1.0&1.0 & 0.59&0.59&0.59&0.58&0.58&0.58&0.57&0.58 \\
W51-E & 0.42&0.26&0.99&0.98 & 0.48&0.48&0.43&0.48&0.47&0.4&0.47&0.48 \\
W51-IRS2 & 0.41&0.96&0.99&0.98 & 0.65&0.65&0.65&0.65&0.65&0.65&0.65&0.98 \\
\hline \noalign {\smallskip}
\end{tabular}
\end{table*}
\section{Known data issues}

Towards several regions that contain bright hot cores, there can be a slight overestimation of the continuum startmodel at the positions of the hot cores, resulting in localised negative features in some of the residuals. This overestimate is likely due to rich line emission (line forests) across the spectrum at those positions. Often the negative residuals are < 10\% of the flux in the restored image, but in some cases with faint (real) continuum and bright line emission, they are larger. We also note that in some instances the sidelobes can remain particularly bright in the final imaged cubes (e.g. G327 in B6).

\section{Special cases}

Several regions required additional special attention in early 2022 because \tclean failed for a variety of reasons related to software configurations.
We were never able to fully determine what the underlying causes of the failures were, but they manifested as \texttt{CASA} like the following:

\noindent 
\texttt{MPICommandServer:: 
command\_request\_handler
\_service:: SynthesisImagerVi2:: CubeMajorCycle::MPIServer-23 (file src/code/synthesis/ImagerObjects/ CubeMajorCycleAlgorithm.cc, line 336)      Exception for chan range [1921, 1921] ---   Error in running Major Cycle: A nasty Visbuffer2 error occurred...wait for CNGI
W51-E\_B3\_fullcube\_12M\_0\_34232805.log: Exception: Error in running Major Cycle: A nasty Visbuffer2 error occurred...wait for CNGI}

or

\noindent
\texttt{SEVERE  tclean::::casa  Task tclean raised an exception of class RuntimeError with the following message: Error in making PSF : One or more  of the cube section failed in de/gridding. Return values for the sections: [1, 1, 1, 1, 1, 1, 1, 0, 0, 1, 1, 1, 1, 1, 1, 1, 1, 1, 1, 1, 1, 1, 1, 1, 1, 1, 1, 1, 1, 1, 1, 1, 1, 1, 1, 1, 1, 1, 1, 1, 1, 1, 1, 1, 1, 1, 1, 1, 1, 1, 1, 1, 1, 1, 1, 1, 1, 1, 1, 1, 1, 1, 1, 1, 1, 1, 1, 1, 1, 1, 1, 1, 1, 1, 1, 1, 1, 1, 1, 1, 1, 1, 1, 1, 1, 1, 1, 1, 1, 1, 1, 1, 1, 0, 1, 1, 1, 1, 1, 1, 1, 1, 1, 1, 1, 1, 1, 1, 1, 1, 1, 1, 1, 1, 1, 1, 1, 1, 1, 1, 1, 1, 1, 1]}
\newline
Two combinations of field and band, W51-E B3 spw0 and G333.60 B6 spw 1, proved particularly difficult to run \texttt{tclean} to completion.
We, therefore, modified the W51-E B3 spw0 imaging parameters to specify the number of channels by hand.
For G333.60, we found that the final model exhibited `zigzag' spectral patterns, in which, at the end of cleaning, adjacent spectral channels oscillated from $-1t$ to $+1t$, where $t$ is the final clean threshold.
These oscillations manifest strongly in the image and spectra and render the data unusable for science, even though technically the clean threshold has been met.
To correct this issue, we re-processed the final \texttt{.model} file by applying a 3-pixel maximum spectral filter, that is, replacing the value at each pixel in the cube with the maximum of the $\pm1$ pixel range surrounding it in the spectral dimension.
This approach has the effect of retaining the spectral peaks, but slightly over-predicting the values to either edge of the peaks.

\section{DCN moment maps \label{appendix:dcn}}
In Figure \ref{fig:appendixmomments} we present the first three moment maps for the remaining 13 protoclusters. G338.93 and G333.60 are presented in the main body of the text in Figure \ref{fig:G333G338_momentmaps}. 
\begin{figure*}
    \centering
    \includegraphics[width=0.98\textwidth]{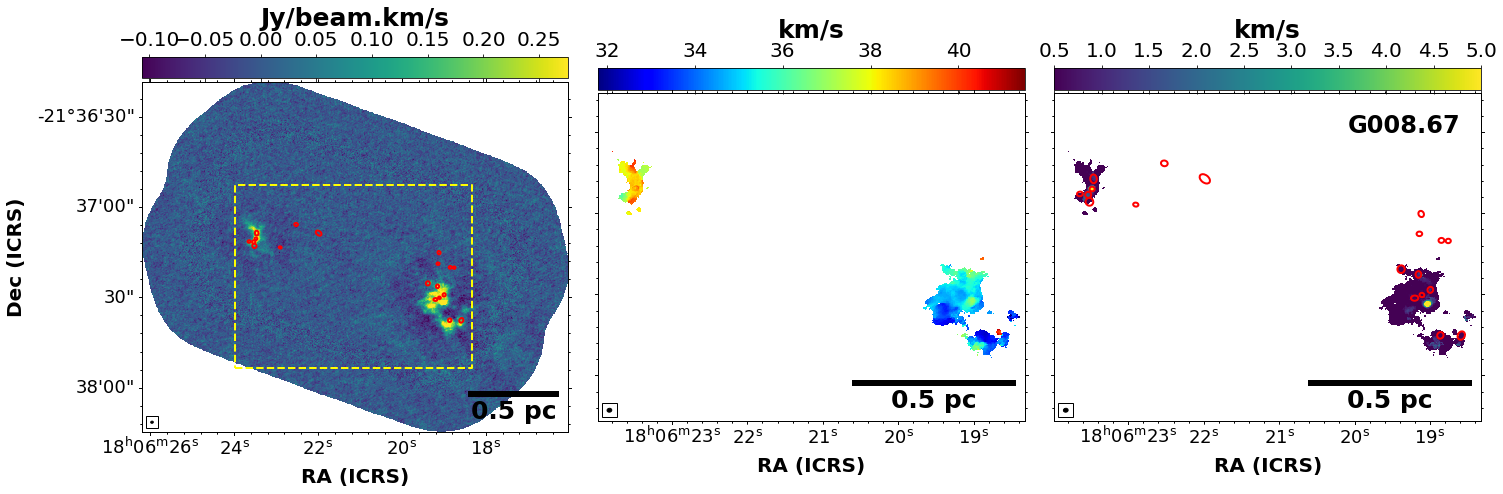}
    \includegraphics[width=0.98\textwidth]{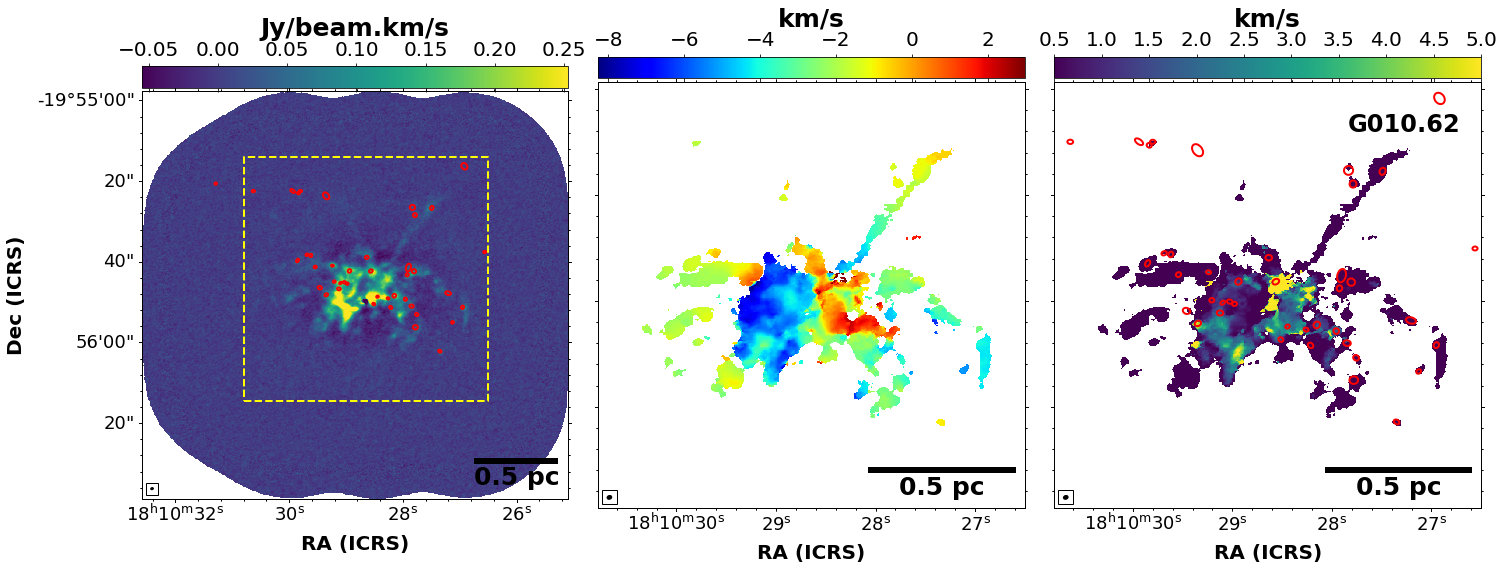}
    \includegraphics[width=0.98\textwidth]{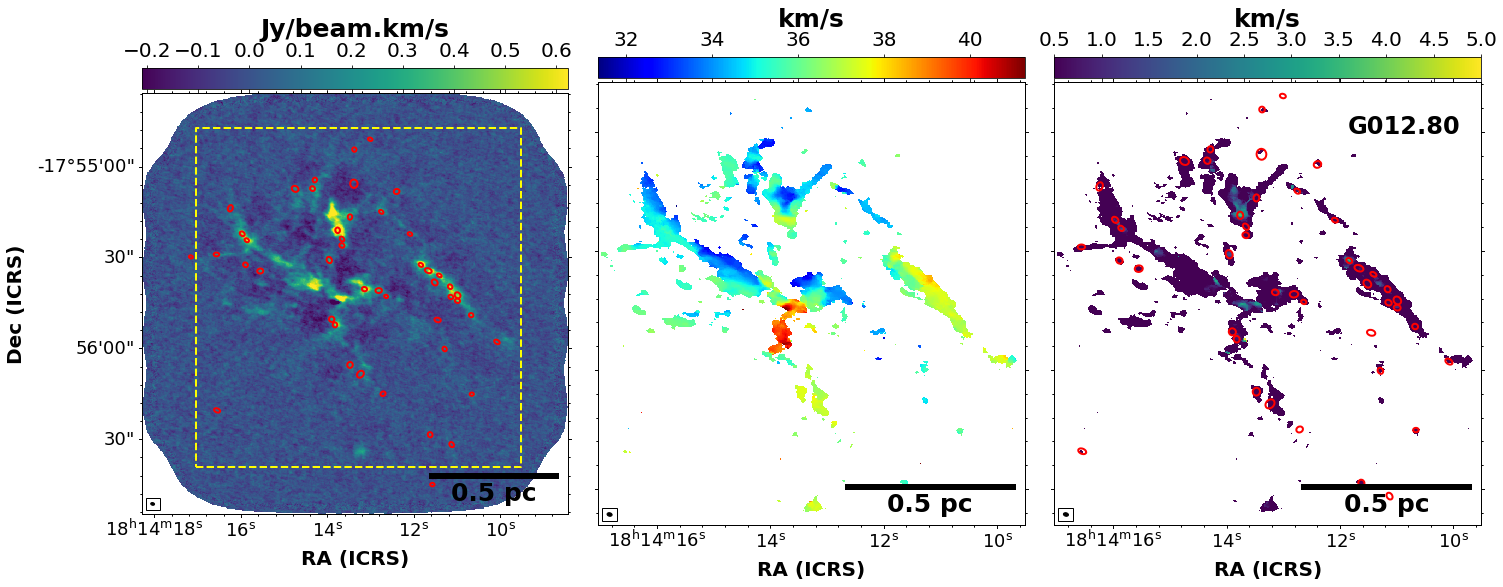}
    \caption{Moment maps (moment 0, 1,  and 2 in the left, centre, and right panels, respectively) of DCN (3-2) emission towards the protoclusters G008.67 (top), G010.62 (middle), and G012.80 (bottom). All three moments have been determined over a velocity range of 31.7 to 43~\kms, $-9.3$ to 5.8~\kms, and 30.7 to 43~\kms for G008.67, G010.62, and G012.80, respectively. For the moment 0 and moment 2 maps, we overlay the core positions and sizes (red ellipses) from the continuum core catalogue described in \cite{louvet22}. For the moment 1 and moment 2 maps, we show a zoom-in of the area highlighted by the yellow dashed box overlaid on the moment 0 map. The moment 1 and moment 2 maps have an additional threshold cut per channel of 4~$\sigma$. The synthesised beam is shown in the bottom left corner of each image.}
    \label{fig:appendixmomments}
\end{figure*}

\begin{figure*}
\ContinuedFloat
    \centering
    \includegraphics[width=0.98\textwidth]{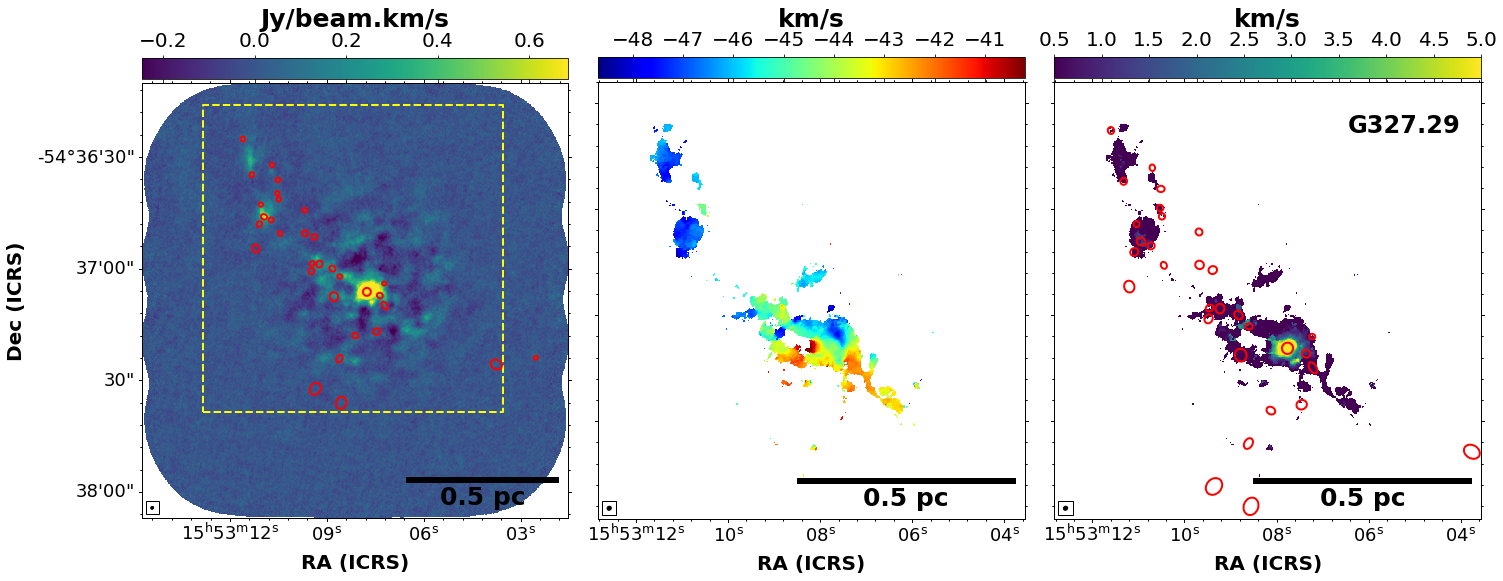}
    \includegraphics[width=0.98\textwidth]{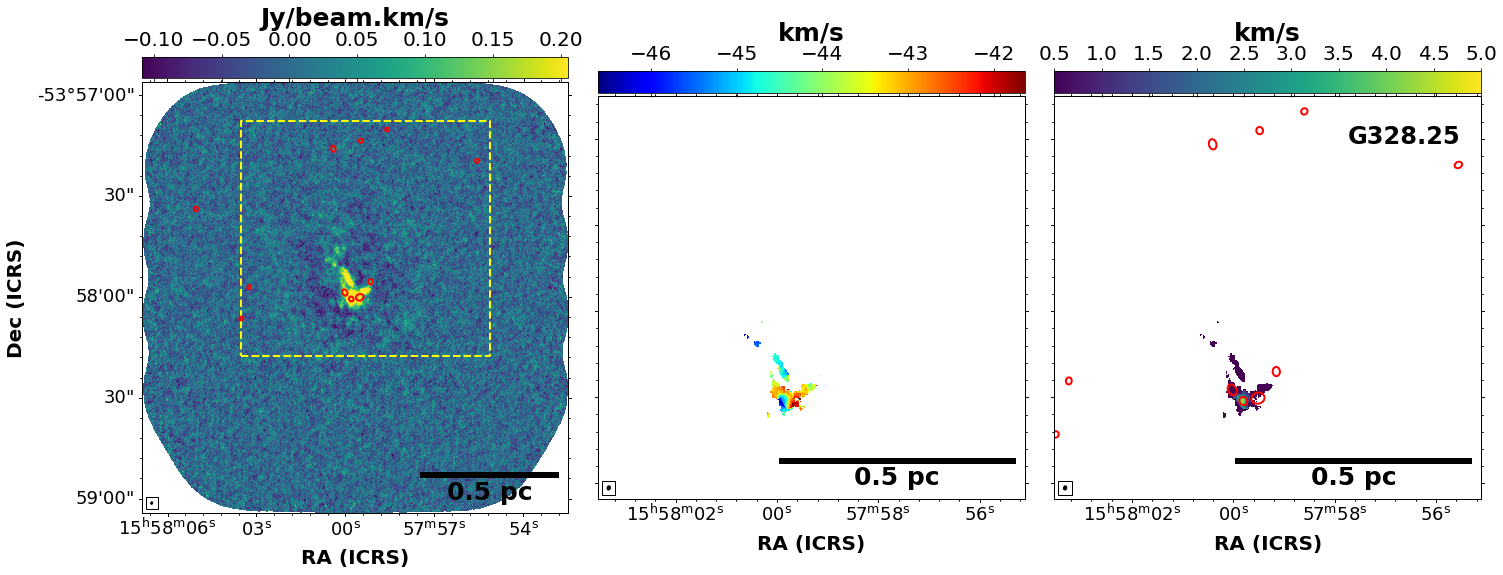}
    \includegraphics[width=0.98\textwidth]{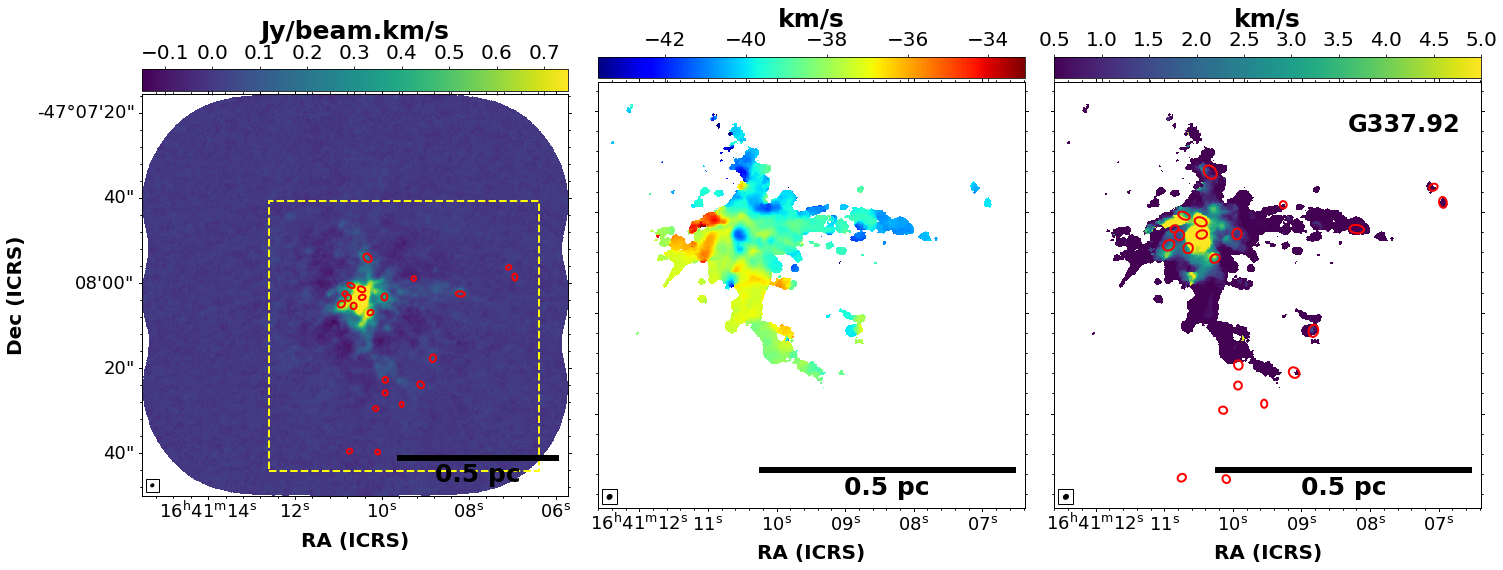}
    \caption{
    Continued. Moment maps (Moment 0, 1, and 2 in the left, centre, and right panels, respectively) of DCN (3-2) for G327.29 (top), G328.25 (middle), and G337.92 (bottom). Maps were calculated over a velocity range of $-50.2$ to $-39.7$~\kms,  $-47.5$ to $-41$~\kms, and $-30.8$ to $-45$~\kms, respectively.
 }
\end{figure*}

\begin{figure*}
\ContinuedFloat
    \centering
    \includegraphics[width=0.98\textwidth]{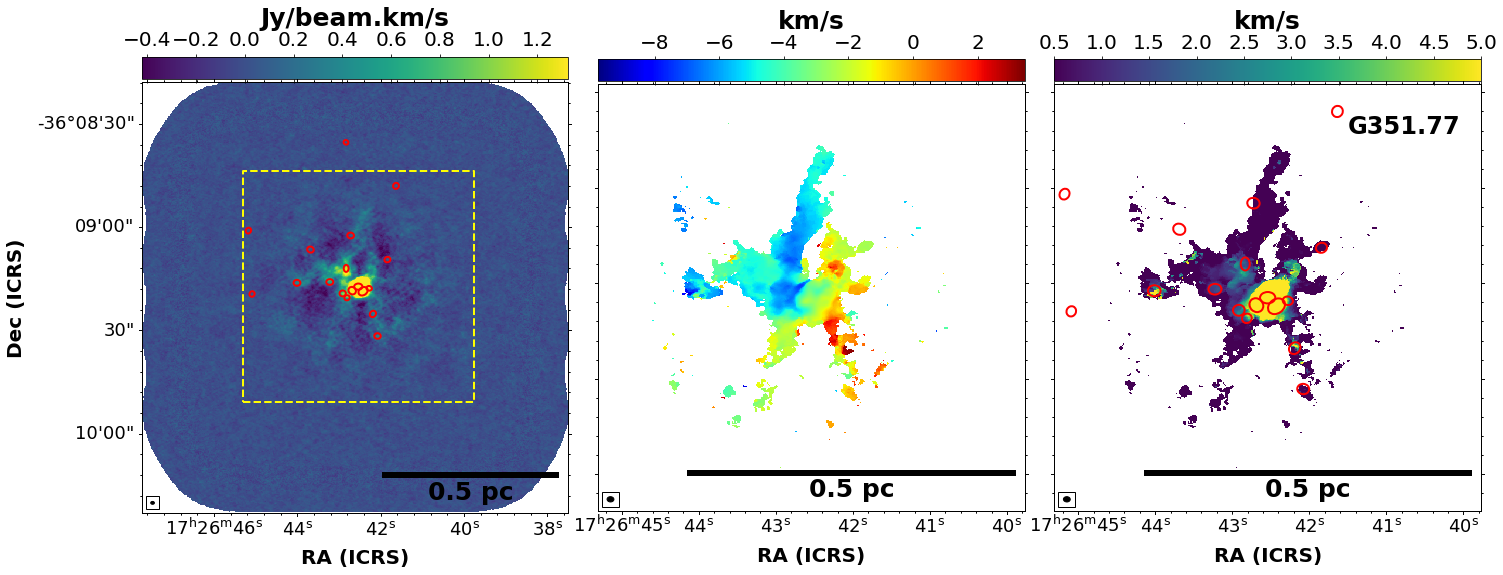}
    \includegraphics[width=0.98\textwidth]{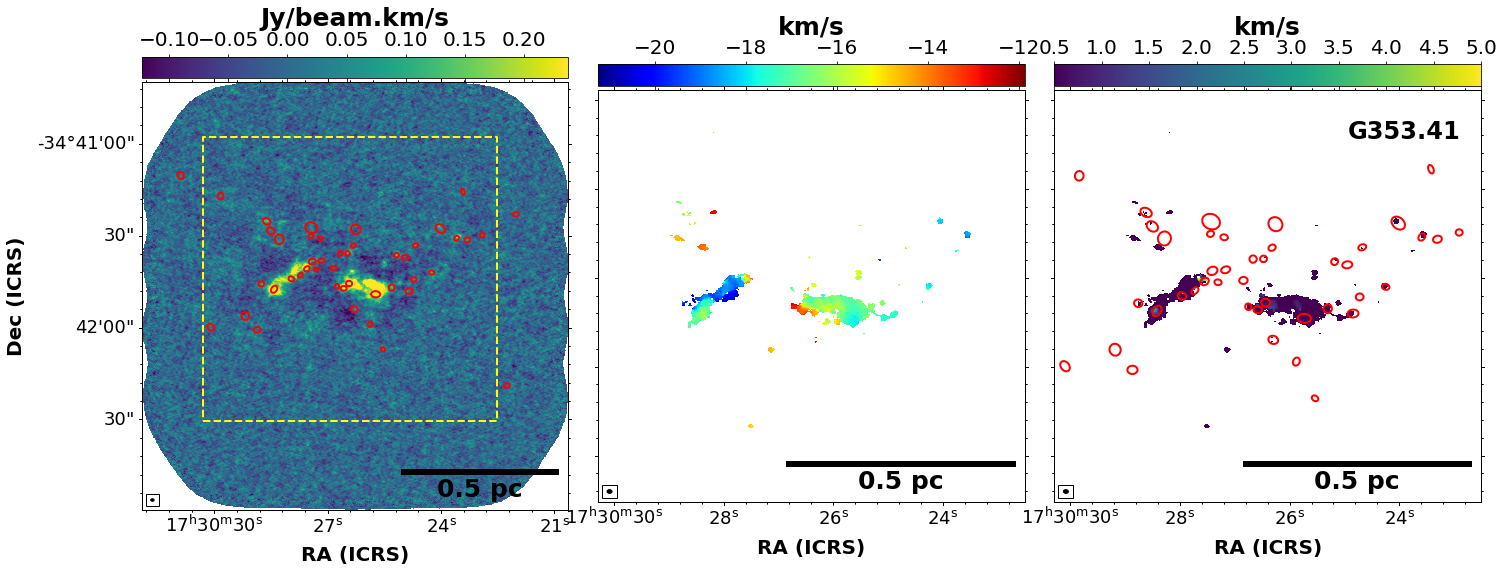}
    \includegraphics[width=0.98\textwidth]{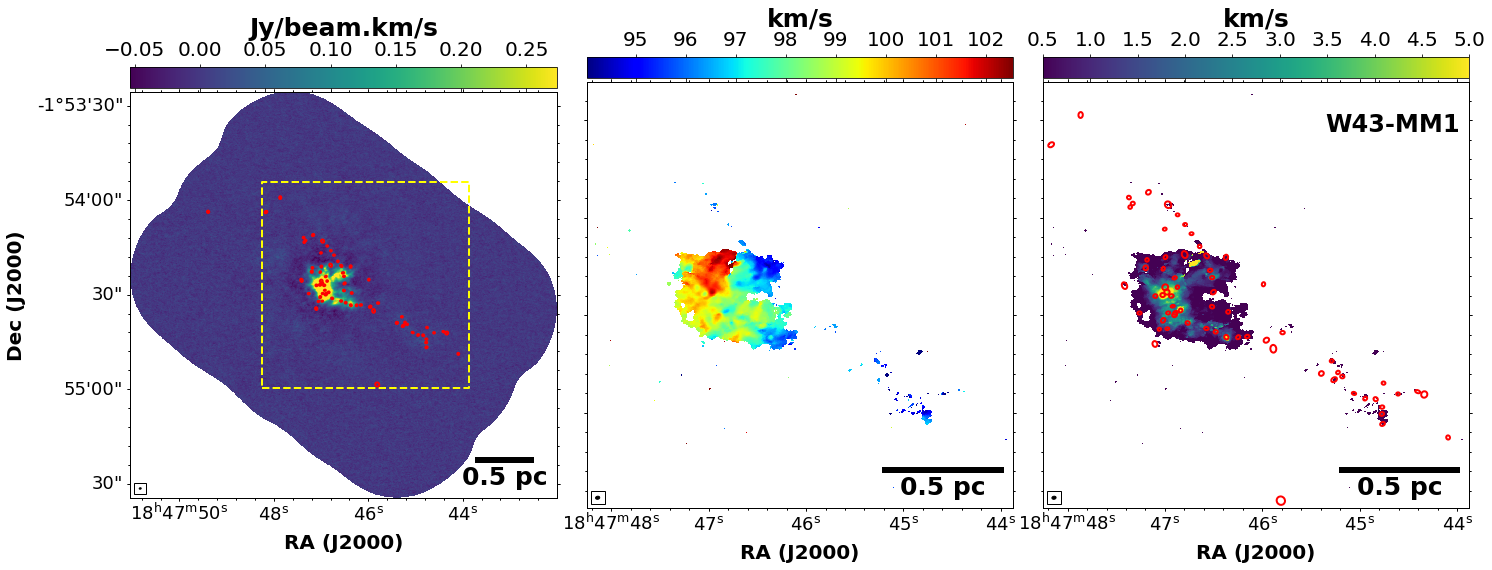}
    \caption{Continued. Moment maps (Moment 0, 1, and 2 in the left, centre, and right panels, respectively) of DCN (3-2) for G351.77 (top), G353.41 (middle), and W43-MM1 (bottom). Maps were calculated over a velocity range of $-9.8$ to  3.2~\kms, $-20.5$ to $-12.7$~\kms, and 92.7 to 105~\kms, respectively. 
 }
\end{figure*}

\begin{figure*}
\ContinuedFloat
    \centering
    \includegraphics[width=0.98\textwidth]{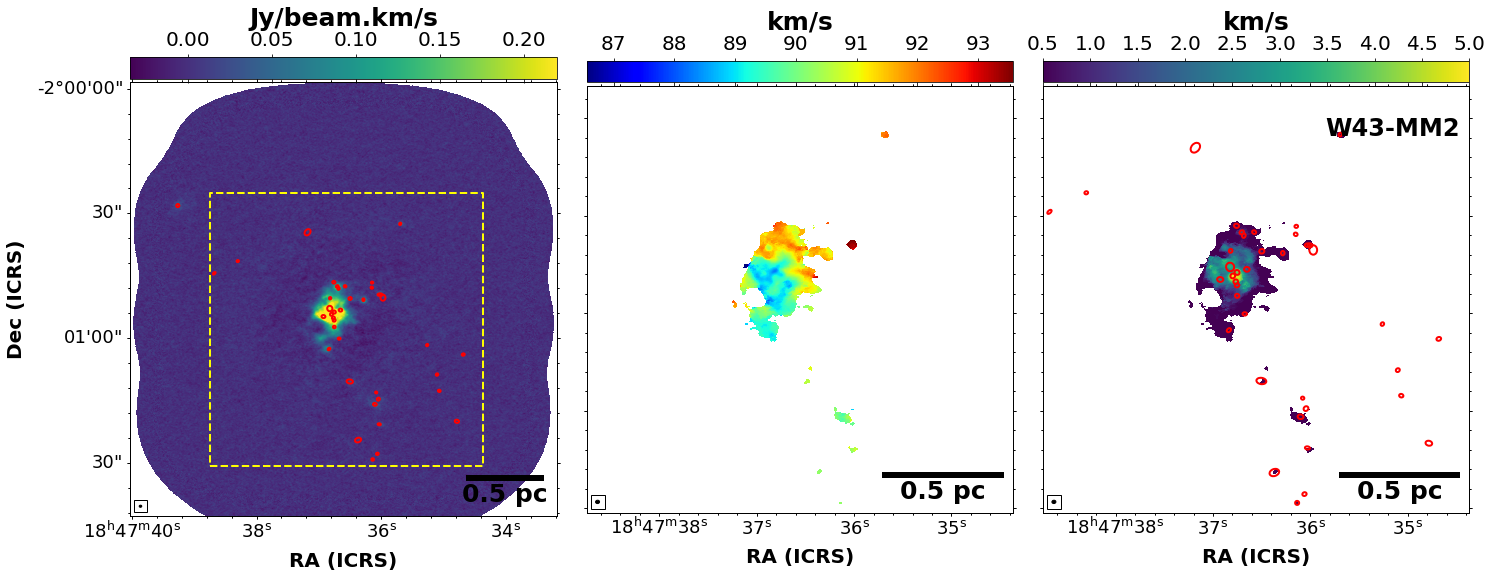}
    \includegraphics[width=0.98\textwidth]{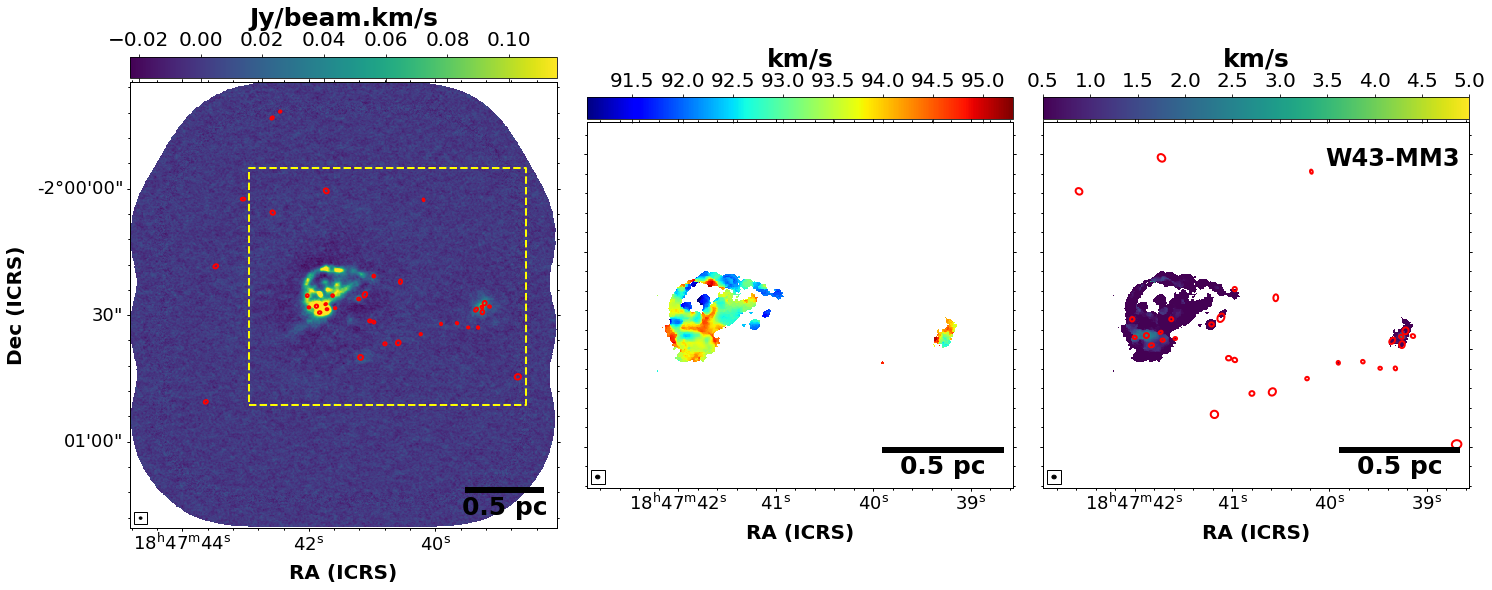}
    \includegraphics[width=0.98\textwidth]{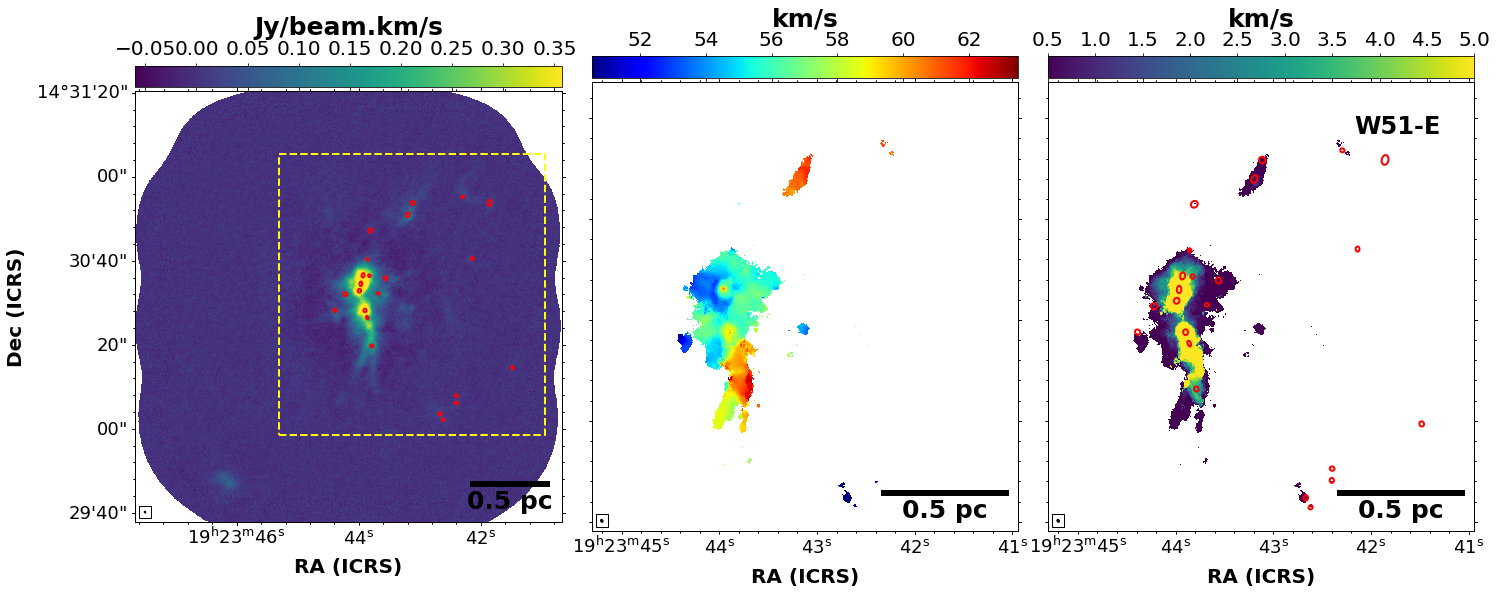}
    \caption{Continued. Moment maps (Moment 0, 1, and 2 in the left, centre, and right panels, respectively) of DCN (3-2) for W43-MM2 (top), W43-MM3 (middle), and W51-E (bottom). Maps were calculated over a velocity range of 85.7 to 94~\kms, 90.4 to 95.8~\kms, and 46 to 64~\kms, respectively. 
 }
\end{figure*}

\begin{figure*}
\ContinuedFloat
    \centering
    \includegraphics[width=0.98\textwidth]{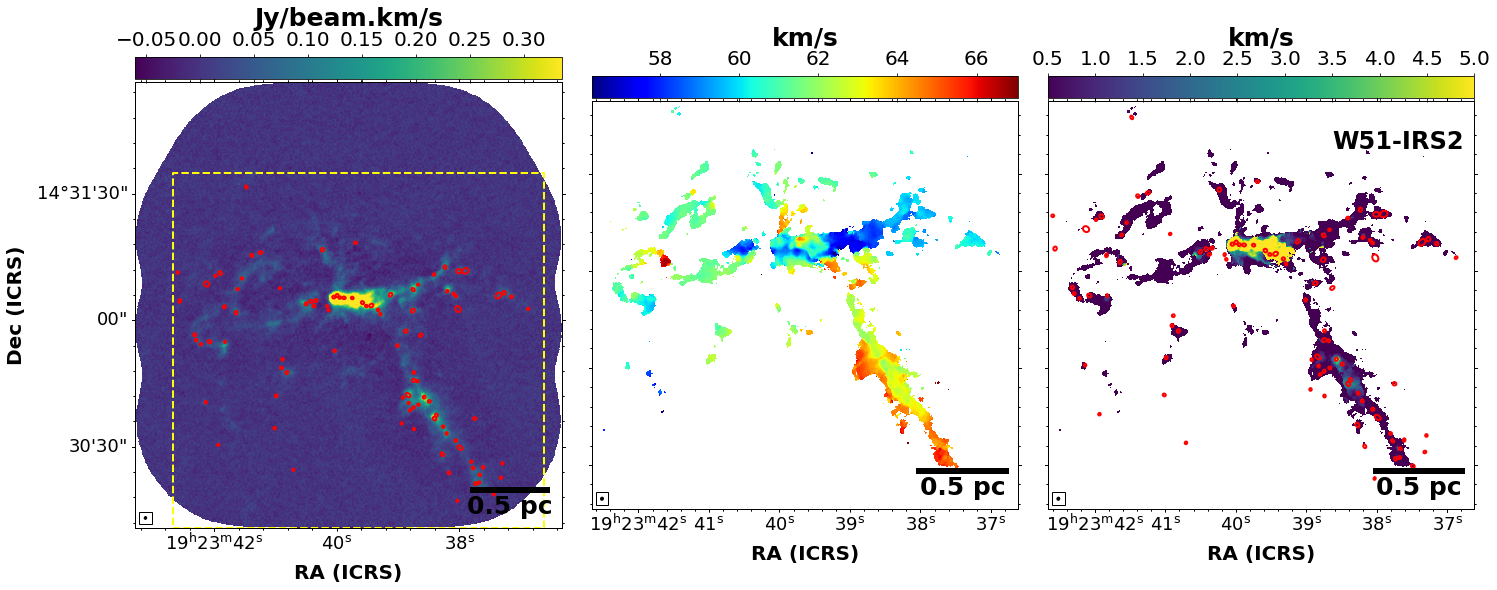}
    \caption{Continued. Moment maps (Moment 0, 1,  and 2 in the left, centre and right panels, respectively) of DCN (3-2) for W51-IRS2. Maps were calculated over a velocity range of 52.4 to 68~\kms. 
     }
\end{figure*}
\section{DCN line extraction, background subtraction, and hyperfine fitting}
\label{appendix:dcn_fitting}

We used \texttt{spectral-cube}\footnote{\url{https://spectral-cube.readthedocs.io/en/latest/}} \citep{Ginsburg2019b} to extract the DCN (3-2) emission from the Band 6 spectral window 1 cube. First, we shifted the cubes into the  frequency rest frame of the DCN (3-2) transition, and then for each protocluster, we extracted a subcube with a 50~\kms width centred on the \vlsr of the protocluster. We then extracted a core-averaged, background-subtracted DCN (3-2) spectrum for each continuum core. We used the smoothed core catalogue from \citet{louvet22} to obtain the positions and core sizes for the 595 continuum cores over the 15 ALMA-IMF protoclusters. We extracted the core-averaged DCN (3-2) spectrum from all pixels within an elliptical aperture with a major (minor) axis length twice that of the continuum source major (minor) FWHM. We subtracted from this an average background spectrum taken from within an elliptical annulus with an inner major (minor) axis size equal to the size of the spectral aperture (i.e. twice the continuum source major (minor) FWHM) and outer radius size 1.5x larger (i.e. three times the continuum source major (minor) FWHM).   
As nearby cores can overlap with the background annulus, we excluded those pixels from neighbouring cores that spatially overlap with the background annulus, creating a background `crown'. An example of the core and background annulus-crown definition can be seen in \cref{appendix:dcn_zoom_masks}.  
The resulting core-averaged, background-subtracted spectrum for each core was then fitted using \texttt{PySpecKit} \citep{Ginsburg2011} with an adapted single component hyperfine model (HFS), to account for the hyperfine structure in the DCN (3-2) line. The values for the specified hyperfine model were taken from the CDMS catalogue to define the frequencies and relative weights of the hyperfine lines \citep{Endres2016,muller2001,muller2005}. They are presented in \cref{tabappendix:hyperfine}. An example of the positions of the hyperfine components overlaid on an example DCN spectrum is presented in \cref{appendix:dcn_hyperfineexample}.
\label{sec:hyperfinemodel}

\subsection{Single- and Complex-type spectral classification}

To classify a spectrum as either detection or non-detection, we require an S/N of at least 4~$\sigma$ between its peak and estimated noise and a velocity dispersion of $> 0.2$~\kms. The noise in the spectrum is estimated from the MAD\footref{note:mad} in 30 continuous channels ($\sim$10\,\kms) from either the lower or upper part of the spectrum (shaded pink regions in the DCN spectra) that were identified by eye to be in a line free part of the spectrum. We then split the detected spectra into Single- or Complex-type. This is done manually by visually inspecting the DCN (3-2) spectra after fitting. Again, we use the higher noise estimates in units of dirty beam. If multiple components are present in the spectrum, we also classify them as Complex-type and  remove them from the analysis. The spectra for each protocluster are shown in \cref{appendix:spectral}. 

\subsection{Comparison of the background and non-background subtracted DCN spectra}

\begin{figure*}
    \centering
\includegraphics[width=0.24\textwidth]{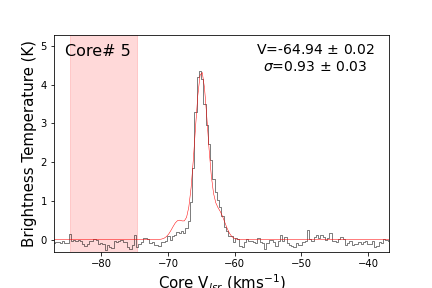}
\includegraphics[width=0.24\textwidth]{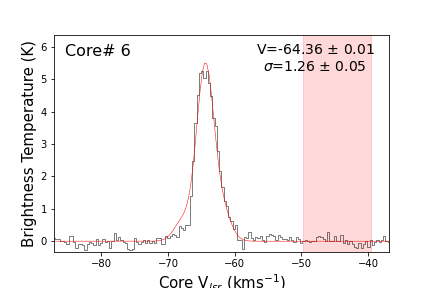}
\includegraphics[width=0.24\textwidth]{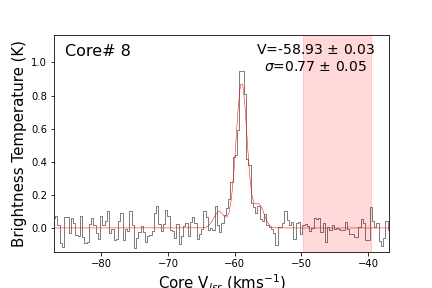}
\includegraphics[width=0.24\textwidth]{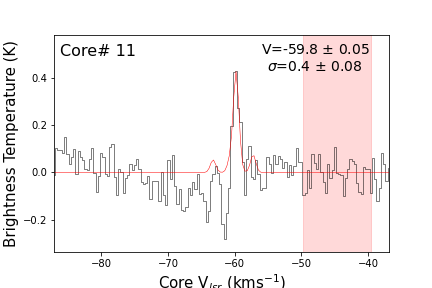}
\includegraphics[width=0.24\textwidth]{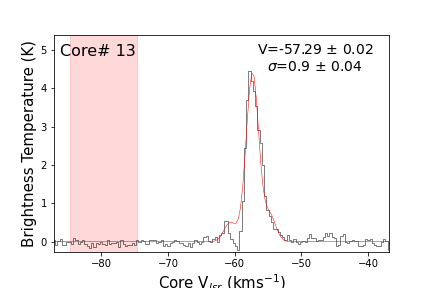}
\includegraphics[width=0.24\textwidth]{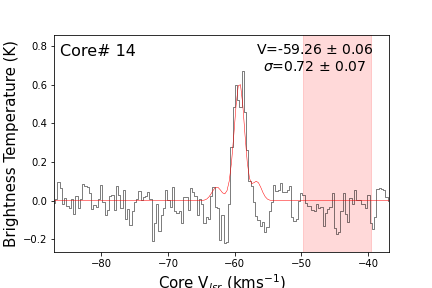}
\includegraphics[width=0.24\textwidth]{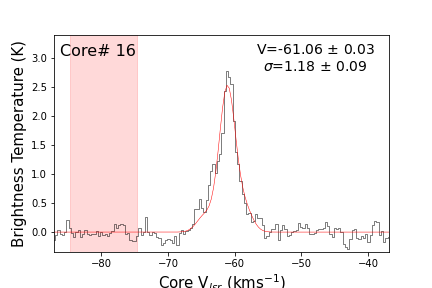}
\includegraphics[width=0.24\textwidth]{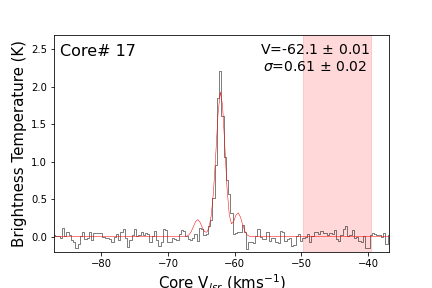}
\includegraphics[width=0.24\textwidth]{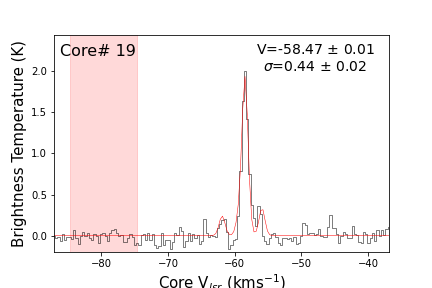}
\includegraphics[width=0.24\textwidth]{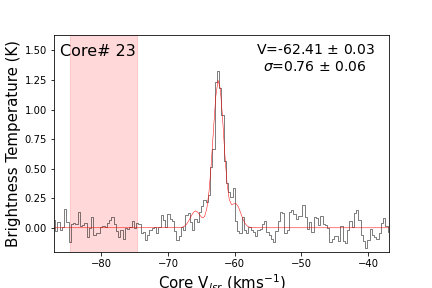}
\includegraphics[width=0.24\textwidth]{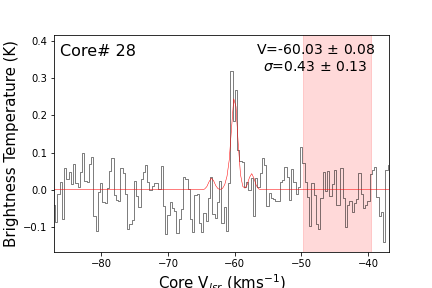}
\includegraphics[width=0.24\textwidth]{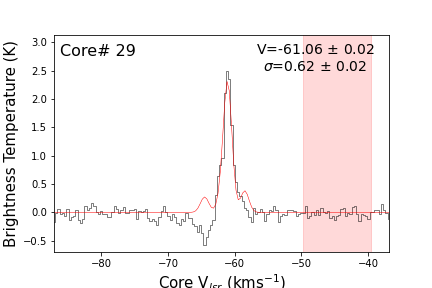}
\includegraphics[width=0.24\textwidth]{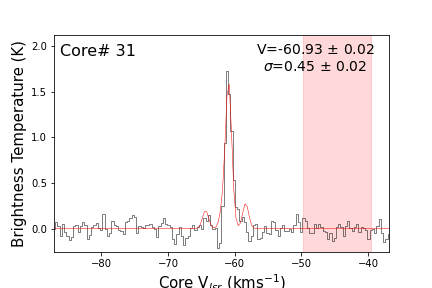}
\includegraphics[width=0.24\textwidth]{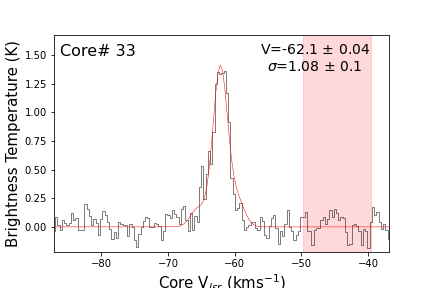}
\includegraphics[width=0.24\textwidth]{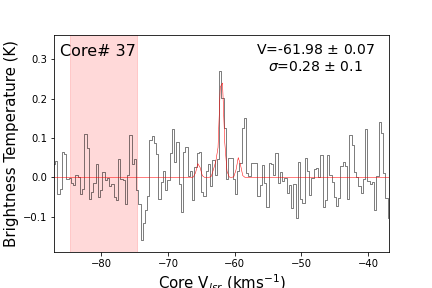}
\includegraphics[width=0.24\textwidth]{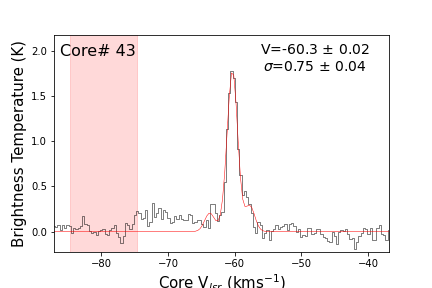}
\includegraphics[width=0.24\textwidth]{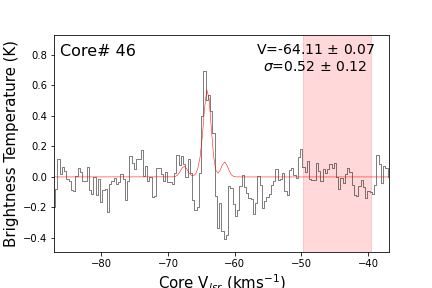}
\includegraphics[width=0.24\textwidth]{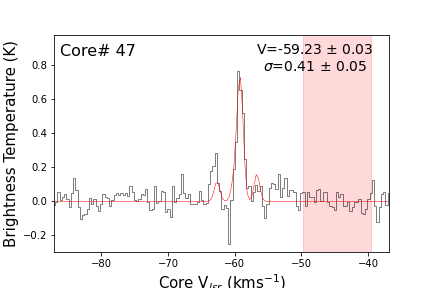}

\caption{Single-type core-averaged, background-subtracted DCN spectra extracted from the cores in the young protocluster G338.93. See \cref{tabappendix:coretables_g338} for the line fit parameters for each core.\label{figspectra:dcnspectra_split_G338_s}}
\end{figure*}

In Figures \ref{figspectra:dcnspectra_split_G338_s}, and \ref{appendicfig:g388_on}, we show the DCN (3-2) detected Single- and Complex-type spectra identified when performing background-subtraction, and non-background subtraction, respectively, for G338.93. In Figure \ref{g338:off}, for reference, we also show the background spectra. Some of the 'absorption features' in the background-subtracted spectra result from the background subtraction (e.g. core 13 and 29). The background subtraction also results in some cores with initially Complex-type spectra due to the background contamination becoming Single-type spectra (e.g. core 16). The background and non-background subtraction detection rates are similar in G338.93, with 18 and 21 detected Single-type cores, respectively (with slight differences between which cores are Single-type in the two sets), and eight cores with Complex-type spectra in both. Furthermore, the average \vlsr extracted from all cores is the same (-61~\kms) when considering either background- or non-background-subtraction. The average linewidth of the fits, however, is smaller ($\sim$1.6~\kms) for the background-subtracted compared with ($\sim$2.1~\kms) for the non-background subtracted, which is due to the removal of the background contribution. These background-subtracted spectra are likely more representative of the gas associated with the cores. In \cref{tabappendix:dcnfits_average}, we show the same table as in \cref{tab:dcnfits_average} but from fitting the non-background subtracted spectra for all protoclusters. This highlights that the general trends across the sample are the same regardless of whether the background or non-background spectra are used.

\begin{table*}[htbp!]
\centering
\begin{threeparttable}[c]
\caption{Characteristic parameters of the DCN (3-2) hyperfine fits for the non-background subtracted.}
\label{tabappendix:dcnfits_average}
 \begin{tabular}{lllllllllllll}
\hline \noalign {\smallskip}
Protocluster         & Number\tnote{1}  & Number\tnote{2}   & Detection\tnote{2} & Velocity\tnote{3} & \vlsr\tnote{3}   & \vlsr\tnote{3}    & Linewidth\tnote{3}  & Linewidth\tnote{3}   &  Linewidth\tnote{3}  \\ 
cloud name & of Cores & detected    & rate [\%]   & range    & mean   & std     & mean  & median &  std \\
&  &(complex)        &  (complex)      & [km~s$^{-1}$]   & [km~s$^{-1}$] & [km~s$^{-1}$]  & [km~s$^{-1}$]& [km~s$^{-1}$] & [km~s$^{-1}$] \\

\hline \noalign {\smallskip}
\multicolumn{10}{c}{Young} \\
G327.29 & 32 & 18 (23)&  56 (72)\% & 5.3 & -44.9 & 1.9  & 1.7 & 1.7 & 0.5 \\
G328.25 & 11 & 3 (4)&  27 (36)\% & 1.4 & -43.1 & 0.6  & 1.9 & 1.6 & 0.9 \\
G338.93 & 42 & 21 (29)&  50 (69)\% & 6.9 & -60.9 & 2.0  & 2.1 & 2.0 & 0.7 \\
G337.92 & 22 & 10 (17)&  45 (77)\% & 4.0 & -39.2 & 1.3  & 2.1 & 2.0 & 1.1 \\
W43-MM1 & 70 & 32 (51)&  46 (73)\% & 7.5 & 97.2 & 2.1  & 1.9 & 1.6 & 1.1 \\
W43-MM2 & 40 & 23 (30)&  58 (75)\% & 4.3 & 90.8 & 1.3  & 1.9 & 1.6 & 0.9 \\
Average &  36 & 18(26) & 49 (71)\% &  4.9 &  & 1.8  & 1.9 & 1.7 & 0.9 \\
Total &  217 & 107 (154)  \\
\hline
\multicolumn{10}{c}{Intermediate} \\
G351.77 & 18 & 5 (14)&  28 (78)\% & 5.9 & -4.5 & 2.1  & 2.0 & 1.6 & 0.6 \\
G353.41 & 45 & 17 (22)&  38 (49)\% & 7.4 & -17.3 & 1.8  & 1.7 & 1.5 & 0.8 \\
G008.67 & 19 & 8 (13)&  42 (68)\% & 6.6 & 35.8 & 2.0  & 2.0 & 1.9 & 0.6 \\
W43-MM3 & 36 & 12 (18)&  33 (50)\% & 2.6 & 93.1 & 0.8  & 1.8 & 2.0 & 0.7 \\
W51-E & 23 & 12 (21)&  52 (91)\% & 11.1 & 56.8 & 3.9  & 2.0 & 1.6 & 1.0 \\
Average &  28 & 11(18) & 38 (62)\% &  6.7 &  & 2.8  & 1.8 & 1.6 & 0.8 \\
Total &  141 & 54 (88)  \\
\hline
\multicolumn{10}{c}{Evolved} \\
G333.60 &  52 & 36 (42) &  69 (81)\% & 10.0 & -47.8 & 1.9  & 1.7 & 1.7 & 0.5 \\
G010.62 &  42 & 28 (37) &  67 (88)\% & 9.3 & -2.5 & 2.1  & 1.4 & 1.3 & 0.6 \\
G012.80 &  46 & 38 (40) &  83 (87)\% & 6.4 & 36.1 & 1.4  & 1.4 & 1.2 & 0.6 \\
W51-IRS2 &  97 & 69 (84) &  71 (87)\% & 18.3 & 61.9 & 2.9  & 1.5 & 1.3 & 0.8 \\
Average &  59 & 43 (51) & 72 (86)\% &  11.0 &  & 2.5  & 1.5 & 1.3 & 0.7 \\
Total &  237 & 171 (203)  \\
\hline \noalign {\smallskip}
\end{tabular}
\begin{tablenotes}
\item[1] The number of continuum cores are taken from the smoothed core catalogue from \citet{louvet22}. 
\item[2] The total count/percentages in the brackets include both Single- and Complex-type DCN (3-2) detections. Where spectra that are described as Complex-type spectra and are not well fit by a single component. We provide the percentages and number of detections, but cores with a Complex-type spectrum are not included in the estimates of the Velocity range, \vlsr, or linewidth.
\item[3] Only Single-type DCN spectra (fit with a single component) and listed as Single in the Tables of the DCN fits (e.g. \cref{tab:coretables_g338}) are included in the estimates of the Velocity ranges, \vlsr and linewidth for each protocluster. 
\end{tablenotes}
\end{threeparttable}
\end{table*}

\begin{figure*}
    \centering
    \includegraphics[width=0.44\textwidth]{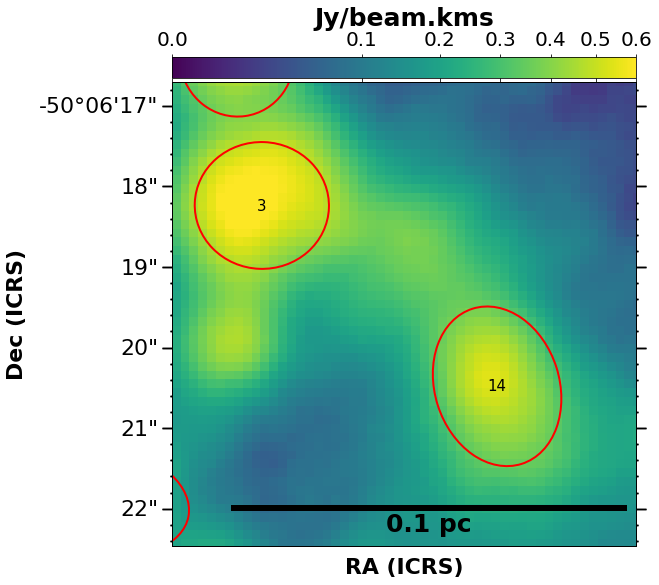}
    \includegraphics[width=0.44\textwidth]{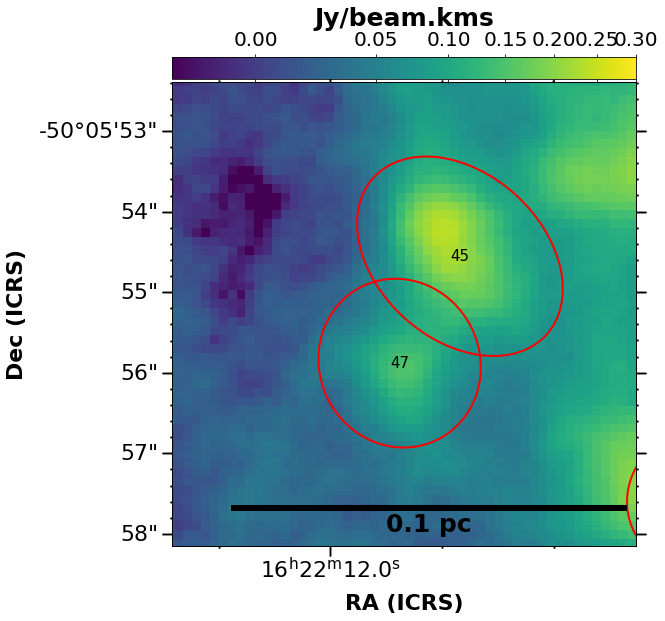}
    \includegraphics[width=0.48\textwidth]{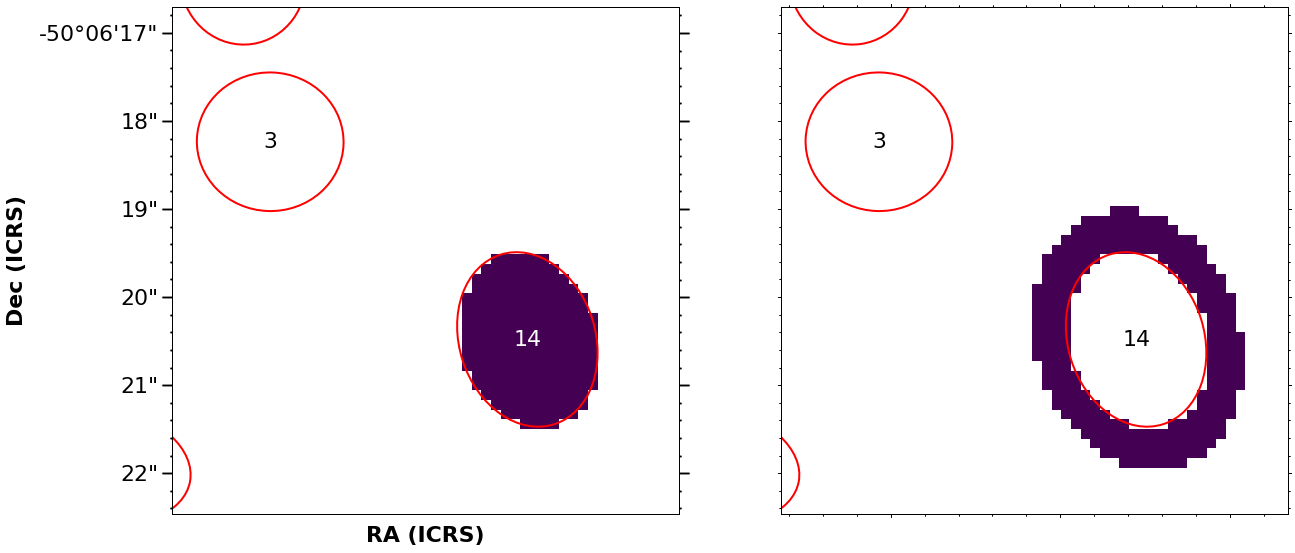}
    \includegraphics[width=0.48\textwidth]{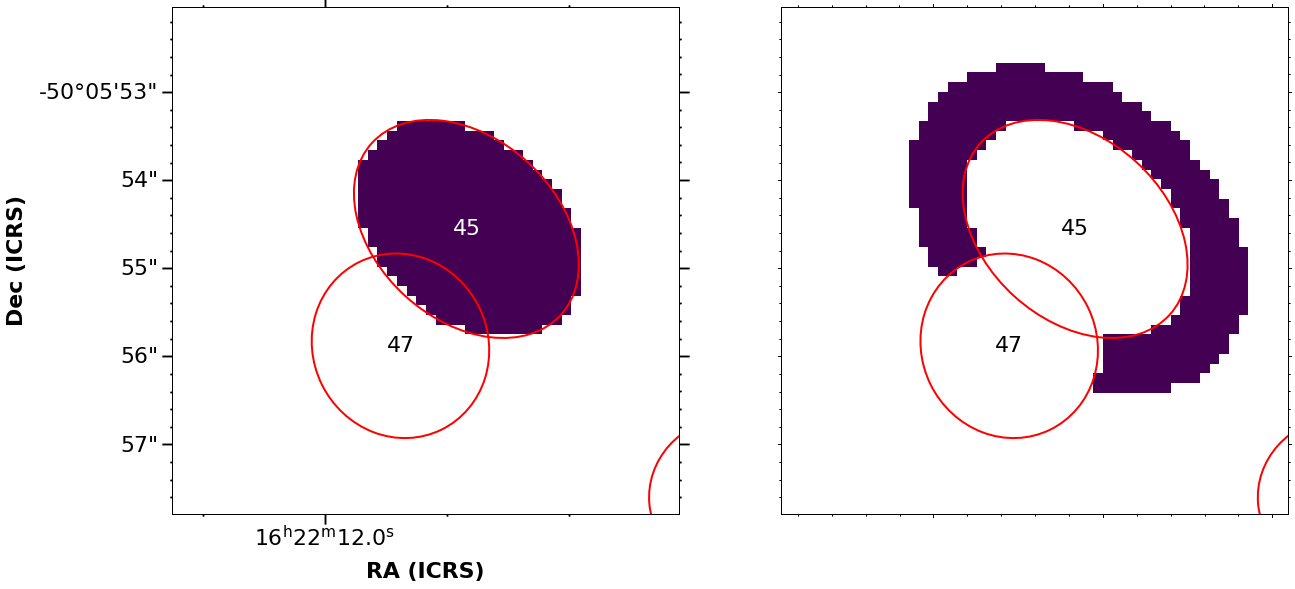}
    \caption{ Zoom in towards cores 14 (top-left panel) and 45 (top-right panel) overlaid on the DCN moment 0 map in the protocluster G333.60. The red ellipses are 2 times the major (minor) FWHM of the continuum cores. In the bottom panels, we show examples of the masks (shaded blue areas) and their respective annulus (for core 14) or crown (for core 45), that is used to estimate the core-averaged, background-subtracted DCN (3-2) spectra. The background annulus is defined as 2-3 times the core major (minor) FWHMs from \citet{louvet22}, with pixels from overlapping cores being masked out (i.e. the resulting crown in core 45). \label{appendix:dcn_zoom_masks}} 
\end{figure*}

\begin{figure}
    \centering
    \includegraphics[width=0.44\textwidth]{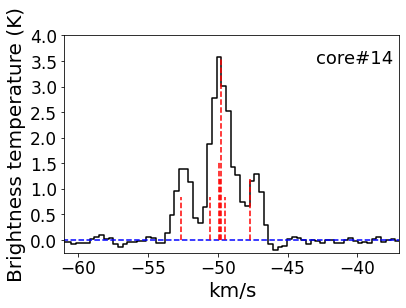}
    \caption{ Example of the DCN (3-2) core-averaged, background-subtracted spectrum extracted for core 14, in G333.60 with the DCN hyperfine model (see \cref{tabappendix:hyperfine}) overlaid in red vertical dashed lines and are set relative to the central core \vlsr. \label{appendix:dcn_hyperfineexample}}
\end{figure}

\begin{figure*}

    \centering
\includegraphics[width=0.24\textwidth]{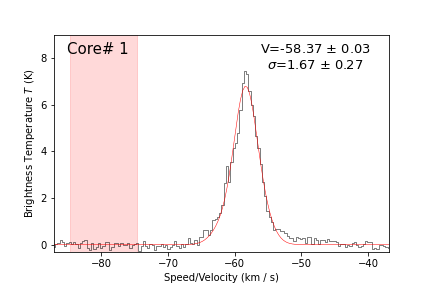}
\includegraphics[width=0.24\textwidth]{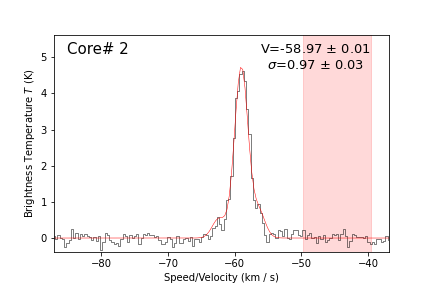}
\includegraphics[width=0.24\textwidth]{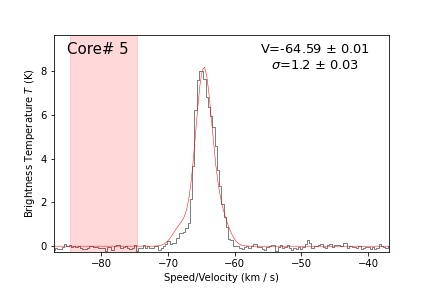}
\includegraphics[width=0.24\textwidth]{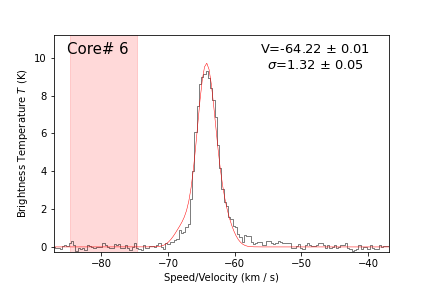}
\includegraphics[width=0.24\textwidth]{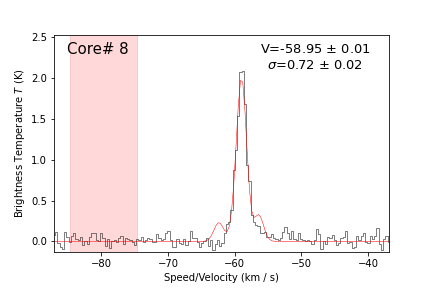}
\includegraphics[width=0.24\textwidth]{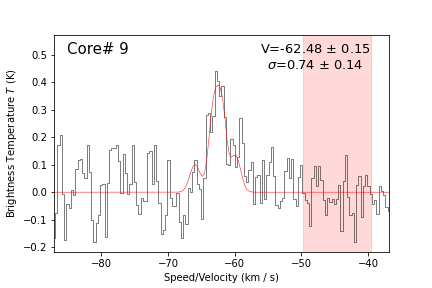}
\includegraphics[width=0.24\textwidth]{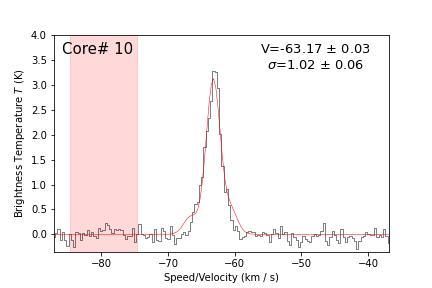}
\includegraphics[width=0.24\textwidth]{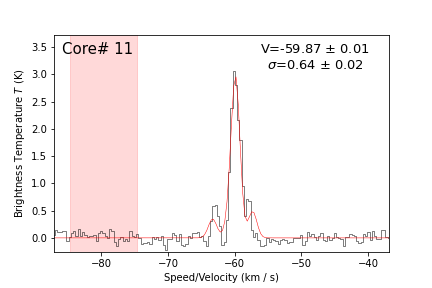}
\includegraphics[width=0.24\textwidth]{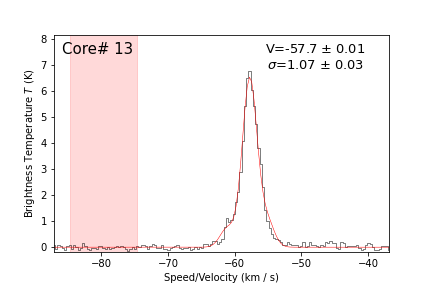}
\includegraphics[width=0.24\textwidth]{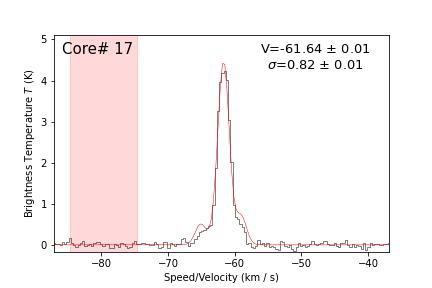}
\includegraphics[width=0.24\textwidth]{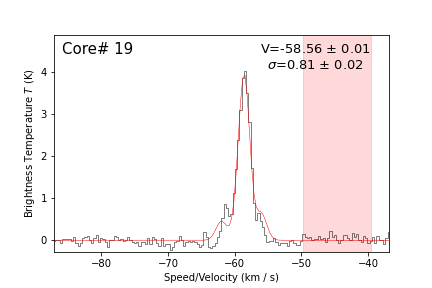}
\includegraphics[width=0.24\textwidth]{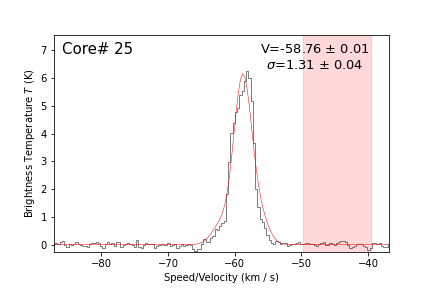}
\includegraphics[width=0.24\textwidth]{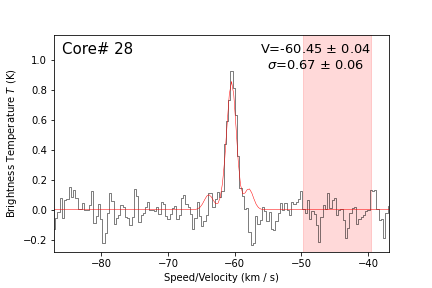}
\includegraphics[width=0.24\textwidth]{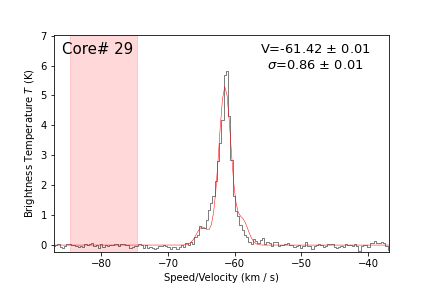}
\includegraphics[width=0.24\textwidth]{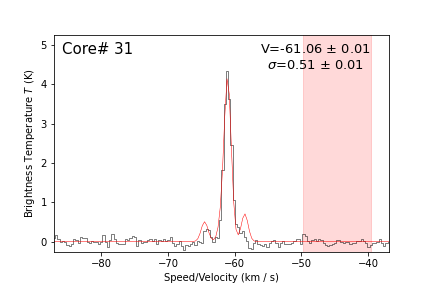}
\includegraphics[width=0.24\textwidth]{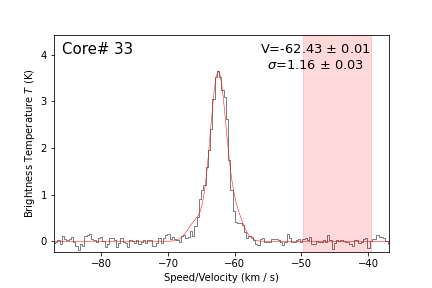}
\includegraphics[width=0.24\textwidth]{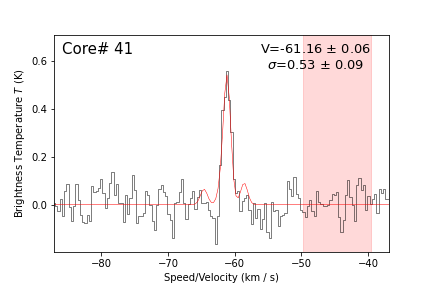}
\includegraphics[width=0.24\textwidth]{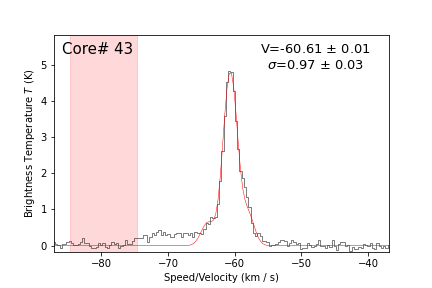}
\includegraphics[width=0.24\textwidth]{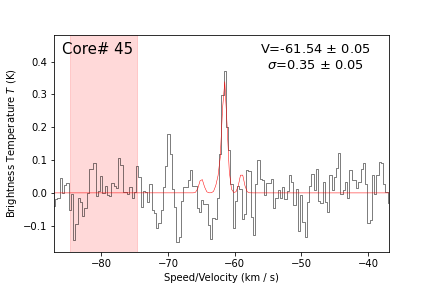}
\includegraphics[width=0.24\textwidth]{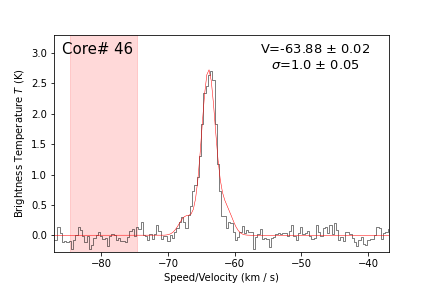}
\includegraphics[width=0.24\textwidth]{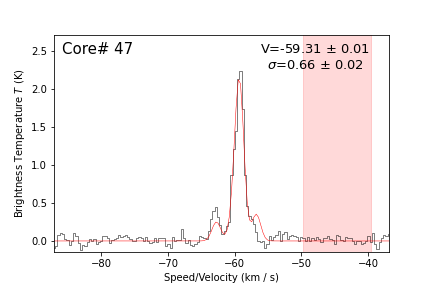}

\caption{Single-type non-background subtracted spectra in the young protocluster G338.93. They are classified in the same way as the background-subtracted spectra. The associated continuum core number is given in the top left of each panel (the core numbering is taken from \citet{louvet22}) the core \vlsr (V) and velocity dispersion ($\sigma$) both in units of \kms from the HSF fit are given in the top right. The pink shaded region represents the part of the spectrum used to estimate the MAD noise. \label{appendicfig:g388_on}}
\end{figure*}

\begin{figure*}\ContinuedFloat
    \centering
\includegraphics[width=0.24\textwidth]{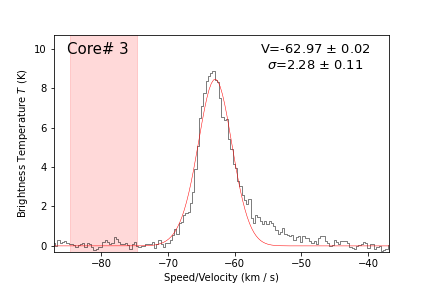}
\includegraphics[width=0.24\textwidth]{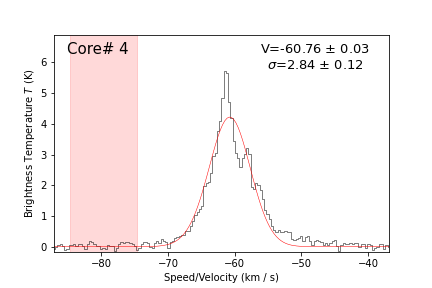}
\includegraphics[width=0.24\textwidth]{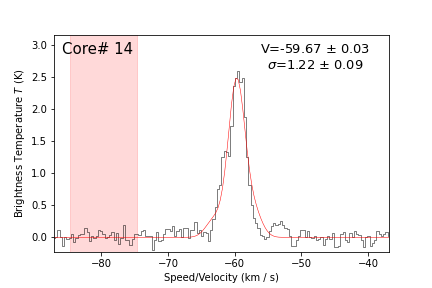}
\includegraphics[width=0.24\textwidth]{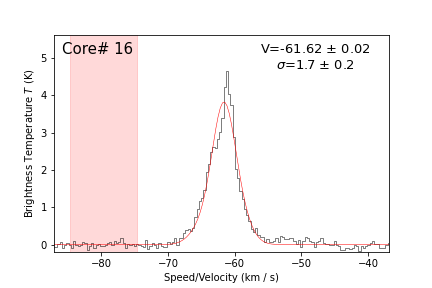}
\includegraphics[width=0.24\textwidth]{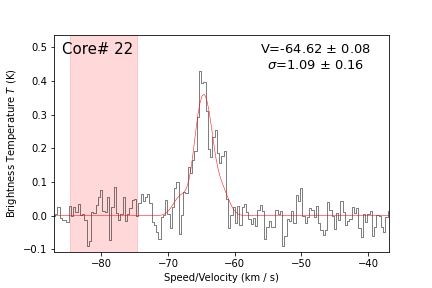}
\includegraphics[width=0.24\textwidth]{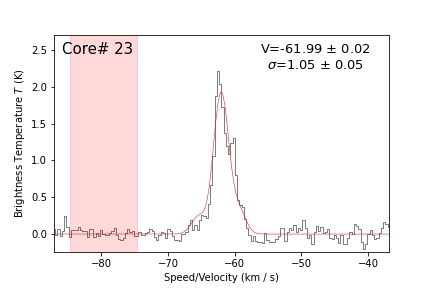}
\includegraphics[width=0.24\textwidth]{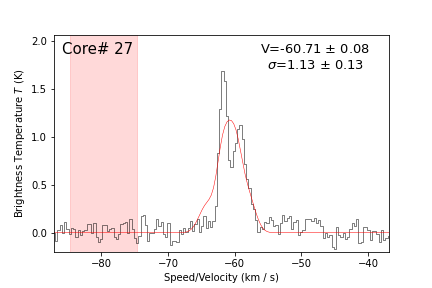}
\includegraphics[width=0.24\textwidth]{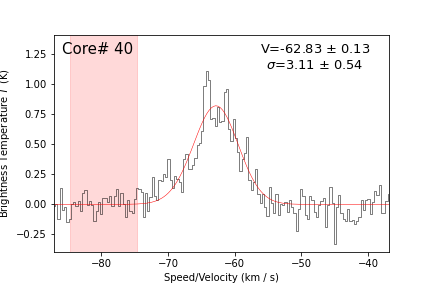}
\caption{Continued: Complex-type, non-background subtracted spectra towards the young protocluster G338.93. \label{appendicfig:g388_on_c}}

\end{figure*}
\begin{figure*}

    \centering
\includegraphics[width=0.24\textwidth]{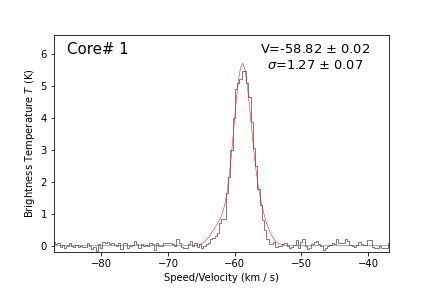}
\includegraphics[width=0.24\textwidth]{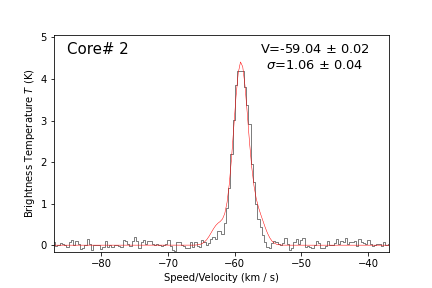}
\includegraphics[width=0.24\textwidth]{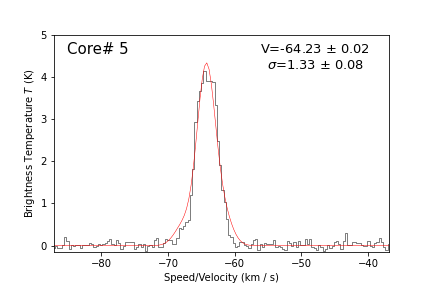}
\includegraphics[width=0.24\textwidth]{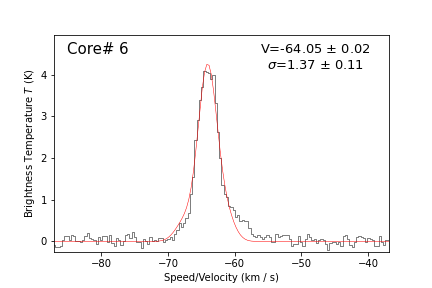}
\includegraphics[width=0.24\textwidth]{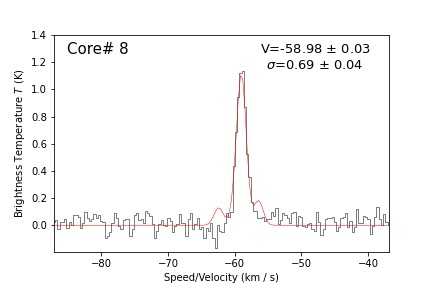}
\includegraphics[width=0.24\textwidth]{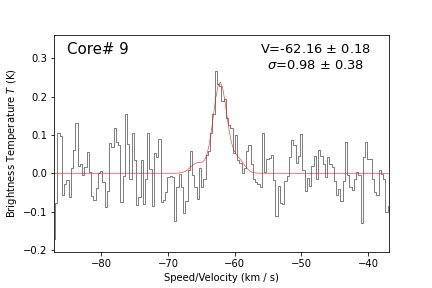}
\includegraphics[width=0.24\textwidth]{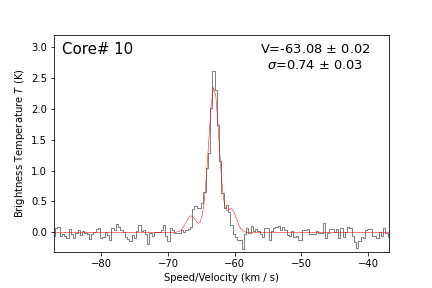}
\includegraphics[width=0.24\textwidth]{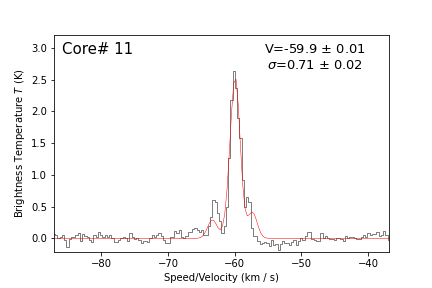}
\includegraphics[width=0.24\textwidth]{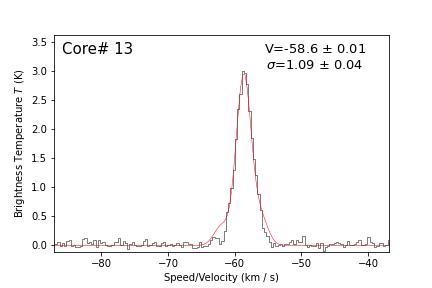}
\includegraphics[width=0.24\textwidth]{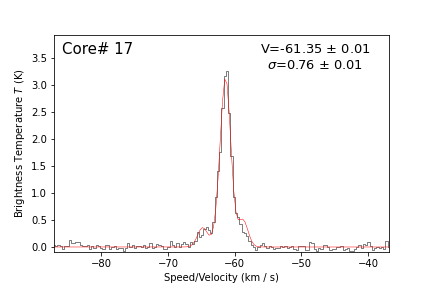}
\includegraphics[width=0.24\textwidth]{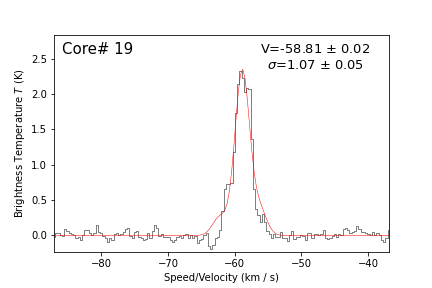}
\includegraphics[width=0.24\textwidth]{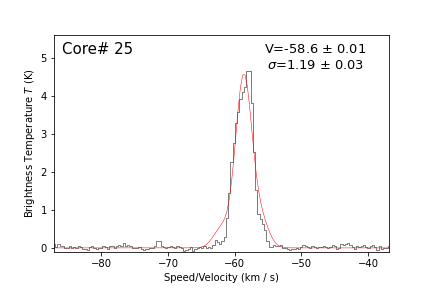}
\includegraphics[width=0.24\textwidth]{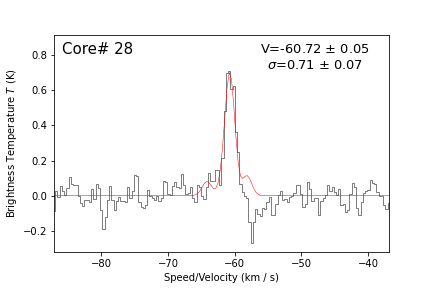}
\includegraphics[width=0.24\textwidth]{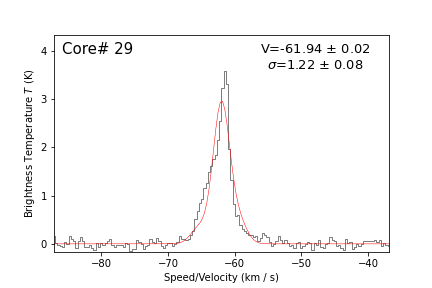}
\includegraphics[width=0.24\textwidth]{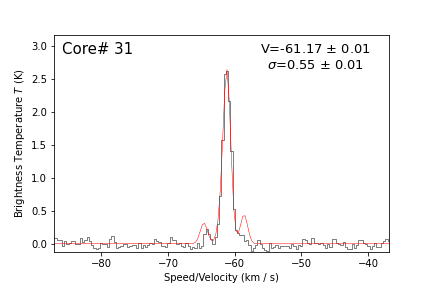}
\includegraphics[width=0.24\textwidth]{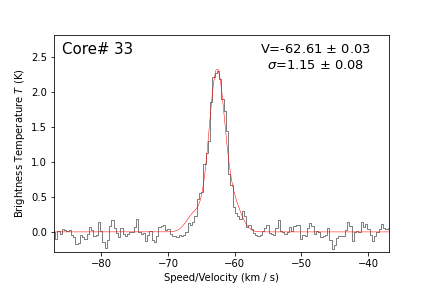}
\includegraphics[width=0.24\textwidth]{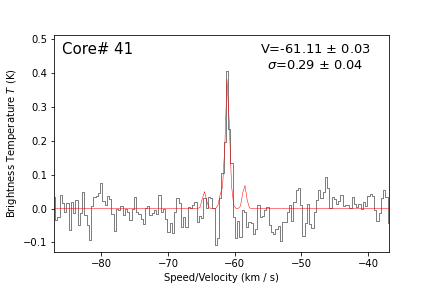}
\includegraphics[width=0.24\textwidth]{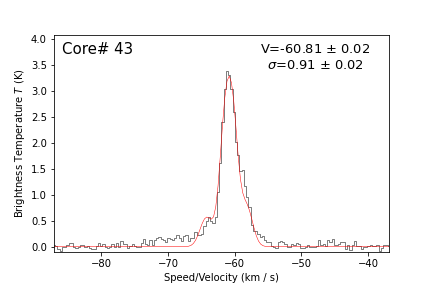}
\includegraphics[width=0.24\textwidth]{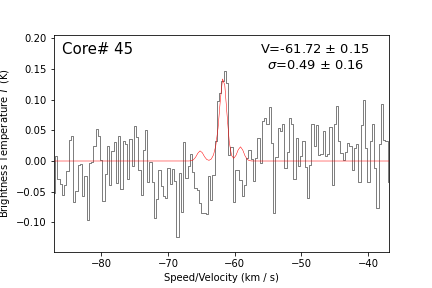}
\includegraphics[width=0.24\textwidth]{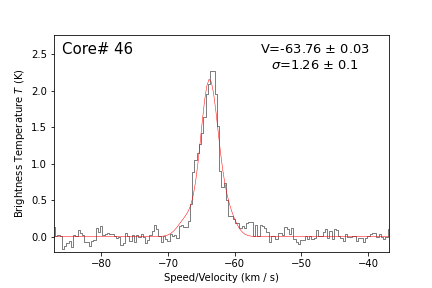}
\includegraphics[width=0.24\textwidth]{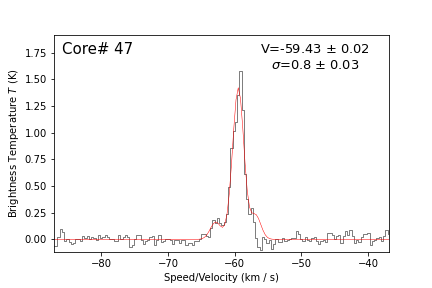}

\caption{Background spectra which are the average spectra from the background annulus towards the positions of the Single-type cores shown in \cref{appendicfig:g388_on} in the young protocluster G338.93. The associated core numbers are provided in the top left of each panel. \label{g338:off}}
\end{figure*}

\begin{figure*}\ContinuedFloat
    \centering

\includegraphics[width=0.24\textwidth]{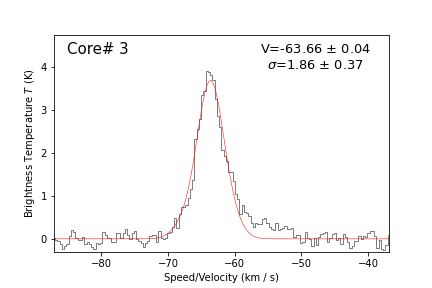}
\includegraphics[width=0.24\textwidth]{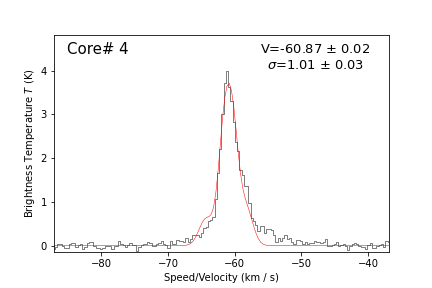}
\includegraphics[width=0.24\textwidth]{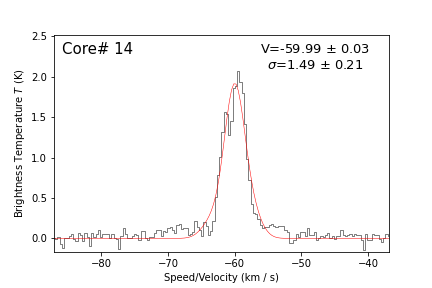}
\includegraphics[width=0.24\textwidth]{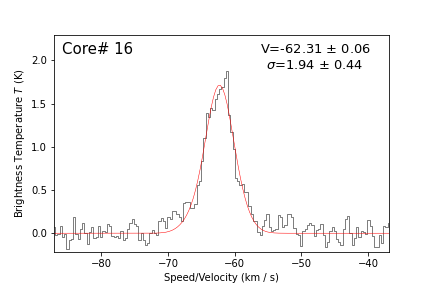}
\includegraphics[width=0.24\textwidth]{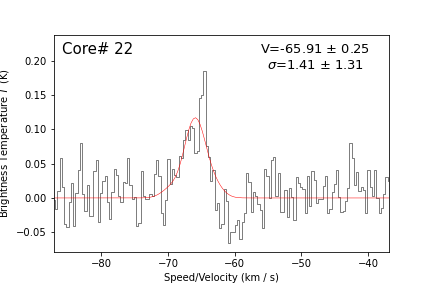}
\includegraphics[width=0.24\textwidth]{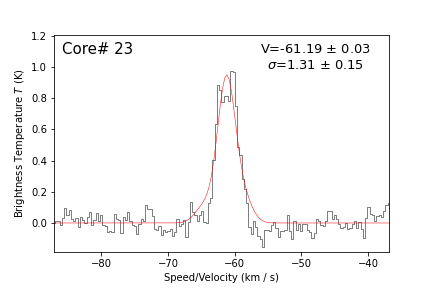}
\includegraphics[width=0.24\textwidth]{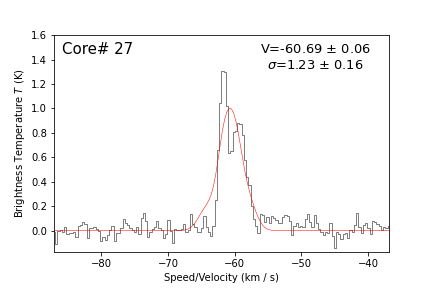}
\includegraphics[width=0.24\textwidth]{Appendix/appendix_figures/spectra_plots/G338_OFF_spectra/G338.93_core_16_hyperfine_fits_off.png}
\includegraphics[width=0.24\textwidth]{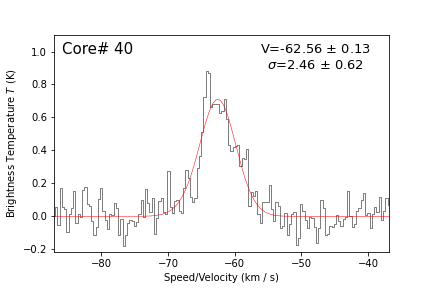}
\caption{Continued: Background spectra which are the average spectra from the background annulus towards the positions of the Complex-type cores shown in \cref{appendicfig:g388_on_c} in the young protocluster G338.93.}

\end{figure*}

\begin{table}[htbp!]
\centering
\caption{DCN hyperfine model.}
\label{tabappendix:hyperfine}
 \begin{tabular}{llll}
\hline \noalign {\smallskip}
Hyperfine    & Frequency   & Relative Velocity &  Relative   \\ 
transition             &  [GHz]      &   [kms$^{-1}$]&        Line Strength  \\
\hline \noalign {\smallskip}
f1  &    217.236999  & 2.12      & 7   \\
f2  &    217.238300  & 0.328     & 5 \\
f3  &    217.238538  & 0.0       & 21 \\
f4  &    217.238555  &  -0.02373 & 7 \\
f5  &    217.238612  & -0.10239  &  9 \\
f6  &    217.239079  & -0.7468   & 5  \\
f7  &    217.240622  &  -2.8762  & 5\\
\hline \noalign {\smallskip}
\end{tabular}
\end{table}

\section{DCN line fits and spectra}
\label{appendix:spectral}
We show the single- and Complex-type DCN (3-2) spectra for each of the 15 protoclusters in the ALMA-IMF survey. We also provide the tables of the associated core number, core name, position, and major and minor axes taken from \citet{louvet22} along with the DCN line fits extracted for each respective core. If no fit parameters are provided for the DCN or the name Single or Complex is not provided, the core was not detected in DCN. 
\begin{table*}[htp]
\centering
\small
\caption{DCN fits towards the core population of the intermediate protocluster G008.67.}
\label{g008coretables}
\begin{tabular}{llllllllccc}

\hline 
n   & Core Name &RA  & DEC & F$_{A}$   & F$_{B}$    & PA & T & \vlsr  & Linewidth   & Spectral \\
 &  & [ICRS]  &  [ICRS]   &  [\arcsec] &[\arcsec] &  [deg] & [K] & [\kms]  & [\kms] & Type \\
\hline 
2  &  271.5978343-21.6195911  &  18:06:23.48  &  -21:37:10.53  &  0.96  &  0.87  &  92  &  100 $\pm$ 50  &  39.5 $\pm$ 0.07  &  --  &  Complex \\
3  &  271.5796091-21.6250348  &  18:06:19.11  &  -21:37:30.13  &  0.9  &  0.79  &  71  &  34 $\pm$ 7  &  34.27 $\pm$ 0.07  &  1.14 $\pm$ 0.16  &  Single \\
4  &  271.580761-21.6237093  &  18:06:19.38  &  -21:37:25.35  &  1.42  &  1.35  &  168  &  27 $\pm$ 6  &  35.4 $\pm$ 0.05  &  1.16 $\pm$ 0.11  &  Single \\
5  &  271.5785345-21.6222306  &  18:06:18.85  &  -21:37:20.03  &  1.06  &  0.91  &  95  &  24 $\pm$ 5  &  --  &  --  &  -- \\
6  &  271.5980489-21.6198889  &  18:06:23.53  &  -21:37:11.60  &  1.13  &  1.11  &  4  &  29 $\pm$ 6  &  38.72 $\pm$ 0.05  &  --  &  Complex \\
7  &  271.5800126-21.6251966  &  18:06:19.20  &  -21:37:30.71  &  1.35  &  0.96  &  92  &  33 $\pm$ 7  &  34.32 $\pm$ 0.03  &  1.69 $\pm$ 0.17  &  Single \\
8  &  271.5954206-21.620392  &  18:06:22.90  &  -21:37:13.41  &  0.97  &  0.73  &  80  &  25 $\pm$ 5  &  --  &  --  &  -- \\
9  &  271.5985087-21.6198427  &  18:06:23.64  &  -21:37:11.43  &  1.09  &  0.85  &  99  &  28 $\pm$ 6  &  37.56 $\pm$ 0.09  &  --  &  Complex \\
10  &  271.5774356-21.6271193  &  18:06:18.58  &  -21:37:37.63  &  1.62  &  1.29  &  153  &  32 $\pm$ 6  &  33.78 $\pm$ 0.08  &  1.53 $\pm$ 0.21  &  Single \\
11  &  271.5791461-21.6247713  &  18:06:19.00  &  -21:37:29.18  &  1.14  &  1.03  &  156  &  32 $\pm$ 7  &  35.12 $\pm$ 0.02  &  1.43 $\pm$ 0.08  &  Single \\
12  &  271.5796516-21.6208729  &  18:06:19.12  &  -21:37:15.14  &  1.18  &  0.98  &  28  &  22 $\pm$ 5  &  39.19 $\pm$ 0.12  &  1.43 $\pm$ 0.45  &  Single \\
13  &  271.5781549-21.622259  &  18:06:18.76  &  -21:37:20.13  &  0.95  &  0.81  &  94  &  24 $\pm$ 5  &  --  &  --  &  -- \\
14  &  271.5798029-21.6239806  &  18:06:19.15  &  -21:37:26.33  &  1.24  &  1.06  &  1  &  28 $\pm$ 6  &  35.07 $\pm$ 0.1  &  --  &  Complex \\
15  &  271.5785913-21.6271084  &  18:06:18.86  &  -21:37:37.59  &  1.25  &  1.15  &  148  &  32 $\pm$ 6  &  32.45 $\pm$ 0.02  &  1.24 $\pm$ 0.08  &  Single \\
16  &  271.5938426-21.6182732  &  18:06:22.52  &  -21:37:5.780  &  1.25  &  1.05  &  77  &  19 $\pm$ 4  &  --  &  --  &  -- \\
17  &  271.5797502-21.6218939  &  18:06:19.14  &  -21:37:18.82  &  1.04  &  0.83  &  91  &  23 $\pm$ 5  &  --  &  --  &  -- \\
19  &  271.5916065-21.6190688  &  18:06:21.99  &  -21:37:8.650  &  2.16  &  1.4  &  52  &  18 $\pm$ 4  &  --  &  --  &  -- \\
21  &  271.5977539-21.6190685  &  18:06:23.46  &  -21:37:8.650  &  1.62  &  1.21  &  4  &  29 $\pm$ 6  &  38.44 $\pm$ 0.05  &  2.87 $\pm$ 1.19  &  Single \\
22  &  271.5979817-21.6202471  &  18:06:23.52  &  -21:37:12.89  &  1.47  &  1.41  &  31  &  29 $\pm$ 6  &  39.7 $\pm$ 0.03  &  1.06 $\pm$ 0.06  &  Single \\
\hline 
\end{tabular}
\end{table*}

\begin{figure*}
    \centering
\includegraphics[width=0.24\textwidth]{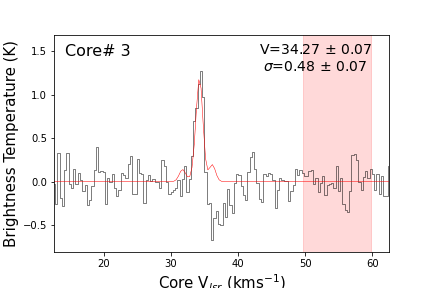}
\includegraphics[width=0.24\textwidth]{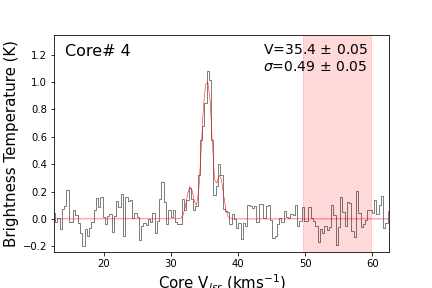}
\includegraphics[width=0.24\textwidth]{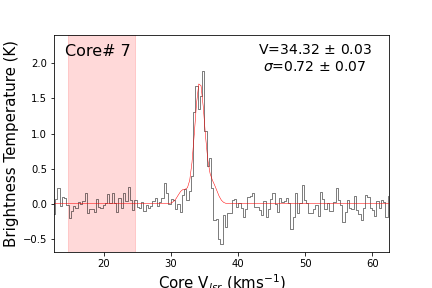}
\includegraphics[width=0.24\textwidth]{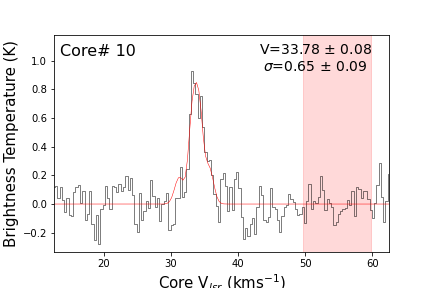}
\includegraphics[width=0.24\textwidth]{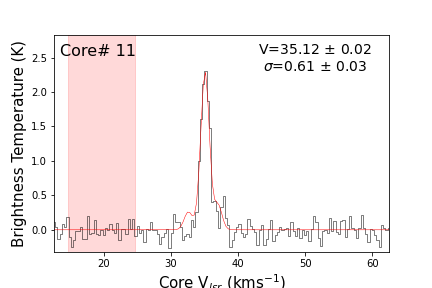}
\includegraphics[width=0.24\textwidth]{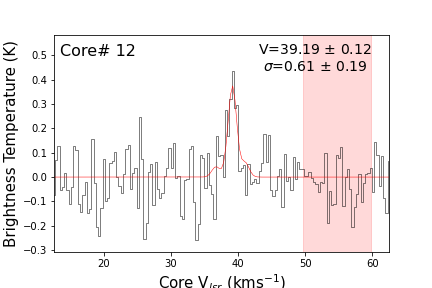}
\includegraphics[width=0.24\textwidth]{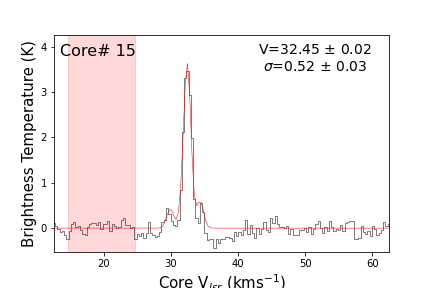}
\includegraphics[width=0.24\textwidth]{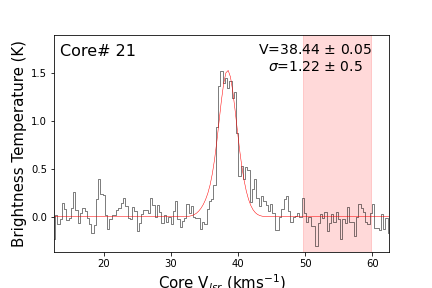}
\includegraphics[width=0.24\textwidth]{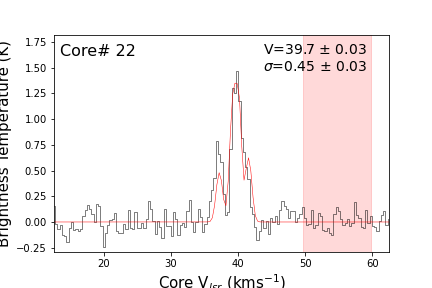}
\caption{Single-type core-averaged, background-subtracted DCN spectra extracted from the cores in the intermediate protocluster G008.67. See \cref{g008coretables} for the line fit parameters for each core. The associated continuum core number is given in the top left of each panel (the core numbering is taken from \citet{louvet22}) the core \vlsr (V) and velocity dispersion ($\sigma$) both in units of \kms from the HSF fit are given in the top right. The pink shaded region represents the part of the spectrum used to estimate the MAD noise.}
\end{figure*}

\begin{figure*}
    \centering

\includegraphics[width=0.24\textwidth]{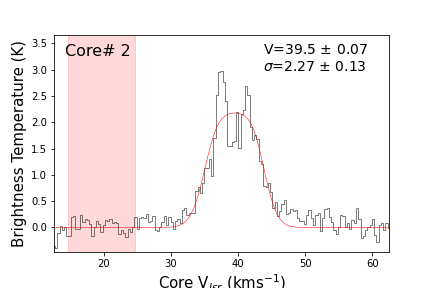} 
\includegraphics[width=0.24\textwidth]{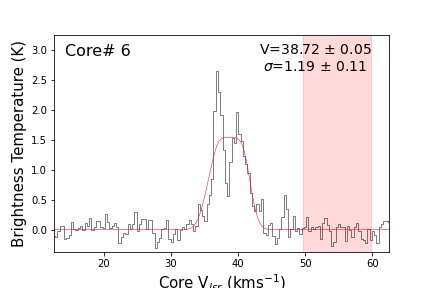} 
\includegraphics[width=0.24\textwidth]{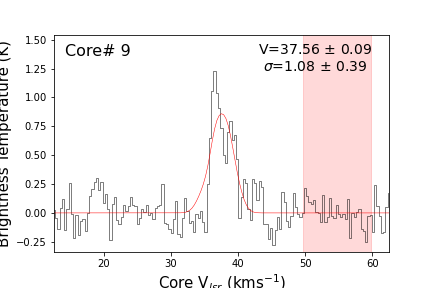} 
\includegraphics[width=0.24\textwidth]{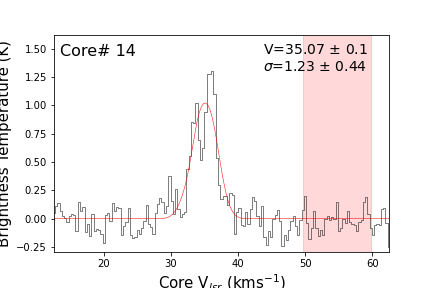}
\caption{Complex-type core-averaged, background-subtracted DCN spectra extracted from cores in the intermediate protocluster G008.67. See \cref{g008coretables} for the line fit parameters for each core.} 
\end{figure*}

\begin{table*}[htbp!]
\centering
\small
\caption{DCN fits towards the core population of the evolved protocluster G010.62.}
\label{tabappendix:coretables_g010}
\begin{tabular}{llllllllccc}

\hline 
n   & Core Name &RA  & DEC & F$_{A}$   & F$_{B}$    & PA & T & \vlsr  & Linewidth   & Spectral \\
 &  & [ICRS]  &  [ICRS]   &  [\arcsec] &[\arcsec] &  [deg] & [K] & [\kms]  & [\kms] & Type \\
\hline 
2  &  272.6133567-19.9299385  &  18:10:27.21  &  -19:55:47.78  &  1.44  &  0.88  &  76  &  29 $\pm$ 6  &  -3.15 $\pm$ 0.04  &  0.98 $\pm$ 0.11  &  Single \\
3  &  272.6158612-19.9284295  &  18:10:27.81  &  -19:55:42.35  &  1.09  &  1.03  &  66  &  29 $\pm$ 6  &  -2.49 $\pm$ 0.02  &  1.28 $\pm$ 0.08  &  Single \\
4  &  272.6216933-19.9291438  &  18:10:29.21  &  -19:55:44.92  &  0.71  &  0.62  &  97  &  33 $\pm$ 7  &  -7.64 $\pm$ 0.02  &  1.27 $\pm$ 0.06  &  Single \\
5  &  272.6106772-19.9271061  &  18:10:26.56  &  -19:55:37.58  &  0.72  &  0.56  &  93  &  25 $\pm$ 5  &  --  &  --  &  -- \\
6  &  272.6218286-19.9280416  &  18:10:29.24  &  -19:55:40.95  &  0.71  &  0.53  &  81  &  100 $\pm$ 50  &  -0.0 $\pm$ 0.01  &  1.22 $\pm$ 0.04  &  Single \\
7  &  272.6243153-19.9230489  &  18:10:29.84  &  -19:55:22.98  &  0.74  &  0.62  &  61  &  24 $\pm$ 5  &  -1.9 $\pm$ 0.09  &  1.13 $\pm$ 0.37  &  Single \\
8  &  272.6185252-19.9301787  &  18:10:28.45  &  -19:55:48.64  &  0.64  &  0.55  &  114  &  38 $\pm$ 8  &  1.24 $\pm$ 0.07  &  --  &  Complex \\
9  &  272.6130348-19.9319451  &  18:10:27.13  &  -19:55:55.00  &  0.72  &  0.6  &  125  &  29 $\pm$ 6  &  -5.16 $\pm$ 0.02  &  0.62 $\pm$ 0.04  &  Single \\
11  &  272.6241652-19.9229069  &  18:10:29.80  &  -19:55:22.46  &  0.76  &  0.57  &  90  &  24 $\pm$ 5  &  -1.74 $\pm$ 0.05  &  1.11 $\pm$ 0.17  &  Single \\
14  &  272.6213499-19.9296433  &  18:10:29.12  &  -19:55:46.72  &  0.92  &  0.72  &  86  &  36 $\pm$ 7  &  -7.43 $\pm$ 0.04  &  --  &  Complex \\
16  &  272.6207417-19.929286  &  18:10:28.98  &  -19:55:45.43  &  0.7  &  0.59  &  101  &  36 $\pm$ 7  &  -5.6 $\pm$ 0.02  &  0.69 $\pm$ 0.08  &  Single \\
17  &  272.6303738-19.922395  &  18:10:31.29  &  -19:55:20.62  &  0.64  &  0.55  &  124  &  19 $\pm$ 4  &  --  &  --  &  -- \\
19  &  272.6234089-19.9273408  &  18:10:29.62  &  -19:55:38.43  &  0.81  &  0.68  &  158  &  29 $\pm$ 6  &  -0.19 $\pm$ 0.02  &  0.86 $\pm$ 0.09  &  Single \\
20  &  272.6212378-19.9292402  &  18:10:29.10  &  -19:55:45.26  &  0.72  &  0.49  &  126  &  35 $\pm$ 7  &  -6.6 $\pm$ 0.07  &  2.03 $\pm$ 0.36  &  Single \\
21  &  272.6230891-19.9281344  &  18:10:29.54  &  -19:55:41.28  &  0.76  &  0.66  &  110  &  29 $\pm$ 6  &  --  &  --  &  -- \\
22  &  272.6162629-19.9281751  &  18:10:27.90  &  -19:55:41.43  &  1.94  &  1.2  &  165  &  29 $\pm$ 6  &  -2.44 $\pm$ 0.01  &  1.39 $\pm$ 0.05  &  Single \\
24  &  272.6222764-19.930054  &  18:10:29.35  &  -19:55:48.19  &  0.93  &  0.73  &  113  &  32 $\pm$ 6  &  -3.32 $\pm$ 0.03  &  --  &  Complex \\
25  &  272.6122967-19.9309094  &  18:10:26.95  &  -19:55:51.27  &  0.86  &  0.73  &  149  &  28 $\pm$ 6  &  -4.05 $\pm$ 0.01  &  1.08 $\pm$ 0.05  &  Single \\
26  &  272.6209449-19.9291949  &  18:10:29.03  &  -19:55:45.10  &  0.79  &  0.59  &  73  &  35 $\pm$ 7  &  -5.6 $\pm$ 0.06  &  0.56 $\pm$ 0.13  &  Single \\
27  &  272.6175606-19.9309193  &  18:10:28.21  &  -19:55:51.31  &  0.86  &  0.68  &  42  &  34 $\pm$ 7  &  -2.32 $\pm$ 0.03  &  1.1 $\pm$ 0.1  &  Single \\
29  &  272.6159709-19.9240334  &  18:10:27.83  &  -19:55:26.52  &  1.3  &  1.26  &  110  &  27 $\pm$ 5  &  -1.36 $\pm$ 0.04  &  0.99 $\pm$ 0.17  &  Single \\
33  &  272.6157835-19.9245689  &  18:10:27.79  &  -19:55:28.45  &  1.02  &  0.97  &  176  &  27 $\pm$ 6  &  -1.03 $\pm$ 0.02  &  0.64 $\pm$ 0.07  &  Single \\
34  &  272.6177509-19.9302971  &  18:10:28.26  &  -19:55:49.07  &  0.72  &  0.53  &  73  &  35 $\pm$ 7  &  2.43 $\pm$ 0.03  &  1.69 $\pm$ 0.07  &  Single \\
37  &  272.6222902-19.9232388  &  18:10:29.35  &  -19:55:23.66  &  1.92  &  1.31  &  38  &  25 $\pm$ 5  &  --  &  --  &  -- \\
38  &  272.6243811-19.9276902  &  18:10:29.85  &  -19:55:39.68  &  0.98  &  0.62  &  149  &  27 $\pm$ 5  &  --  &  --  &  -- \\
39  &  272.6205793-19.9284073  &  18:10:28.94  &  -19:55:42.27  &  0.94  &  0.84  &  137  &  32 $\pm$ 7  &  -6.94 $\pm$ 0.01  &  1.17 $\pm$ 0.03  &  Single \\
40  &  272.6276189-19.922909  &  18:10:30.63  &  -19:55:22.47  &  0.82  &  0.6  &  88  &  21 $\pm$ 4  &  --  &  --  &  -- \\
42  &  272.6163507-19.9286735  &  18:10:27.92  &  -19:55:43.22  &  0.86  &  0.77  &  165  &  29 $\pm$ 6  &  -2.72 $\pm$ 0.02  &  1.64 $\pm$ 0.08  &  Single \\
43  &  272.6139595-19.9339321  &  18:10:27.35  &  -19:56:2.160  &  0.86  &  0.62  &  58  &  28 $\pm$ 6  &  -0.85 $\pm$ 0.02  &  0.72 $\pm$ 0.09  &  Single \\
44  &  272.6157525-19.9322865  &  18:10:27.78  &  -19:55:56.23  &  1.22  &  1.08  &  92  &  32 $\pm$ 6  &  -3.07 $\pm$ 0.02  &  1.19 $\pm$ 0.09  &  Single \\
45  &  272.6172993-19.9301129  &  18:10:28.15  &  -19:55:48.41  &  1.0  &  0.88  &  141  &  33 $\pm$ 7  &  2.04 $\pm$ 0.02  &  --  &  Complex \\
46  &  272.6164851-19.9303554  &  18:10:27.96  &  -19:55:49.28  &  0.82  &  0.78  &  18  &  33 $\pm$ 7  &  1.27 $\pm$ 0.02  &  --  &  Complex \\
47  &  272.6188053-19.9306895  &  18:10:28.51  &  -19:55:50.48  &  0.73  &  0.62  &  54  &  38 $\pm$ 8  &  -1.97 $\pm$ 0.07  &  1.19 $\pm$ 0.12  &  Single \\
48  &  272.6121657-19.9212008  &  18:10:26.92  &  -19:55:16.32  &  1.73  &  1.36  &  37  &  23 $\pm$ 5  &  --  &  --  &  -- \\
49  &  272.6190151-19.9284127  &  18:10:28.56  &  -19:55:42.29  &  1.0  &  0.74  &  119  &  33 $\pm$ 7  &  -1.09 $\pm$ 0.02  &  1.12 $\pm$ 0.06  &  Single \\
50  &  272.6193111-19.9274685  &  18:10:28.63  &  -19:55:38.89  &  0.93  &  0.79  &  92  &  31 $\pm$ 6  &  0.32 $\pm$ 0.02  &  1.45 $\pm$ 0.06  &  Single \\
52  &  272.6160332-19.9308294  &  18:10:27.85  &  -19:55:50.99  &  1.07  &  0.8  &  81  &  32 $\pm$ 7  &  -2.33 $\pm$ 0.12  &  --  &  Complex \\
53  &  272.6236904-19.9272934  &  18:10:29.69  &  -19:55:38.26  &  0.78  &  0.58  &  111  &  28 $\pm$ 6  &  -0.79 $\pm$ 0.06  &  1.3 $\pm$ 0.24  &  Single \\
56  &  272.6156542-19.9314086  &  18:10:27.76  &  -19:55:53.07  &  0.87  &  0.74  &  59  &  32 $\pm$ 6  &  -3.92 $\pm$ 0.03  &  0.81 $\pm$ 0.11  &  Single \\
58  &  272.6247409-19.9229049  &  18:10:29.94  &  -19:55:22.46  &  1.33  &  0.69  &  55  &  24 $\pm$ 5  &  --  &  --  &  -- \\
60  &  272.6227533-19.9295616  &  18:10:29.46  &  -19:55:46.42  &  1.08  &  0.84  &  69  &  30 $\pm$ 6  &  -5.29 $\pm$ 0.05  &  1.08 $\pm$ 0.17  &  Single \\
62  &  272.6145473-19.924069  &  18:10:27.49  &  -19:55:26.65  &  1.04  &  0.83  &  156  &  25 $\pm$ 5  &  -1.65 $\pm$ 0.03  &  0.76 $\pm$ 0.12  &  Single \\
\hline \noalign {\smallskip}
\end{tabular}
\end{table*}

\begin{figure*}
    \centering
\includegraphics[width=0.24\textwidth]{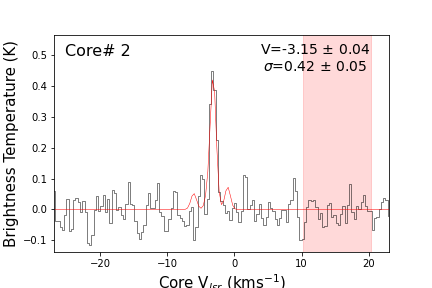}
\includegraphics[width=0.24\textwidth]{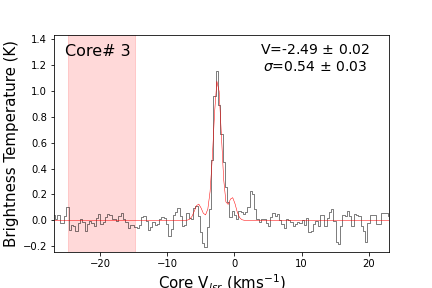}
\includegraphics[width=0.24\textwidth]{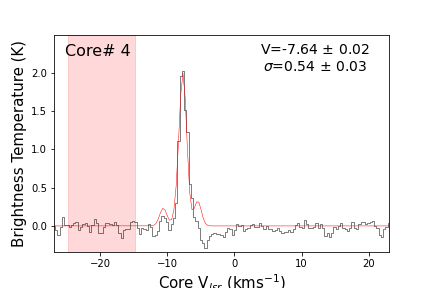}
\includegraphics[width=0.24\textwidth]{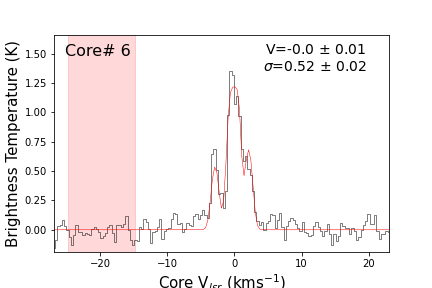}
\includegraphics[width=0.24\textwidth]{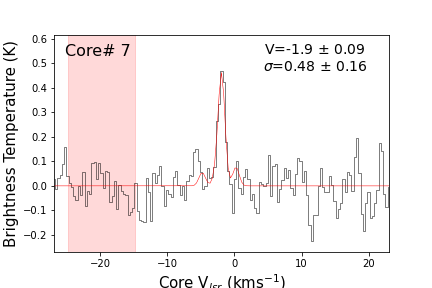}
\includegraphics[width=0.24\textwidth]{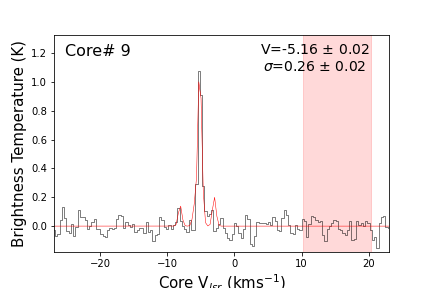}
\includegraphics[width=0.24\textwidth]{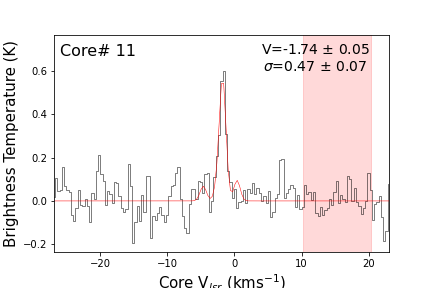}
\includegraphics[width=0.24\textwidth]{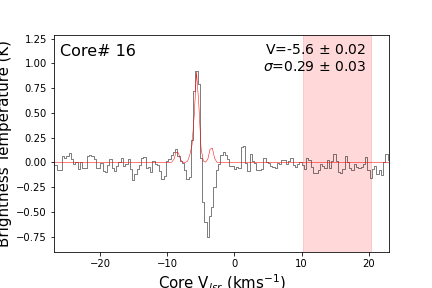}
\includegraphics[width=0.24\textwidth]{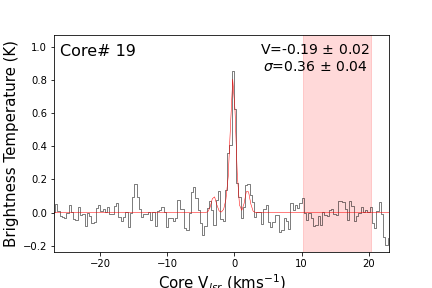}
\includegraphics[width=0.24\textwidth]{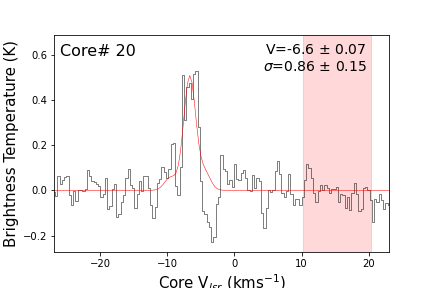}
\includegraphics[width=0.24\textwidth]{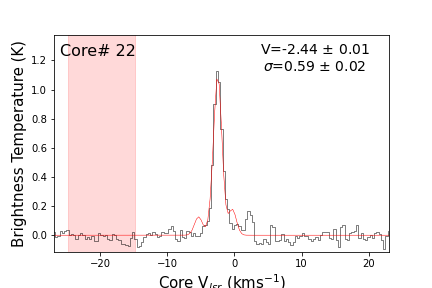}
\includegraphics[width=0.24\textwidth]{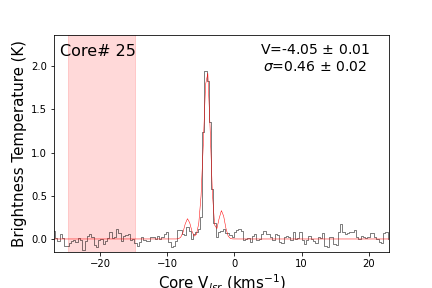}
\includegraphics[width=0.24\textwidth]{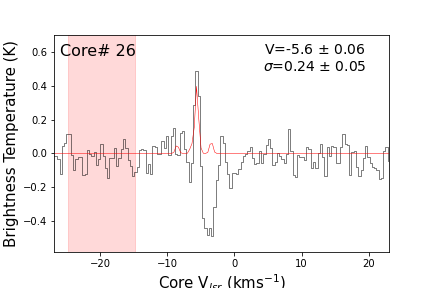}
\includegraphics[width=0.24\textwidth]{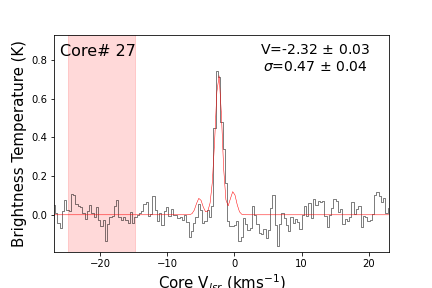}
\includegraphics[width=0.24\textwidth]{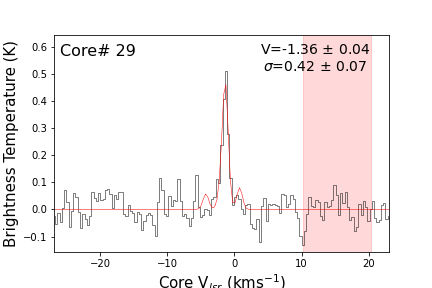}
\includegraphics[width=0.24\textwidth]{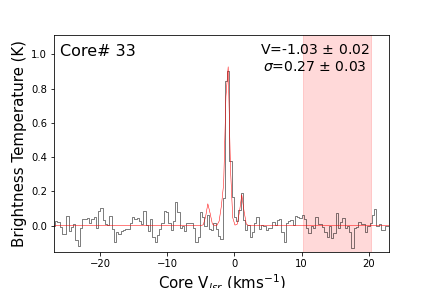}
\includegraphics[width=0.24\textwidth]{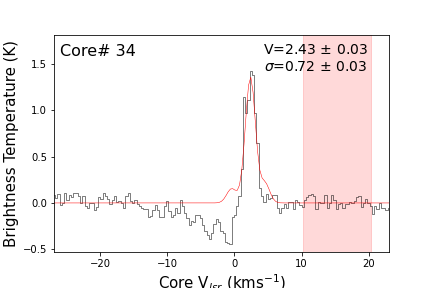}
\includegraphics[width=0.24\textwidth]{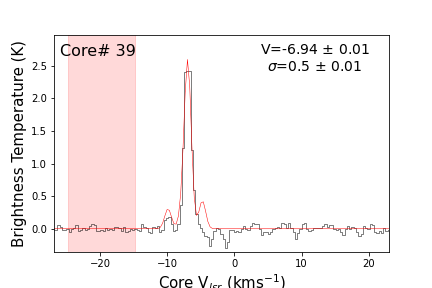}
\includegraphics[width=0.24\textwidth]{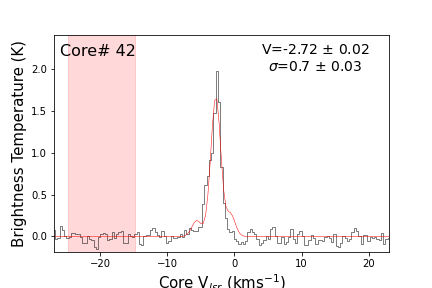}
\includegraphics[width=0.24\textwidth]{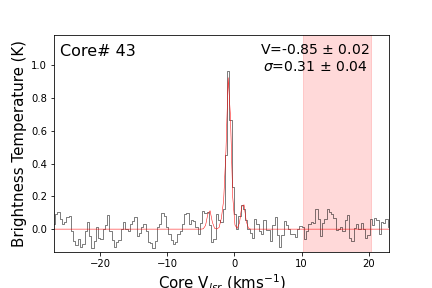}
\includegraphics[width=0.24\textwidth]{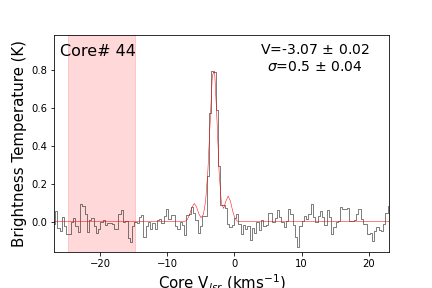}
\includegraphics[width=0.24\textwidth]{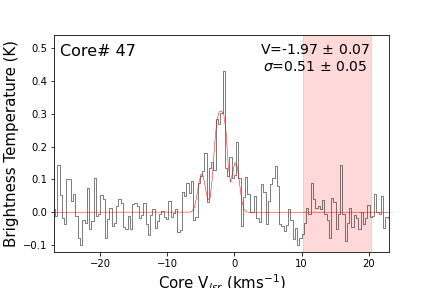}
\includegraphics[width=0.24\textwidth]{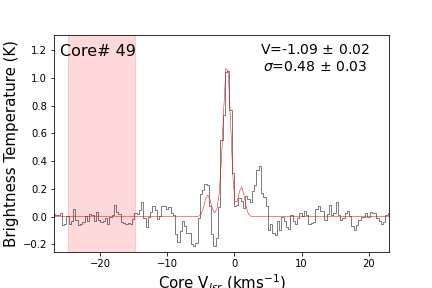}
\includegraphics[width=0.24\textwidth]{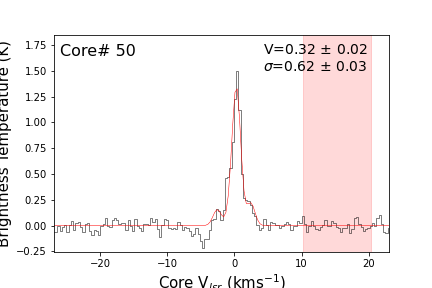}
\includegraphics[width=0.24\textwidth]{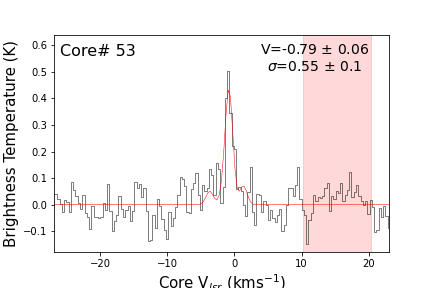}
\includegraphics[width=0.24\textwidth]{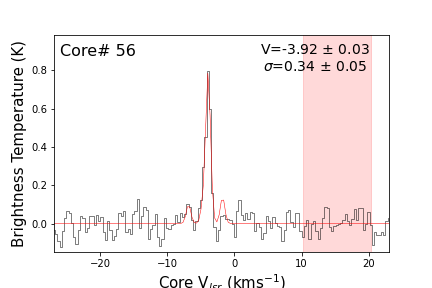}
\includegraphics[width=0.24\textwidth]{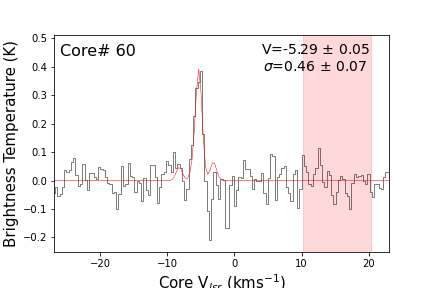}
\includegraphics[width=0.24\textwidth]{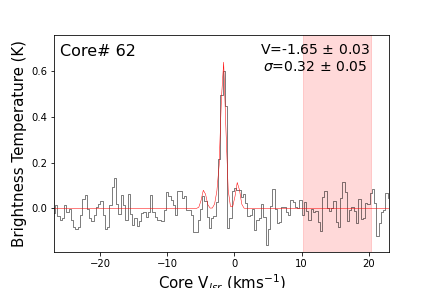}

\caption{Single-type core-averaged, background-subtracted DCN spectra extracted from the cores in the evolved protocluster G010.62. See \cref{tabappendix:coretables_g010} for the line fit parameters for each core.\label{figspectra:dcnspectra_split_G010_s}}
\end{figure*}

\begin{figure*}\ContinuedFloat
    \centering
\includegraphics[width=0.24\textwidth]{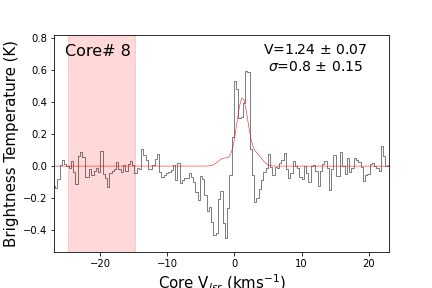}
\includegraphics[width=0.24\textwidth]{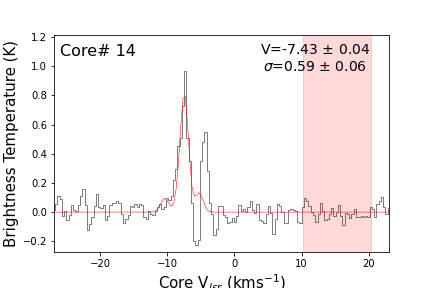}
\includegraphics[width=0.24\textwidth]{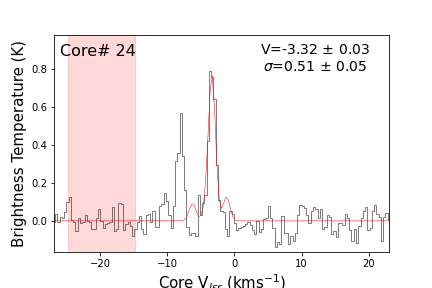}
\includegraphics[width=0.24\textwidth]{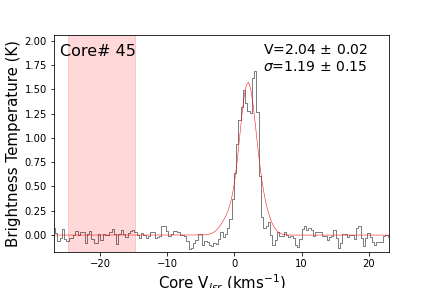}
\includegraphics[width=0.24\textwidth]{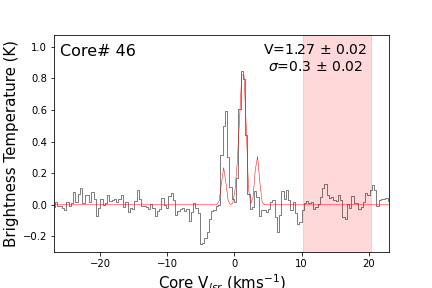}
\includegraphics[width=0.24\textwidth]{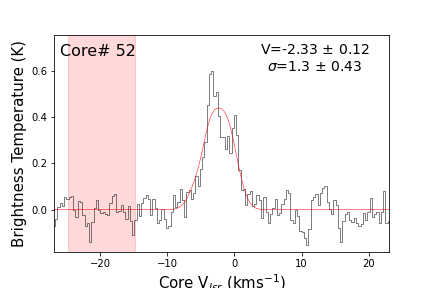}
\caption{Complex-type core-averaged, background-subtracted DCN spectra extracted from the cores in the evolved protocluster G010.62. See \cref{tabappendix:coretables_g010} for the line fit parameters for each core.\label{figspectra:dcnspectra_split_G010_c}} 
\end{figure*}

\begin{table*}[htbp!]
\centering
\small
\caption{DCN fits towards the core population of the evolved protocluster G012.80.}
\label{tabappendix:coretables_g012}
\begin{tabular}{llllllllccc}

\hline 
n   & Core Name &RA  & DEC & F$_{A}$   & F$_{B}$    & PA & T & \vlsr  & Linewidth   & Spectral \\
 &  & [ICRS]  &  [ICRS]   &  [\arcsec] &[\arcsec] &  [deg] & [K] & [\kms]  & [\kms] & Type \\
\hline 
1  &  273.5493292-17.9256817  &  18:14:11.84  &  -17:55:32.45  &  1.6  &  1.17  &  57  &  100 $\pm$ 50  &  37.06 $\pm$ 0.06  &  4.1 $\pm$ 0.7  &  Single \\
3  &  273.5573504-17.9225106  &  18:14:13.76  &  -17:55:21.04  &  1.84  &  1.52  &  8  &  100 $\pm$ 50  &  36.16 $\pm$ 0.04  &  --  &  Complex \\
4  &  273.5444266-17.9375329  &  18:14:10.66  &  -17:56:15.12  &  1.41  &  1.09  &  92  &  25 $\pm$ 5  &  36.16 $\pm$ 0.08  &  1.54 $\pm$ 0.27  &  Single \\
5  &  273.5482204-17.9458195  &  18:14:11.57  &  -17:56:44.95  &  1.44  &  1.14  &  89  &  23 $\pm$ 5  &  --  &  --  &  -- \\
6  &  273.5531484-17.9208183  &  18:14:12.76  &  -17:55:14.95  &  1.59  &  1.11  &  54  &  29 $\pm$ 6  &  32.6 $\pm$ 0.12  &  2.9 $\pm$ 2.8  &  Single \\
8  &  273.548615-17.9262103  &  18:14:11.67  &  -17:55:34.36  &  2.35  &  1.54  &  69  &  31 $\pm$ 6  &  36.5 $\pm$ 0.03  &  1.81 $\pm$ 0.16  &  Single \\
9  &  273.5557419-17.9151082  &  18:14:13.38  &  -17:54:54.39  &  1.46  &  1.4  &  26  &  25 $\pm$ 5  &  --  &  --  &  -- \\
10  &  273.554211-17.9141627  &  18:14:13.01  &  -17:54:50.99  &  1.53  &  1.09  &  73  &  24 $\pm$ 5  &  --  &  --  &  -- \\
11  &  273.5547682-17.9278997  &  18:14:13.14  &  -17:55:40.44  &  1.72  &  1.42  &  77  &  38 $\pm$ 8  &  35.82 $\pm$ 0.03  &  1.65 $\pm$ 0.15  &  Single \\
12  &  273.5444858-17.9302895  &  18:14:10.68  &  -17:55:49.04  &  1.5  &  1.37  &  72  &  27 $\pm$ 5  &  36.87 $\pm$ 0.04  &  1.22 $\pm$ 0.17  &  Single \\
13  &  273.5484365-17.9412478  &  18:14:11.62  &  -17:56:28.49  &  1.75  &  1.54  &  17  &  27 $\pm$ 5  &  35.4 $\pm$ 0.05  &  0.84 $\pm$ 0.17  &  Single \\
16  &  273.5526754-17.9285675  &  18:14:12.64  &  -17:55:42.84  &  1.35  &  1.07  &  67  &  38 $\pm$ 8  &  33.87 $\pm$ 0.03  &  1.26 $\pm$ 0.13  &  Single \\
19  &  273.5561582-17.9212927  &  18:14:13.48  &  -17:55:16.65  &  1.78  &  1.46  &  176  &  33 $\pm$ 7  &  34.78 $\pm$ 0.06  &  1.43 $\pm$ 0.26  &  Single \\
20  &  273.5569282-17.9233073  &  18:14:13.66  &  -17:55:23.91  &  1.82  &  1.27  &  41  &  35 $\pm$ 7  &  36.27 $\pm$ 0.08  &  1.28 $\pm$ 0.33  &  Single \\
22  &  273.5661028-17.9234199  &  18:14:15.86  &  -17:55:24.31  &  1.56  &  1.18  &  54  &  28 $\pm$ 6  &  35.44 $\pm$ 0.02  &  1.06 $\pm$ 0.05  &  Single \\
23  &  273.5464417-17.9286205  &  18:14:11.15  &  -17:55:43.03  &  1.68  &  1.35  &  61  &  29 $\pm$ 6  &  37.05 $\pm$ 0.02  &  1.05 $\pm$ 0.07  &  Single \\
24  &  273.5689656-17.9390182  &  18:14:16.55  &  -17:56:20.47  &  2.1  &  1.31  &  68  &  32 $\pm$ 6  &  37.31 $\pm$ 0.08  &  0.92 $\pm$ 0.27  &  Single \\
25  &  273.5690243-17.9247231  &  18:14:16.57  &  -17:55:29.00  &  1.85  &  1.23  &  90  &  26 $\pm$ 5  &  36.09 $\pm$ 0.03  &  0.71 $\pm$ 0.09  &  Single \\
26  &  273.5419973-17.932744  &  18:14:10.08  &  -17:55:57.88  &  1.98  &  1.29  &  57  &  25 $\pm$ 5  &  --  &  --  &  -- \\
27  &  273.5561592-17.9348499  &  18:14:13.48  &  -17:56:5.460  &  1.97  &  1.82  &  171  &  30 $\pm$ 6  &  37.53 $\pm$ 0.02  &  0.79 $\pm$ 0.09  &  Single \\
29  &  273.5516652-17.9189507  &  18:14:12.40  &  -17:55:8.220  &  1.91  &  1.67  &  112  &  27 $\pm$ 5  &  35.87 $\pm$ 0.09  &  1.31 $\pm$ 0.27  &  Single \\
30  &  273.5576077-17.9311763  &  18:14:13.83  &  -17:55:52.23  &  1.68  &  1.49  &  16  &  32 $\pm$ 7  &  40.0 $\pm$ 0.07  &  2.2 $\pm$ 0.75  &  Single \\
31  &  273.5676865-17.9204887  &  18:14:16.24  &  -17:55:13.76  &  2.18  &  1.55  &  158  &  26 $\pm$ 5  &  33.52 $\pm$ 0.07  &  0.98 $\pm$ 0.19  &  Single \\
32  &  273.5470383-17.9333909  &  18:14:11.29  &  -17:56:0.210  &  1.6  &  1.32  &  30  &  27 $\pm$ 5  &  35.39 $\pm$ 0.08  &  1.19 $\pm$ 0.28  &  Single \\
33  &  273.5581592-17.9252346  &  18:14:13.96  &  -17:55:30.84  &  2.16  &  1.68  &  25  &  34 $\pm$ 7  &  35.31 $\pm$ 0.02  &  0.89 $\pm$ 0.07  &  Single \\
34  &  273.5465089-17.9276804  &  18:14:11.16  &  -17:55:39.65  &  1.63  &  1.42  &  37  &  29 $\pm$ 6  &  37.57 $\pm$ 0.02  &  1.1 $\pm$ 0.06  &  Single \\
36  &  273.5569394-17.9239001  &  18:14:13.67  &  -17:55:26.04  &  1.71  &  1.45  &  68  &  35 $\pm$ 7  &  35.82 $\pm$ 0.03  &  1.39 $\pm$ 0.14  &  Single \\
39  &  273.5665414-17.9228278  &  18:14:15.97  &  -17:55:22.18  &  1.8  &  1.36  &  50  &  27 $\pm$ 5  &  35.17 $\pm$ 0.02  &  1.02 $\pm$ 0.05  &  Single \\
40  &  273.5457926-17.9284661  &  18:14:10.99  &  -17:55:42.48  &  2.12  &  1.76  &  40  &  28 $\pm$ 6  &  37.31 $\pm$ 0.02  &  1.53 $\pm$ 0.09  &  Single \\
41  &  273.5662283-17.9256745  &  18:14:15.89  &  -17:55:32.43  &  1.69  &  1.47  &  48  &  29 $\pm$ 6  &  35.02 $\pm$ 0.07  &  1.15 $\pm$ 0.28  &  Single \\
42  &  273.564807-17.9262439  &  18:14:15.55  &  -17:55:34.48  &  2.01  &  1.65  &  98  &  32 $\pm$ 6  &  36.09 $\pm$ 0.06  &  1.34 $\pm$ 0.17  &  Single \\
44  &  273.5714882-17.9249465  &  18:14:17.16  &  -17:55:29.81  &  1.37  &  1.0  &  73  &  26 $\pm$ 5  &  --  &  --  &  -- \\
45  &  273.5557839-17.9182388  &  18:14:13.39  &  -17:55:5.660  &  2.75  &  2.44  &  7  &  28 $\pm$ 6  &  35.99 $\pm$ 0.08  &  0.74 $\pm$ 0.24  &  Single \\
46  &  273.5579243-17.9306652  &  18:14:13.90  &  -17:55:50.39  &  1.83  &  1.75  &  45  &  33 $\pm$ 7  &  39.71 $\pm$ 0.05  &  1.33 $\pm$ 0.1  &  Single \\
47  &  273.5479984-17.9272905  &  18:14:11.52  &  -17:55:38.25  &  2.22  &  1.72  &  44  &  30 $\pm$ 6  &  35.98 $\pm$ 0.02  &  0.62 $\pm$ 0.06  &  Single \\
51  &  273.5458037-17.9289719  &  18:14:10.99  &  -17:55:44.30  &  1.75  &  1.43  &  77  &  28 $\pm$ 6  &  37.23 $\pm$ 0.02  &  1.19 $\pm$ 0.07  &  Single \\
52  &  273.5503813-17.9228494  &  18:14:12.09  &  -17:55:22.26  &  1.6  &  1.12  &  67  &  30 $\pm$ 6  &  34.41 $\pm$ 0.05  &  0.72 $\pm$ 0.17  &  Single \\
53  &  273.5463654-17.942157  &  18:14:11.13  &  -17:56:31.77  &  1.84  &  1.37  &  34  &  26 $\pm$ 5  &  36.48 $\pm$ 0.07  &  0.6 $\pm$ 0.24  &  Single \\
55  &  273.5551571-17.9357044  &  18:14:13.24  &  -17:56:8.540  &  2.55  &  1.99  &  136  &  29 $\pm$ 6  &  --  &  --  &  -- \\
57  &  273.5614471-17.9187094  &  18:14:14.75  &  -17:55:7.350  &  2.34  &  1.89  &  61  &  30 $\pm$ 6  &  35.71 $\pm$ 0.02  &  0.74 $\pm$ 0.09  &  Single \\
58  &  273.559544-17.9178954  &  18:14:14.29  &  -17:55:4.420  &  1.59  &  1.45  &  158  &  30 $\pm$ 6  &  34.27 $\pm$ 0.07  &  1.59 $\pm$ 0.33  &  Single \\
59  &  273.5529774-17.9374942  &  18:14:12.71  &  -17:56:14.98  &  1.71  &  1.46  &  117  &  29 $\pm$ 6  &  --  &  --  &  -- \\
61  &  273.5475497-17.92664  &  18:14:11.41  &  -17:55:35.90  &  1.69  &  1.04  &  60  &  30 $\pm$ 6  &  37.8 $\pm$ 0.02  &  1.32 $\pm$ 0.09  &  Single \\
62  &  273.5533917-17.9280329  &  18:14:12.81  &  -17:55:40.92  &  2.13  &  1.63  &  94  &  38 $\pm$ 8  &  34.03 $\pm$ 0.02  &  0.87 $\pm$ 0.07  &  Single \\
65  &  273.5597707-17.918686  &  18:14:14.34  &  -17:55:7.270  &  1.7  &  1.52  &  51  &  32 $\pm$ 6  &  34.85 $\pm$ 0.06  &  1.65 $\pm$ 0.25  &  Single \\
66  &  273.5477174-17.9307289  &  18:14:11.45  &  -17:55:50.62  &  2.13  &  1.44  &  75  &  29 $\pm$ 6  &  --  &  --  &  -- \\
\hline \noalign {\smallskip}
\end{tabular}
\end{table*}

\begin{figure*}
    \centering
\includegraphics[width=0.24\textwidth]{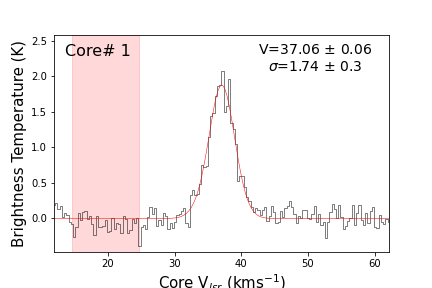}
\includegraphics[width=0.24\textwidth]{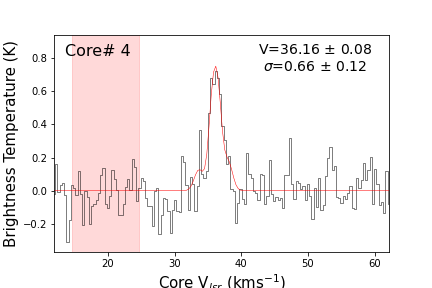}
\includegraphics[width=0.24\textwidth]{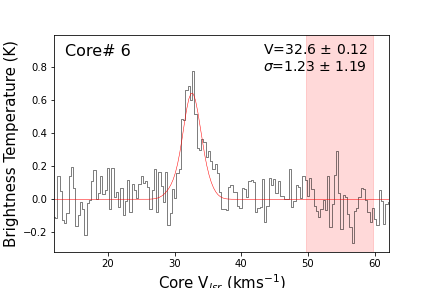}
\includegraphics[width=0.24\textwidth]{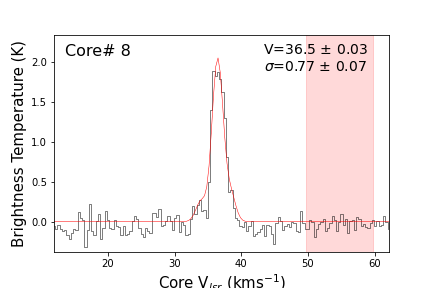}
\includegraphics[width=0.24\textwidth]{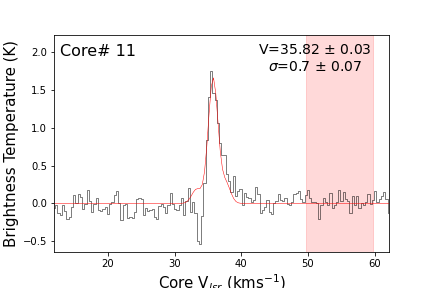}
\includegraphics[width=0.24\textwidth]{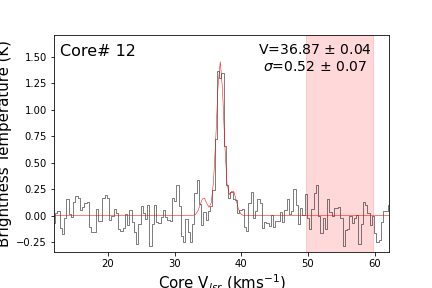}
\includegraphics[width=0.24\textwidth]{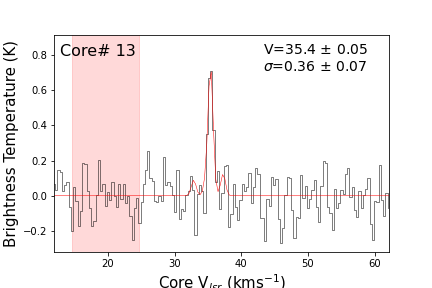}
\includegraphics[width=0.24\textwidth]{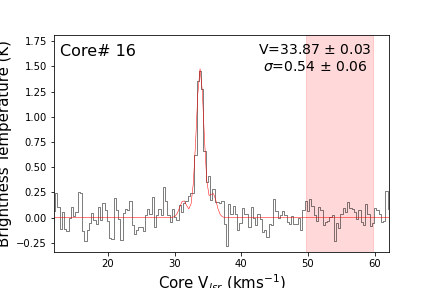}
\includegraphics[width=0.24\textwidth]{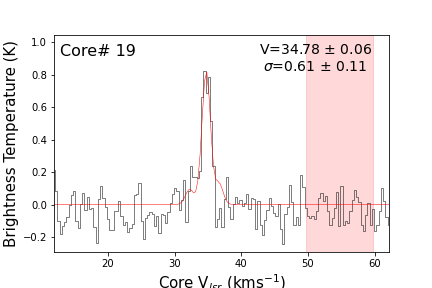}
\includegraphics[width=0.24\textwidth]{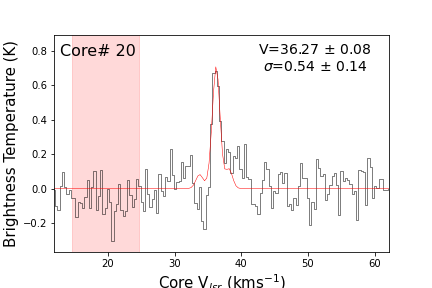}
\includegraphics[width=0.24\textwidth]{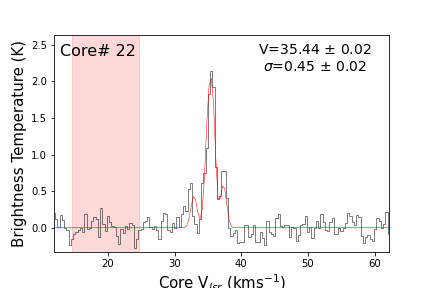}
\includegraphics[width=0.24\textwidth]{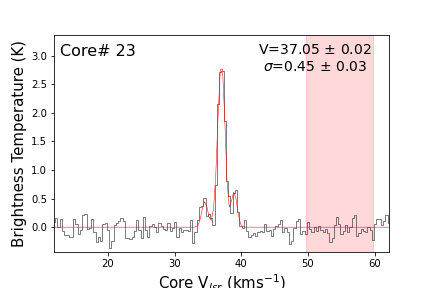}
\includegraphics[width=0.24\textwidth]{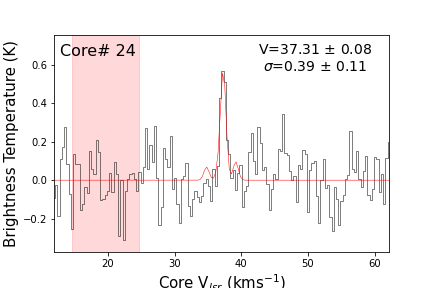}
\includegraphics[width=0.24\textwidth]{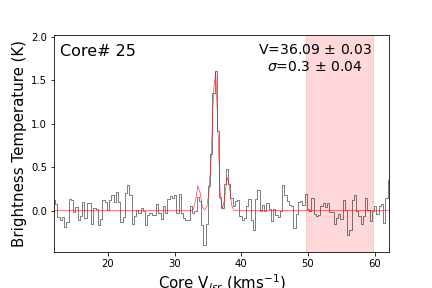}
\includegraphics[width=0.24\textwidth]{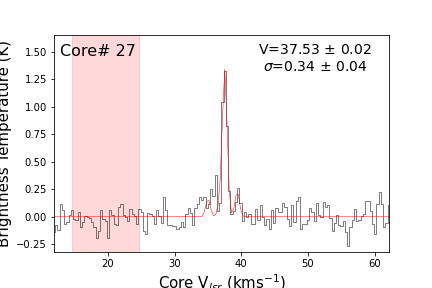}
\includegraphics[width=0.24\textwidth]{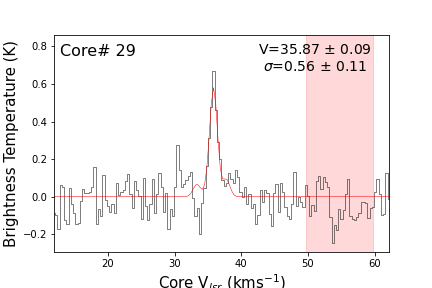}
\includegraphics[width=0.24\textwidth]{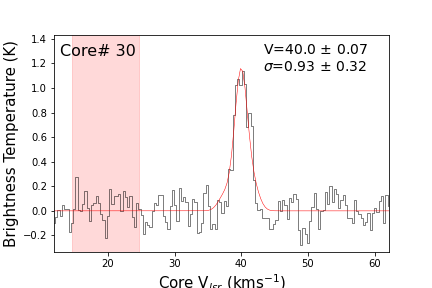}
\includegraphics[width=0.24\textwidth]{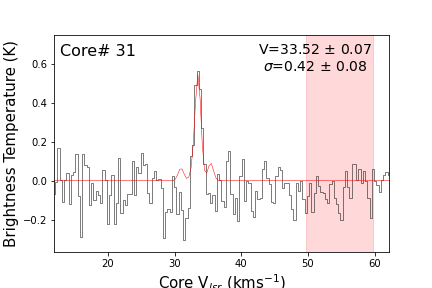}
\includegraphics[width=0.24\textwidth]{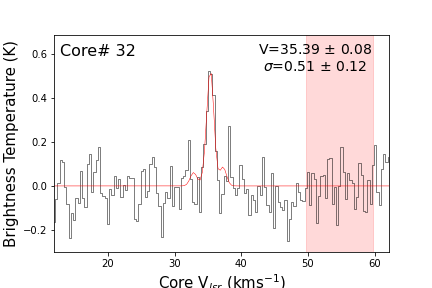}
\includegraphics[width=0.24\textwidth]{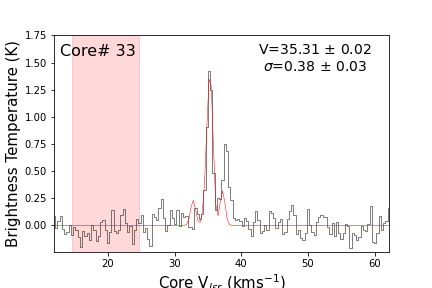}
\includegraphics[width=0.24\textwidth]{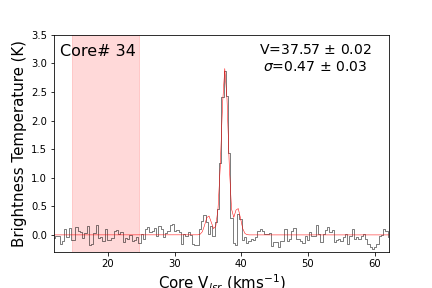}
\includegraphics[width=0.24\textwidth]{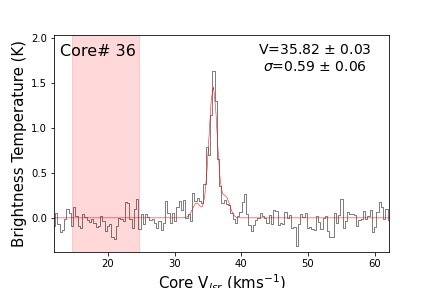}
\includegraphics[width=0.24\textwidth]{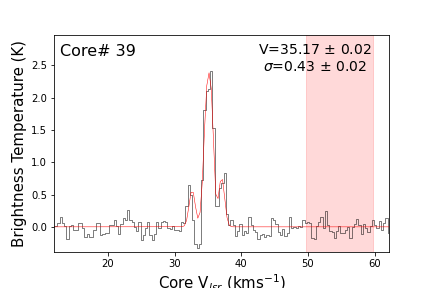}
\includegraphics[width=0.24\textwidth]{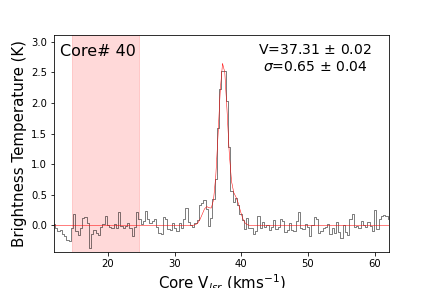}
\includegraphics[width=0.24\textwidth]{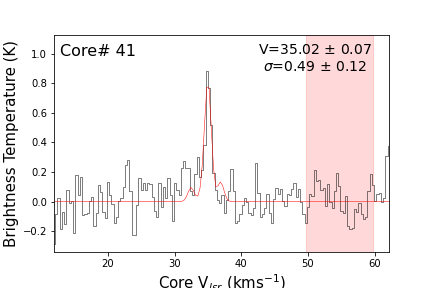}
\includegraphics[width=0.24\textwidth]{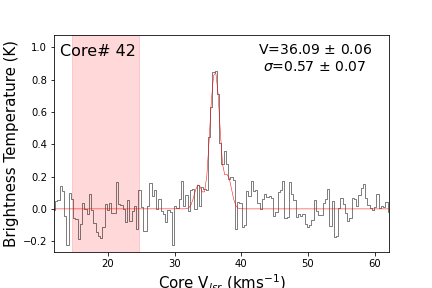}
\includegraphics[width=0.24\textwidth]{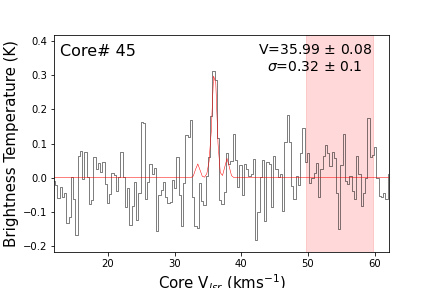}
\includegraphics[width=0.24\textwidth]{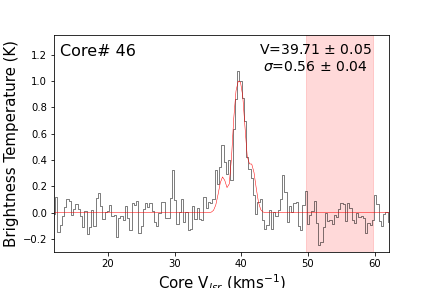}

\caption{Single-type core-averaged, background-subtracted DCN spectra extracted from the cores in the evolved protocluster G012.80. See \cref{tabappendix:coretables_g012} for the line fit parameters for each core.\label{figspectra:dcnspectra_split_G012_s}}
\end{figure*}

\begin{figure*}\ContinuedFloat
    \centering
\includegraphics[width=0.24\textwidth]{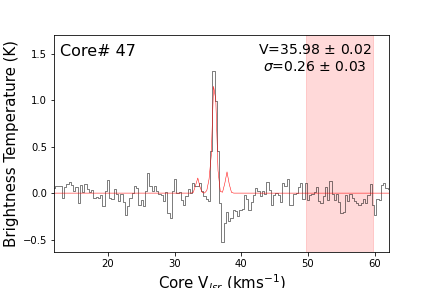}
\includegraphics[width=0.24\textwidth]{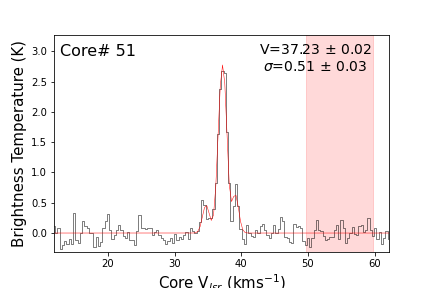}
\includegraphics[width=0.24\textwidth]{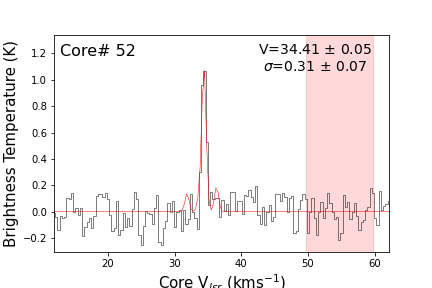}
\includegraphics[width=0.24\textwidth]{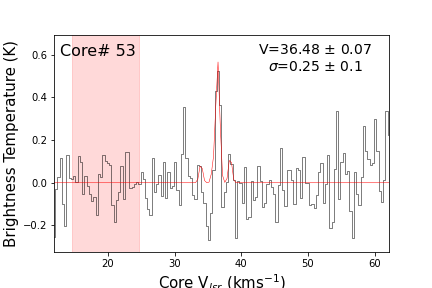}

\includegraphics[width=0.24\textwidth]{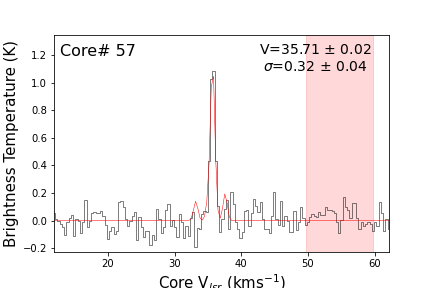}
\includegraphics[width=0.24\textwidth]{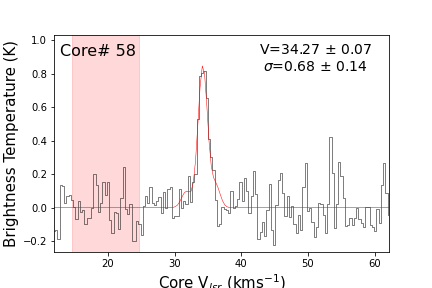}
\includegraphics[width=0.24\textwidth]{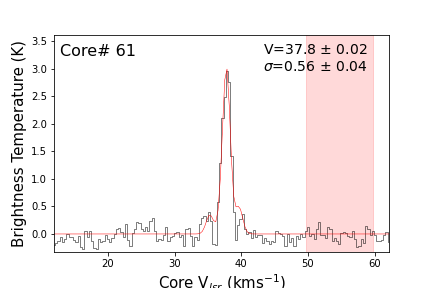}
\includegraphics[width=0.24\textwidth]{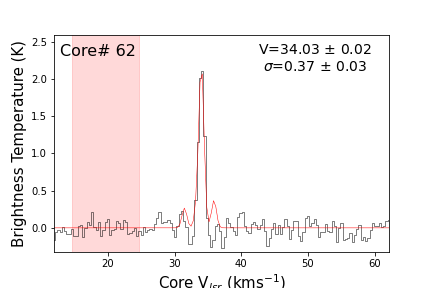}
\includegraphics[width=0.24\textwidth]{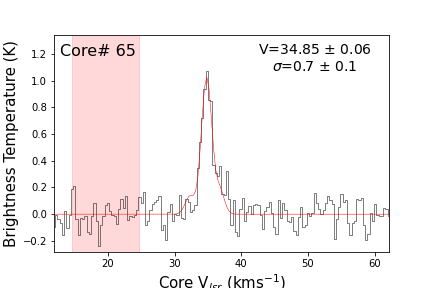}

\caption{Continued: Single-type core-averaged, background-subtracted DCN spectra extracted from the cores in the evolved protocluster G012.80. See \cref{tabappendix:coretables_g012} for the line fit parameters for each core.}
\end{figure*}

\begin{figure*}
    \centering

\includegraphics[width=0.24\textwidth]{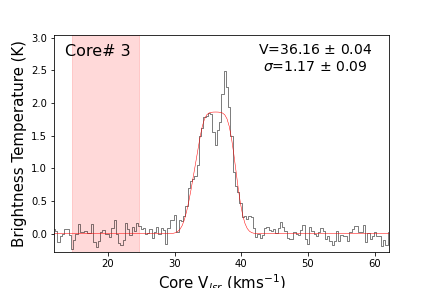}

\caption{Complex-type core-averaged, background-subtracted DCN spectra extracted from the cores in the evolved protocluster G012.80. See \cref{tabappendix:coretables_g012} for the line fit parameters for each core.\label{figspectra:dcnspectra_split_G012_c}} 
\end{figure*}

\begin{table*}[htbp!]
\centering
\small
\caption{DCN fits towards the core population of the young protocluster G327.29.}
\label{tabappendix:coretables_g327}
\begin{tabular}{llllllllccc}

\hline 
n   & Core Name &RA  & DEC & F$_{A}$   & F$_{B}$    & PA & T & \vlsr  & Linewidth   & Spectral \\
 &  & [ICRS]  &  [ICRS]   &  [\arcsec] &[\arcsec] &  [deg] & [K] & [\kms]  & [\kms] & Type \\
\hline 
1  &  238.282335-54.618403  &  15:53:7.760  &  -54:37:6.250  &  2.14  &  2.09  &  164  &  100 $\pm$ 50  &  -42.87 $\pm$ 0.13  &  --  &  Complex \\
2  &  238.2894812-54.6168557  &  15:53:9.480  &  -54:37:0.680  &  1.78  &  1.5  &  152  &  28 $\pm$ 6  &  --  &  --  &  -- \\
3  &  238.280656-54.6186811  &  15:53:7.360  &  -54:37:7.250  &  1.62  &  1.47  &  62  &  36 $\pm$ 7  &  -42.18 $\pm$ 0.05  &  --  &  Complex \\
4  &  238.2936982-54.6114916  &  15:53:10.49  &  -54:36:41.37  &  1.2  &  1.19  &  7  &  25 $\pm$ 5  &  -44.26 $\pm$ 0.1  &  1.36 $\pm$ 0.4  &  Single \\
5  &  238.2955904-54.6128047  &  15:53:10.94  &  -54:36:46.10  &  1.7  &  1.45  &  64  &  27 $\pm$ 5  &  -47.08 $\pm$ 0.03  &  1.76 $\pm$ 0.09  &  Single \\
6  &  238.2938644-54.6110507  &  15:53:10.53  &  -54:36:39.78  &  1.24  &  1.19  &  53  &  25 $\pm$ 5  &  -44.57 $\pm$ 0.06  &  1.06 $\pm$ 0.21  &  Single \\
7  &  238.2961992-54.6133521  &  15:53:11.09  &  -54:36:48.07  &  1.6  &  1.41  &  129  &  26 $\pm$ 5  &  -47.79 $\pm$ 0.05  &  1.98 $\pm$ 0.22  &  Single \\
8  &  238.2983074-54.6069915  &  15:53:11.59  &  -54:36:25.17  &  1.32  &  1.19  &  16  &  26 $\pm$ 5  &  -47.41 $\pm$ 0.08  &  1.21 $\pm$ 0.28  &  Single \\
9  &  238.2858254-54.6172552  &  15:53:8.600  &  -54:37:2.120  &  1.39  &  1.13  &  73  &  29 $\pm$ 6  &  -44.42 $\pm$ 0.04  &  0.95 $\pm$ 0.15  &  Single \\
10  &  238.2605463-54.6233229  &  15:53:2.530  &  -54:37:23.96  &  1.16  &  1.04  &  134  &  24 $\pm$ 5  &  --  &  --  &  -- \\
11  &  238.2800319-54.6194279  &  15:53:7.210  &  -54:37:9.940  &  2.35  &  1.32  &  32  &  33 $\pm$ 7  &  -42.11 $\pm$ 0.02  &  1.1 $\pm$ 0.06  &  Single \\
12  &  238.2838451-54.6216681  &  15:53:8.120  &  -54:37:18.01  &  1.72  &  1.38  &  59  &  24 $\pm$ 5  &  --  &  --  &  -- \\
13  &  238.2903446-54.6123086  &  15:53:9.680  &  -54:36:44.31  &  1.35  &  1.26  &  36  &  24 $\pm$ 5  &  --  &  --  &  -- \\
14  &  238.2971584-54.609653  &  15:53:11.32  &  -54:36:34.75  &  1.34  &  1.26  &  163  &  25 $\pm$ 5  &  -47.17 $\pm$ 0.06  &  1.26 $\pm$ 0.21  &  Single \\
15  &  238.2867749-54.6166625  &  15:53:8.830  &  -54:36:59.98  &  1.76  &  1.44  &  47  &  28 $\pm$ 6  &  -43.64 $\pm$ 0.03  &  1.76 $\pm$ 0.15  &  Single \\
16  &  238.2891016-54.6143036  &  15:53:9.380  &  -54:36:51.49  &  1.64  &  1.47  &  114  &  25 $\pm$ 5  &  --  &  --  &  -- \\
17  &  238.2893864-54.6162879  &  15:53:9.450  &  -54:36:58.64  &  1.49  &  1.21  &  114  &  27 $\pm$ 6  &  -45.32 $\pm$ 0.2  &  --  &  Complex \\
18  &  238.2800971-54.6177887  &  15:53:7.220  &  -54:37:4.040  &  1.14  &  1.03  &  76  &  32 $\pm$ 7  &  --  &  --  &  -- \\
19  &  238.2810557-54.62136  &  15:53:7.450  &  -54:37:16.90  &  1.99  &  1.76  &  113  &  25 $\pm$ 5  &  -42.75 $\pm$ 0.07  &  1.01 $\pm$ 0.23  &  Single \\
20  &  238.2935252-54.6140562  &  15:53:10.45  &  -54:36:50.60  &  1.36  &  1.1  &  26  &  24 $\pm$ 5  &  --  &  --  &  -- \\
21  &  238.2959968-54.6118908  &  15:53:11.04  &  -54:36:42.81  &  1.18  &  1.12  &  15  &  25 $\pm$ 5  &  -46.74 $\pm$ 0.04  &  0.9 $\pm$ 0.15  &  Single \\
22  &  238.2884448-54.6163242  &  15:53:9.230  &  -54:36:58.77  &  1.83  &  1.64  &  171  &  27 $\pm$ 6  &  -45.19 $\pm$ 0.05  &  1.49 $\pm$ 0.15  &  Single \\
23  &  238.2946778-54.6130334  &  15:53:10.72  &  -54:36:46.92  &  1.37  &  1.29  &  113  &  26 $\pm$ 5  &  -46.49 $\pm$ 0.13  &  --  &  Complex \\
24  &  238.2656528-54.6238211  &  15:53:3.760  &  -54:37:25.76  &  3.17  &  2.47  &  59  &  24 $\pm$ 5  &  --  &  --  &  -- \\
25  &  238.2865705-54.6187657  &  15:53:8.780  &  -54:37:7.560  &  2.44  &  2.21  &  17  &  28 $\pm$ 6  &  -42.85 $\pm$ 0.05  &  1.62 $\pm$ 0.13  &  Single \\
26  &  238.2966528-54.615169  &  15:53:11.20  &  -54:36:54.61  &  2.24  &  1.92  &  16  &  23 $\pm$ 5  &  --  &  --  &  -- \\
27  &  238.2945683-54.6089436  &  15:53:10.70  &  -54:36:32.20  &  1.24  &  1.02  &  8  &  26 $\pm$ 5  &  --  &  --  &  -- \\
28  &  238.2889899-54.6256413  &  15:53:9.360  &  -54:37:32.31  &  3.44  &  2.78  &  138  &  22 $\pm$ 5  &  --  &  --  &  -- \\
29  &  238.2902959-54.6140325  &  15:53:9.670  &  -54:36:50.52  &  1.66  &  1.56  &  59  &  24 $\pm$ 5  &  --  &  --  &  -- \\
30  &  238.2937964-54.610052  &  15:53:10.51  &  -54:36:36.19  &  1.43  &  1.18  &  88  &  25 $\pm$ 5  &  --  &  --  &  -- \\
31  &  238.2856418-54.6266832  &  15:53:8.550  &  -54:37:36.06  &  3.37  &  2.73  &  159  &  23 $\pm$ 5  &  --  &  --  &  -- \\
34  &  238.285875-54.6233919  &  15:53:8.610  &  -54:37:24.21  &  2.18  &  1.52  &  149  &  23 $\pm$ 5  &  --  &  --  &  -- \\

\hline \noalign {\smallskip}
\end{tabular}
\end{table*}

\begin{figure*}
    \centering
\includegraphics[width=0.24\textwidth]{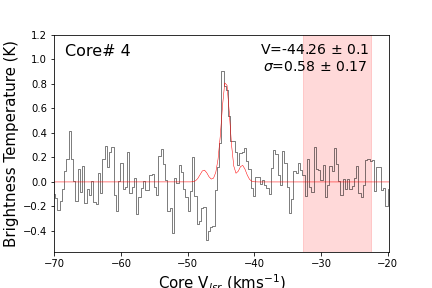}
\includegraphics[width=0.24\textwidth]{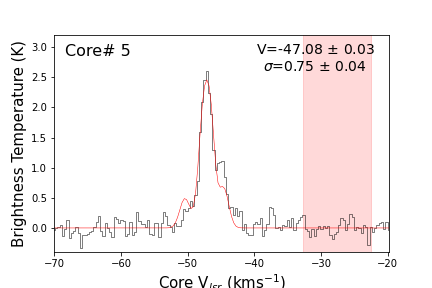}
\includegraphics[width=0.24\textwidth]{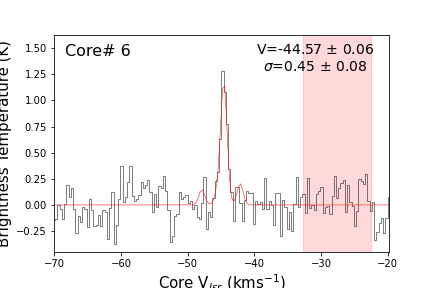}
\includegraphics[width=0.24\textwidth]{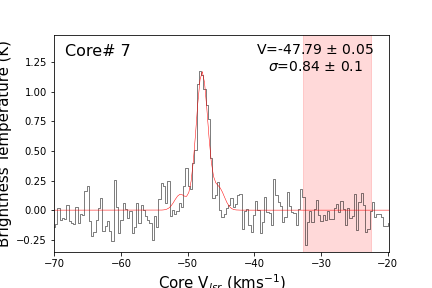}
\includegraphics[width=0.24\textwidth]{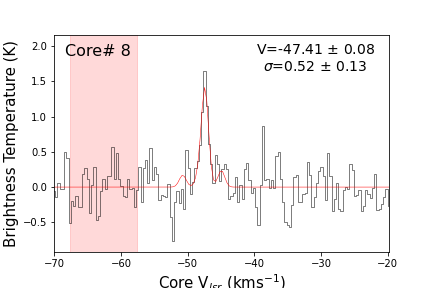}
\includegraphics[width=0.24\textwidth]{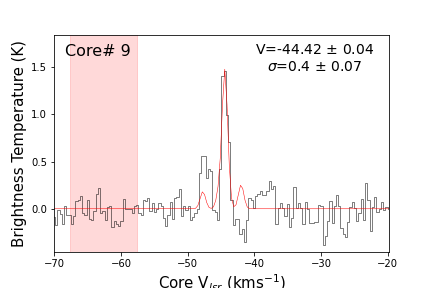}
\includegraphics[width=0.24\textwidth]{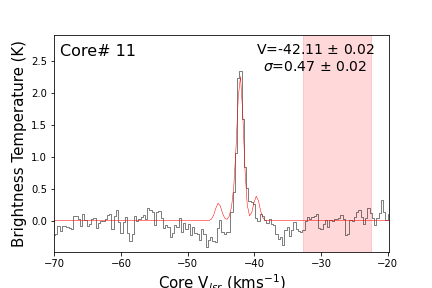}
\includegraphics[width=0.24\textwidth]{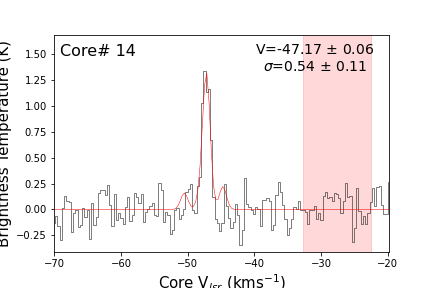}
\includegraphics[width=0.24\textwidth]{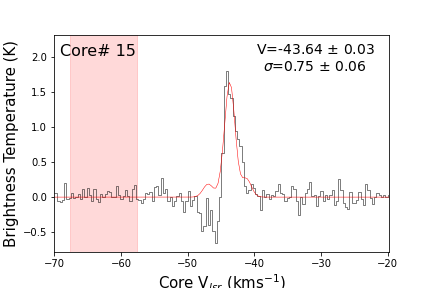}
\includegraphics[width=0.24\textwidth]{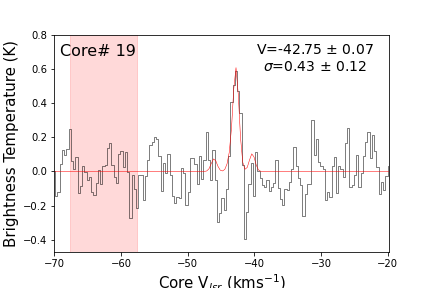}
\includegraphics[width=0.24\textwidth]{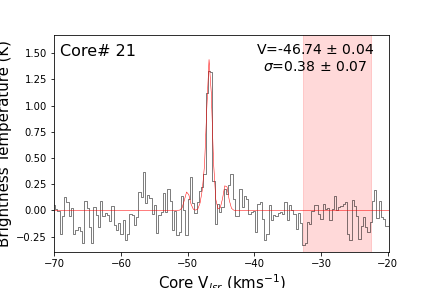}
\includegraphics[width=0.24\textwidth]{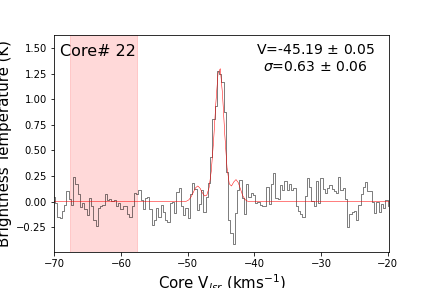}
\includegraphics[width=0.24\textwidth]{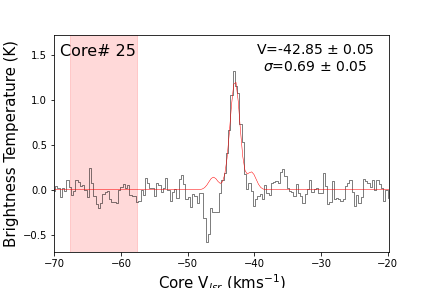}
\caption{ Single-type core-averaged, background-subtracted DCN spectra extracted from the cores in the young protocluster G327.29. See \cref{tabappendix:coretables_g327} for the line fit parameters for each core.\label{figspectra:dcnspectra_split_G327_s}}
\end{figure*}

\begin{figure*}\ContinuedFloat
    \centering

\includegraphics[width=0.24\textwidth]{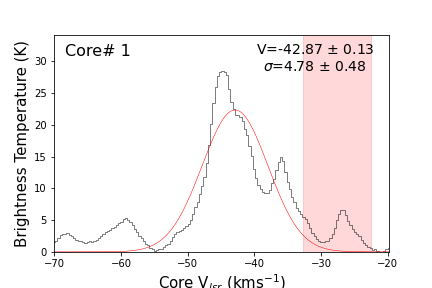}
\includegraphics[width=0.24\textwidth]{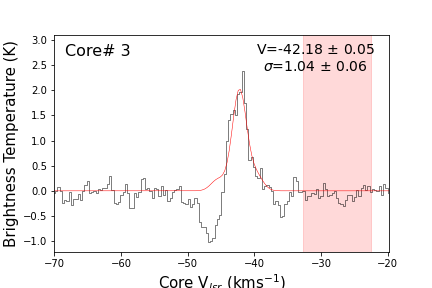}
\includegraphics[width=0.24\textwidth]{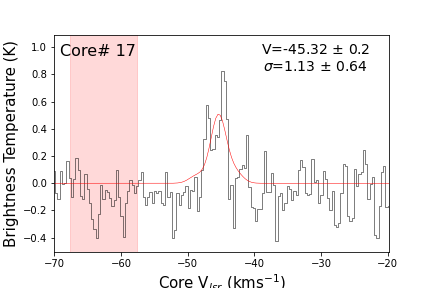}
\includegraphics[width=0.24\textwidth]{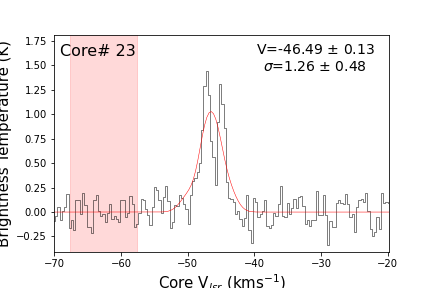}
\caption{Complex-type core-averaged, background-subtracted DCN spectra extracted from the cores in the young protocluster G327.29. See \cref{tabappendix:coretables_g327} for the line fit parameters for each core.\label{figspectra:dcnspectra_split_G327_c}} 
\end{figure*}

\begin{table*}[htbp!]
\centering
\small
\caption{DCN fits towards the core population of the young protocluster G328.25.}
\label{tabappendix:coretables_g328}
\begin{tabular}{llllllllccc}

\hline 
n   & Core Name &RA  & DEC & F$_{A}$   & F$_{B}$    & PA & T & \vlsr  & Linewidth   & Spectral \\
 &  & [ICRS]  &  [ICRS]   &  [\arcsec] &[\arcsec] &  [deg] & [K] & [\kms]  & [\kms] & Type \\
\hline 
1  &  239.4991778-53.9668493  &  15:57:59.80  &  -53:58:0.660  &  1.49  &  1.32  &  29  &  100 $\pm$ 50  &  -43.09 $\pm$ 0.05  &  --  &  Complex \\
2  &  239.4978364-53.9537718  &  15:57:59.48  &  -53:57:13.58  &  1.28  &  1.16  &  12  &  23 $\pm$ 5  &  --  &  --  &  -- \\
3  &  239.5016973-53.954434  &  15:58:0.410  &  -53:57:15.96  &  1.81  &  1.3  &  14  &  23 $\pm$ 5  &  --  &  --  &  -- \\
4  &  239.4941628-53.9528437  &  15:57:58.60  &  -53:57:10.24  &  1.15  &  1.07  &  135  &  22 $\pm$ 4  &  --  &  --  &  -- \\
5  &  239.5209961-53.959408  &  15:58:5.040  &  -53:57:33.87  &  1.32  &  1.19  &  145  &  23 $\pm$ 5  &  --  &  --  &  -- \\
6  &  239.4815006-53.9554302  &  15:57:55.56  &  -53:57:19.55  &  1.34  &  1.07  &  118  &  21 $\pm$ 4  &  --  &  --  &  -- \\
7  &  239.5135243-53.965875  &  15:58:3.250  &  -53:57:57.15  &  1.23  &  1.01  &  175  &  23 $\pm$ 5  &  --  &  --  &  -- \\
8  &  239.5146161-53.968461  &  15:58:3.510  &  -53:58:6.460  &  1.14  &  1.12  &  108  &  23 $\pm$ 5  &  --  &  --  &  -- \\
9  &  239.5000898-53.9663118  &  15:58:0.020  &  -53:57:58.72  &  2.02  &  1.41  &  30  &  36 $\pm$ 7  &  -43.0 $\pm$ 0.02  &  1.21 $\pm$ 0.06  &  Single \\
10  &  239.4979744-53.9667109  &  15:57:59.51  &  -53:58:0.160  &  2.31  &  1.93  &  102  &  37 $\pm$ 8  &  -41.27 $\pm$ 0.09  &  3.69 $\pm$ 1.24  &  Single \\
11  &  239.4964686-53.9654253  &  15:57:59.15  &  -53:57:55.53  &  1.6  &  1.26  &  3  &  33 $\pm$ 7  &  -43.77 $\pm$ 0.06  &  1.11 $\pm$ 0.2  &  Single \\
\hline \noalign {\smallskip}
\end{tabular}
\end{table*}

\begin{figure*}
    \centering
\includegraphics[width=0.24\textwidth]{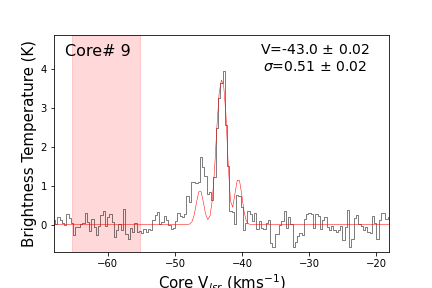}
\includegraphics[width=0.24\textwidth]{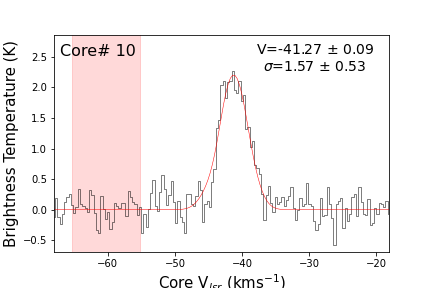}
\includegraphics[width=0.24\textwidth]{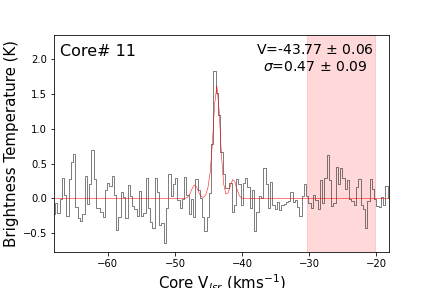}
\caption{Single-type core-averaged, background-subtracted DCN spectra extracted from the cores in the young protocluster G328.25. See \cref{tabappendix:coretables_g328} for the line fit parameters for each core.\label{figspectra:dcnspectra_split_G328_s}}
\end{figure*}

\begin{figure*}
    \centering
\includegraphics[width=0.24\textwidth]{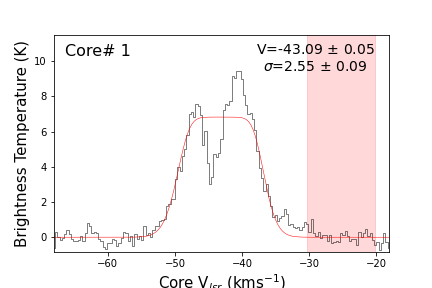}
\caption{Complex-type core-averaged, background-subtracted DCN spectra extracted from the cores in the young protocluster G328.25. See \cref{tabappendix:coretables_g328} for the line fit parameters for each core.\label{figspectra:dcnspectra_split_G328_c}} 
\end{figure*}

\begin{table*}[htbp!]
\centering
\small
\caption{DCN fits towards the core population of the young protocluster G337.92.}
\label{tabappendix:coretables_g337}
\begin{tabular}{llllllllccc}

\hline 
n   & Core Name &RA  & DEC & F$_{A}$   & F$_{B}$    & PA & T & \vlsr  & Linewidth   & Spectral \\
 &  & [ICRS]  &  [ICRS]   &  [\arcsec] &[\arcsec] &  [deg] & [K] & [\kms]  & [\kms] & Type \\
\hline 
1  &  250.2935548-47.1342727  &  16:41:10.45  &  -47:08:3.380  &  1.61  &  1.23  &  105  &  100 $\pm$ 50  &  -41.92 $\pm$ 0.02  &  --  &  Complex \\
2  &  250.2936172-47.133748  &  16:41:10.47  &  -47:08:1.490  &  1.85  &  1.27  &  69  &  35 $\pm$ 10  &  -38.58 $\pm$ 0.01  &  --  &  Complex \\
3  &  250.2949021-47.1343006  &  16:41:10.78  &  -47:08:3.480  &  1.51  &  1.35  &  170  &  33 $\pm$ 7  &  0.0 $\pm$ 0.0  &  --  &  Complex \\
4  &  250.2922563-47.1415274  &  16:41:10.14  &  -47:08:29.50  &  1.22  &  1.06  &  73  &  18 $\pm$ 4  &  --  &  --  &  -- \\
5  &  250.295556-47.134715  &  16:41:10.93  &  -47:08:4.970  &  1.8  &  1.46  &  124  &  30 $\pm$ 6  &  -36.83 $\pm$ 0.02  &  3.61 $\pm$ 0.34  &  Single \\
6  &  250.2913536-47.1405094  &  16:41:9.920  &  -47:08:25.83  &  1.19  &  1.15  &  8  &  20 $\pm$ 4  &  --  &  --  &  -- \\
7  &  250.2794944-47.1323197  &  16:41:7.080  &  -47:07:56.35  &  1.3  &  1.01  &  115  &  19 $\pm$ 4  &  -40.4 $\pm$ 0.07  &  1.66 $\pm$ 0.29  &  Single \\
8  &  250.2927529-47.1352673  &  16:41:10.26  &  -47:08:6.960  &  1.46  &  1.21  &  122  &  31 $\pm$ 6  &  -39.36 $\pm$ 0.01  &  --  &  Complex \\
9  &  250.294383-47.1348356  &  16:41:10.65  &  -47:08:5.410  &  1.52  &  1.41  &  162  &  33 $\pm$ 7  &  -39.91 $\pm$ 0.07  &  --  &  Complex \\
10  &  250.2841444-47.1340494  &  16:41:8.190  &  -47:08:2.580  &  2.14  &  1.24  &  83  &  22 $\pm$ 5  &  -40.69 $\pm$ 0.03  &  1.16 $\pm$ 0.09  &  Single \\
12  &  250.2947667-47.1443172  &  16:41:10.74  &  -47:08:39.54  &  1.29  &  1.09  &  122  &  18 $\pm$ 4  &  --  &  --  &  -- \\
13  &  250.293043-47.1316977  &  16:41:10.33  &  -47:07:54.11  &  2.23  &  1.68  &  43  &  28 $\pm$ 6  &  -40.04 $\pm$ 0.04  &  3.09 $\pm$ 0.59  &  Single \\
14  &  250.2952101-47.1340394  &  16:41:10.85  &  -47:08:2.540  &  1.23  &  1.06  &  111  &  31 $\pm$ 6  &  -37.78 $\pm$ 0.04  &  --  &  Complex \\
15  &  250.2920694-47.1443736  &  16:41:10.10  &  -47:08:39.74  &  1.22  &  1.06  &  37  &  18 $\pm$ 4  &  --  &  --  &  -- \\
16  &  250.2886082-47.1330445  &  16:41:9.270  &  -47:07:58.96  &  1.09  &  1.03  &  15  &  26 $\pm$ 5  &  --  &  --  &  -- \\
17  &  250.294635-47.1334957  &  16:41:10.71  &  -47:08:0.580  &  1.81  &  1.05  &  66  &  31 $\pm$ 6  &  -38.12 $\pm$ 0.03  &  4.5 $\pm$ 0.5  &  Single \\
18  &  250.2879462-47.1399751  &  16:41:9.110  &  -47:08:23.91  &  1.71  &  1.44  &  46  &  21 $\pm$ 4  &  -38.22 $\pm$ 0.16  &  2.18 $\pm$ 0.75  &  Single \\
19  &  250.2789162-47.1329451  &  16:41:6.940  &  -47:07:58.60  &  1.63  &  1.14  &  8  &  20 $\pm$ 4  &  -39.91 $\pm$ 0.04  &  0.54 $\pm$ 0.14  &  Single \\
20  &  250.2867777-47.1382501  &  16:41:8.830  &  -47:08:17.70  &  1.82  &  1.41  &  165  &  22 $\pm$ 5  &  -38.4 $\pm$ 0.13  &  --  &  Complex \\
21  &  250.2897651-47.1412616  &  16:41:9.540  &  -47:08:28.54  &  1.21  &  0.92  &  2  &  19 $\pm$ 4  &  --  &  --  &  -- \\
23  &  250.2913342-47.1396608  &  16:41:9.920  &  -47:08:22.78  &  1.4  &  1.26  &  19  &  22 $\pm$ 4  &  --  &  --  &  -- \\
25  &  250.2914162-47.134249  &  16:41:9.940  &  -47:08:3.300  &  1.61  &  1.36  &  174  &  30 $\pm$ 6  &  -39.43 $\pm$ 0.02  &  1.8 $\pm$ 0.07  &  Single \\
\hline \noalign {\smallskip}
\end{tabular}
\end{table*}

\begin{figure*}
    \centering
\includegraphics[width=0.24\textwidth]{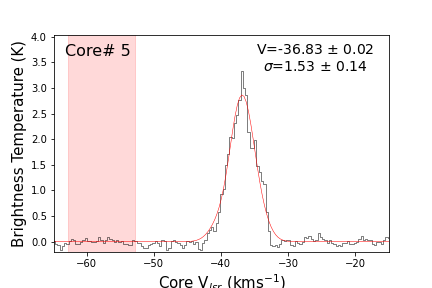}
\includegraphics[width=0.24\textwidth]{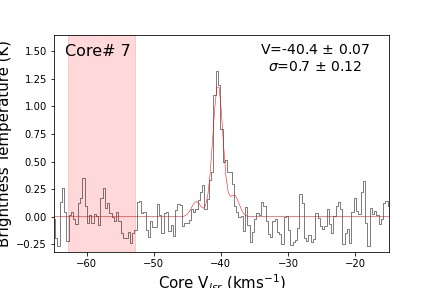}
\includegraphics[width=0.24\textwidth]{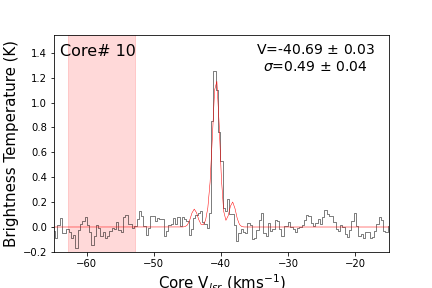}
\includegraphics[width=0.24\textwidth]{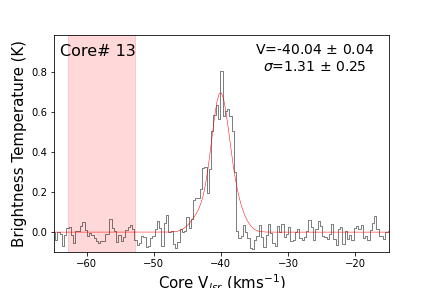}
\includegraphics[width=0.24\textwidth]{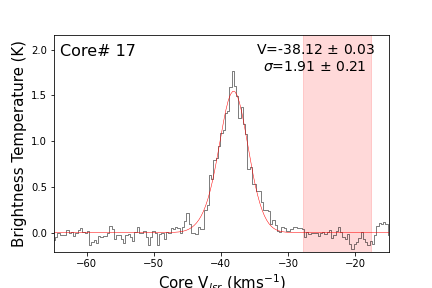}
\includegraphics[width=0.24\textwidth]{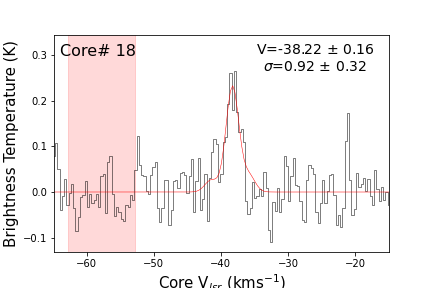}
\includegraphics[width=0.24\textwidth]{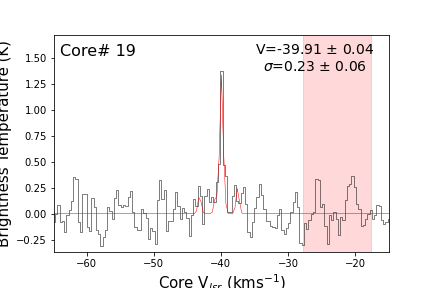}
\includegraphics[width=0.24\textwidth]{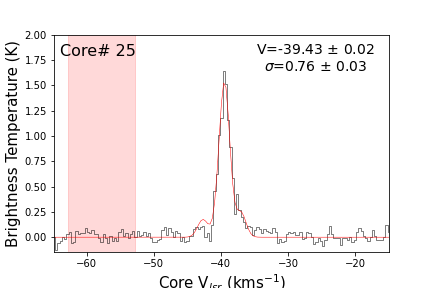}

\caption{Single-type core-averaged, background-subtracted DCN spectra extracted from the cores in the young protocluster G337.92. See \cref{tabappendix:coretables_g337} for the line fit parameters for each core.\label{figspectra:dcnspectra_split_G337_s}}
\end{figure*}

\begin{figure*}
    \centering
\includegraphics[width=0.24\textwidth]{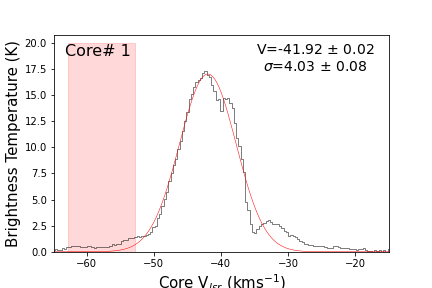}
\includegraphics[width=0.24\textwidth]{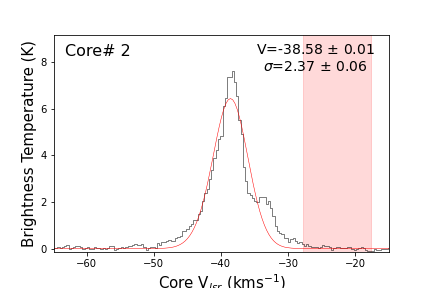}
\includegraphics[width=0.24\textwidth]{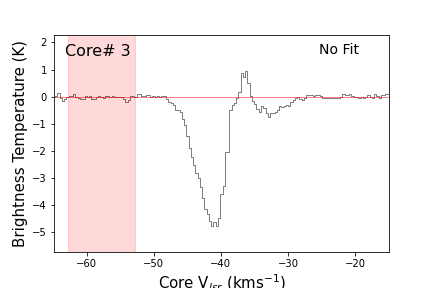}
\includegraphics[width=0.24\textwidth]{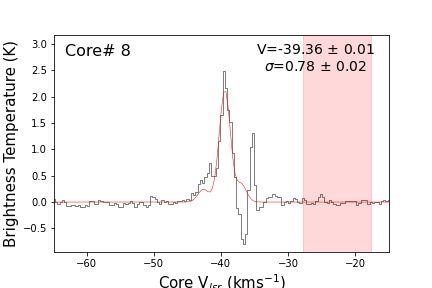}
\includegraphics[width=0.24\textwidth]{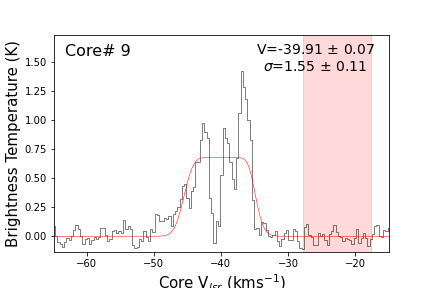}
\includegraphics[width=0.24\textwidth]{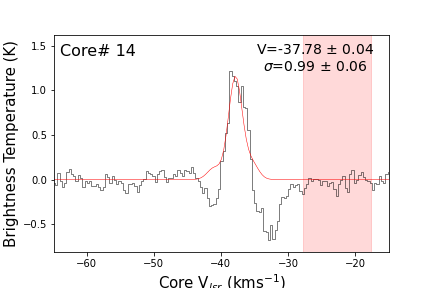}
\includegraphics[width=0.24\textwidth]{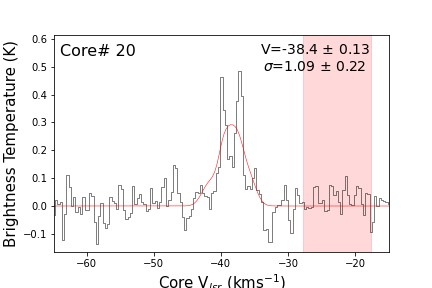}

\caption{Complex-type core-averaged, background-subtracted DCN spectra extracted from the cores in the young protocluster G337.92. See \cref{tabappendix:coretables_g337} for the line fit parameters for each core.\label{figspectra:dcnspectra_split_G337_c}} 
\end{figure*}
\clearpage
\begin{table*}[htbp!]
\centering
\small
\caption{DCN fits towards the core population of the young protocluster G338.93.}
\label{tabappendix:coretables_g338}
\begin{tabular}{llllllllccc}

\hline 
n   & Core Name &RA  & DEC & F$_{A}$   & F$_{B}$    & PA & T & \vlsr  & Linewidth   & Spectral \\
 &  & [ICRS]  &  [ICRS]   &  [\arcsec] &[\arcsec] &  [deg] & [K] & [\kms]  & [\kms] & Type \\
\hline 
1  &  250.1422011-45.6934028  &  16:40:34.13  &  -45:41:36.25  &  0.89  &  0.79  &  84  &  100 $\pm$ 50  &  -57.34 $\pm$ 0.08  &  --  &  Complex \\
2  &  250.1427268-45.6936217  &  16:40:34.25  &  -45:41:37.04  &  0.97  &  0.73  &  124  &  100 $\pm$ 50  &  -58.71 $\pm$ 0.08  &  --  &  Complex \\
3  &  250.1417044-45.7020244  &  16:40:34.01  &  -45:42:7.290  &  0.86  &  0.79  &  92  &  100 $\pm$ 50  &  -62.49 $\pm$ 0.03  &  --  &  Complex \\
4  &  250.1397543-45.6936984  &  16:40:33.54  &  -45:41:37.31  &  0.86  &  0.82  &  7  &  100 $\pm$ 50  &  -60.37 $\pm$ 0.04  &  --  &  Complex \\
5  &  250.1417392-45.7025004  &  16:40:34.02  &  -45:42:9.000  &  0.86  &  0.79  &  89  &  36 $\pm$ 7  &  -64.94 $\pm$ 0.02  &  2.19 $\pm$ 0.07  &  Single \\
6  &  250.1419183-45.7022799  &  16:40:34.06  &  -45:42:8.210  &  1.0  &  0.86  &  65  &  35 $\pm$ 10  &  -64.36 $\pm$ 0.01  &  2.98 $\pm$ 0.12  &  Single \\
7  &  250.1378537-45.7040233  &  16:40:33.08  &  -45:42:14.48  &  0.81  &  0.76  &  56  &  30 $\pm$ 6  &  --  &  --  &  -- \\
8  &  250.1433977-45.6938542  &  16:40:34.42  &  -45:41:37.88  &  1.42  &  0.88  &  119  &  27 $\pm$ 6  &  -58.93 $\pm$ 0.03  &  1.81 $\pm$ 0.12  &  Single \\
9  &  250.1497218-45.6929092  &  16:40:35.93  &  -45:41:34.47  &  0.98  &  0.74  &  29  &  24 $\pm$ 5  &  --  &  --  &  -- \\
10  &  250.1403862-45.7027264  &  16:40:33.69  &  -45:42:9.820  &  1.02  &  0.88  &  63  &  100 $\pm$ 50  &  -63.32 $\pm$ 0.07  &  --  &  Complex \\
11  &  250.1409277-45.6954519  &  16:40:33.82  &  -45:41:43.63  &  0.87  &  0.74  &  15  &  27 $\pm$ 6  &  -59.8 $\pm$ 0.05  &  0.95 $\pm$ 0.2  &  Single \\
12  &  250.1444989-45.69665  &  16:40:34.68  &  -45:41:47.94  &  1.99  &  1.75  &  16  &  24 $\pm$ 5  &  --  &  --  &  -- \\
13  &  250.1423231-45.6931039  &  16:40:34.16  &  -45:41:35.17  &  1.06  &  1.04  &  129  &  29 $\pm$ 6  &  -57.29 $\pm$ 0.02  &  2.12 $\pm$ 0.09  &  Single \\
14  &  250.1394346-45.6928199  &  16:40:33.46  &  -45:41:34.15  &  0.87  &  0.7  &  72  &  28 $\pm$ 6  &  -59.26 $\pm$ 0.06  &  1.69 $\pm$ 0.17  &  Single \\
15  &  250.1461922-45.6894135  &  16:40:35.09  &  -45:41:21.89  &  1.68  &  1.26  &  124  &  24 $\pm$ 5  &  --  &  --  &  -- \\
16  &  250.1418854-45.701708  &  16:40:34.05  &  -45:42:6.150  &  1.06  &  0.98  &  72  &  33 $\pm$ 7  &  -61.06 $\pm$ 0.03  &  2.78 $\pm$ 0.2  &  Single \\
17  &  250.139044-45.6942019  &  16:40:33.37  &  -45:41:39.13  &  1.25  &  1.05  &  20  &  30 $\pm$ 6  &  -62.1 $\pm$ 0.01  &  1.44 $\pm$ 0.04  &  Single \\
18  &  250.1508635-45.6953106  &  16:40:36.21  &  -45:41:43.12  &  1.28  &  1.1  &  63  &  24 $\pm$ 5  &  --  &  --  &  -- \\
19  &  250.1409412-45.693164  &  16:40:33.83  &  -45:41:35.39  &  0.92  &  0.89  &  118  &  26 $\pm$ 5  &  -58.47 $\pm$ 0.01  &  1.03 $\pm$ 0.04  &  Single \\
21  &  250.1528564-45.6839538  &  16:40:36.69  &  -45:41:2.230  &  1.23  &  0.99  &  114  &  24 $\pm$ 5  &  --  &  --  &  -- \\
22  &  250.1435778-45.6996258  &  16:40:34.46  &  -45:41:58.65  &  2.09  &  1.82  &  21  &  25 $\pm$ 5  &  -63.63 $\pm$ 0.19  &  --  &  Complex \\
23  &  250.1390889-45.6957652  &  16:40:33.38  &  -45:41:44.75  &  1.01  &  0.89  &  170  &  28 $\pm$ 6  &  -62.41 $\pm$ 0.03  &  1.79 $\pm$ 0.14  &  Single \\
24  &  250.1386357-45.6860028  &  16:40:33.27  &  -45:41:9.610  &  0.78  &  0.75  &  86  &  24 $\pm$ 5  &  --  &  --  &  -- \\
25  &  250.1417571-45.6936655  &  16:40:34.02  &  -45:41:37.20  &  1.12  &  1.03  &  40  &  28 $\pm$ 6  &  -59.3 $\pm$ 0.03  &  --  &  Complex \\
27  &  250.1396701-45.6923769  &  16:40:33.52  &  -45:41:32.56  &  0.92  &  0.73  &  32  &  26 $\pm$ 5  &  -60.67 $\pm$ 0.28  &  --  &  Complex \\
28  &  250.138712-45.6950579  &  16:40:33.29  &  -45:41:42.21  &  0.93  &  0.81  &  172  &  29 $\pm$ 6  &  -60.03 $\pm$ 0.08  &  1.02 $\pm$ 0.32  &  Single \\
29  &  250.1412979-45.7015713  &  16:40:33.91  &  -45:42:5.660  &  1.33  &  1.1  &  129  &  32 $\pm$ 7  &  -61.06 $\pm$ 0.02  &  1.46 $\pm$ 0.04  &  Single \\
30  &  250.1517326-45.6946186  &  16:40:36.42  &  -45:41:40.63  &  1.06  &  0.88  &  180  &  24 $\pm$ 5  &  --  &  --  &  -- \\
31  &  250.1380669-45.6941846  &  16:40:33.14  &  -45:41:39.06  &  1.0  &  0.91  &  112  &  30 $\pm$ 6  &  -60.93 $\pm$ 0.02  &  1.05 $\pm$ 0.05  &  Single \\
33  &  250.1415524-45.7011978  &  16:40:33.97  &  -45:42:4.310  &  1.09  &  1.0  &  21  &  29 $\pm$ 6  &  -62.1 $\pm$ 0.04  &  2.53 $\pm$ 0.25  &  Single \\
34  &  250.1519045-45.6953464  &  16:40:36.46  &  -45:41:43.25  &  1.21  &  1.05  &  144  &  24 $\pm$ 5  &  --  &  --  &  -- \\
35  &  250.1430564-45.7006254  &  16:40:34.33  &  -45:42:2.250  &  1.52  &  1.39  &  30  &  26 $\pm$ 5  &  --  &  --  &  -- \\
37  &  250.1366979-45.6914341  &  16:40:32.81  &  -45:41:29.16  &  1.57  &  1.47  &  155  &  27 $\pm$ 5  &  -61.98 $\pm$ 0.07  &  0.66 $\pm$ 0.24  &  Single \\
38  &  250.1355573-45.7039553  &  16:40:32.53  &  -45:42:14.24  &  1.53  &  1.29  &  180  &  25 $\pm$ 5  &  --  &  --  &  -- \\
39  &  250.1511653-45.6876105  &  16:40:36.28  &  -45:41:15.40  &  2.35  &  1.71  &  179  &  24 $\pm$ 5  &  --  &  --  &  -- \\
40  &  250.1425257-45.7017414  &  16:40:34.21  &  -45:42:6.270  &  0.83  &  0.57  &  71  &  31 $\pm$ 6  &  --  &  --  &  -- \\
41  &  250.1363471-45.6947624  &  16:40:32.72  &  -45:41:41.14  &  1.33  &  1.24  &  80  &  27 $\pm$ 6  &  --  &  --  &  -- \\
42  &  250.1380589-45.697165  &  16:40:33.13  &  -45:41:49.79  &  1.5  &  1.26  &  90  &  26 $\pm$ 5  &  --  &  --  &  -- \\
43  &  250.1392522-45.6932893  &  16:40:33.42  &  -45:41:35.84  &  1.2  &  0.8  &  134  &  30 $\pm$ 6  &  -60.3 $\pm$ 0.02  &  1.76 $\pm$ 0.09  &  Single \\
45  &  250.1373942-45.6927488  &  16:40:32.97  &  -45:41:33.90  &  1.44  &  0.97  &  157  &  28 $\pm$ 6  &  --  &  --  &  -- \\
46  &  250.1406937-45.7022641  &  16:40:33.77  &  -45:42:8.150  &  1.11  &  0.82  &  38  &  34 $\pm$ 7  &  -64.11 $\pm$ 0.07  &  1.22 $\pm$ 0.29  &  Single \\
47  &  250.1421862-45.6953419  &  16:40:34.12  &  -45:41:43.23  &  1.47  &  1.17  &  106  &  26 $\pm$ 5  &  -59.23 $\pm$ 0.03  &  0.96 $\pm$ 0.11  &  Single \\
\hline \noalign {\smallskip}

\end{tabular}
\end{table*}

\begin{figure*}
    \centering

\includegraphics[width=0.24\textwidth]{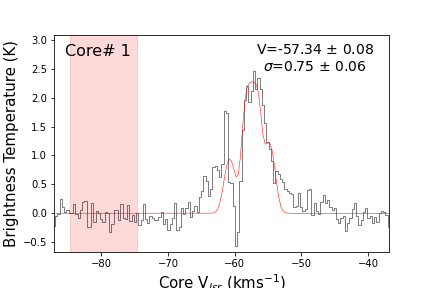}
\includegraphics[width=0.24\textwidth]{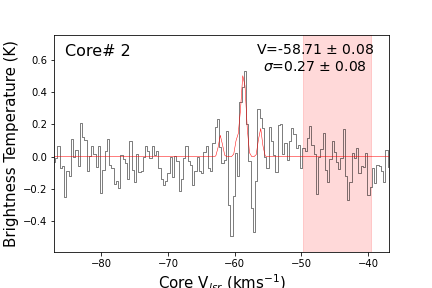}
\includegraphics[width=0.24\textwidth]{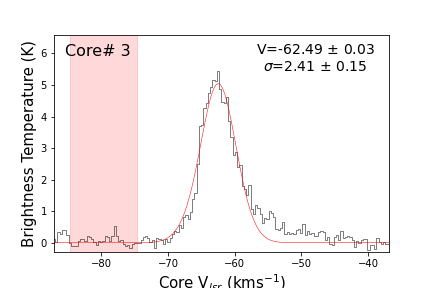}
\includegraphics[width=0.24\textwidth]{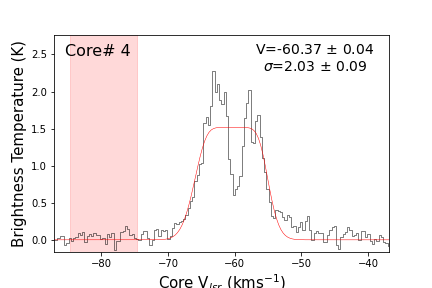}
\includegraphics[width=0.24\textwidth]{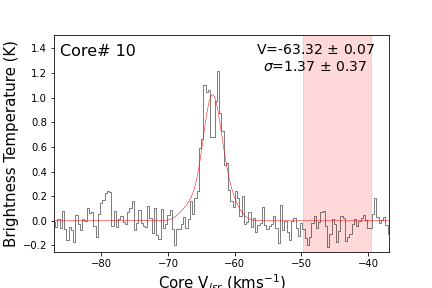}
\includegraphics[width=0.24\textwidth]{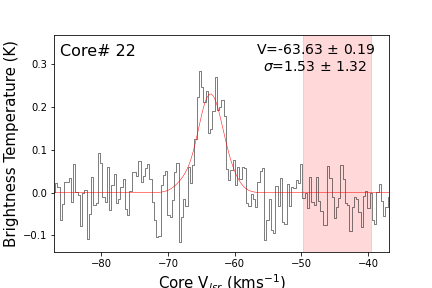}
\includegraphics[width=0.24\textwidth]{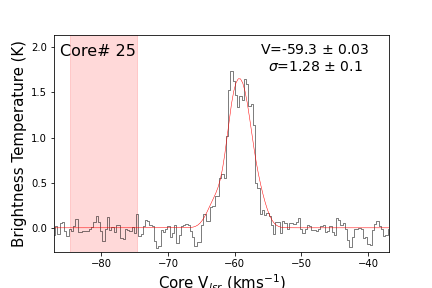}
\includegraphics[width=0.24\textwidth]{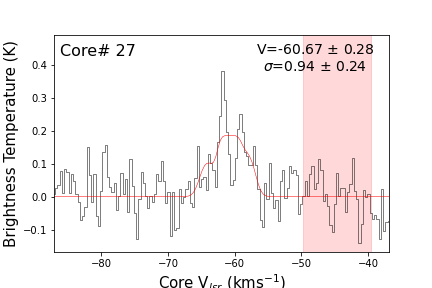}

\caption{Complex-type core-averaged, background-subtracted DCN spectra extracted from the cores in the young protocluster G338.93. See \cref{tabappendix:coretables_g338} for the line fit parameters for each core.\label{figspectra:dcnspectra_split_G338_c}} 
\end{figure*}

\begin{table*}[htbp!]
\centering
\small
\caption{DCN fits towards the core population of the evolved protocluster G333.60.}
\label{tabappendix:coretables_g333}
\begin{tabular}{llllllllccc}

\hline 
n   & Core Name &RA  & DEC & F$_{A}$   & F$_{B}$    & PA & T & \vlsr  & Linewidth   & Spectral \\
 &  & [ICRS]  &  [ICRS]   &  [\arcsec] &[\arcsec] &  [deg] & [K] & [\kms]  & [\kms] & Type \\
\hline 
1  &  245.5468006-50.0988073  &  16:22:11.23  &  -50:05:55.71  &  1.41  &  0.9  &  67  &  38 $\pm$ 8  &  -49.51 $\pm$ 0.02  &  1.99 $\pm$ 0.1  &  Single \\
2  &  245.5356169-50.104813  &  16:22:8.550  &  -50:06:17.33  &  0.94  &  0.75  &  67  &  37 $\pm$ 8  &  -48.48 $\pm$ 0.03  &  --  &  Complex \\
3  &  245.528392-50.1050657  &  16:22:6.810  &  -50:06:18.24  &  0.83  &  0.79  &  88  &  33 $\pm$ 7  &  -49.87 $\pm$ 0.02  &  --  &  Complex \\
4  &  245.5460571-50.0990531  &  16:22:11.05  &  -50:05:56.59  &  0.85  &  0.63  &  47  &  100 $\pm$ 50  &  -51.35 $\pm$ 0.07  &  --  &  Complex \\
5  &  245.5285258-50.1045541  &  16:22:6.850  &  -50:06:16.39  &  0.74  &  0.71  &  8  &  32 $\pm$ 7  &  -48.75 $\pm$ 0.04  &  --  &  Complex \\
7  &  245.5370048-50.1027976  &  16:22:8.880  &  -50:06:10.07  &  1.3  &  0.96  &  24  &  38 $\pm$ 8  &  -46.37 $\pm$ 0.04  &  2.76 $\pm$ 0.31  &  Single \\
8  &  245.5282954-50.1068134  &  16:22:6.790  &  -50:06:24.53  &  1.0  &  0.77  &  50  &  30 $\pm$ 6  &  -50.16 $\pm$ 0.01  &  1.22 $\pm$ 0.03  &  Single \\
11  &  245.5357042-50.1033887  &  16:22:8.570  &  -50:06:12.20  &  1.33  &  1.13  &  36  &  100 $\pm$ 50  &  -47.33 $\pm$ 0.02  &  2.02 $\pm$ 0.09  &  Single \\
14  &  245.5271272-50.1056891  &  16:22:6.510  &  -50:06:20.48  &  1.01  &  0.77  &  18  &  31 $\pm$ 6  &  -49.83 $\pm$ 0.01  &  1.25 $\pm$ 0.02  &  Single \\
18  &  245.5296271-50.1064415  &  16:22:7.110  &  -50:06:23.19  &  0.69  &  0.6  &  43  &  31 $\pm$ 6  &  -50.22 $\pm$ 0.07  &  --  &  Complex \\
19  &  245.5270406-50.1089457  &  16:22:6.490  &  -50:06:32.20  &  1.25  &  1.07  &  151  &  28 $\pm$ 6  &  -47.89 $\pm$ 0.01  &  1.15 $\pm$ 0.04  &  Single \\
21  &  245.5311806-50.1042724  &  16:22:7.480  &  -50:06:15.38  &  1.17  &  1.04  &  16  &  34 $\pm$ 7  &  -50.18 $\pm$ 0.01  &  1.25 $\pm$ 0.03  &  Single \\
22  &  245.5170221-50.1089966  &  16:22:4.090  &  -50:06:32.39  &  0.86  &  0.7  &  54  &  28 $\pm$ 6  &  -46.04 $\pm$ 0.04  &  0.86 $\pm$ 0.11  &  Single \\
23  &  245.5474175-50.0959979  &  16:22:11.38  &  -50:05:45.59  &  1.08  &  0.82  &  150  &  36 $\pm$ 7  &  -44.09 $\pm$ 0.01  &  1.22 $\pm$ 0.03  &  Single \\
25  &  245.5167915-50.108572  &  16:22:4.030  &  -50:06:30.86  &  0.76  &  0.62  &  174  &  28 $\pm$ 6  &  -47.89 $\pm$ 0.05  &  1.24 $\pm$ 0.21  &  Single \\
26  &  245.535353-50.1041745  &  16:22:8.480  &  -50:06:15.03  &  0.95  &  0.75  &  148  &  38 $\pm$ 8  &  -47.69 $\pm$ 0.11  &  --  &  Complex \\
28  &  245.5508287-50.0914106  &  16:22:12.20  &  -50:05:29.08  &  1.34  &  0.9  &  169  &  32 $\pm$ 6  &  -45.93 $\pm$ 0.04  &  1.05 $\pm$ 0.16  &  Single \\
29  &  245.5475066-50.0997527  &  16:22:11.40  &  -50:05:59.11  &  0.95  &  0.84  &  19  &  37 $\pm$ 7  &  -50.1 $\pm$ 0.06  &  0.98 $\pm$ 0.21  &  Single \\
30  &  245.5475313-50.0989706  &  16:22:11.41  &  -50:05:56.29  &  1.68  &  1.32  &  117  &  38 $\pm$ 8  &  -49.63 $\pm$ 0.02  &  --  &  Complex \\
31  &  245.547985-50.0993306  &  16:22:11.52  &  -50:05:57.59  &  0.97  &  0.91  &  95  &  37 $\pm$ 8  &  -49.78 $\pm$ 0.04  &  --  &  Complex \\
33  &  245.5496874-50.0921633  &  16:22:11.92  &  -50:05:31.79  &  1.7  &  1.11  &  168  &  31 $\pm$ 6  &  -47.01 $\pm$ 0.03  &  1.02 $\pm$ 0.08  &  Single \\
35  &  245.5174577-50.1082967  &  16:22:4.190  &  -50:06:29.87  &  0.96  &  0.82  &  129  &  28 $\pm$ 6  &  -48.53 $\pm$ 0.02  &  1.1 $\pm$ 0.07  &  Single \\
37  &  245.530804-50.1034521  &  16:22:7.390  &  -50:06:12.43  &  1.29  &  1.16  &  132  &  34 $\pm$ 7  &  -49.7 $\pm$ 0.05  &  --  &  Complex \\
43  &  245.5223011-50.1105471  &  16:22:5.350  &  -50:06:37.97  &  0.81  &  0.69  &  113  &  25 $\pm$ 5  &  --  &  --  &  -- \\
44  &  245.5165096-50.1098853  &  16:22:3.960  &  -50:06:35.59  &  1.57  &  1.26  &  44  &  27 $\pm$ 6  &  --  &  --  &  -- \\
45  &  245.549303-50.0984873  &  16:22:11.83  &  -50:05:54.55  &  1.45  &  1.03  &  48  &  35 $\pm$ 7  &  -49.0 $\pm$ 0.01  &  1.19 $\pm$ 0.05  &  Single \\
46  &  245.5530669-50.0886225  &  16:22:12.74  &  -50:05:19.04  &  1.24  &  1.08  &  42  &  28 $\pm$ 6  &  -47.58 $\pm$ 0.07  &  1.1 $\pm$ 0.29  &  Single \\
47  &  245.5496272-50.0988558  &  16:22:11.91  &  -50:05:55.88  &  1.06  &  1.0  &  26  &  35 $\pm$ 7  &  -49.75 $\pm$ 0.02  &  1.32 $\pm$ 0.09  &  Single \\
49  &  245.5471944-50.0982171  &  16:22:11.33  &  -50:05:53.58  &  1.33  &  0.94  &  38  &  37 $\pm$ 8  &  -46.03 $\pm$ 0.02  &  1.41 $\pm$ 0.07  &  Single \\
61  &  245.551584-50.0920068  &  16:22:12.38  &  -50:05:31.22  &  1.23  &  1.03  &  3  &  32 $\pm$ 6  &  -46.71 $\pm$ 0.07  &  0.89 $\pm$ 0.19  &  Single \\
63  &  245.5332111-50.0860966  &  16:22:7.970  &  -50:05:9.950  &  1.86  &  1.14  &  92  &  24 $\pm$ 5  &  -39.83 $\pm$ 0.05  &  1.25 $\pm$ 0.18  &  Single \\
65  &  245.5394054-50.1081881  &  16:22:9.460  &  -50:06:29.48  &  1.17  &  0.9  &  113  &  35 $\pm$ 7  &  --  &  --  &  -- \\
66  &  245.5292379-50.1060849  &  16:22:7.020  &  -50:06:21.91  &  1.06  &  0.67  &  80  &  31 $\pm$ 6  &  -49.46 $\pm$ 0.02  &  1.24 $\pm$ 0.06  &  Single \\
71  &  245.5554581-50.1093706  &  16:22:13.31  &  -50:06:33.73  &  1.52  &  1.21  &  45  &  26 $\pm$ 5  &  --  &  --  &  -- \\
74  &  245.526505-50.0998535  &  16:22:6.360  &  -50:05:59.47  &  1.11  &  0.9  &  159  &  30 $\pm$ 6  &  -46.71 $\pm$ 0.02  &  0.95 $\pm$ 0.1  &  Single \\
76  &  245.5321185-50.1024642  &  16:22:7.710  &  -50:06:8.870  &  0.86  &  0.7  &  20  &  36 $\pm$ 7  &  -49.77 $\pm$ 0.01  &  1.51 $\pm$ 0.04  &  Single \\
77  &  245.509509-50.108116  &  16:22:2.280  &  -50:06:29.22  &  0.78  &  0.57  &  18  &  22 $\pm$ 4  &  --  &  --  &  -- \\
80  &  245.5338592-50.1044306  &  16:22:8.130  &  -50:06:15.95  &  1.47  &  1.15  &  59  &  36 $\pm$ 7  &  -48.01 $\pm$ 0.01  &  1.12 $\pm$ 0.04  &  Single \\
81  &  245.5157057-50.1118428  &  16:22:3.770  &  -50:06:42.63  &  1.42  &  1.16  &  106  &  25 $\pm$ 5  &  --  &  --  &  -- \\
84  &  245.5171073-50.1084609  &  16:22:4.110  &  -50:06:30.46  &  0.95  &  0.67  &  159  &  28 $\pm$ 6  &  -48.05 $\pm$ 0.03  &  0.97 $\pm$ 0.08  &  Single \\
85  &  245.514463-50.1066733  &  16:22:3.470  &  -50:06:24.02  &  1.27  &  1.18  &  97  &  23 $\pm$ 5  &  --  &  --  &  -- \\
88  &  245.5487618-50.1045099  &  16:22:11.70  &  -50:06:16.24  &  0.94  &  0.78  &  17  &  36 $\pm$ 7  &  --  &  --  &  -- \\
90  &  245.533779-50.1030615  &  16:22:8.110  &  -50:06:11.02  &  0.96  &  0.8  &  173  &  36 $\pm$ 7  &  --  &  --  &  -- \\
91  &  245.5563744-50.1099885  &  16:22:13.53  &  -50:06:35.96  &  1.05  &  0.8  &  42  &  25 $\pm$ 5  &  --  &  --  &  -- \\
94  &  245.5546166-50.1076554  &  16:22:13.11  &  -50:06:27.56  &  1.17  &  1.08  &  49  &  25 $\pm$ 5  &  --  &  --  &  -- \\
98  &  245.5313456-50.1018072  &  16:22:7.520  &  -50:06:6.510  &  0.83  &  0.73  &  68  &  35 $\pm$ 7  &  -47.83 $\pm$ 0.02  &  1.57 $\pm$ 0.08  &  Single \\
99  &  245.5583826-50.1029587  &  16:22:14.01  &  -50:06:10.65  &  0.85  &  0.82  &  18  &  23 $\pm$ 5  &  --  &  --  &  -- \\
101  &  245.5267009-50.1004679  &  16:22:6.410  &  -50:06:1.680  &  1.52  &  1.14  &  39  &  30 $\pm$ 6  &  --  &  --  &  -- \\
103  &  245.5326259-50.0866914  &  16:22:7.830  &  -50:05:12.09  &  0.8  &  0.7  &  68  &  24 $\pm$ 5  &  --  &  --  &  -- \\
104  &  245.535577-50.1053101  &  16:22:8.540  &  -50:06:19.12  &  0.93  &  0.8  &  178  &  36 $\pm$ 7  &  -48.39 $\pm$ 0.06  &  --  &  Complex \\
108  &  245.5246254-50.103576  &  16:22:5.910  &  -50:06:12.87  &  1.71  &  1.39  &  46  &  29 $\pm$ 6  &  -48.41 $\pm$ 0.04  &  0.8 $\pm$ 0.1  &  Single \\
116  &  245.526364-50.1034004  &  16:22:6.330  &  -50:06:12.24  &  1.24  &  0.99  &  78  &  30 $\pm$ 6  &  -49.36 $\pm$ 0.05  &  1.14 $\pm$ 0.2  &  Single \\
\hline \noalign {\smallskip}

\end{tabular}
\end{table*}

\begin{figure*}
    \centering
\includegraphics[width=0.24\textwidth]{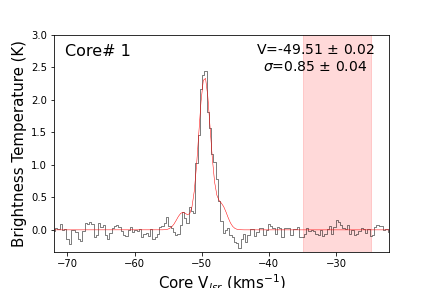}
\includegraphics[width=0.24\textwidth]{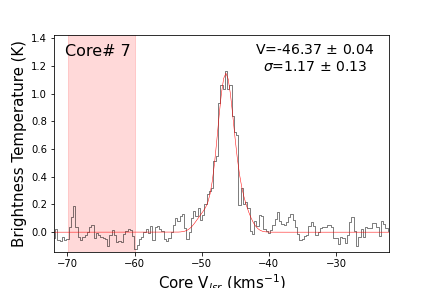}
\includegraphics[width=0.24\textwidth]{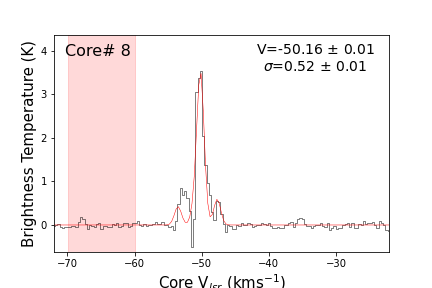}
\includegraphics[width=0.24\textwidth]{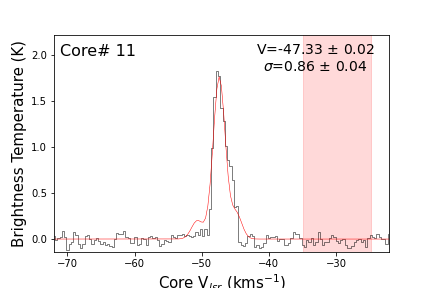}
\includegraphics[width=0.24\textwidth]{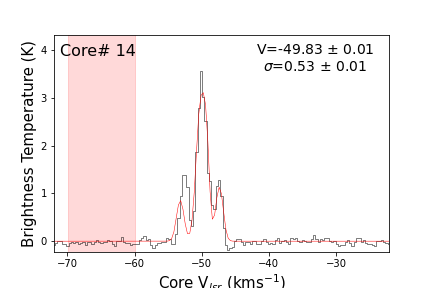}
\includegraphics[width=0.24\textwidth]{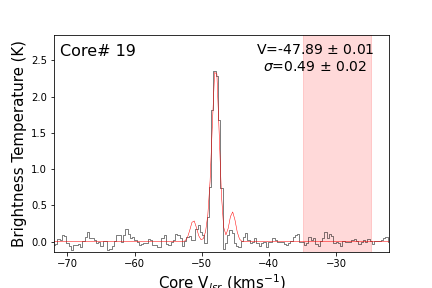}
\includegraphics[width=0.24\textwidth]{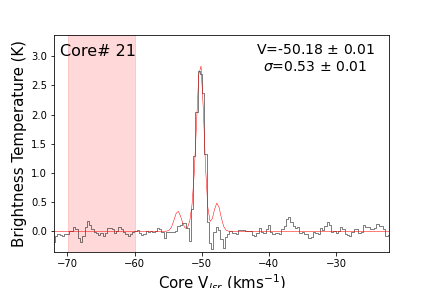}
\includegraphics[width=0.24\textwidth]{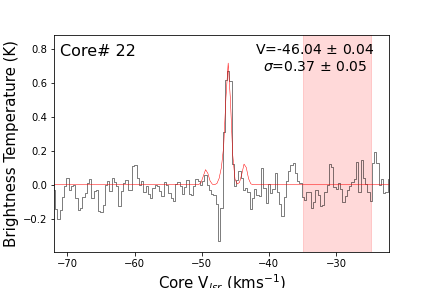}
\includegraphics[width=0.24\textwidth]{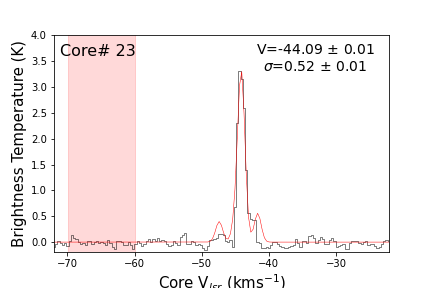}
\includegraphics[width=0.24\textwidth]{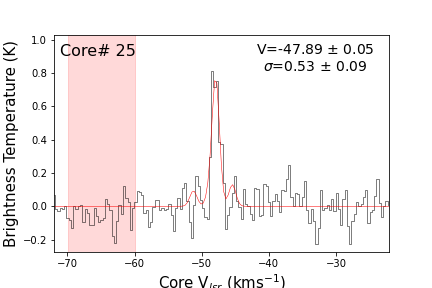}
\includegraphics[width=0.24\textwidth]{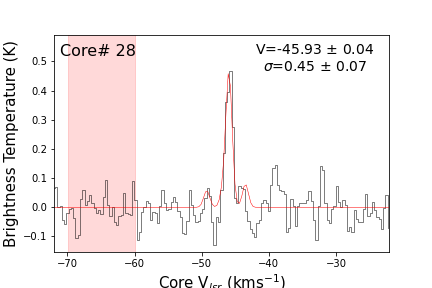}
\includegraphics[width=0.24\textwidth]{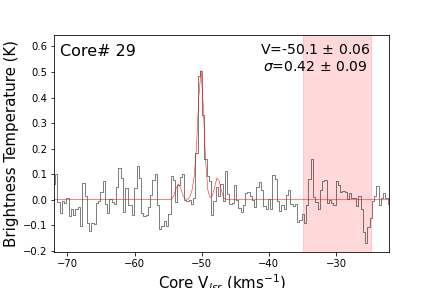}
\includegraphics[width=0.24\textwidth]{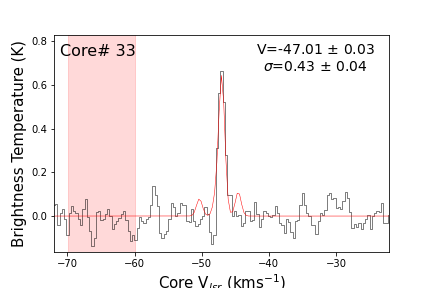}
\includegraphics[width=0.24\textwidth]{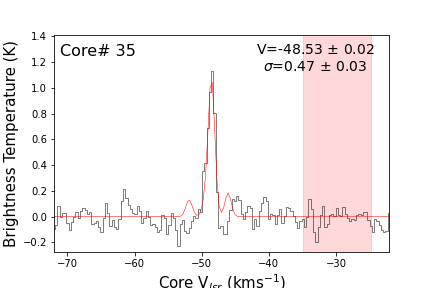}
\includegraphics[width=0.24\textwidth]{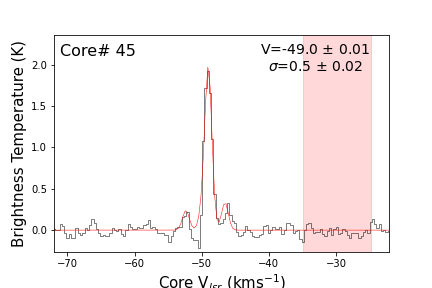}
\includegraphics[width=0.24\textwidth]{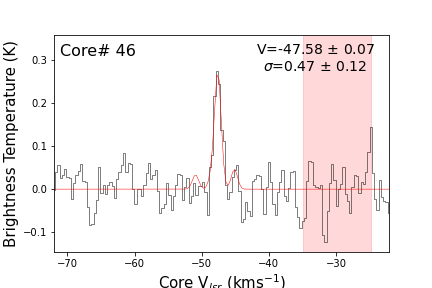}
\includegraphics[width=0.24\textwidth]{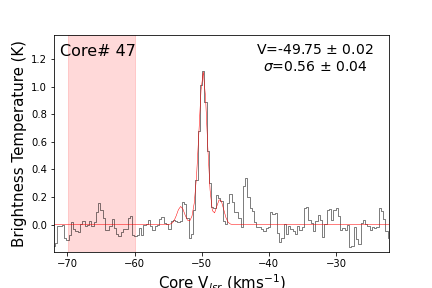}
\includegraphics[width=0.24\textwidth]{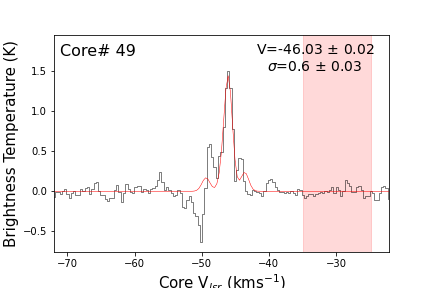}
\includegraphics[width=0.24\textwidth]{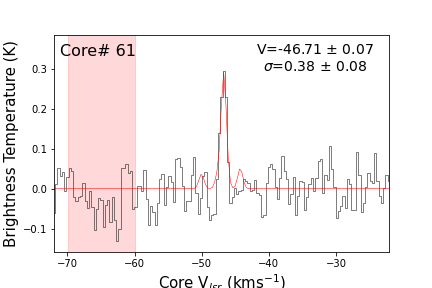}
\includegraphics[width=0.24\textwidth]{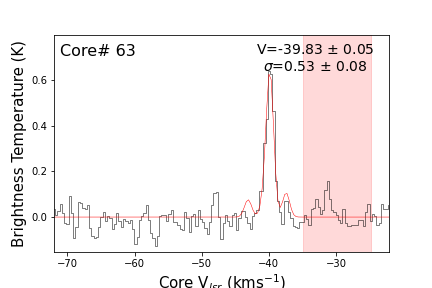}
\includegraphics[width=0.24\textwidth]{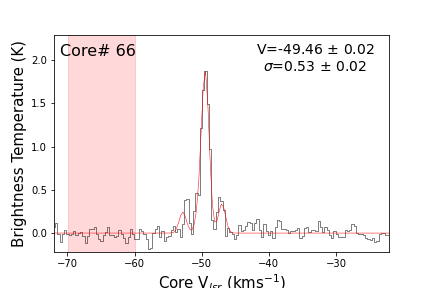}
\includegraphics[width=0.24\textwidth]{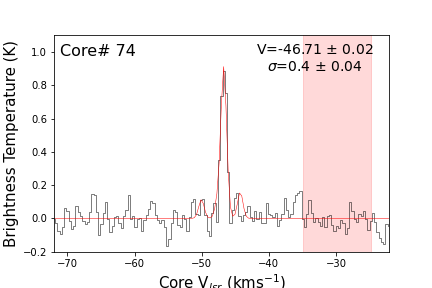}
\includegraphics[width=0.24\textwidth]{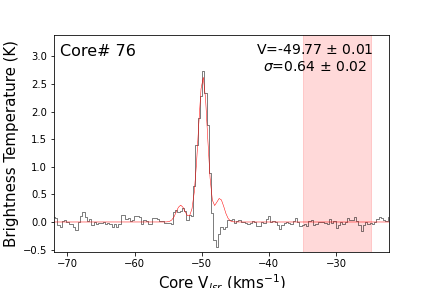}
\includegraphics[width=0.24\textwidth]{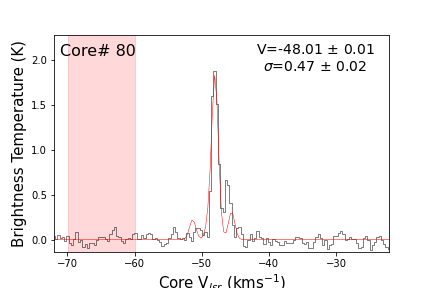}
\includegraphics[width=0.24\textwidth]{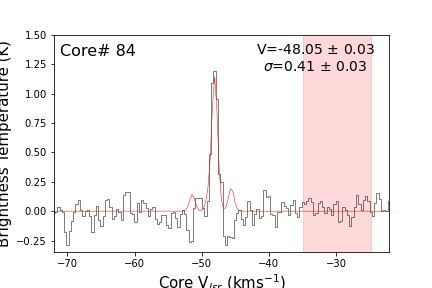}
\includegraphics[width=0.24\textwidth]{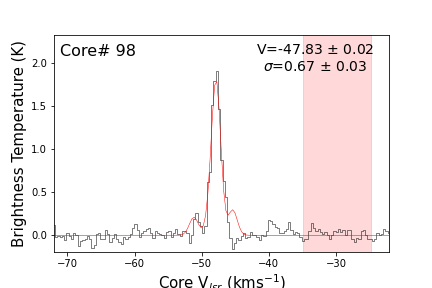}
\includegraphics[width=0.24\textwidth]{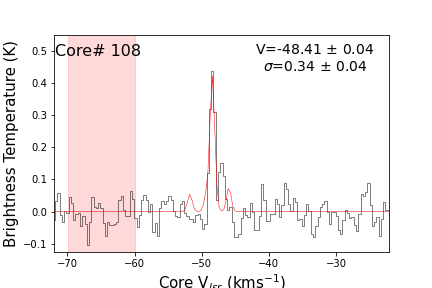}
\includegraphics[width=0.24\textwidth]{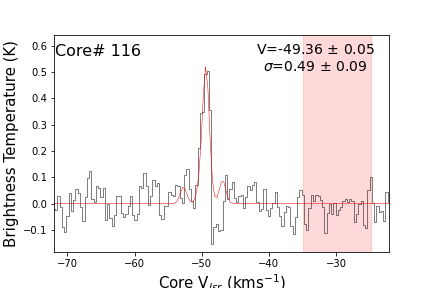}

\caption{Single-type core-averaged, background-subtracted DCN spectra extracted from the cores in the evolved protocluster G333.60. See \cref{tabappendix:coretables_g333} for the line fit parameters extracted for each core.\label{figspectra:dcnspectra_split_G333_s}}
\end{figure*}

\begin{figure*}

    \centering
    
\includegraphics[width=0.24\textwidth]{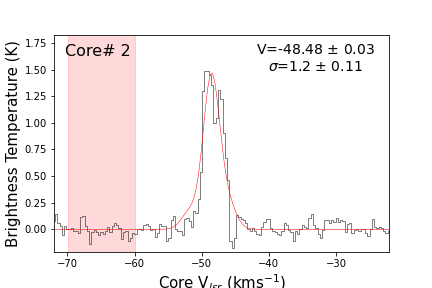}
\includegraphics[width=0.24\textwidth]{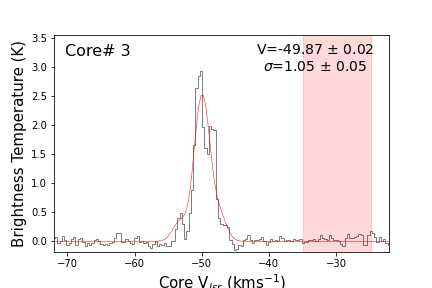}
\includegraphics[width=0.24\textwidth]{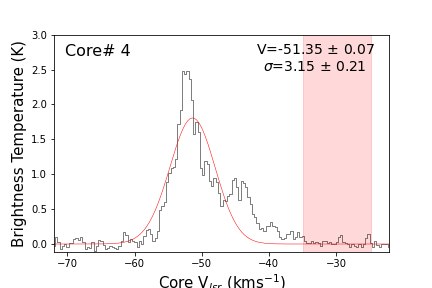}
\includegraphics[width=0.24\textwidth]{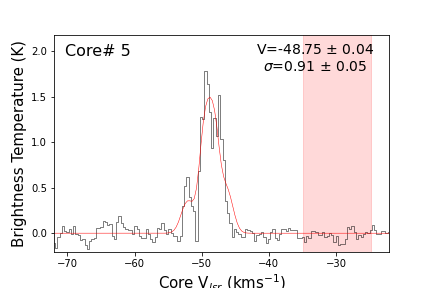}
\includegraphics[width=0.24\textwidth]{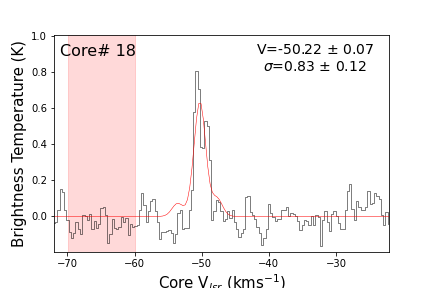}
\includegraphics[width=0.24\textwidth]{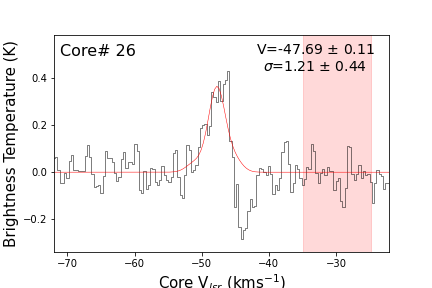}
\includegraphics[width=0.24\textwidth]{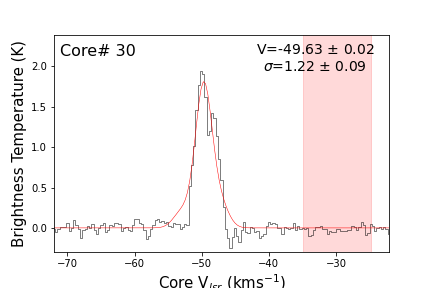}
\includegraphics[width=0.24\textwidth]{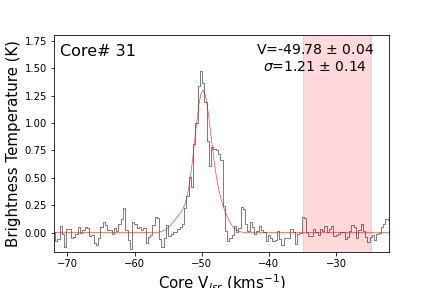}
\includegraphics[width=0.24\textwidth]{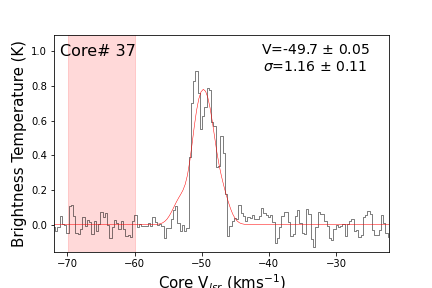}
\includegraphics[width=0.24\textwidth]{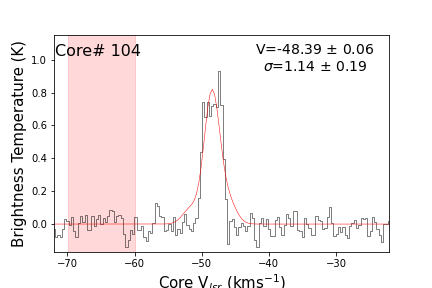}

\caption{Complex-type core-averaged, background-subtracted DCN spectra extracted from the cores in the evolved protocluster G333.60. See \cref{tabappendix:coretables_g333} for the line fit parameters extracted for each core.\label{figspectra:dcnspectra_split_G333_c}} 
\end{figure*}

\begin{table*}[htbp!]
\centering
\small
\caption{DCN fits towards the core population of the intermediate protocluster G351.77.}
\label{tabappendix:coretables_g351}
\begin{tabular}{llllllllccc}

\hline 
n   & Core Name &RA  & DEC & F$_{A}$   & F$_{B}$    & PA & T & \vlsr  & Linewidth   & Spectral \\
 &  & [ICRS]  &  [ICRS]   &  [\arcsec] &[\arcsec] &  [deg] & [K] & [\kms]  & [\kms] & Type \\
\hline 
1  &  261.677251-36.1547987  &  17:26:42.54  &  -36:09:17.28  &  2.4  &  1.79  &  91  &  100 $\pm$ 50  &  -2.44 $\pm$ 0.04  &  --  &  Complex \\
2  &  261.6767723-36.1551725  &  17:26:42.43  &  -36:09:18.62  &  2.88  &  2.22  &  126  &  100 $\pm$ 50  &  -1.96 $\pm$ 0.03  &  --  &  Complex \\
4  &  261.6761591-36.1549329  &  17:26:42.28  &  -36:09:17.76  &  1.62  &  1.33  &  78  &  35 $\pm$ 10  &  -1.06 $\pm$ 0.02  &  --  &  Complex \\
5  &  261.6783746-36.1556996  &  17:26:42.81  &  -36:09:20.52  &  1.56  &  1.46  &  122  &  100 $\pm$ 50  &  -4.84 $\pm$ 0.04  &  --  &  Complex \\
6  &  261.6778603-36.1551246  &  17:26:42.69  &  -36:09:18.45  &  2.3  &  2.05  &  45  &  35 $\pm$ 10  &  -3.02 $\pm$ 0.05  &  --  &  Complex \\
7  &  261.6788098-36.1553482  &  17:26:42.91  &  -36:09:19.25  &  1.84  &  1.72  &  99  &  38 $\pm$ 8  &  -4.51 $\pm$ 0.1  &  --  &  Complex \\
8  &  261.67848-36.1431525  &  17:26:42.84  &  -36:08:35.35  &  1.42  &  1.28  &  61  &  29 $\pm$ 6  &  --  &  --  &  -- \\
10  &  261.6882344-36.1502759  &  17:26:45.18  &  -36:09:0.990  &  1.76  &  1.51  &  149  &  26 $\pm$ 5  &  --  &  --  &  -- \\
11  &  261.6753281-36.1587854  &  17:26:42.08  &  -36:09:31.63  &  1.88  &  1.62  &  84  &  33 $\pm$ 7  &  -0.73 $\pm$ 0.06  &  1.32 $\pm$ 0.21  &  Single \\
12  &  261.6743392-36.1526192  &  17:26:41.84  &  -36:09:9.430  &  1.76  &  1.55  &  111  &  37 $\pm$ 8  &  -2.42 $\pm$ 0.02  &  0.89 $\pm$ 0.07  &  Single \\
13  &  261.6878699-36.1553934  &  17:26:45.09  &  -36:09:19.42  &  1.64  &  1.48  &  155  &  29 $\pm$ 6  &  --  &  --  &  -- \\
14  &  261.6800995-36.1544281  &  17:26:43.22  &  -36:09:15.94  &  2.04  &  1.68  &  92  &  38 $\pm$ 8  &  -4.0 $\pm$ 0.02  &  --  &  Complex \\
15  &  261.6757897-36.1570022  &  17:26:42.19  &  -36:09:25.21  &  1.97  &  1.67  &  141  &  36 $\pm$ 7  &  0.37 $\pm$ 0.18  &  --  &  Complex \\
16  &  261.6734791-36.1466672  &  17:26:41.63  &  -36:08:48.00  &  1.78  &  1.68  &  153  &  30 $\pm$ 6  &  --  &  --  &  -- \\
18  &  261.6780185-36.1506715  &  17:26:42.72  &  -36:09:2.420  &  1.95  &  1.68  &  82  &  35 $\pm$ 7  &  -5.73 $\pm$ 0.12  &  1.83 $\pm$ 0.3  &  Single \\
19  &  261.6820301-36.1518086  &  17:26:43.69  &  -36:09:6.510  &  1.95  &  1.72  &  65  &  33 $\pm$ 7  &  --  &  --  &  -- \\
20  &  261.6784625-36.153325  &  17:26:42.83  &  -36:09:11.97  &  2.05  &  1.42  &  180  &  39 $\pm$ 8  &  -6.69 $\pm$ 0.01  &  1.59 $\pm$ 0.06  &  Single \\
23  &  261.6833811-36.1544942  &  17:26:44.01  &  -36:09:16.18  &  1.99  &  1.73  &  88  &  34 $\pm$ 7  &  -4.6 $\pm$ 0.08  &  --  &  Complex \\
\hline \noalign {\smallskip}
\end{tabular}
\end{table*}

\begin{figure*}
    \centering

\includegraphics[width=0.24\textwidth]{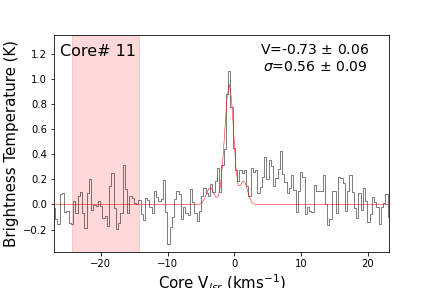}
\includegraphics[width=0.24\textwidth]{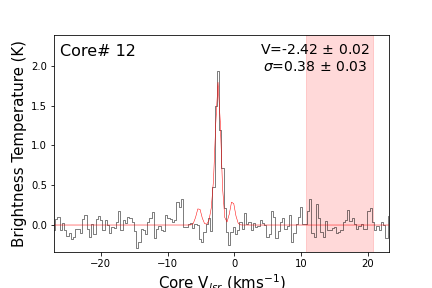}
\includegraphics[width=0.24\textwidth]{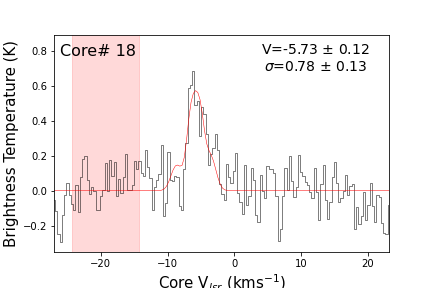}
\includegraphics[width=0.24\textwidth]{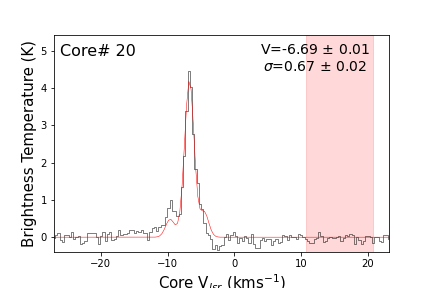}

\caption{Single-type core-averaged, background-subtracted DCN spectra extracted from the cores in the intermediate protocluster G351.77. See \cref{tabappendix:coretables_g351} for the line fit parameters for each core.\label{figspectra:dcnspectra_split_G351_s}}
\end{figure*}

\begin{figure*}
    \centering

\includegraphics[width=0.24\textwidth]{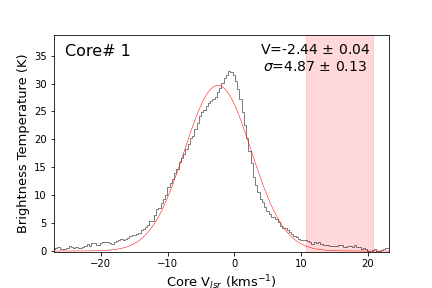}
\includegraphics[width=0.24\textwidth]{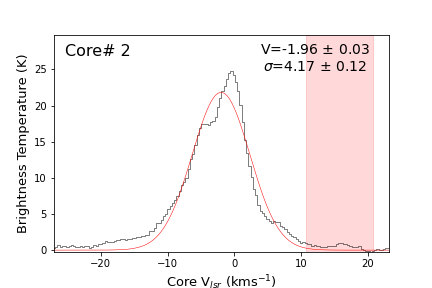}
\includegraphics[width=0.24\textwidth]{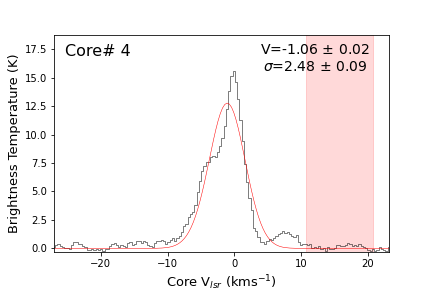}
\includegraphics[width=0.24\textwidth]{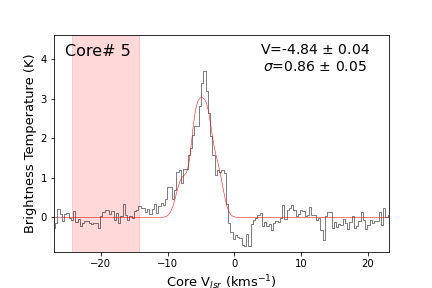}
\includegraphics[width=0.24\textwidth]{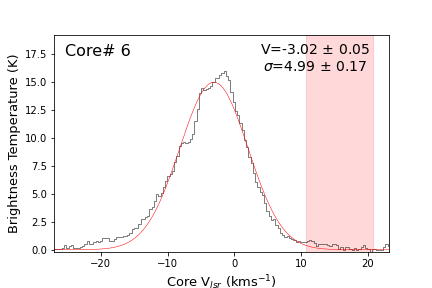}
\includegraphics[width=0.24\textwidth]{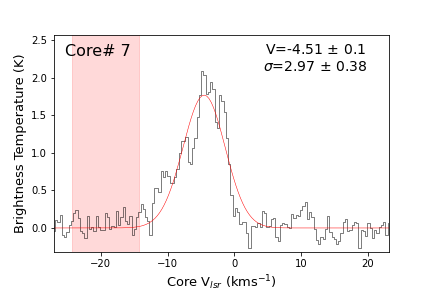}
\includegraphics[width=0.24\textwidth]{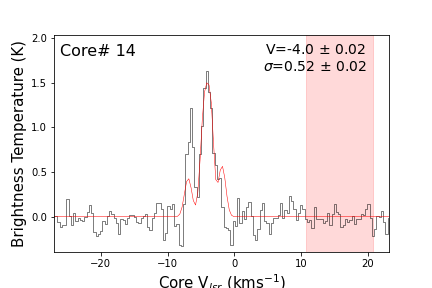}
\includegraphics[width=0.24\textwidth]{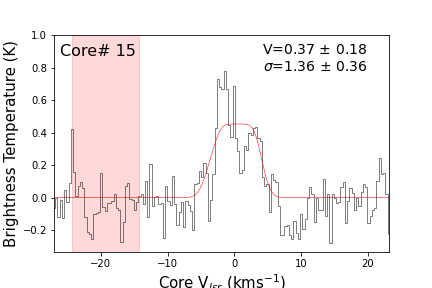}
\includegraphics[width=0.24\textwidth]{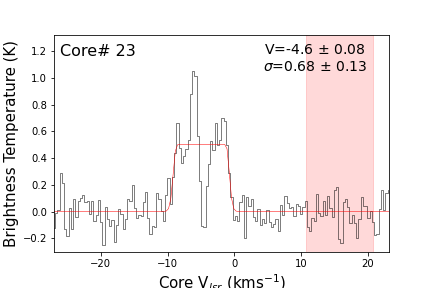}

\caption{Complex-type core-averaged, background-subtracted DCN spectra extracted from the cores in the intermediate protocluster G351.77. See \cref{tabappendix:coretables_g351} for the line fit parameters for each core.\label{figspectra:dcnspectra_split_G351_c}} 
\end{figure*}

\begin{table*}[htbp!]
\centering
\small
\caption{DCN fits towards the core population of the intermediate protocluster G353.41.}
\label{tabappendix:coretables_g353}
\begin{tabular}{llllllllccc}

\hline 
n   & Core Name &RA  & DEC & F$_{A}$   & F$_{B}$    & PA & T & \vlsr  & Linewidth   & Spectral \\
 &  & [ICRS]  &  [ICRS]   &  [\arcsec] &[\arcsec] &  [deg] & [K] & [\kms]  & [\kms] & Type \\
\hline 
2  &  262.6165032-34.6955865  &  17:30:27.96  &  -34:41:44.11  &  1.98  &  1.59  &  64  &  25 $\pm$ 5  &  -20.45 $\pm$ 0.06  &  1.2 $\pm$ 0.25  &  Single \\
3  &  262.6184156-34.696524  &  17:30:28.42  &  -34:41:47.49  &  2.59  &  1.79  &  146  &  100 $\pm$ 50  &  -16.49 $\pm$ 0.04  &  2.86 $\pm$ 0.5  &  Single \\
4  &  262.6103159-34.6932659  &  17:30:26.48  &  -34:41:35.76  &  1.56  &  1.46  &  104  &  31 $\pm$ 6  &  -16.54 $\pm$ 0.16  &  1.47 $\pm$ 0.66  &  Single \\
5  &  262.6101515-34.6960014  &  17:30:26.44  &  -34:41:45.61  &  2.03  &  1.75  &  79  &  30 $\pm$ 6  &  -16.94 $\pm$ 0.09  &  2.63 $\pm$ 0.75  &  Single \\
6  &  262.6049155-34.6934384  &  17:30:25.18  &  -34:41:36.38  &  1.6  &  1.48  &  129  &  27 $\pm$ 5  &  -18.68 $\pm$ 0.13  &  2.13 $\pm$ 0.65  &  Single \\
7  &  262.6137738-34.6947298  &  17:30:27.31  &  -34:41:41.03  &  1.63  &  1.28  &  81  &  26 $\pm$ 5  &  --  &  --  &  -- \\
8  &  262.6039531-34.6936374  &  17:30:24.95  &  -34:41:37.09  &  2.26  &  1.57  &  97  &  26 $\pm$ 5  &  --  &  --  &  -- \\
9  &  262.6192359-34.690365  &  17:30:28.62  &  -34:41:25.31  &  2.71  &  1.93  &  62  &  26 $\pm$ 5  &  --  &  --  &  -- \\
11  &  262.6243189-34.688078  &  17:30:29.84  &  -34:41:17.08  &  2.16  &  1.95  &  171  &  29 $\pm$ 6  &  --  &  --  &  -- \\
12  &  262.6072148-34.6969795  &  17:30:25.73  &  -34:41:49.13  &  2.98  &  2.02  &  86  &  28 $\pm$ 6  &  -17.84 $\pm$ 0.02  &  1.02 $\pm$ 0.06  &  Single \\
13  &  262.6078228-34.6996836  &  17:30:25.88  &  -34:41:58.86  &  1.91  &  1.47  &  154  &  26 $\pm$ 5  &  --  &  --  &  -- \\
14  &  262.6147937-34.6946762  &  17:30:27.55  &  -34:41:40.83  &  2.0  &  1.58  &  124  &  26 $\pm$ 5  &  -14.36 $\pm$ 0.09  &  1.18 $\pm$ 0.29  &  Single \\
15  &  262.6107433-34.6964412  &  17:30:26.58  &  -34:41:47.19  &  1.96  &  1.59  &  94  &  28 $\pm$ 6  &  -14.79 $\pm$ 0.09  &  --  &  Complex \\
16  &  262.6215941-34.6989408  &  17:30:29.18  &  -34:41:56.19  &  2.7  &  2.5  &  19  &  22 $\pm$ 5  &  --  &  --  &  -- \\
17  &  262.5954514-34.6916168  &  17:30:22.91  &  -34:41:29.82  &  1.57  &  1.43  &  88  &  19 $\pm$ 4  &  --  &  --  &  -- \\
18  &  262.5927434-34.7052494  &  17:30:22.26  &  -34:42:18.90  &  1.95  &  1.53  &  90  &  18 $\pm$ 4  &  --  &  --  &  -- \\
19  &  262.6064012-34.7019756  &  17:30:25.54  &  -34:42:7.110  &  1.55  &  1.2  &  58  &  24 $\pm$ 5  &  --  &  --  &  -- \\
20  &  262.6111096-34.6932787  &  17:30:26.67  &  -34:41:35.80  &  1.66  &  1.62  &  178  &  29 $\pm$ 6  &  --  &  --  &  -- \\
21  &  262.613191-34.6939495  &  17:30:27.17  &  -34:41:38.22  &  2.1  &  1.48  &  108  &  26 $\pm$ 5  &  -18.56 $\pm$ 0.06  &  0.69 $\pm$ 0.23  &  Single \\
22  &  262.6118441-34.694615  &  17:30:26.84  &  -34:41:40.61  &  1.89  &  1.6  &  97  &  28 $\pm$ 6  &  --  &  --  &  -- \\
23  &  262.6028175-34.6925438  &  17:30:24.68  &  -34:41:33.16  &  1.84  &  1.31  &  113  &  24 $\pm$ 5  &  --  &  --  &  -- \\
24  &  262.6198349-34.6960383  &  17:30:28.76  &  -34:41:45.74  &  1.88  &  1.72  &  76  &  24 $\pm$ 5  &  -19.51 $\pm$ 0.09  &  1.03 $\pm$ 0.35  &  Single \\
25  &  262.6155222-34.6952591  &  17:30:27.73  &  -34:41:42.93  &  1.88  &  1.2  &  137  &  25 $\pm$ 5  &  --  &  --  &  -- \\
26  &  262.6143434-34.6917027  &  17:30:27.44  &  -34:41:30.13  &  1.62  &  1.35  &  110  &  25 $\pm$ 5  &  --  &  --  &  -- \\
27  &  262.6000802-34.6910324  &  17:30:24.02  &  -34:41:27.72  &  3.39  &  2.5  &  48  &  22 $\pm$ 4  &  --  &  --  &  -- \\
28  &  262.6253977-34.6999713  &  17:30:30.10  &  -34:41:59.90  &  2.53  &  1.88  &  38  &  20 $\pm$ 4  &  --  &  --  &  -- \\
29  &  262.6133074-34.6919187  &  17:30:27.19  &  -34:41:30.91  &  1.7  &  1.32  &  79  &  25 $\pm$ 5  &  --  &  --  &  -- \\
30  &  262.6114686-34.6962602  &  17:30:26.75  &  -34:41:46.54  &  1.52  &  1.41  &  31  &  27 $\pm$ 6  &  -12.81 $\pm$ 0.07  &  1.77 $\pm$ 0.3  &  Single \\
31  &  262.6096651-34.692568  &  17:30:26.32  &  -34:41:33.24  &  1.71  &  1.33  &  119  &  30 $\pm$ 6  &  --  &  --  &  -- \\
32  &  262.6094126-34.6910985  &  17:30:26.26  &  -34:41:27.95  &  3.31  &  2.99  &  43  &  28 $\pm$ 6  &  --  &  --  &  -- \\
33  &  262.6287106-34.6862068  &  17:30:30.89  &  -34:41:10.34  &  2.41  &  2.0  &  24  &  24 $\pm$ 5  &  --  &  --  &  -- \\
34  &  262.5982914-34.6919006  &  17:30:23.59  &  -34:41:30.84  &  1.81  &  1.36  &  139  &  22 $\pm$ 5  &  -18.77 $\pm$ 0.08  &  1.1 $\pm$ 0.16  &  Single \\
35  &  262.6142011-34.6940134  &  17:30:27.41  &  -34:41:38.45  &  2.32  &  1.82  &  111  &  25 $\pm$ 5  &  --  &  --  &  -- \\
36  &  262.6202758-34.7001995  &  17:30:28.87  &  -34:42:0.720  &  2.29  &  1.87  &  92  &  23 $\pm$ 5  &  --  &  --  &  -- \\
37  &  262.6010398-34.6950114  &  17:30:24.25  &  -34:41:42.04  &  1.74  &  1.33  &  75  &  25 $\pm$ 5  &  -17.83 $\pm$ 0.05  &  0.97 $\pm$ 0.21  &  Single \\
38  &  262.5971034-34.6920396  &  17:30:23.30  &  -34:41:31.34  &  2.01  &  1.6  &  105  &  21 $\pm$ 4  &  --  &  --  &  -- \\
39  &  262.6054437-34.6963773  &  17:30:25.31  &  -34:41:46.96  &  2.03  &  1.88  &  159  &  29 $\pm$ 6  &  --  &  --  &  -- \\
40  &  262.5917454-34.6897316  &  17:30:22.02  &  -34:41:23.03  &  1.87  &  1.47  &  96  &  18 $\pm$ 4  &  --  &  --  &  -- \\
41  &  262.6095777-34.6983259  &  17:30:26.30  &  -34:41:53.97  &  2.24  &  1.83  &  72  &  26 $\pm$ 5  &  --  &  --  &  -- \\
42  &  262.5975992-34.6876666  &  17:30:23.42  &  -34:41:15.60  &  2.01  &  1.14  &  21  &  18 $\pm$ 4  &  --  &  --  &  -- \\
43  &  262.6035166-34.6966807  &  17:30:24.84  &  -34:41:48.05  &  2.48  &  1.78  &  95  &  27 $\pm$ 5  &  -17.34 $\pm$ 0.06  &  0.59 $\pm$ 0.19  &  Single \\
44  &  262.6030115-34.6956424  &  17:30:24.72  &  -34:41:44.31  &  1.76  &  1.55  &  97  &  27 $\pm$ 5  &  -17.0 $\pm$ 0.09  &  1.1 $\pm$ 0.31  &  Single \\
45  &  262.6143008-34.6909376  &  17:30:27.43  &  -34:41:27.38  &  4.09  &  3.38  &  67  &  25 $\pm$ 5  &  --  &  --  &  -- \\
46  &  262.6187648-34.6912377  &  17:30:28.50  &  -34:41:28.46  &  2.76  &  2.12  &  53  &  25 $\pm$ 5  &  --  &  --  &  -- \\
47  &  262.6178453-34.6919943  &  17:30:28.28  &  -34:41:31.18  &  3.17  &  2.9  &  168  &  25 $\pm$ 5  &  --  &  --  &  -- \\
\hline \noalign {\smallskip}
\end{tabular}
\end{table*}

\begin{figure*}
    \centering

\includegraphics[width=0.24\textwidth]{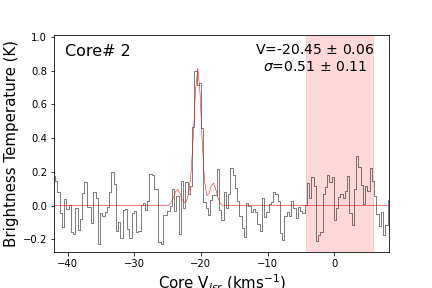}
\includegraphics[width=0.24\textwidth]{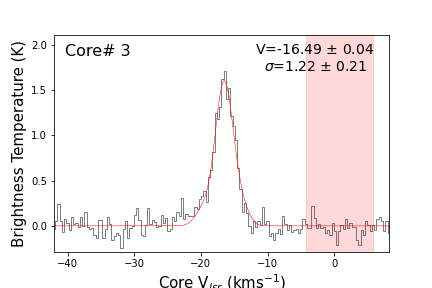}
\includegraphics[width=0.24\textwidth]{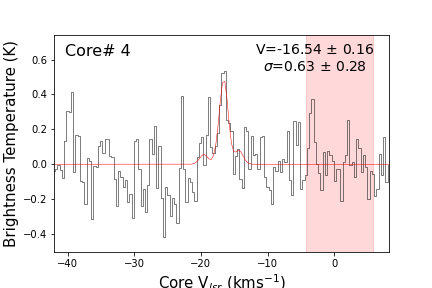}
\includegraphics[width=0.24\textwidth]{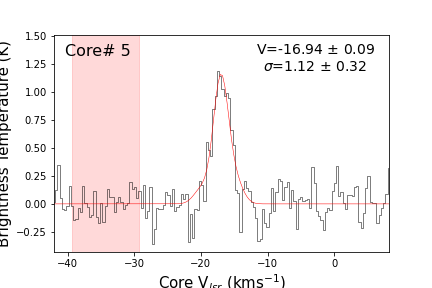}
\includegraphics[width=0.24\textwidth]{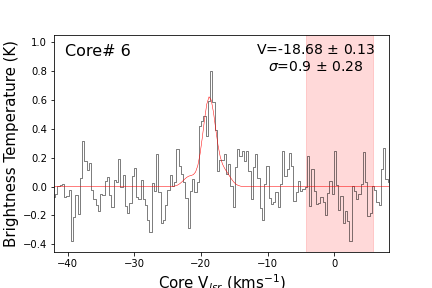}
\includegraphics[width=0.24\textwidth]{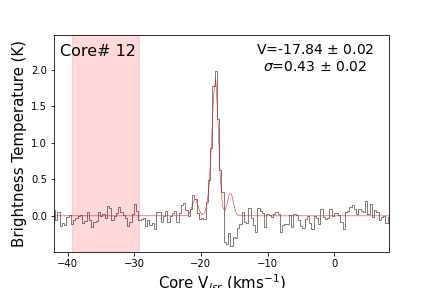}
\includegraphics[width=0.24\textwidth]{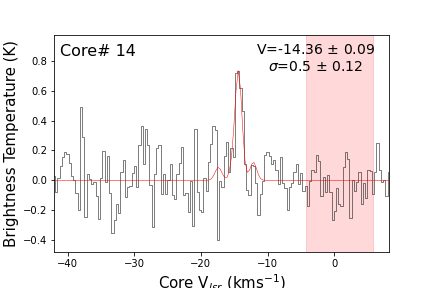}
\includegraphics[width=0.24\textwidth]{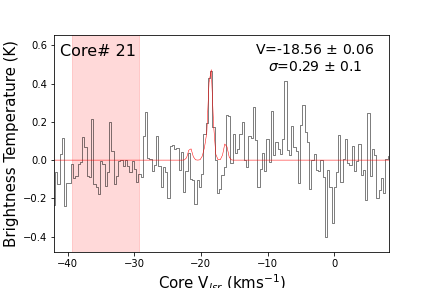}
\includegraphics[width=0.24\textwidth]{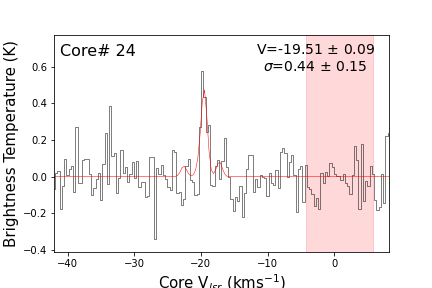}
\includegraphics[width=0.24\textwidth]{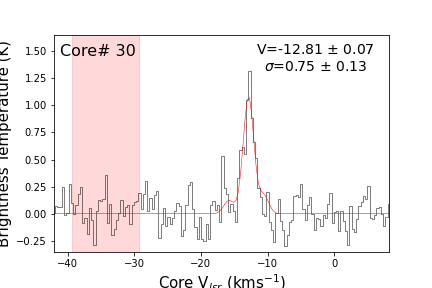}
\includegraphics[width=0.24\textwidth]{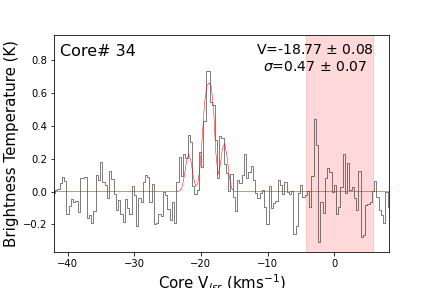}
\includegraphics[width=0.24\textwidth]{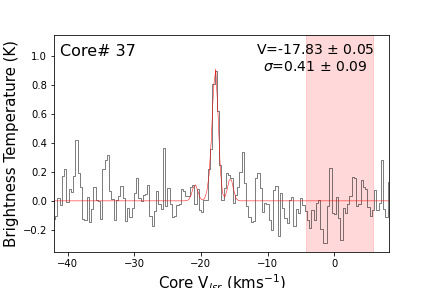}
\includegraphics[width=0.24\textwidth]{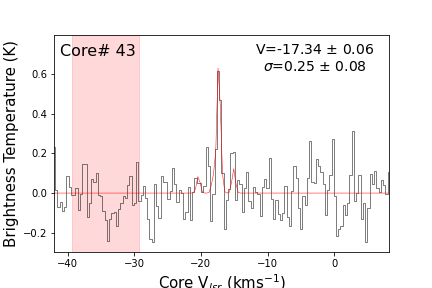}
\includegraphics[width=0.24\textwidth]{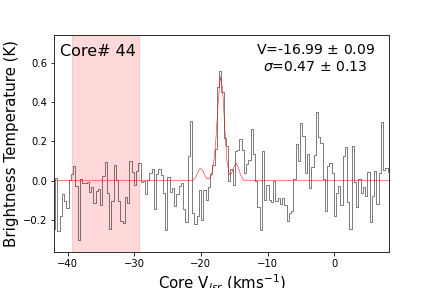}

\caption{Single-type core-averaged, background-subtracted DCN spectra extracted from the cores in the intermediate protocluster G353.41. See \cref{tabappendix:coretables_g353} for the line fit parameters for each core.\label{figspectra:dcnspectra_split_G353_s}}
\end{figure*}

\begin{figure*}
    \centering
\includegraphics[width=0.24\textwidth]{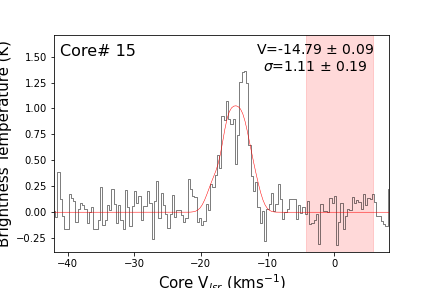}
\caption{Complex-type core-averaged, background-subtracted DCN spectra extracted from the cores in the intermediate protocluster G353.41. See \cref{tabappendix:coretables_g353} for the line fit parameters for each core.\label{figspectra:dcnspectra_split_G353_c}} 
\end{figure*}
\clearpage

\begin{table*}[htbp!]
\centering
\small
\caption{DCN fits towards the core population of the young protocluster W43-MM1.}
\label{tabappendix:coretables_w43_mm1}
\begin{tabular}{llllllllccc}

\hline 
n   & Core Name &RA  & DEC & F$_{A}$   & F$_{B}$    & PA & T & \vlsr  & Linewidth   & Spectral \\
 &  & [ICRS]  &  [ICRS]   &  [\arcsec] &[\arcsec] &  [deg] & [K] & [\kms]  & [\kms] & Type \\
\hline 
1  &  281.9459334-1.9074736  &  18:47:47.02  &  -01:54:26.90  &  0.71  &  0.62  &  88  &  100 $\pm$ 50  &  102.22 $\pm$ 0.31  &  --  &  Complex \\
2  &  281.9451648-1.9081343  &  18:47:46.84  &  -01:54:29.28  &  0.61  &  0.56  &  143  &  100 $\pm$ 50  &  96.01 $\pm$ 0.04  &  --  &  Complex \\
3  &  281.9432069-1.9092736  &  18:47:46.37  &  -01:54:33.38  &  0.72  &  0.63  &  37  &  100 $\pm$ 50  &  96.9 $\pm$ 0.08  &  --  &  Complex \\
4  &  281.945748-1.9073463  &  18:47:46.98  &  -01:54:26.45  &  0.82  &  0.77  &  62  &  100 $\pm$ 50  &  103.56 $\pm$ 0.36  &  --  &  Complex \\
5  &  281.9469228-1.908243  &  18:47:47.26  &  -01:54:29.67  &  0.65  &  0.49  &  96  &  18 $\pm$ 4  &  99.13 $\pm$ 0.07  &  1.33 $\pm$ 0.34  &  Single \\
6  &  281.9423152-1.909246  &  18:47:46.16  &  -01:54:33.29  &  0.7  &  0.54  &  104  &  22 $\pm$ 4  &  --  &  --  &  -- \\
7  &  281.9438998-1.9064175  &  18:47:46.54  &  -01:54:23.10  &  0.72  &  0.5  &  101  &  100 $\pm$ 50  &  95.99 $\pm$ 0.02  &  0.73 $\pm$ 0.05  &  Single \\
8  &  281.938703-1.9102889  &  18:47:45.29  &  -01:54:37.04  &  0.6  &  0.49  &  99  &  21 $\pm$ 4  &  --  &  --  &  -- \\
9  &  281.9448606-1.9086666  &  18:47:46.77  &  -01:54:31.20  &  0.66  &  0.54  &  67  &  100 $\pm$ 50  &  95.9 $\pm$ 0.13  &  --  &  Complex \\
10  &  281.9458236-1.9071389  &  18:47:47.00  &  -01:54:25.70  &  0.93  &  0.86  &  131  &  35 $\pm$ 10  &  102.92 $\pm$ 0.25  &  --  &  Complex \\
11  &  281.9365573-1.9125569  &  18:47:44.77  &  -01:54:45.20  &  0.69  &  0.58  &  101  &  100 $\pm$ 50  &  94.74 $\pm$ 0.05  &  2.49 $\pm$ 0.95  &  Single \\
12  &  281.9436641-1.9090461  &  18:47:46.48  &  -01:54:32.57  &  0.65  &  0.54  &  91  &  100 $\pm$ 50  &  95.0 $\pm$ 0.14  &  --  &  Complex \\
13  &  281.9427046-1.9092745  &  18:47:46.25  &  -01:54:33.39  &  0.69  &  0.55  &  132  &  22 $\pm$ 4  &  94.52 $\pm$ 0.07  &  --  &  Complex \\
14  &  281.944974-1.9044595  &  18:47:46.79  &  -01:54:16.05  &  0.6  &  0.47  &  97  &  100 $\pm$ 50  &  --  &  --  &  -- \\
15  &  281.9438148-1.9067276  &  18:47:46.52  &  -01:54:24.22  &  0.66  &  0.51  &  96  &  100 $\pm$ 50  &  93.98 $\pm$ 0.04  &  --  &  Complex \\
16  &  281.9372707-1.9118969  &  18:47:44.94  &  -01:54:42.83  &  0.66  &  0.54  &  134  &  21 $\pm$ 4  &  --  &  --  &  -- \\
17  &  281.9454215-1.9083192  &  18:47:46.90  &  -01:54:29.95  &  0.78  &  0.67  &  121  &  100 $\pm$ 50  &  --  &  --  &  -- \\
18  &  281.9462451-1.9075171  &  18:47:47.10  &  -01:54:27.06  &  0.64  &  0.57  &  92  &  35 $\pm$ 10  &  99.79 $\pm$ 0.1  &  --  &  Complex \\
20  &  281.9456983-1.9036124  &  18:47:46.97  &  -01:54:13.00  &  1.07  &  0.95  &  39  &  21 $\pm$ 4  &  --  &  --  &  -- \\
21  &  281.9453834-1.9082357  &  18:47:46.89  &  -01:54:29.65  &  0.74  &  0.58  &  121  &  35 $\pm$ 10  &  --  &  --  &  -- \\
22  &  281.9384338-1.9107821  &  18:47:45.22  &  -01:54:38.82  &  0.6  &  0.51  &  122  &  21 $\pm$ 4  &  --  &  --  &  -- \\
23  &  281.9452927-1.9040458  &  18:47:46.87  &  -01:54:14.56  &  0.57  &  0.45  &  86  &  21 $\pm$ 4  &  --  &  --  &  -- \\
24  &  281.9440629-1.9089  &  18:47:46.58  &  -01:54:32.04  &  0.69  &  0.49  &  82  &  22 $\pm$ 5  &  --  &  --  &  -- \\
25  &  281.9459111-1.9085541  &  18:47:47.02  &  -01:54:30.79  &  0.78  &  0.6  &  146  &  100 $\pm$ 50  &  98.37 $\pm$ 0.04  &  1.39 $\pm$ 0.13  &  Single \\
26  &  281.94313-1.9081925  &  18:47:46.35  &  -01:54:29.49  &  0.62  &  0.57  &  105  &  23 $\pm$ 5  &  --  &  --  &  -- \\
27  &  281.9472105-1.9035597  &  18:47:47.33  &  -01:54:12.81  &  0.62  &  0.55  &  102  &  20 $\pm$ 4  &  --  &  --  &  -- \\
28  &  281.9453005-1.9071369  &  18:47:46.87  &  -01:54:25.69  &  0.57  &  0.52  &  133  &  100 $\pm$ 50  &  99.91 $\pm$ 0.33  &  --  &  Complex \\
29  &  281.9460739-1.9089351  &  18:47:47.06  &  -01:54:32.17  &  0.54  &  0.44  &  115  &  19 $\pm$ 4  &  --  &  --  &  -- \\
30  &  281.9456971-1.9082389  &  18:47:46.97  &  -01:54:29.66  &  0.63  &  0.47  &  97  &  25 $\pm$ 5  &  --  &  --  &  -- \\
31  &  281.9337243-1.9135618  &  18:47:44.09  &  -01:54:48.82  &  0.65  &  0.57  &  18  &  21 $\pm$ 4  &  --  &  --  &  -- \\
32  &  281.9455609-1.9075139  &  18:47:46.93  &  -01:54:27.05  &  0.71  &  0.56  &  99  &  35 $\pm$ 10  &  103.86 $\pm$ 0.14  &  --  &  Complex \\
33  &  281.9438197-1.9079708  &  18:47:46.52  &  -01:54:28.69  &  0.68  &  0.56  &  73  &  23 $\pm$ 5  &  97.55 $\pm$ 0.02  &  0.87 $\pm$ 0.07  &  Single \\
34  &  281.9358647-1.9117019  &  18:47:44.61  &  -01:54:42.13  &  0.57  &  0.49  &  88  &  21 $\pm$ 4  &  96.08 $\pm$ 0.05  &  --  &  Complex \\
35  &  281.9385991-1.9110901  &  18:47:45.26  &  -01:54:39.92  &  0.95  &  0.66  &  124  &  21 $\pm$ 4  &  --  &  --  &  -- \\
36  &  281.9443508-1.9053944  &  18:47:46.64  &  -01:54:19.42  &  0.55  &  0.53  &  72  &  24 $\pm$ 5  &  --  &  --  &  -- \\
37  &  281.9454183-1.9067383  &  18:47:46.90  &  -01:54:24.26  &  0.68  &  0.48  &  109  &  100 $\pm$ 50  &  101.59 $\pm$ 0.02  &  0.79 $\pm$ 0.05  &  Single \\
38  &  281.9507008-1.9010492  &  18:47:48.17  &  -01:54:3.780  &  0.96  &  0.68  &  123  &  18 $\pm$ 4  &  --  &  --  &  -- \\
39  &  281.9365275-1.9129829  &  18:47:44.77  &  -01:54:46.74  &  0.76  &  0.58  &  116  &  21 $\pm$ 4  &  96.97 $\pm$ 0.05  &  0.68 $\pm$ 0.15  &  Single \\
40  &  281.9440532-1.9057986  &  18:47:46.57  &  -01:54:20.87  &  0.86  &  0.64  &  40  &  25 $\pm$ 5  &  95.79 $\pm$ 0.04  &  0.48 $\pm$ 0.1  &  Single \\
41  &  281.945843-1.9046555  &  18:47:47.00  &  -01:54:16.76  &  0.59  &  0.46  &  104  &  21 $\pm$ 4  &  --  &  --  &  -- \\
42  &  281.9455026-1.9079514  &  18:47:46.92  &  -01:54:28.63  &  0.64  &  0.38  &  104  &  26 $\pm$ 5  &  98.68 $\pm$ 0.08  &  --  &  Complex \\
43  &  281.9558069-1.9010298  &  18:47:49.39  &  -01:54:3.710  &  0.76  &  0.69  &  28  &  20 $\pm$ 4  &  --  &  --  &  -- \\
44  &  281.9465996-1.9059703  &  18:47:47.18  &  -01:54:21.49  &  0.56  &  0.54  &  153  &  20 $\pm$ 4  &  --  &  --  &  -- \\
45  &  281.9446978-1.9048517  &  18:47:46.73  &  -01:54:17.47  &  0.62  &  0.44  &  92  &  22 $\pm$ 5  &  --  &  --  &  -- \\
46  &  281.9473216-1.9037125  &  18:47:47.36  &  -01:54:13.36  &  0.57  &  0.54  &  120  &  20 $\pm$ 4  &  --  &  --  &  -- \\
47  &  281.9350259-1.9116027  &  18:47:44.41  &  -01:54:41.77  &  0.79  &  0.44  &  71  &  21 $\pm$ 4  &  --  &  --  &  -- \\
48  &  281.9377497-1.9116788  &  18:47:45.06  &  -01:54:42.04  &  0.68  &  0.45  &  74  &  21 $\pm$ 4  &  --  &  --  &  -- \\
49  &  281.9408041-1.9090791  &  18:47:45.79  &  -01:54:32.68  &  0.69  &  0.52  &  89  &  21 $\pm$ 4  &  --  &  --  &  -- \\
50  &  281.9494405-1.8997773  &  18:47:47.87  &  -01:53:59.20  &  0.91  &  0.67  &  176  &  18 $\pm$ 4  &  --  &  --  &  -- \\
51  &  281.938249-1.9109571  &  18:47:45.18  &  -01:54:39.45  &  0.66  &  0.55  &  85  &  21 $\pm$ 4  &  --  &  --  &  -- \\
52  &  281.9365434-1.9122622  &  18:47:44.77  &  -01:54:44.14  &  0.65  &  0.52  &  92  &  22 $\pm$ 4  &  --  &  --  &  -- \\
53  &  281.9457328-1.9089106  &  18:47:46.98  &  -01:54:32.08  &  0.6  &  0.42  &  97  &  20 $\pm$ 4  &  99.83 $\pm$ 0.06  &  0.79 $\pm$ 0.21  &  Single \\
54  &  281.936822-1.9119242  &  18:47:44.84  &  -01:54:42.93  &  0.69  &  0.6  &  54  &  21 $\pm$ 4  &  --  &  --  &  -- \\
55  &  281.9414962-1.9094029  &  18:47:45.96  &  -01:54:33.85  &  0.89  &  0.72  &  124  &  21 $\pm$ 4  &  --  &  --  &  -- \\
56  &  281.9391449-1.9108264  &  18:47:45.39  &  -01:54:38.98  &  0.77  &  0.69  &  123  &  21 $\pm$ 4  &  --  &  --  &  -- \\
58  &  281.9465409-1.9030795  &  18:47:47.17  &  -01:54:11.09  &  0.8  &  0.66  &  130  &  20 $\pm$ 4  &  --  &  --  &  -- \\
59  &  281.9466558-1.9063105  &  18:47:47.20  &  -01:54:22.72  &  0.74  &  0.69  &  15  &  20 $\pm$ 4  &  97.32 $\pm$ 0.02  &  1.26 $\pm$ 0.08  &  Single \\
60  &  281.9347438-1.9117236  &  18:47:44.34  &  -01:54:42.20  &  1.0  &  0.93  &  152  &  21 $\pm$ 4  &  --  &  --  &  -- \\
61  &  281.9473742-1.9033061  &  18:47:47.37  &  -01:54:11.90  &  0.61  &  0.54  &  64  &  20 $\pm$ 4  &  --  &  --  &  -- \\
62  &  281.9364851-1.9112375  &  18:47:44.76  &  -01:54:40.45  &  0.58  &  0.5  &  84  &  21 $\pm$ 4  &  --  &  --  &  -- \\
63  &  281.9408778-1.9162621  &  18:47:45.81  &  -01:54:58.54  &  1.33  &  1.25  &  38  &  25 $\pm$ 5  &  --  &  --  &  -- \\
64  &  281.9416193-1.9070122  &  18:47:45.99  &  -01:54:25.24  &  0.64  &  0.51  &  166  &  23 $\pm$ 5  &  --  &  --  &  -- \\
66  &  281.9411991-1.9097694  &  18:47:45.89  &  -01:54:35.17  &  1.19  &  0.91  &  4  &  21 $\pm$ 4  &  --  &  --  &  -- \\
67  &  281.9459087-1.9063496  &  18:47:47.02  &  -01:54:22.86  &  0.64  &  0.46  &  149  &  21 $\pm$ 4  &  101.17 $\pm$ 0.03  &  1.13 $\pm$ 0.11  &  Single \\
68  &  281.9475634-1.907071  &  18:47:47.42  &  -01:54:25.46  &  1.07  &  0.77  &  23  &  19 $\pm$ 4  &  98.57 $\pm$ 0.05  &  1.14 $\pm$ 0.2  &  Single \\
\hline \noalign {\smallskip}
\end{tabular}
\end{table*}

\begin{table*}[htbp!]
\ContinuedFloat
\centering
\small
\caption{Continued DCN fits towards the core population of the young protocluster W43-MM1.}
\begin{tabular}{llllllllccc}

\hline 
n   & Core Name &RA  & DEC & F$_{A}$   & F$_{B}$    & PA & T & \vlsr  & Linewidth   & Spectral \\
 &  & [ICRS]  &  [ICRS]   &  [\arcsec] &[\arcsec] &  [deg] & [K] & [\kms]  & [\kms] & Type \\
\hline 
69  &  281.946263-1.9095657  &  18:47:47.10  &  -01:54:34.44  &  0.9  &  0.73  &  17  &  18 $\pm$ 4  &  97.8 $\pm$ 0.07  &  0.63 $\pm$ 0.26  &  Single \\
70  &  281.9449909-1.9057425  &  18:47:46.80  &  -01:54:20.67  &  1.09  &  0.86  &  23  &  22 $\pm$ 5  &  101.7 $\pm$ 0.02  &  1.74 $\pm$ 0.25  &  Single \\
73  &  281.9458121-1.905854  &  18:47:46.99  &  -01:54:21.07  &  0.66  &  0.53  &  143  &  20 $\pm$ 4  &  100.74 $\pm$ 0.02  &  0.83 $\pm$ 0.09  &  Single \\
74  &  281.9432263-1.9058611  &  18:47:46.37  &  -01:54:21.10  &  0.82  &  0.65  &  161  &  25 $\pm$ 5  &  --  &  --  &  -- \\
75  &  281.9437613-1.9073486  &  18:47:46.50  &  -01:54:26.45  &  0.82  &  0.61  &  129  &  25 $\pm$ 5  &  --  &  --  &  -- \\
\hline \noalign {\smallskip}
\end{tabular}
\end{table*}

\begin{figure*}
    \centering
\includegraphics[width=0.24\textwidth]{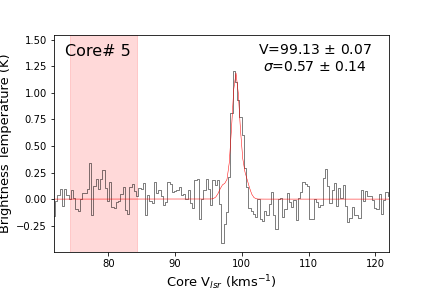}
\includegraphics[width=0.24\textwidth]{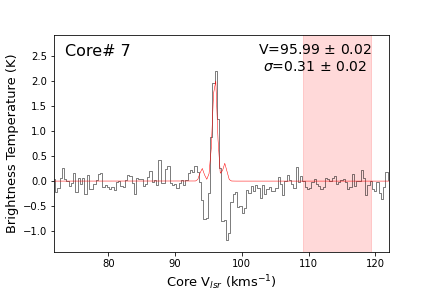}
\includegraphics[width=0.24\textwidth]{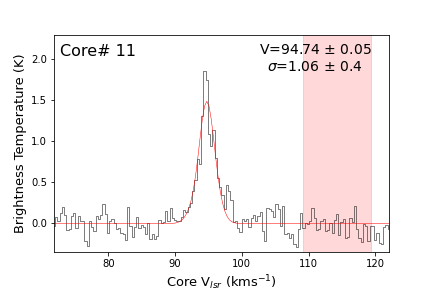}
\includegraphics[width=0.24\textwidth]{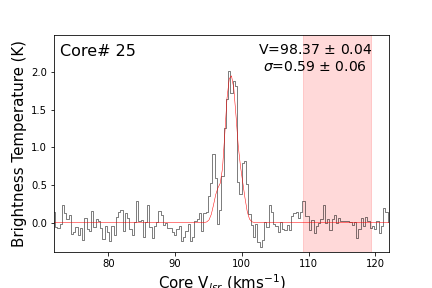}
\includegraphics[width=0.24\textwidth]{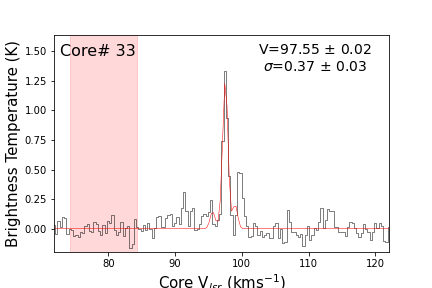}
\includegraphics[width=0.24\textwidth]{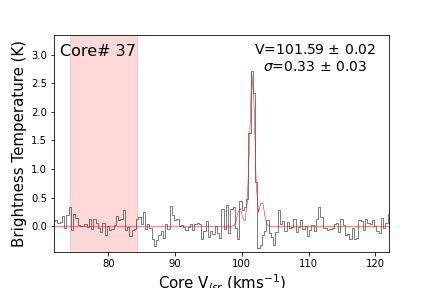}
\includegraphics[width=0.24\textwidth]{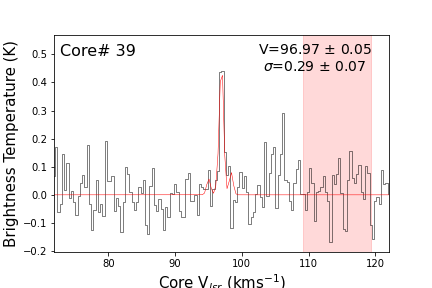}
\includegraphics[width=0.24\textwidth]{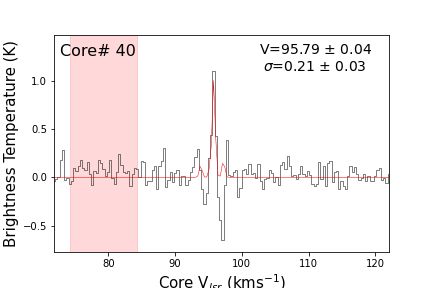}
\includegraphics[width=0.24\textwidth]{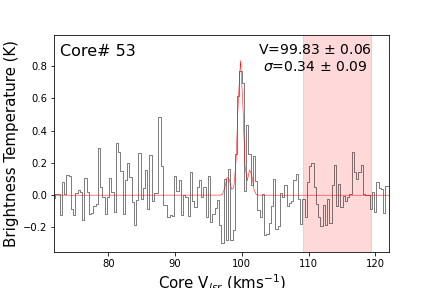}
\includegraphics[width=0.24\textwidth]{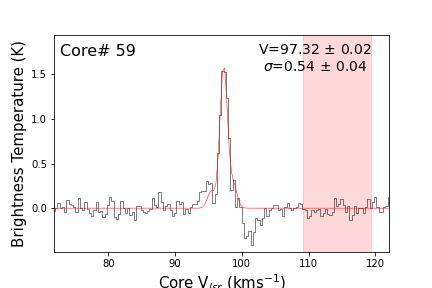}
\includegraphics[width=0.24\textwidth]{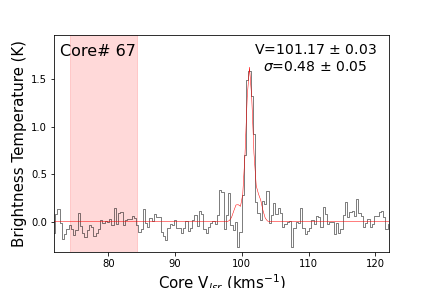}
\includegraphics[width=0.24\textwidth]{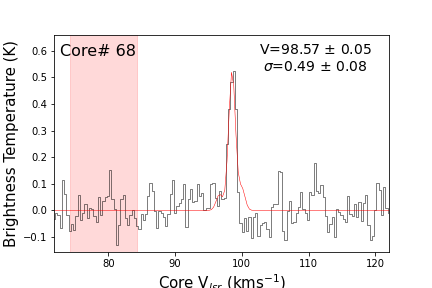}
\includegraphics[width=0.24\textwidth]{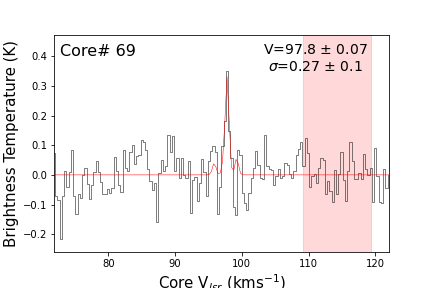}
\includegraphics[width=0.24\textwidth]{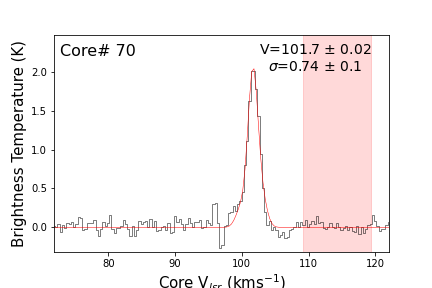}
\includegraphics[width=0.24\textwidth]{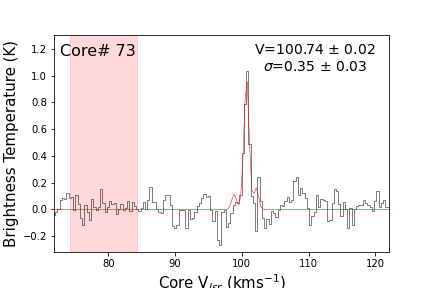}
\caption{ Single-type core-averaged, background-subtracted DCN spectra extracted from the cores in the young protocluster W43-MM1. Readers can refer to \cref{tabappendix:coretables_w43_mm1} for the line fit parameters for each core.}
\end{figure*}

\begin{figure*}
    \centering
\includegraphics[width=0.24\textwidth]{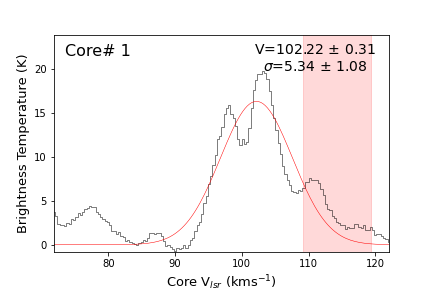}
\includegraphics[width=0.24\textwidth]{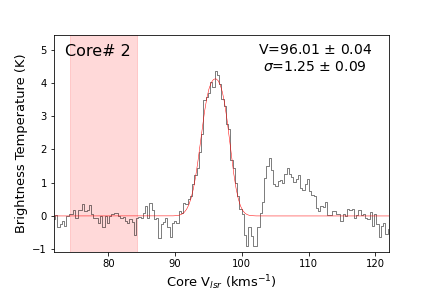}
\includegraphics[width=0.24\textwidth]{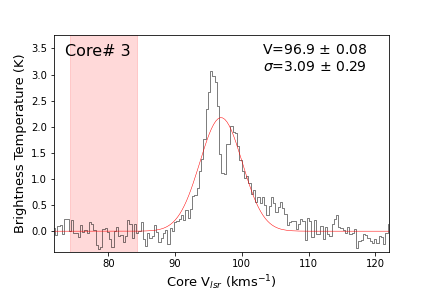}
\includegraphics[width=0.24\textwidth]{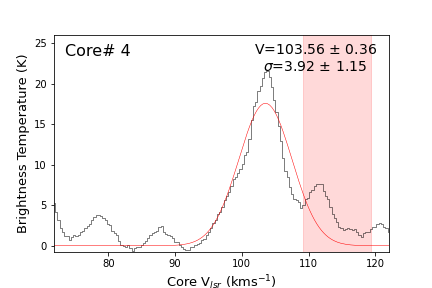}
\includegraphics[width=0.24\textwidth]{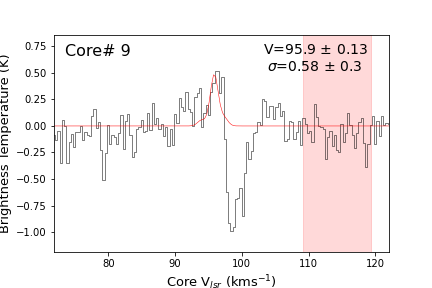}
\includegraphics[width=0.24\textwidth]{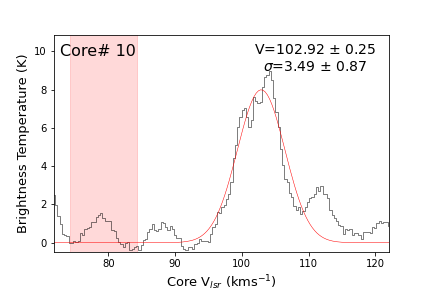}
\includegraphics[width=0.24\textwidth]{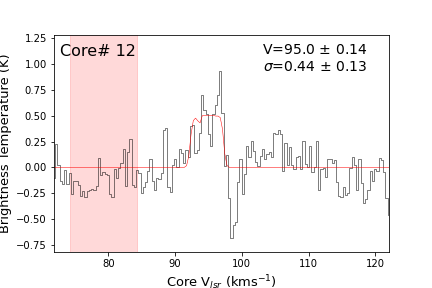}
\includegraphics[width=0.24\textwidth]{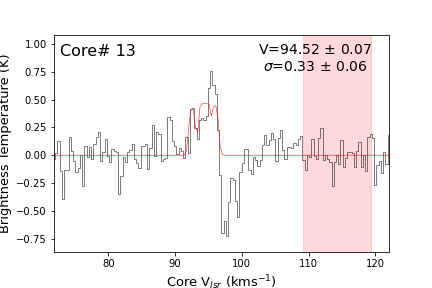}
\includegraphics[width=0.24\textwidth]{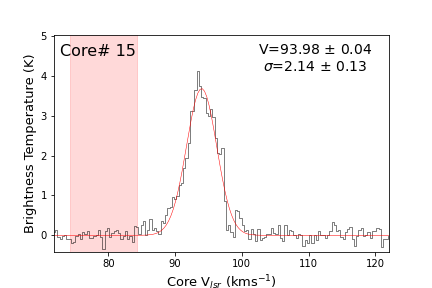}
\includegraphics[width=0.24\textwidth]{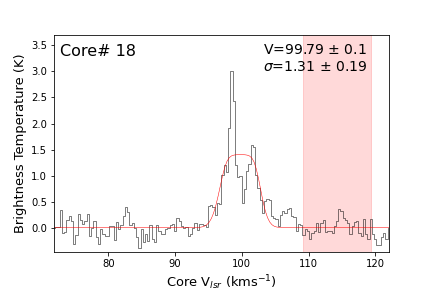}
\includegraphics[width=0.24\textwidth]{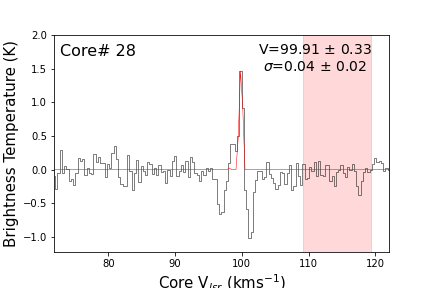}
\includegraphics[width=0.24\textwidth]{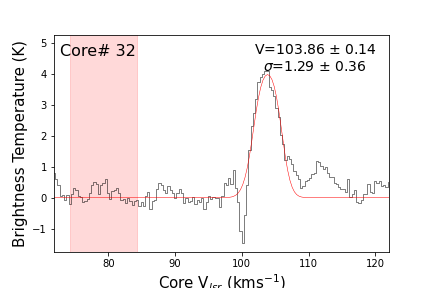}
\includegraphics[width=0.24\textwidth]{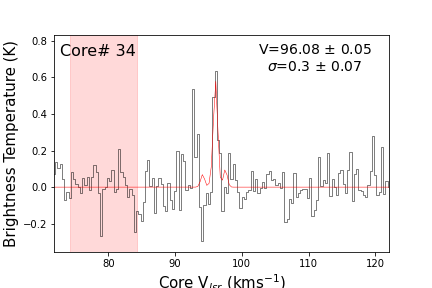}
\includegraphics[width=0.24\textwidth]{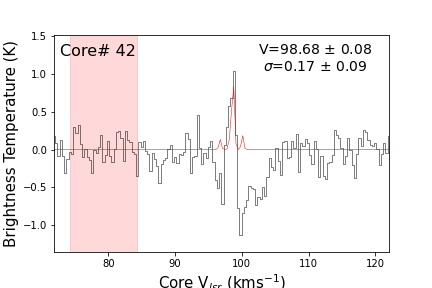}
\caption{Complex-type core-averaged, background-subtracted DCN spectra extracted from the cores in the young protocluster W43-MM1. Readers can refer to \cref{tabappendix:coretables_w43_mm1} for the line fit parameters for each core.} 
\end{figure*}

\begin{table*}[htp]
\centering
\small
\caption{DCN fits towards the core population of the young protocluster W43-MM2.}
\label{tabappendix:coretables_w43_mm2}
\begin{tabular}{llllllllccc}
\hline 
n   & Core Name &RA  & DEC & F$_{A}$   & F$_{B}$    & PA & T & \vlsr  & Linewidth   & Spectral \\
 &  & [ICRS]  &  [ICRS]   &  [\arcsec] &[\arcsec] &  [deg] & [K] & [\kms]  & [\kms] & Type \\
\hline 
1  &  281.9033208-2.0150752  &  18:47:36.80  &  -02:00:54.27  &  0.76  &  0.7  &  147  &  100 $\pm$ 50  &  89.59 $\pm$ 0.08  &  --  &  Complex \\
2  &  281.9001273-2.0224281  &  18:47:36.03  &  -02:01:20.74  &  0.78  &  0.48  &  84  &  19 $\pm$ 4  &  90.33 $\pm$ 0.07  &  1.25 $\pm$ 0.33  &  Single \\
3  &  281.9004197-2.0210949  &  18:47:36.10  &  -02:01:15.94  &  0.9  &  0.7  &  80  &  18 $\pm$ 4  &  90.14 $\pm$ 0.06  &  2.21 $\pm$ 1.43  &  Single \\
4  &  281.9006289-2.0132967  &  18:47:36.15  &  -02:00:47.87  &  0.59  &  0.52  &  64  &  18 $\pm$ 4  &  --  &  --  &  -- \\
5  &  281.9011837-2.0141012  &  18:47:36.28  &  -02:00:50.76  &  0.67  &  0.54  &  175  &  19 $\pm$ 4  &  91.13 $\pm$ 0.03  &  2.11 $\pm$ 0.6  &  Single \\
6  &  281.9031379-2.0149264  &  18:47:36.75  &  -02:00:53.74  &  0.79  &  0.73  &  56  &  35 $\pm$ 10  &  88.33 $\pm$ 0.03  &  --  &  Complex \\
7  &  281.9029377-2.0132083  &  18:47:36.71  &  -02:00:47.55  &  0.72  &  0.55  &  118  &  100 $\pm$ 50  &  91.59 $\pm$ 0.02  &  1.84 $\pm$ 0.13  &  Single \\
8  &  281.8962619-2.0191026  &  18:47:35.10  &  -02:01:8.770  &  0.64  &  0.55  &  128  &  18 $\pm$ 4  &  --  &  --  &  -- \\
9  &  281.9034863-2.0173945  &  18:47:36.84  &  -02:01:2.620  &  0.74  &  0.54  &  132  &  20 $\pm$ 4  &  89.57 $\pm$ 0.03  &  1.47 $\pm$ 0.16  &  Single \\
10  &  281.9135971-2.0078124  &  18:47:39.26  &  -02:00:28.12  &  0.79  &  0.74  &  148  &  23 $\pm$ 5  &  93.99 $\pm$ 0.09  &  --  &  Complex \\
11  &  281.8987232-2.0090321  &  18:47:35.69  &  -02:00:32.52  &  0.62  &  0.49  &  148  &  21 $\pm$ 4  &  92.08 $\pm$ 0.02  &  1.42 $\pm$ 0.1  &  Single \\
12  &  281.9027193-2.014794  &  18:47:36.65  &  -02:00:53.26  &  0.86  &  0.72  &  86  &  29 $\pm$ 6  &  91.32 $\pm$ 0.03  &  2.01 $\pm$ 0.58  &  Single \\
13  &  281.9028491-2.0133518  &  18:47:36.68  &  -02:00:48.07  &  0.69  &  0.56  &  168  &  21 $\pm$ 4  &  91.49 $\pm$ 0.03  &  2.16 $\pm$ 0.41  &  Single \\
14  &  281.9002537-2.0244012  &  18:47:36.06  &  -02:01:27.84  &  0.66  &  0.57  &  111  &  20 $\pm$ 4  &  --  &  --  &  -- \\
15  &  281.9034339-2.0146926  &  18:47:36.82  &  -02:00:52.89  &  1.33  &  1.24  &  22  &  35 $\pm$ 10  &  88.08 $\pm$ 0.02  &  --  &  Complex \\
16  &  281.9006053-2.0129566  &  18:47:36.15  &  -02:00:46.64  &  0.56  &  0.48  &  98  &  18 $\pm$ 4  &  --  &  --  &  -- \\
17  &  281.9095872-2.011522  &  18:47:38.30  &  -02:00:41.48  &  0.57  &  0.48  &  104  &  18 $\pm$ 4  &  --  &  --  &  -- \\
19  &  281.8998797-2.0139649  &  18:47:35.97  &  -02:00:50.27  &  1.46  &  1.2  &  1  &  18 $\pm$ 4  &  93.14 $\pm$ 0.07  &  1.34 $\pm$ 0.25  &  Single \\
20  &  281.9031931-2.0153229  &  18:47:36.77  &  -02:00:55.16  &  0.72  &  0.69  &  110  &  35 $\pm$ 10  &  90.52 $\pm$ 0.05  &  --  &  Complex \\
21  &  281.9005682-2.0247778  &  18:47:36.14  &  -02:01:29.20  &  0.62  &  0.62  &  48  &  20 $\pm$ 4  &  --  &  --  &  -- \\
22  &  281.9034115-2.0140036  &  18:47:36.82  &  -02:00:50.41  &  0.6  &  0.54  &  154  &  24 $\pm$ 5  &  92.85 $\pm$ 0.03  &  0.8 $\pm$ 0.09  &  Single \\
24  &  281.9031397-2.0154806  &  18:47:36.75  &  -02:00:55.73  &  0.71  &  0.61  &  107  &  30 $\pm$ 6  &  88.59 $\pm$ 0.02  &  1.29 $\pm$ 0.08  &  Single \\
25  &  281.8945099-2.0177668  &  18:47:34.68  &  -02:01:3.960  &  0.73  &  0.57  &  108  &  19 $\pm$ 4  &  --  &  --  &  -- \\
26  &  281.9038579-2.0152273  &  18:47:36.93  &  -02:00:54.82  &  0.94  &  0.74  &  83  &  27 $\pm$ 6  &  90.1 $\pm$ 0.02  &  --  &  Complex \\
27  &  281.9003313-2.0203002  &  18:47:36.08  &  -02:01:13.08  &  0.51  &  0.47  &  74  &  18 $\pm$ 4  &  --  &  --  &  -- \\
28  &  281.9031601-2.0129333  &  18:47:36.76  &  -02:00:46.56  &  0.77  &  0.65  &  80  &  21 $\pm$ 4  &  92.15 $\pm$ 0.03  &  1.0 $\pm$ 0.11  &  Single \\
29  &  281.9000187-2.0137672  &  18:47:36.00  &  -02:00:49.56  &  0.78  &  0.62  &  80  &  18 $\pm$ 4  &  92.89 $\pm$ 0.1  &  1.12 $\pm$ 0.19  &  Single \\
30  &  281.9001893-2.0207471  &  18:47:36.05  &  -02:01:14.69  &  0.78  &  0.68  &  136  &  18 $\pm$ 4  &  88.97 $\pm$ 0.1  &  1.17 $\pm$ 0.42  &  Single \\
31  &  281.9020978-2.0195665  &  18:47:36.50  &  -02:01:10.44  &  1.53  &  0.97  &  82  &  18 $\pm$ 4  &  90.38 $\pm$ 0.07  &  0.86 $\pm$ 0.11  &  Single \\
32  &  281.9001361-2.0137613  &  18:47:36.03  &  -02:00:49.54  &  0.77  &  0.54  &  97  &  18 $\pm$ 4  &  93.28 $\pm$ 0.04  &  2.05 $\pm$ 0.83  &  Single \\
33  &  281.9028092-2.0166989  &  18:47:36.67  &  -02:01:0.120  &  0.69  &  0.52  &  92  &  21 $\pm$ 4  &  --  &  --  &  -- \\
35  &  281.8969217-2.0171335  &  18:47:35.26  &  -02:01:1.680  &  0.58  &  0.5  &  140  &  18 $\pm$ 4  &  --  &  --  &  -- \\
36  &  281.8949343-2.0222271  &  18:47:34.78  &  -02:01:20.02  &  1.03  &  0.75  &  78  &  19 $\pm$ 4  &  --  &  --  &  -- \\
37  &  281.904921-2.0095957  &  18:47:37.18  &  -02:00:34.54  &  1.69  &  1.23  &  139  &  20 $\pm$ 4  &  --  &  --  &  -- \\
38  &  281.9024073-2.0132046  &  18:47:36.58  &  -02:00:47.54  &  0.64  &  0.55  &  63  &  21 $\pm$ 4  &  --  &  --  &  -- \\
39  &  281.9020797-2.0140221  &  18:47:36.50  &  -02:00:50.48  &  0.79  &  0.68  &  104  &  21 $\pm$ 4  &  90.3 $\pm$ 0.03  &  0.76 $\pm$ 0.13  &  Single \\
40  &  281.8961141-2.0201944  &  18:47:35.07  &  -02:01:12.70  &  0.66  &  0.54  &  75  &  18 $\pm$ 4  &  --  &  --  &  -- \\
44  &  281.9031339-2.0159252  &  18:47:36.75  &  -02:00:57.33  &  0.66  &  0.62  &  78  &  26 $\pm$ 5  &  88.68 $\pm$ 0.04  &  --  &  Complex \\
46  &  281.911161-2.0123366  &  18:47:38.68  &  -02:00:44.41  &  0.8  &  0.43  &  132  &  18 $\pm$ 4  &  --  &  --  &  -- \\
47  &  281.9015429-2.0234787  &  18:47:36.37  &  -02:01:24.52  &  1.52  &  1.07  &  110  &  20 $\pm$ 4  &  90.33 $\pm$ 0.05  &  0.61 $\pm$ 0.15  &  Single \\
\hline
\end{tabular}
\end{table*}

\begin{figure*}
    \centering
\includegraphics[width=0.24\textwidth]{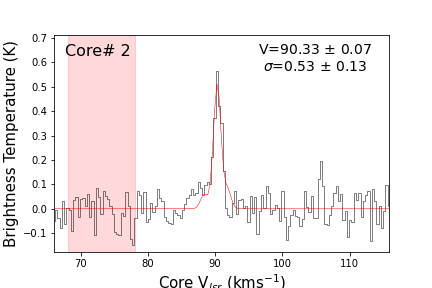}
\includegraphics[width=0.24\textwidth]{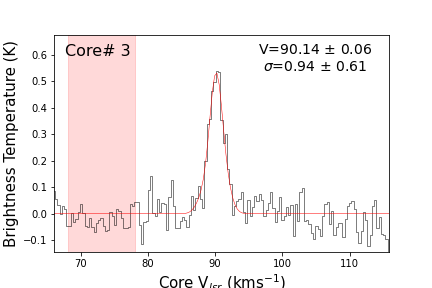}
\includegraphics[width=0.24\textwidth]{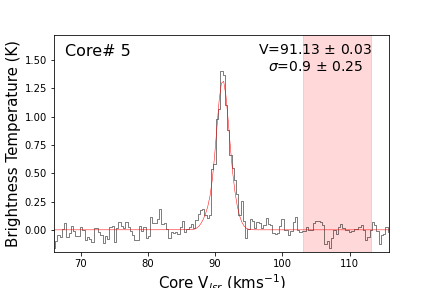}
\includegraphics[width=0.24\textwidth]{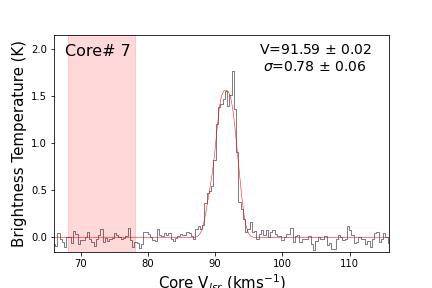}
\includegraphics[width=0.24\textwidth]{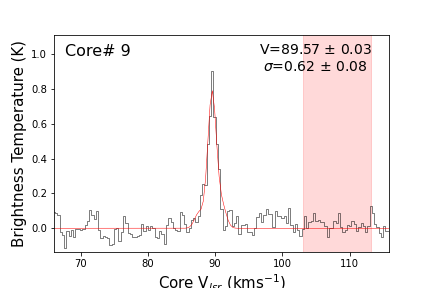}
\includegraphics[width=0.24\textwidth]{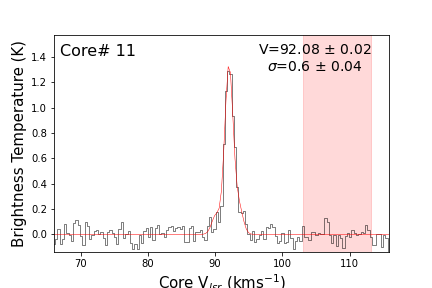}
\includegraphics[width=0.24\textwidth]{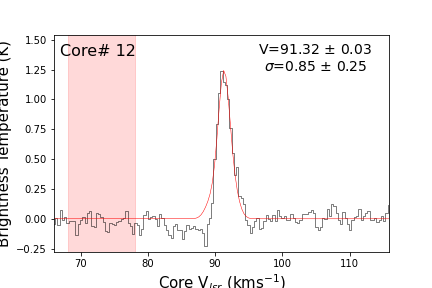}
\includegraphics[width=0.24\textwidth]{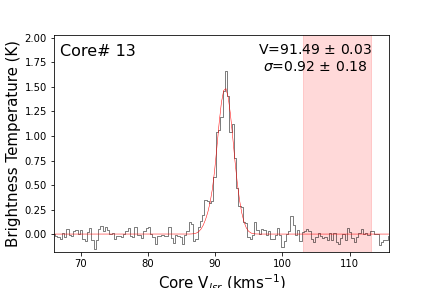}
\includegraphics[width=0.24\textwidth]{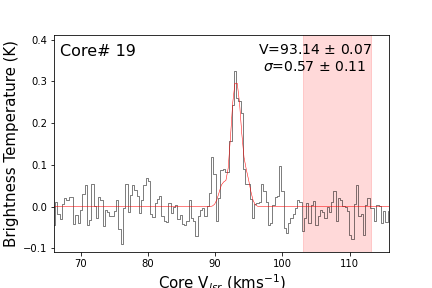}
\includegraphics[width=0.24\textwidth]{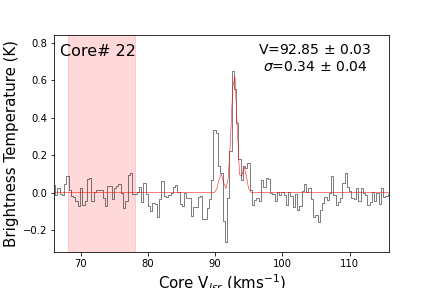}
\includegraphics[width=0.24\textwidth]{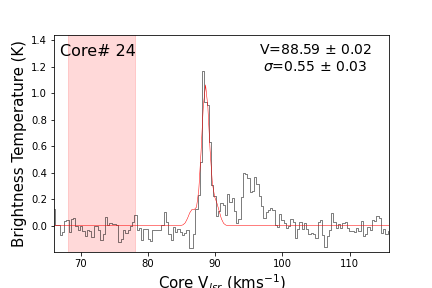}
\includegraphics[width=0.24\textwidth]{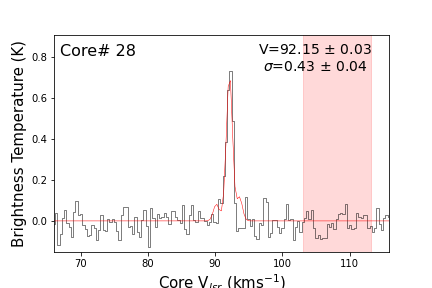}
\includegraphics[width=0.24\textwidth]{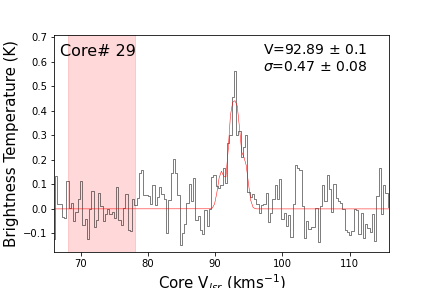}
\includegraphics[width=0.24\textwidth]{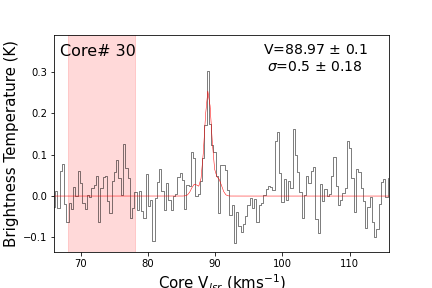}
\includegraphics[width=0.24\textwidth]{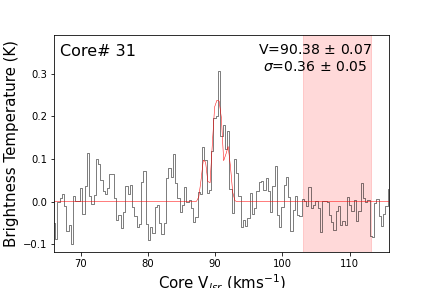}
\includegraphics[width=0.24\textwidth]{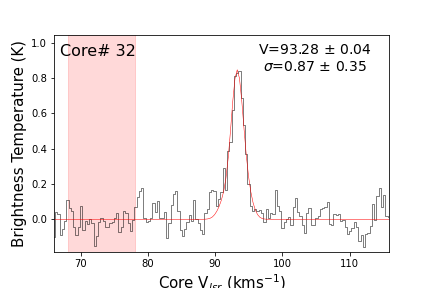}
\includegraphics[width=0.24\textwidth]{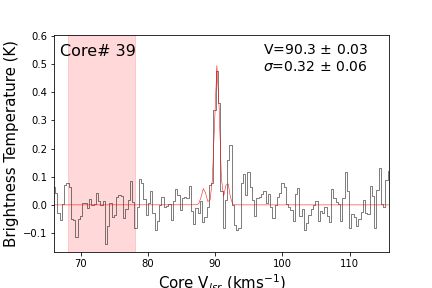}
\includegraphics[width=0.24\textwidth]{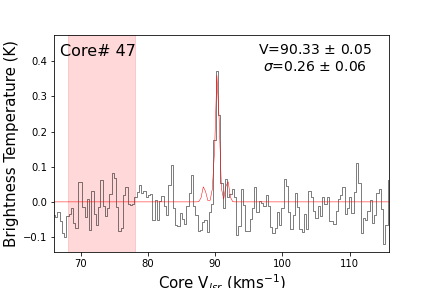}
\caption{ Single-type core-averaged, background-subtracted DCN spectra extracted from the cores in the young protocluster W43-MM2. Readers can refer to \cref{tabappendix:coretables_w43_mm2} for the line fit parameters for each core.}
\end{figure*}

\begin{figure*}
    \centering
\includegraphics[width=0.24\textwidth]{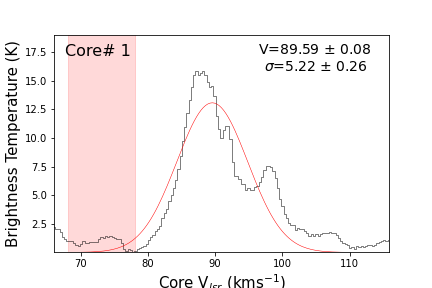}
\includegraphics[width=0.24\textwidth]{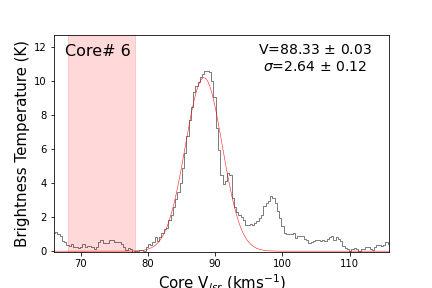}
\includegraphics[width=0.24\textwidth]{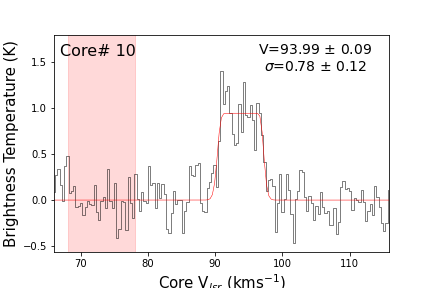}
\includegraphics[width=0.24\textwidth]{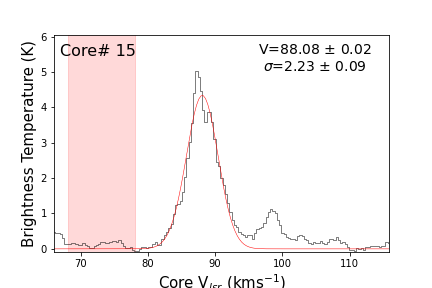}
\includegraphics[width=0.24\textwidth]{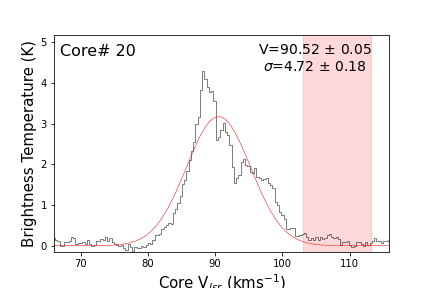}
\includegraphics[width=0.24\textwidth]{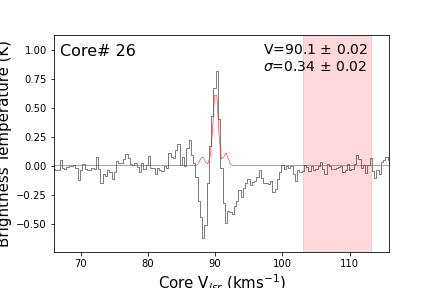}
\includegraphics[width=0.24\textwidth]{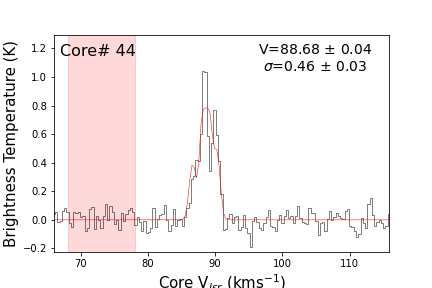}
\caption{Complex-type core-averaged, background-subtracted DCN spectra extracted from the cores in the young protocluster W43-MM2. Readers can refer to \cref{tabappendix:coretables_w43_mm2} for the line fit parameters for each core.} 
\end{figure*}

\begin{table*}[htbp!]
\centering
\small
\caption{DCN fits towards the core population of the intermediate protocluster W43-MM3.}
\label{w43mm3tabappendix:coretables}
\begin{tabular}{llllllllccc}

\hline 
n   & Core Name &RA  & DEC & F$_{A}$   & F$_{B}$    & PA & T & \vlsr  & Linewidth   & Spectral \\
 &  & [ICRS]  &  [ICRS]   &  [\arcsec] &[\arcsec] &  [deg] & [K] & [\kms]  & [\kms] & Type \\
\hline 
1  &  281.9238036-2.0079434  &  18:47:41.71  &  -02:00:28.60  &  0.65  &  0.5  &  90  &  100 $\pm$ 50  &  92.46 $\pm$ 0.02  &  1.22 $\pm$ 0.09  &  Single \\
2  &  281.9135958-2.0078027  &  18:47:39.26  &  -02:00:28.09  &  0.73  &  0.57  &  57  &  100 $\pm$ 50  &  93.38 $\pm$ 0.12  &  --  &  Complex \\
3  &  281.9207197-2.0057611  &  18:47:40.97  &  -02:00:20.74  &  0.67  &  0.65  &  20  &  25 $\pm$ 5  &  --  &  --  &  -- \\
4  &  281.9238914-2.0076105  &  18:47:41.73  &  -02:00:27.40  &  0.64  &  0.45  &  118  &  31 $\pm$ 6  &  95.3 $\pm$ 0.09  &  1.63 $\pm$ 0.73  &  Single \\
5  &  281.9176265-2.0095864  &  18:47:40.23  &  -02:00:34.51  &  0.56  &  0.53  &  111  &  20 $\pm$ 4  &  --  &  --  &  -- \\
6  &  281.9234345-2.007048  &  18:47:41.62  &  -02:00:25.37  &  0.64  &  0.61  &  44  &  32 $\pm$ 7  &  92.98 $\pm$ 0.07  &  1.11 $\pm$ 0.15  &  Single \\
8  &  281.9145005-2.0091444  &  18:47:39.48  &  -02:00:32.92  &  0.55  &  0.47  &  82  &  22 $\pm$ 5  &  --  &  --  &  -- \\
9  &  281.9232633-2.0078731  &  18:47:41.58  &  -02:00:28.34  &  0.55  &  0.46  &  88  &  29 $\pm$ 6  &  --  &  --  &  -- \\
10  &  281.924284-2.008162  &  18:47:41.83  &  -02:00:29.38  &  0.82  &  0.62  &  98  &  29 $\pm$ 6  &  93.9 $\pm$ 0.05  &  --  &  Complex \\
11  &  281.9273799-2.0015818  &  18:47:42.57  &  -02:00:5.690  &  1.1  &  0.94  &  52  &  22 $\pm$ 5  &  --  &  --  &  -- \\
12  &  281.9249972-2.0078222  &  18:47:42.00  &  -02:00:28.16  &  0.63  &  0.51  &  81  &  28 $\pm$ 6  &  --  &  --  &  -- \\
13  &  281.9134182-2.0075463  &  18:47:39.22  &  -02:00:27.17  &  1.09  &  0.93  &  158  &  26 $\pm$ 5  &  93.7 $\pm$ 0.11  &  --  &  Complex \\
14  &  281.921711-2.00727  &  18:47:41.21  &  -02:00:26.17  &  0.73  &  0.64  &  82  &  25 $\pm$ 5  &  92.44 $\pm$ 0.03  &  0.96 $\pm$ 0.09  &  Single \\
16  &  281.9244991-2.0077454  &  18:47:41.88  &  -02:00:27.88  &  0.81  &  0.76  &  110  &  30 $\pm$ 6  &  90.66 $\pm$ 0.01  &  1.4 $\pm$ 0.06  &  Single \\
17  &  281.9293338-2.0006956  &  18:47:43.04  &  -02:00:2.500  &  1.05  &  0.66  &  92  &  20 $\pm$ 4  &  --  &  --  &  -- \\
18  &  281.9209802-2.0087135  &  18:47:41.04  &  -02:00:31.37  &  0.81  &  0.69  &  89  &  23 $\pm$ 5  &  --  &  --  &  -- \\
19  &  281.9215884-2.0111158  &  18:47:41.18  &  -02:00:40.02  &  1.15  &  1.12  &  64  &  21 $\pm$ 4  &  --  &  --  &  -- \\
20  &  281.9191116-2.0101555  &  18:47:40.59  &  -02:00:36.56  &  1.19  &  1.04  &  140  &  21 $\pm$ 4  &  --  &  --  &  -- \\
21  &  281.9199849-2.0102219  &  18:47:40.80  &  -02:00:36.80  &  0.77  &  0.72  &  99  &  21 $\pm$ 4  &  --  &  --  &  -- \\
22  &  281.9189643-2.0061312  &  18:47:40.55  &  -02:00:22.07  &  1.05  &  0.73  &  176  &  22 $\pm$ 5  &  --  &  --  &  -- \\
23  &  281.9268886-1.9949305  &  18:47:42.45  &  -01:59:41.75  &  0.59  &  0.5  &  114  &  21 $\pm$ 4  &  --  &  --  &  -- \\
24  &  281.9139833-2.0079715  &  18:47:39.36  &  -02:00:28.70  &  1.1  &  0.75  &  140  &  25 $\pm$ 5  &  93.59 $\pm$ 0.14  &  --  &  Complex \\
25  &  281.9207192-2.0087896  &  18:47:40.97  &  -02:00:31.64  &  0.76  &  0.64  &  59  &  22 $\pm$ 5  &  --  &  --  &  -- \\
26  &  281.9174417-2.0007412  &  18:47:40.19  &  -02:00:2.670  &  0.64  &  0.4  &  14  &  21 $\pm$ 4  &  --  &  --  &  -- \\
27  &  281.9317875-2.0140444  &  18:47:43.63  &  -02:00:50.56  &  1.0  &  0.79  &  119  &  18 $\pm$ 4  &  --  &  --  &  -- \\
28  &  281.9138487-2.00915  &  18:47:39.32  &  -02:00:32.94  &  0.61  &  0.51  &  22  &  22 $\pm$ 5  &  --  &  --  &  -- \\
29  &  281.9130975-2.0077682  &  18:47:39.14  &  -02:00:27.97  &  0.69  &  0.57  &  68  &  26 $\pm$ 5  &  --  &  --  &  -- \\
31  &  281.9162905-2.0089111  &  18:47:39.91  &  -02:00:32.08  &  0.58  &  0.5  &  14  &  21 $\pm$ 4  &  94.97 $\pm$ 0.07  &  0.95 $\pm$ 0.15  &  Single \\
32  &  281.9238534-2.0001514  &  18:47:41.72  &  -02:00:0.550  &  1.28  &  1.05  &  45  &  22 $\pm$ 4  &  --  &  --  &  -- \\
34  &  281.9251055-2.0070496  &  18:47:42.03  &  -02:00:25.38  &  0.7  &  0.65  &  146  &  29 $\pm$ 6  &  91.92 $\pm$ 0.04  &  1.82 $\pm$ 0.45  &  Single \\
36  &  281.9274195-1.9953527  &  18:47:42.58  &  -01:59:43.27  &  0.9  &  0.6  &  129  &  20 $\pm$ 4  &  --  &  --  &  -- \\
37  &  281.9112287-2.0123947  &  18:47:38.69  &  -02:00:44.62  &  1.47  &  1.27  &  97  &  21 $\pm$ 4  &  --  &  --  &  -- \\
38  &  281.9152398-2.0088625  &  18:47:39.66  &  -02:00:31.91  &  0.57  &  0.52  &  49  &  22 $\pm$ 4  &  --  &  --  &  -- \\
39  &  281.9135666-2.0081333  &  18:47:39.26  &  -02:00:29.28  &  1.06  &  0.88  &  174  &  26 $\pm$ 5  &  92.88 $\pm$ 0.09  &  3.1 $\pm$ 1.13  &  Single \\
40  &  281.9213102-2.0069861  &  18:47:41.11  &  -02:00:25.15  &  1.36  &  1.0  &  153  &  25 $\pm$ 5  &  91.73 $\pm$ 0.03  &  0.67 $\pm$ 0.13  &  Single \\
42  &  281.9311334-2.005116  &  18:47:43.47  &  -02:00:18.42  &  1.17  &  0.88  &  119  &  21 $\pm$ 4  &  --  &  --  &  -- \\
\hline \noalign {\smallskip}
\end{tabular}
\end{table*}

\begin{figure*}
    \centering
\includegraphics[width=0.24\textwidth]{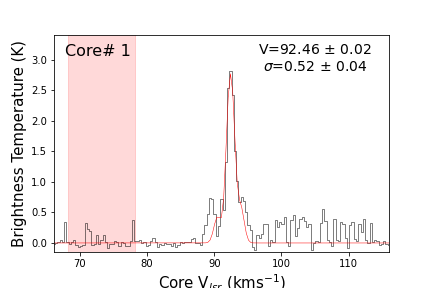}
\includegraphics[width=0.24\textwidth]{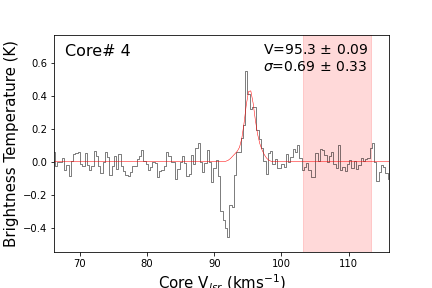}
\includegraphics[width=0.24\textwidth]{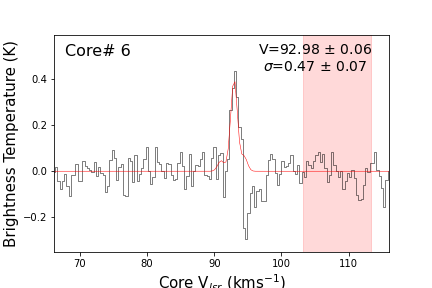}
\includegraphics[width=0.24\textwidth]{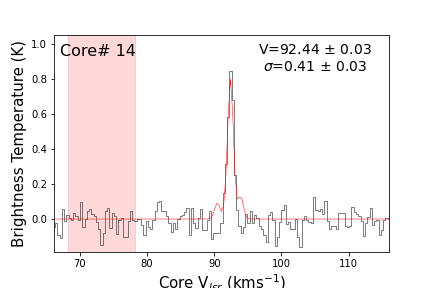}
\includegraphics[width=0.24\textwidth]{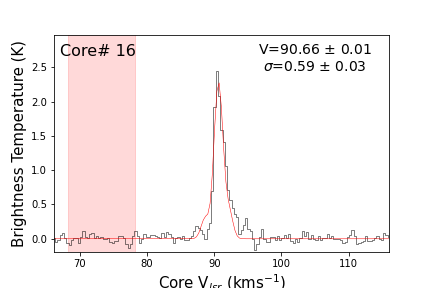}
\includegraphics[width=0.24\textwidth]{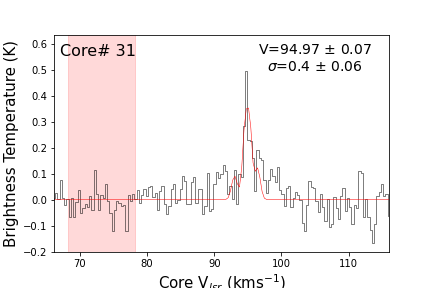}
\includegraphics[width=0.24\textwidth]{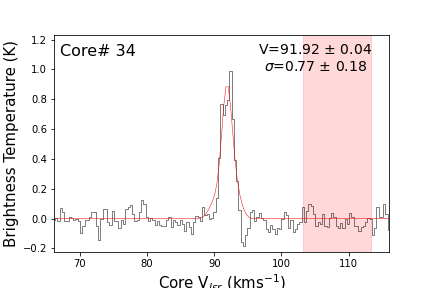}
\includegraphics[width=0.24\textwidth]{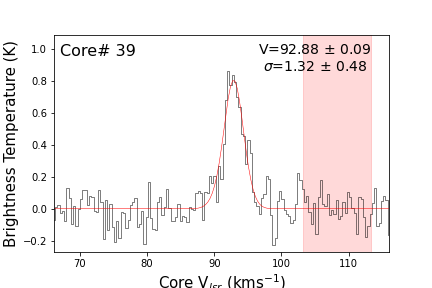}
\includegraphics[width=0.24\textwidth]{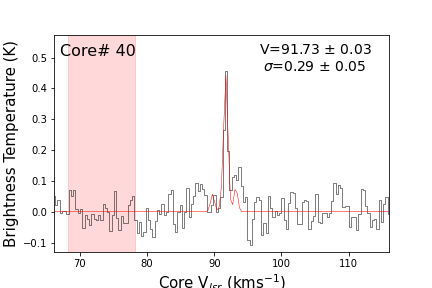}
\caption{Single-type core-averaged, background-subtracted DCN spectra extracted from the cores in the intermediate protocluster W43-MM3. Readers can refer to \cref{w43mm3tabappendix:coretables} for the line fit parameters for each core.}
\end{figure*}

\begin{figure*}
    \centering
\includegraphics[width=0.24\textwidth]{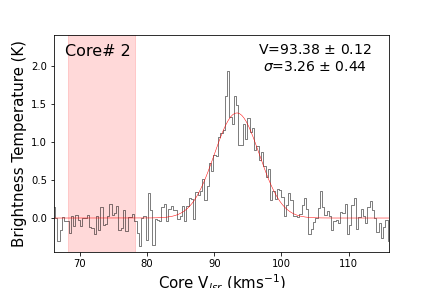}
\includegraphics[width=0.24\textwidth]{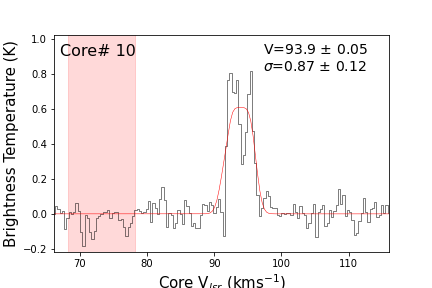}
\includegraphics[width=0.24\textwidth]{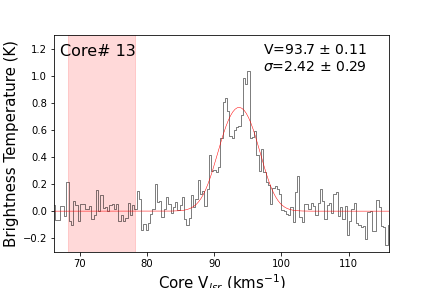}
\includegraphics[width=0.24\textwidth]{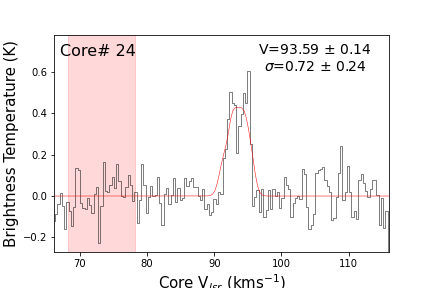}
\caption{Complex-type core-averaged, background-subtracted DCN spectra extracted from the cores in the intermediate protocluster W43-MM3. Readers can refer to \cref{w43mm3tabappendix:coretables} for the line fit parameters for each core.} 
\end{figure*}

\begin{table*}[htbp!]
\centering
\small
\caption{DCN fits towards the core population of the intermediate protocluster W51-E.}
\label{w51tabappendix:coretables}
\begin{tabular}{llllllllccc}
\hline 
n   & Core Name &RA  & DEC & F$_{A}$   & F$_{B}$    & PA & T & \vlsr  & Linewidth   & Spectral \\
 &  & [ICRS]  &  [ICRS]   &  [\arcsec] &[\arcsec] &  [deg] & [K] & [\kms]  & [\kms] & Type \\
\hline 
2  &  290.933185414.5095896  &  19:23:43.96  &  14:30:34.52  &  1.11  &  0.66  &  177  &  300 $\pm$ 100  &  63.97 $\pm$ 0.42  &  --  &  Complex \\
4  &  290.932907614.5078281  &  19:23:43.90  &  14:30:28.18  &  0.85  &  0.73  &  6  &  100 $\pm$ 50  &  63.56 $\pm$ 0.74  &  --  &  Complex \\
8  &  290.933036914.5101522  &  19:23:43.93  &  14:30:36.55  &  1.07  &  0.71  &  167  &  35 $\pm$ 10  &  52.79 $\pm$ 0.07  &  1.97 $\pm$ 0.62  &  Single \\
11  &  290.922836614.5040393  &  19:23:41.48  &  14:30:14.54  &  0.79  &  0.68  &  20  &  25 $\pm$ 5  &  --  &  --  &  -- \\
12  &  290.93328814.5091267  &  19:23:43.99  &  14:30:32.86  &  0.9  &  0.81  &  153  &  35 $\pm$ 10  &  --  &  --  &  -- \\
13  &  290.932447914.5054808  &  19:23:43.79  &  14:30:19.73  &  0.77  &  0.68  &  27  &  31 $\pm$ 6  &  62.86 $\pm$ 0.23  &  --  &  Complex \\
14  &  290.932607114.510126  &  19:23:43.83  &  14:30:36.45  &  0.65  &  0.61  &  174  &  35 $\pm$ 10  &  --  &  --  &  -- \\
15  &  290.931505514.5099588  &  19:23:43.56  &  14:30:35.85  &  0.92  &  0.8  &  26  &  28 $\pm$ 6  &  --  &  --  &  -- \\
16  &  290.932754414.5073581  &  19:23:43.86  &  14:30:26.49  &  0.91  &  0.48  &  21  &  35 $\pm$ 10  &  --  &  --  &  -- \\
17  &  290.929976714.5141521  &  19:23:43.19  &  14:30:50.95  &  1.24  &  0.83  &  150  &  28 $\pm$ 6  &  60.95 $\pm$ 0.07  &  2.84 $\pm$ 1.53  &  Single \\
18  &  290.926656814.502195  &  19:23:42.40  &  14:30:7.900  &  0.72  &  0.64  &  68  &  23 $\pm$ 5  &  --  &  --  &  -- \\
19  &  290.932731514.5112086  &  19:23:43.86  &  14:30:40.35  &  0.6  &  0.51  &  11  &  27 $\pm$ 5  &  --  &  --  &  -- \\
20  &  290.925572814.5112608  &  19:23:42.14  &  14:30:40.54  &  0.77  &  0.57  &  2  &  26 $\pm$ 5  &  --  &  --  &  -- \\
21  &  290.927575714.5005965  &  19:23:42.62  &  14:30:2.150  &  0.64  &  0.57  &  142  &  23 $\pm$ 5  &  51.38 $\pm$ 0.05  &  0.7 $\pm$ 0.18  &  Single \\
22  &  290.927781414.500986  &  19:23:42.67  &  14:30:3.550  &  0.8  &  0.69  &  174  &  24 $\pm$ 5  &  49.9 $\pm$ 0.11  &  1.49 $\pm$ 0.5  &  Single \\
26  &  290.924399714.5149412  &  19:23:41.86  &  14:30:53.79  &  1.47  &  1.0  &  168  &  29 $\pm$ 6  &  60.07 $\pm$ 0.14  &  2.05 $\pm$ 1.02  &  Single \\
28  &  290.931994614.5089561  &  19:23:43.68  &  14:30:32.24  &  0.63  &  0.54  &  65  &  30 $\pm$ 6  &  --  &  --  &  -- \\
29  &  290.929639414.5149377  &  19:23:43.11  &  14:30:53.78  &  1.02  &  0.93  &  44  &  27 $\pm$ 5  &  61.58 $\pm$ 0.15  &  2.55 $\pm$ 3.69  &  Single \\
30  &  290.926221314.5153389  &  19:23:42.29  &  14:30:55.22  &  0.65  &  0.57  &  70  &  27 $\pm$ 6  &  --  &  --  &  -- \\
33  &  290.926670414.5017074  &  19:23:42.40  &  14:30:6.150  &  0.74  &  0.64  &  8  &  24 $\pm$ 5  &  --  &  --  &  -- \\
34  &  290.932534914.5131101  &  19:23:43.81  &  14:30:47.20  &  1.08  &  0.94  &  134  &  25 $\pm$ 5  &  56.6 $\pm$ 0.07  &  1.12 $\pm$ 0.21  &  Single \\
38  &  290.93495214.5078333  &  19:23:44.39  &  14:30:28.20  &  0.84  &  0.71  &  48  &  28 $\pm$ 6  &  --  &  --  &  -- \\
39  &  290.93425614.5089017  &  19:23:44.22  &  14:30:32.05  &  1.02  &  0.89  &  78  &  28 $\pm$ 6  &  --  &  --  &  -- \\
\hline \noalign {\smallskip}
\end{tabular}
\end{table*}

\begin{figure*}
\centering
\includegraphics[width=0.24\textwidth]{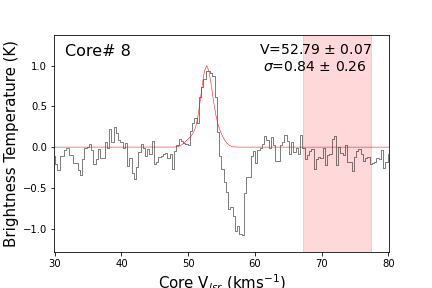}
\includegraphics[width=0.24\textwidth]{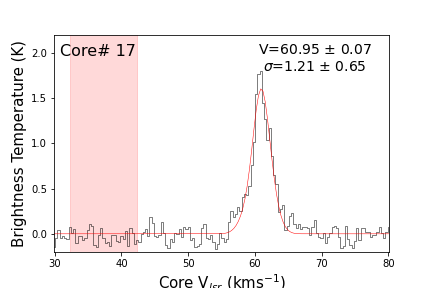}
\includegraphics[width=0.24\textwidth]{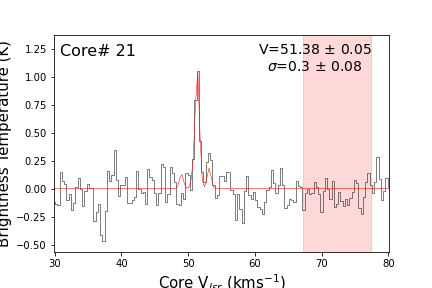}
\includegraphics[width=0.24\textwidth]{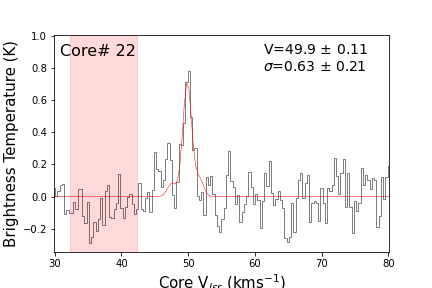}
\includegraphics[width=0.24\textwidth]{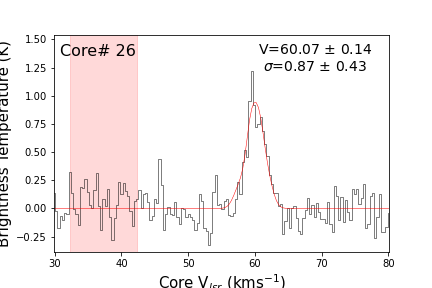}
\includegraphics[width=0.24\textwidth]{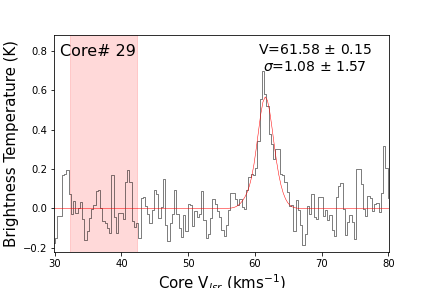}
\includegraphics[width=0.24\textwidth]{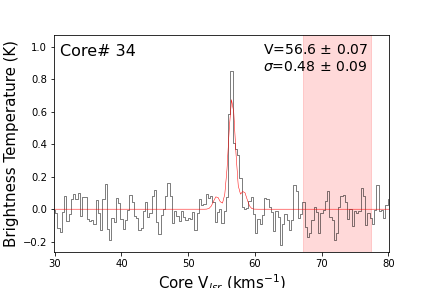}
\caption{Single-type core-averaged, background-subtracted DCN spectra extracted from the cores in the intermediate protocluster W51-E. Readers can refer to \cref{w51tabappendix:coretables} for the line fit parameters for each core.}
\end{figure*}

\begin{figure*}
\centering
\includegraphics[width=0.24\textwidth]{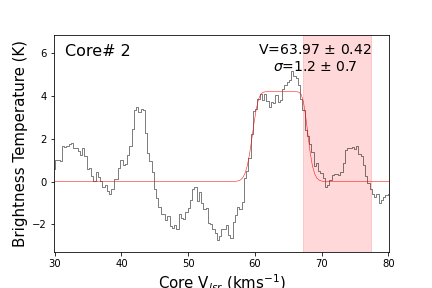}
\includegraphics[width=0.24\textwidth]{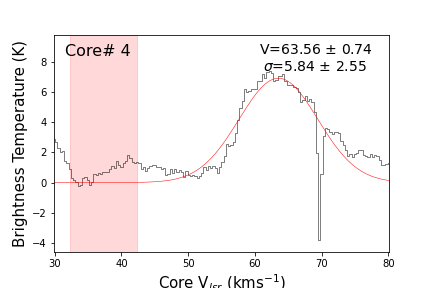}
\includegraphics[width=0.24\textwidth]{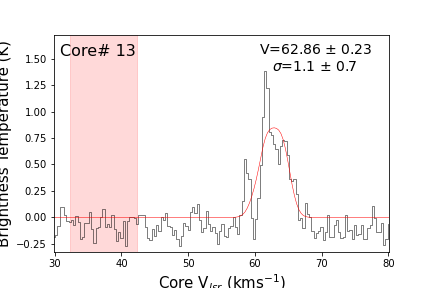}
\caption{Complex-type core-averaged, background-subtracted DCN spectra extracted from the cores in the intermediate protocluster W51-E. Readers can refer to \cref{w51tabappendix:coretables} for the line fit parameters for each core.} 
\end{figure*}
\clearpage
\begin{table*}[htbp!]
\centering
\small
\caption{DCN fits towards the core population of the evolved protocluster W51-IRS2.}
\label{w51irs2_corefits}
\begin{tabular}{llllllllccc}
\hline 
n   & Core Name &RA  & DEC & F$_{A}$   & F$_{B}$    & PA & T & \vlsr  & Linewidth   & Spectral \\
 &  & [ICRS]  &  [ICRS]   &  [\arcsec] &[\arcsec] &  [deg] & [K] & [\kms]  & [\kms] & Type \\
\hline 
1  &  290.916869914.5181896  &  19:23:40.05  &  14:31:5.480  &  0.72  &  0.66  &  89  &  300 $\pm$ 100  &  62.68 $\pm$ 0.19  &  --  &  Complex \\
2  &  290.916470414.5181596  &  19:23:39.95  &  14:31:5.370  &  0.63  &  0.59  &  106  &  100 $\pm$ 50  &  60.72 $\pm$ 0.1  &  4.47 $\pm$ 0.95  &  Single \\
4  &  290.910723514.5116053  &  19:23:38.57  &  14:30:41.78  &  0.74  &  0.61  &  150  &  100 $\pm$ 50  &  66.63 $\pm$ 0.39  &  --  &  Complex \\
6  &  290.91664714.5182923  &  19:23:40.00  &  14:31:5.850  &  0.74  &  0.68  &  166  &  100 $\pm$ 50  &  63.28 $\pm$ 0.17  &  --  &  Complex \\
7  &  290.915614814.5181323  &  19:23:39.75  &  14:31:5.280  &  0.67  &  0.58  &  69  &  100 $\pm$ 50  &  64.98 $\pm$ 0.11  &  --  &  Complex \\
8  &  290.909155314.5185781  &  19:23:38.20  &  14:31:6.880  &  0.69  &  0.55  &  145  &  30 $\pm$ 6  &  61.05 $\pm$ 0.02  &  1.77 $\pm$ 0.12  &  Single \\
9  &  290.910372314.5113251  &  19:23:38.49  &  14:30:40.77  &  0.71  &  0.66  &  84  &  31 $\pm$ 6  &  62.12 $\pm$ 0.02  &  1.12 $\pm$ 0.07  &  Single \\
10  &  290.911183614.5126596  &  19:23:38.68  &  14:30:45.57  &  0.61  &  0.54  &  11  &  30 $\pm$ 6  &  62.31 $\pm$ 0.01  &  1.17 $\pm$ 0.06  &  Single \\
11  &  290.918758714.5177443  &  19:23:40.50  &  14:31:3.880  &  0.82  &  0.67  &  139  &  37 $\pm$ 7  &  57.59 $\pm$ 0.02  &  1.59 $\pm$ 0.12  &  Single \\
12  &  290.90857514.5182523  &  19:23:38.06  &  14:31:5.710  &  0.57  &  0.56  &  103  &  29 $\pm$ 6  &  59.32 $\pm$ 0.01  &  0.95 $\pm$ 0.04  &  Single \\
13  &  290.905996614.5052314  &  19:23:37.44  &  14:30:18.83  &  0.71  &  0.59  &  11  &  28 $\pm$ 6  &  --  &  --  &  -- \\
14  &  290.909192414.5092115  &  19:23:38.21  &  14:30:33.16  &  0.65  &  0.55  &  8  &  29 $\pm$ 6  &  62.37 $\pm$ 0.01  &  1.37 $\pm$ 0.04  &  Single \\
15  &  290.910018914.510189  &  19:23:38.40  &  14:30:36.68  &  0.95  &  0.76  &  47  &  30 $\pm$ 6  &  63.35 $\pm$ 0.04  &  --  &  Complex \\
16  &  290.908730214.5184137  &  19:23:38.10  &  14:31:6.290  &  0.6  &  0.54  &  109  &  30 $\pm$ 6  &  59.11 $\pm$ 0.02  &  0.9 $\pm$ 0.04  &  Single \\
17  &  290.924596814.5197752  &  19:23:41.90  &  14:31:11.19  &  1.01  &  0.78  &  46  &  32 $\pm$ 7  &  61.37 $\pm$ 0.02  &  1.05 $\pm$ 0.08  &  Single \\
19  &  290.924926214.5195877  &  19:23:41.98  &  14:31:10.52  &  0.56  &  0.49  &  150  &  33 $\pm$ 7  &  61.34 $\pm$ 0.03  &  0.75 $\pm$ 0.13  &  Single \\
20  &  290.905668114.5182935  &  19:23:37.36  &  14:31:5.860  &  1.64  &  1.12  &  133  &  27 $\pm$ 6  &  60.33 $\pm$ 0.02  &  1.14 $\pm$ 0.09  &  Single \\
22  &  290.925572314.5112676  &  19:23:42.14  &  14:30:40.56  &  0.62  &  0.48  &  7  &  35 $\pm$ 7  &  --  &  --  &  -- \\
23  &  290.921855114.5211248  &  19:23:41.25  &  14:31:16.05  &  0.95  &  0.64  &  106  &  32 $\pm$ 6  &  63.42 $\pm$ 0.03  &  1.05 $\pm$ 0.1  &  Single \\
24  &  290.911473814.5186878  &  19:23:38.75  &  14:31:7.280  &  0.92  &  0.64  &  38  &  32 $\pm$ 6  &  58.48 $\pm$ 0.02  &  1.64 $\pm$ 0.1  &  Single \\
25  &  290.91465614.5176008  &  19:23:39.52  &  14:31:3.360  &  0.66  &  0.56  &  72  &  100 $\pm$ 50  &  64.65 $\pm$ 0.03  &  --  &  Complex \\
26  &  290.920895914.5095599  &  19:23:41.02  &  14:30:34.42  &  0.63  &  0.56  &  63  &  31 $\pm$ 6  &  --  &  --  &  -- \\
27  &  290.908368814.5198952  &  19:23:38.01  &  14:31:11.62  &  1.26  &  0.99  &  90  &  28 $\pm$ 6  &  59.96 $\pm$ 0.02  &  1.37 $\pm$ 0.1  &  Single \\
28  &  290.905288814.5184717  &  19:23:37.27  &  14:31:6.500  &  0.94  &  0.82  &  146  &  27 $\pm$ 6  &  59.99 $\pm$ 0.09  &  1.14 $\pm$ 0.17  &  Single \\
29  &  290.911773714.5112175  &  19:23:38.83  &  14:30:40.38  &  0.58  &  0.5  &  167  &  30 $\pm$ 6  &  65.3 $\pm$ 0.03  &  1.17 $\pm$ 0.12  &  Single \\
31  &  290.912252514.5098749  &  19:23:38.94  &  14:30:35.55  &  0.56  &  0.5  &  174  &  28 $\pm$ 6  &  --  &  --  &  -- \\
32  &  290.906946714.5064897  &  19:23:37.67  &  14:30:23.36  &  0.69  &  0.54  &  50  &  29 $\pm$ 6  &  63.76 $\pm$ 0.05  &  0.67 $\pm$ 0.16  &  Single \\
33  &  290.923508714.5171682  &  19:23:41.64  &  14:31:1.810  &  0.86  &  0.68  &  170  &  33 $\pm$ 7  &  66.68 $\pm$ 0.04  &  1.61 $\pm$ 0.21  &  Single \\
34  &  290.925906214.5150857  &  19:23:42.22  &  14:30:54.31  &  0.56  &  0.54  &  8  &  36 $\pm$ 7  &  61.08 $\pm$ 0.03  &  0.92 $\pm$ 0.08  &  Single \\
35  &  290.912178514.5115741  &  19:23:38.92  &  14:30:41.67  &  0.59  &  0.51  &  25  &  29 $\pm$ 6  &  --  &  --  &  -- \\
37  &  290.911971914.5159522  &  19:23:38.87  &  14:30:57.43  &  0.93  &  0.77  &  107  &  31 $\pm$ 6  &  64.96 $\pm$ 0.01  &  1.0 $\pm$ 0.05  &  Single \\
38  &  290.926226214.515357  &  19:23:42.29  &  14:30:55.29  &  0.62  &  0.58  &  34  &  35 $\pm$ 7  &  60.66 $\pm$ 0.2  &  --  &  Complex \\
40  &  290.920787414.5116943  &  19:23:40.99  &  14:30:42.10  &  0.66  &  0.62  &  126  &  31 $\pm$ 6  &  62.62 $\pm$ 0.03  &  1.04 $\pm$ 0.09  &  Single \\
41  &  290.916181414.5181418  &  19:23:39.88  &  14:31:5.310  &  0.82  &  0.61  &  85  &  35 $\pm$ 10  &  61.21 $\pm$ 0.41  &  --  &  Complex \\
42  &  290.918079114.5179626  &  19:23:40.34  &  14:31:4.670  &  0.78  &  0.67  &  104  &  38 $\pm$ 8  &  59.15 $\pm$ 0.1  &  1.16 $\pm$ 0.16  &  Single \\
43  &  290.911424714.5132286  &  19:23:38.74  &  14:30:47.62  &  0.66  &  0.6  &  13  &  29 $\pm$ 6  &  --  &  --  &  -- \\
44  &  290.923126314.5194174  &  19:23:41.55  &  14:31:9.900  &  0.56  &  0.54  &  148  &  32 $\pm$ 7  &  60.52 $\pm$ 0.04  &  0.96 $\pm$ 0.16  &  Single \\
45  &  290.918415614.5179099  &  19:23:40.42  &  14:31:4.480  &  0.86  &  0.72  &  65  &  38 $\pm$ 8  &  --  &  --  &  -- \\
46  &  290.907916614.5199136  &  19:23:37.90  &  14:31:11.69  &  1.58  &  1.27  &  117  &  28 $\pm$ 6  &  59.65 $\pm$ 0.02  &  1.23 $\pm$ 0.1  &  Single \\
47  &  290.910052214.5196773  &  19:23:38.41  &  14:31:10.84  &  0.64  &  0.57  &  133  &  30 $\pm$ 6  &  --  &  --  &  -- \\
48  &  290.913699614.517067  &  19:23:39.29  &  14:31:1.440  &  0.65  &  0.5  &  22  &  36 $\pm$ 7  &  64.77 $\pm$ 0.01  &  0.88 $\pm$ 0.05  &  Single \\
49  &  290.911438314.5127072  &  19:23:38.75  &  14:30:45.75  &  0.84  &  0.76  &  102  &  30 $\pm$ 6  &  62.12 $\pm$ 0.01  &  1.22 $\pm$ 0.05  &  Single \\
51  &  290.911693914.5107325  &  19:23:38.81  &  14:30:38.64  &  0.63  &  0.5  &  124  &  29 $\pm$ 6  &  64.92 $\pm$ 0.02  &  0.87 $\pm$ 0.06  &  Single \\
52  &  290.9168114.514653  &  19:23:40.03  &  14:30:52.75  &  0.66  &  0.57  &  57  &  30 $\pm$ 6  &  62.72 $\pm$ 0.04  &  1.48 $\pm$ 0.18  &  Single \\
53  &  290.919614614.5068189  &  19:23:40.71  &  14:30:24.55  &  0.55  &  0.5  &  156  &  31 $\pm$ 6  &  52.93 $\pm$ 0.07  &  0.62 $\pm$ 0.18  &  Single \\
54  &  290.924726714.5084581  &  19:23:41.93  &  14:30:30.45  &  0.55  &  0.49  &  129  &  38 $\pm$ 8  &  --  &  --  &  -- \\
55  &  290.906983914.5059755  &  19:23:37.68  &  14:30:21.51  &  0.62  &  0.57  &  143  &  29 $\pm$ 6  &  63.62 $\pm$ 0.06  &  1.06 $\pm$ 0.16  &  Single \\
57  &  290.911114214.5111058  &  19:23:38.67  &  14:30:39.98  &  0.63  &  0.55  &  62  &  31 $\pm$ 6  &  --  &  --  &  -- \\
59  &  290.908574314.5087426  &  19:23:38.06  &  14:30:31.47  &  0.65  &  0.63  &  158  &  29 $\pm$ 6  &  63.36 $\pm$ 0.02  &  0.88 $\pm$ 0.09  &  Single \\
60  &  290.91181614.5117357  &  19:23:38.84  &  14:30:42.25  &  1.29  &  0.82  &  52  &  30 $\pm$ 6  &  65.49 $\pm$ 0.01  &  1.21 $\pm$ 0.04  &  Single \\
61  &  290.911130914.519012  &  19:23:38.67  &  14:31:8.440  &  0.58  &  0.49  &  113  &  31 $\pm$ 6  &  --  &  --  &  -- \\
62  &  290.905494814.5062954  &  19:23:37.32  &  14:30:22.66  &  0.62  &  0.51  &  116  &  27 $\pm$ 6  &  --  &  --  &  -- \\
63  &  290.920546914.5187726  &  19:23:40.93  &  14:31:7.580  &  0.54  &  0.46  &  127  &  33 $\pm$ 7  &  --  &  --  &  -- \\
64  &  290.905401114.5072415  &  19:23:37.30  &  14:30:26.07  &  0.55  &  0.51  &  1  &  27 $\pm$ 5  &  --  &  --  &  -- \\
65  &  290.920362614.5140959  &  19:23:40.89  &  14:30:50.75  &  0.6  &  0.57  &  120  &  30 $\pm$ 6  &  --  &  --  &  -- \\
66  &  290.927350314.5179344  &  19:23:42.56  &  14:31:4.560  &  0.86  &  0.69  &  161  &  31 $\pm$ 6  &  --  &  --  &  -- \\
67  &  290.924275114.5152428  &  19:23:41.83  &  14:30:54.87  &  0.64  &  0.4  &  174  &  100 $\pm$ 50  &  --  &  --  &  -- \\
68  &  290.914918514.5178211  &  19:23:39.58  &  14:31:4.160  &  0.81  &  0.69  &  47  &  38 $\pm$ 8  &  65.83 $\pm$ 0.03  &  --  &  Complex \\
69  &  290.918235114.5176041  &  19:23:40.38  &  14:31:3.370  &  0.74  &  0.57  &  113  &  37 $\pm$ 8  &  61.65 $\pm$ 0.02  &  1.12 $\pm$ 0.07  &  Single \\
\hline \noalign {\smallskip}
\end{tabular}
\end{table*}

\begin{table*}[htbp!]
\ContinuedFloat
\centering
\small
\caption{Continued DCN fits towards the core population of the evolved protocluster W51-IRS2.}
\begin{tabular}{llllllllccc}
\hline 
n   & Core Name &RA  & DEC & F$_{A}$   & F$_{B}$    & PA & T & \vlsr  & Linewidth   & Spectral \\
 &  & [ICRS]  &  [ICRS]   &  [\arcsec] &[\arcsec] &  [deg] & [K] & [\kms]  & [\kms] & Type \\
\hline 
70  &  290.927487514.5198119  &  19:23:42.60  &  14:31:11.32  &  0.6  &  0.54  &  85  &  31 $\pm$ 6  &  --  &  --  &  -- \\
71  &  290.907521914.5073516  &  19:23:37.81  &  14:30:26.47  &  0.59  &  0.57  &  115  &  28 $\pm$ 6  &  --  &  --  &  -- \\
72  &  290.923421514.5187151  &  19:23:41.62  &  14:31:7.370  &  0.63  &  0.53  &  113  &  33 $\pm$ 7  &  61.58 $\pm$ 0.02  &  1.04 $\pm$ 0.09  &  Single \\
73  &  290.917326114.5175912  &  19:23:40.16  &  14:31:3.330  &  0.63  &  0.47  &  172  &  37 $\pm$ 8  &  --  &  --  &  -- \\
74  &  290.911425114.5094999  &  19:23:38.74  &  14:30:34.20  &  0.59  &  0.52  &  11  &  28 $\pm$ 6  &  --  &  --  &  -- \\
75  &  290.917639714.5213077  &  19:23:40.23  &  14:31:16.71  &  0.91  &  0.75  &  173  &  35 $\pm$ 7  &  60.37 $\pm$ 0.02  &  1.3 $\pm$ 0.06  &  Single \\
76  &  290.907401114.506935  &  19:23:37.78  &  14:30:24.97  &  0.77  &  0.51  &  37  &  28 $\pm$ 6  &  64.01 $\pm$ 0.07  &  1.18 $\pm$ 0.21  &  Single \\
77  &  290.913007314.5183427  &  19:23:39.12  &  14:31:6.030  &  1.01  &  0.71  &  141  &  35 $\pm$ 7  &  57.54 $\pm$ 0.01  &  1.25 $\pm$ 0.03  &  Single \\
78  &  290.903646914.5174172  &  19:23:36.88  &  14:31:2.700  &  0.63  &  0.5  &  129  &  27 $\pm$ 5  &  --  &  --  &  -- \\
79  &  290.908463914.5047802  &  19:23:38.03  &  14:30:17.21  &  0.57  &  0.44  &  159  &  27 $\pm$ 6  &  --  &  --  &  -- \\
80  &  290.906207614.50549  &  19:23:37.49  &  14:30:19.76  &  0.86  &  0.69  &  51  &  28 $\pm$ 6  &  64.99 $\pm$ 0.08  &  0.79 $\pm$ 0.24  &  Single \\
81  &  290.908283514.5082754  &  19:23:37.99  &  14:30:29.79  &  1.38  &  0.85  &  54  &  28 $\pm$ 6  &  63.48 $\pm$ 0.02  &  1.1 $\pm$ 0.06  &  Single \\
84  &  290.922458614.5209362  &  19:23:41.39  &  14:31:15.37  &  0.79  &  0.65  &  39  &  31 $\pm$ 6  &  61.03 $\pm$ 0.08  &  1.07 $\pm$ 0.25  &  Single \\
85  &  290.909046214.5066179  &  19:23:38.17  &  14:30:23.82  &  0.77  &  0.57  &  29  &  27 $\pm$ 6  &  --  &  --  &  -- \\
86  &  290.911497414.5173017  &  19:23:38.76  &  14:31:2.290  &  1.05  &  0.83  &  133  &  31 $\pm$ 6  &  62.72 $\pm$ 0.04  &  1.95 $\pm$ 0.32  &  Single \\
87  &  290.907289414.510199  &  19:23:37.75  &  14:30:36.72  &  0.78  &  0.69  &  81  &  27 $\pm$ 5  &  --  &  --  &  -- \\
88  &  290.904790314.5182133  &  19:23:37.15  &  14:31:5.570  &  0.7  &  0.64  &  3  &  27 $\pm$ 6  &  60.19 $\pm$ 0.05  &  1.41 $\pm$ 0.22  &  Single \\
89  &  290.920063614.5132211  &  19:23:40.82  &  14:30:47.60  &  0.77  &  0.65  &  154  &  30 $\pm$ 6  &  61.49 $\pm$ 0.02  &  1.03 $\pm$ 0.07  &  Single \\
90  &  290.926329814.515708  &  19:23:42.32  &  14:30:56.55  &  0.76  &  0.65  &  9  &  35 $\pm$ 7  &  61.34 $\pm$ 0.05  &  1.05 $\pm$ 0.22  &  Single \\
91  &  290.912545314.5149959  &  19:23:39.01  &  14:30:53.99  &  0.69  &  0.59  &  174  &  30 $\pm$ 6  &  62.23 $\pm$ 0.01  &  0.7 $\pm$ 0.02  &  Single \\
92  &  290.922819714.5254406  &  19:23:41.48  &  14:31:31.59  &  0.78  &  0.53  &  27  &  28 $\pm$ 6  &  --  &  --  &  -- \\
94  &  290.917227714.5173194  &  19:23:40.13  &  14:31:2.350  &  0.57  &  0.5  &  14  &  37 $\pm$ 7  &  --  &  --  &  -- \\
95  &  290.906717114.506982  &  19:23:37.61  &  14:30:25.14  &  0.7  &  0.54  &  171  &  28 $\pm$ 6  &  --  &  --  &  -- \\
97  &  290.924309514.5175529  &  19:23:41.83  &  14:31:3.190  &  0.8  &  0.6  &  166  &  34 $\pm$ 7  &  62.15 $\pm$ 0.05  &  0.87 $\pm$ 0.17  &  Single \\
99  &  290.920432814.5135282  &  19:23:40.90  &  14:30:48.70  &  0.66  &  0.63  &  65  &  31 $\pm$ 6  &  61.25 $\pm$ 0.08  &  1.13 $\pm$ 0.34  &  Single \\
100  &  290.909435414.5096602  &  19:23:38.26  &  14:30:34.78  &  0.64  &  0.59  &  53  &  29 $\pm$ 6  &  62.85 $\pm$ 0.02  &  1.19 $\pm$ 0.06  &  Single \\
101  &  290.913856914.5173454  &  19:23:39.33  &  14:31:2.440  &  0.66  &  0.65  &  103  &  37 $\pm$ 7  &  64.52 $\pm$ 0.01  &  0.91 $\pm$ 0.03  &  Single \\
102  &  290.915392114.5217669  &  19:23:39.69  &  14:31:18.36  &  0.72  &  0.58  &  86  &  35 $\pm$ 7  &  61.13 $\pm$ 0.02  &  0.75 $\pm$ 0.07  &  Single \\
104  &  290.909308914.5201626  &  19:23:38.23  &  14:31:12.59  &  0.86  &  0.67  &  4  &  29 $\pm$ 6  &  60.02 $\pm$ 0.02  &  1.32 $\pm$ 0.06  &  Single \\
105  &  290.92551914.5190497  &  19:23:42.12  &  14:31:8.580  &  1.35  &  1.13  &  53  &  32 $\pm$ 7  &  --  &  --  &  -- \\
106  &  290.925325214.5152483  &  19:23:42.08  &  14:30:54.89  &  0.92  &  0.74  &  69  &  36 $\pm$ 7  &  62.16 $\pm$ 0.04  &  0.96 $\pm$ 0.14  &  Single \\
107  &  290.907227114.5059095  &  19:23:37.73  &  14:30:21.27  &  0.74  &  0.63  &  16  &  29 $\pm$ 6  &  66.19 $\pm$ 0.1  &  --  &  Complex \\
109  &  290.910963614.5156795  &  19:23:38.63  &  14:30:56.45  &  0.98  &  0.68  &  125  &  30 $\pm$ 6  &  --  &  --  &  -- \\
111  &  290.908421814.5174026  &  19:23:38.02  &  14:31:2.650  &  1.53  &  1.17  &  29  &  28 $\pm$ 6  &  60.33 $\pm$ 0.07  &  0.8 $\pm$ 0.17  &  Single \\
114  &  290.909888814.5104222  &  19:23:38.37  &  14:30:37.52  &  0.76  &  0.68  &  32  &  30 $\pm$ 6  &  63.23 $\pm$ 0.06  &  --  &  Complex \\
117  &  290.911454714.5109119  &  19:23:38.75  &  14:30:39.28  &  0.75  &  0.56  &  139  &  30 $\pm$ 6  &  64.98 $\pm$ 0.04  &  0.93 $\pm$ 0.16  &  Single \\
119  &  290.909482214.5078441  &  19:23:38.28  &  14:30:28.24  &  1.03  &  0.72  &  37  &  28 $\pm$ 6  &  65.4 $\pm$ 0.06  &  1.49 $\pm$ 0.26  &  Single \\
122  &  290.914317414.517634  &  19:23:39.44  &  14:31:3.480  &  1.04  &  0.79  &  121  &  38 $\pm$ 8  &  61.13 $\pm$ 0.08  &  --  &  Complex \\
\hline \noalign {\smallskip}
\end{tabular}
\end{table*}

\begin{figure*}
    \centering
\includegraphics[width=0.24\textwidth]{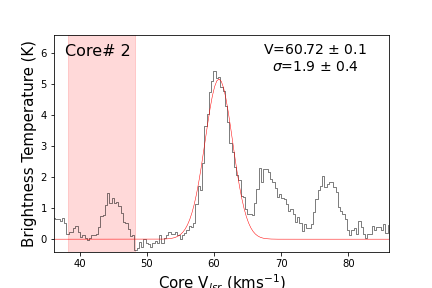}
\includegraphics[width=0.24\textwidth]{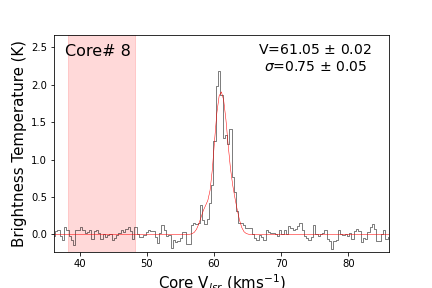}
\includegraphics[width=0.24\textwidth]{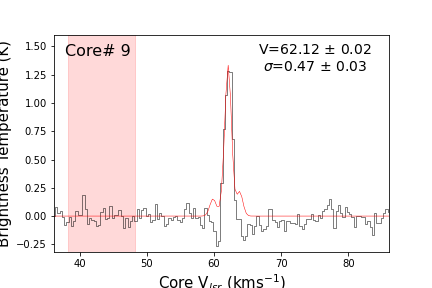}
\includegraphics[width=0.24\textwidth]{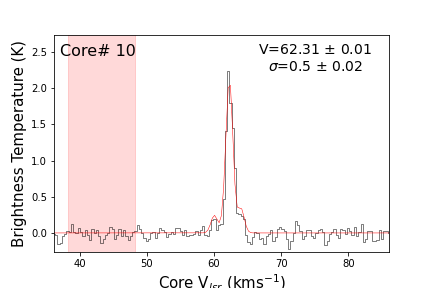}
\includegraphics[width=0.24\textwidth]{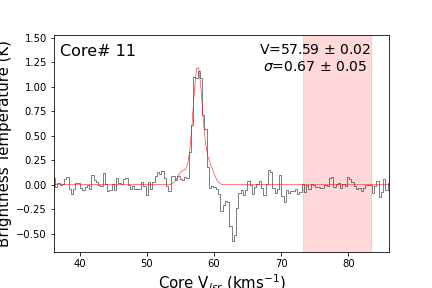}
\includegraphics[width=0.24\textwidth]{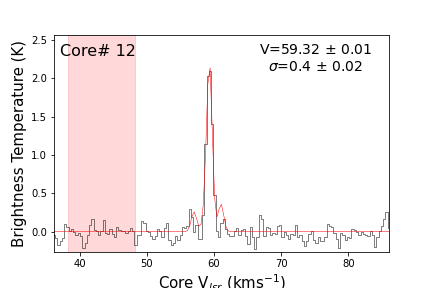}
\includegraphics[width=0.24\textwidth]{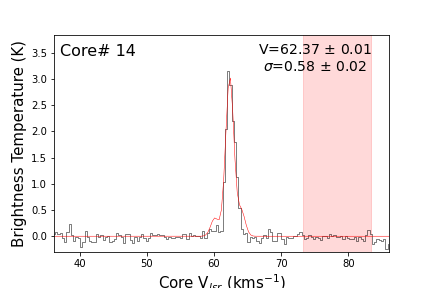}
\includegraphics[width=0.24\textwidth]{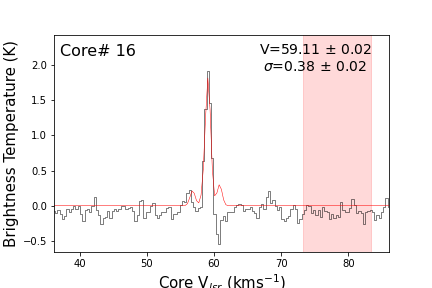}
\includegraphics[width=0.24\textwidth]{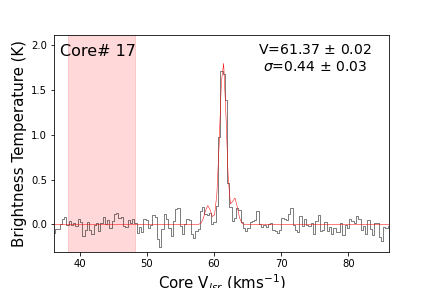}
\includegraphics[width=0.24\textwidth]{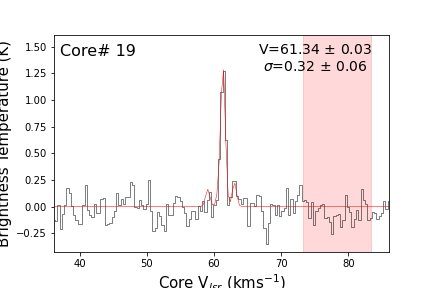}
\includegraphics[width=0.24\textwidth]{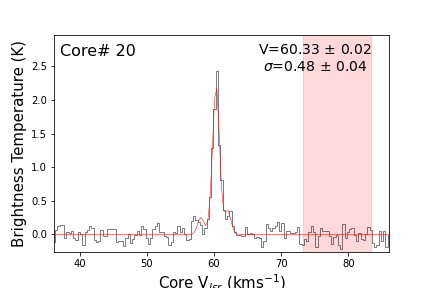}
\includegraphics[width=0.24\textwidth]{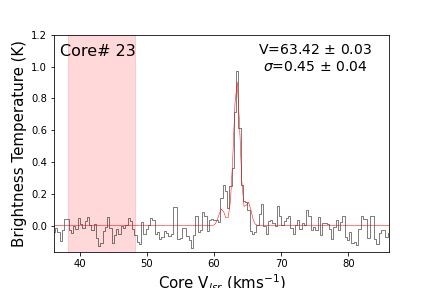}
\includegraphics[width=0.24\textwidth]{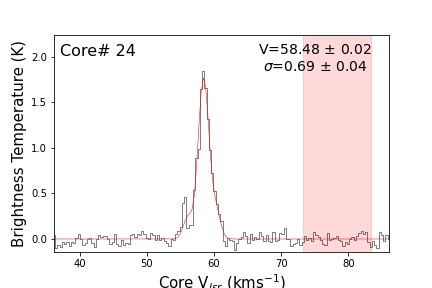}
\includegraphics[width=0.24\textwidth]{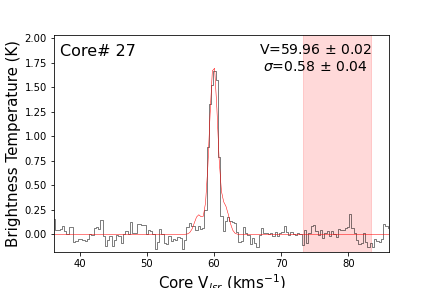}
\includegraphics[width=0.24\textwidth]{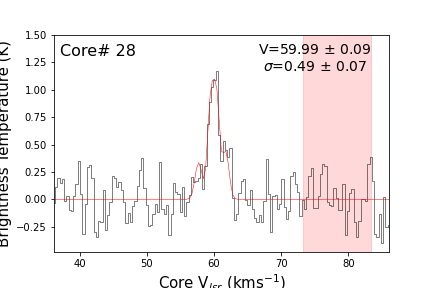}
\includegraphics[width=0.24\textwidth]{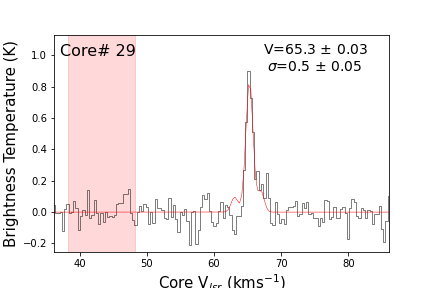}
\includegraphics[width=0.24\textwidth]{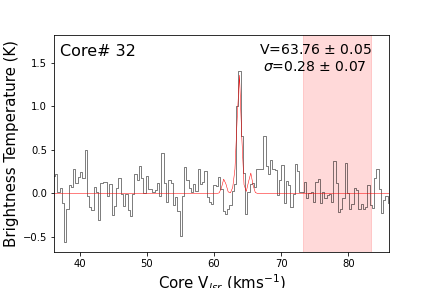}
\includegraphics[width=0.24\textwidth]{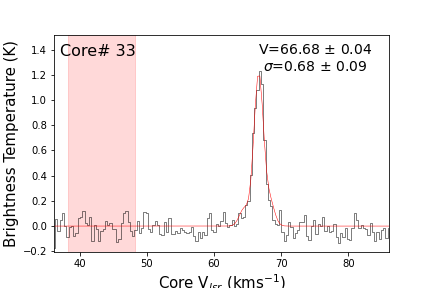}
\includegraphics[width=0.24\textwidth]{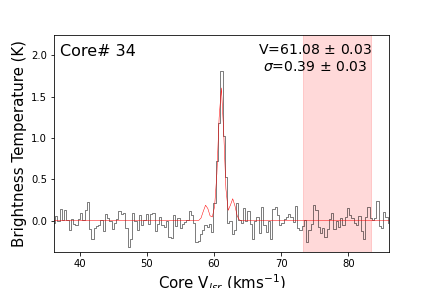}
\includegraphics[width=0.24\textwidth]{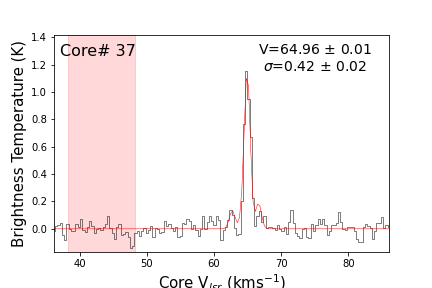}
\includegraphics[width=0.24\textwidth]{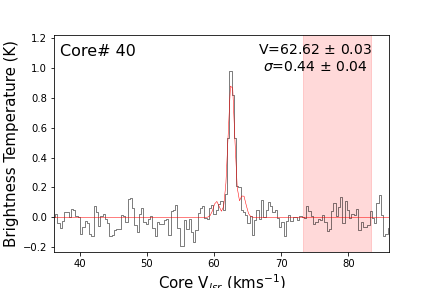}
\includegraphics[width=0.24\textwidth]{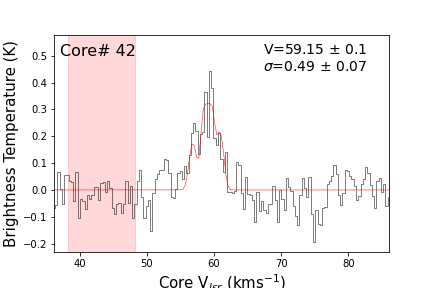}
\includegraphics[width=0.24\textwidth]{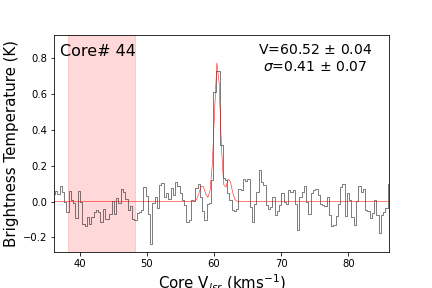}
\includegraphics[width=0.24\textwidth]{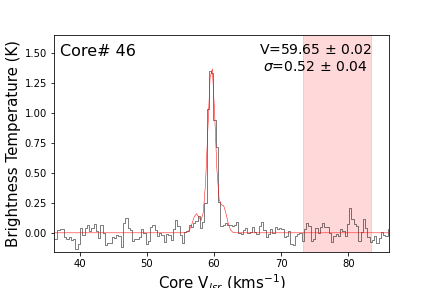}
\includegraphics[width=0.24\textwidth]{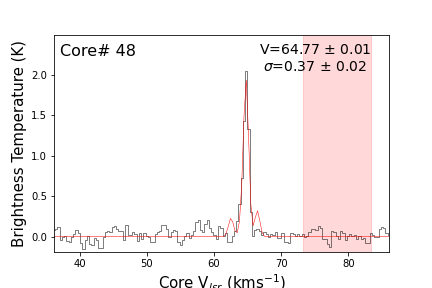}
\includegraphics[width=0.24\textwidth]{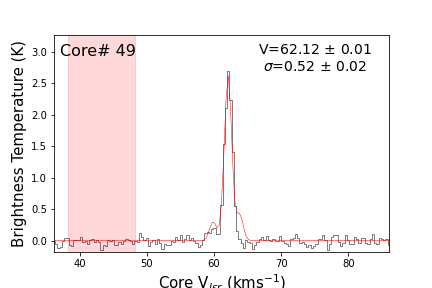}
\includegraphics[width=0.24\textwidth]{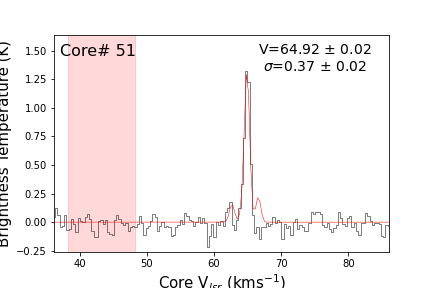}
\includegraphics[width=0.24\textwidth]{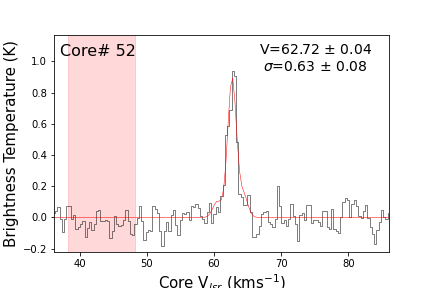}

\caption{Single-type core-averaged, background-subtracted DCN spectra extracted from the cores in the evolved protocluster W51-IRS2. Readers can refer to \cref{w51irs2_corefits} for the line fit parameters for each core.}
\end{figure*}

\begin{figure*}\ContinuedFloat
    \centering
\includegraphics[width=0.24\textwidth]{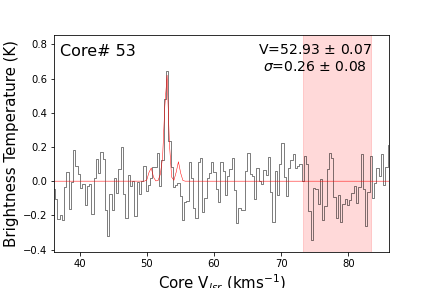}
\includegraphics[width=0.24\textwidth]{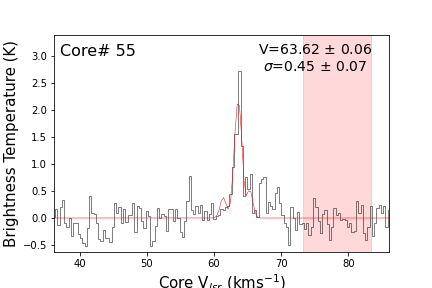}
\includegraphics[width=0.24\textwidth]{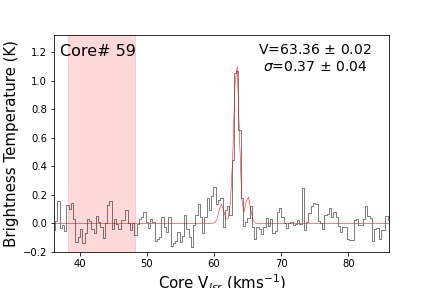}
\includegraphics[width=0.24\textwidth]{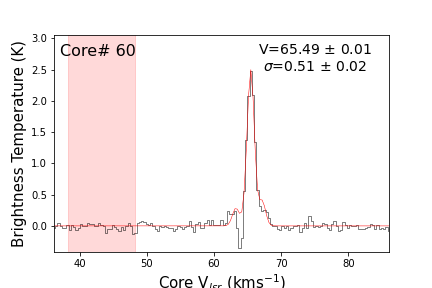}
\includegraphics[width=0.24\textwidth]{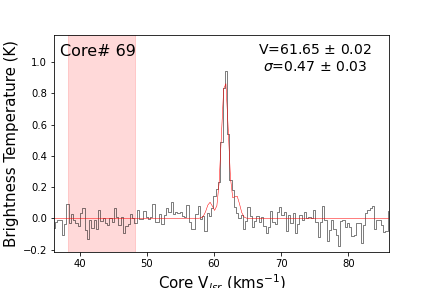}
\includegraphics[width=0.24\textwidth]{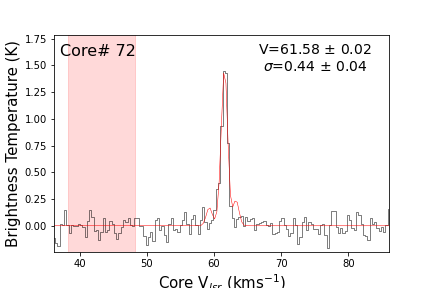}
\includegraphics[width=0.24\textwidth]{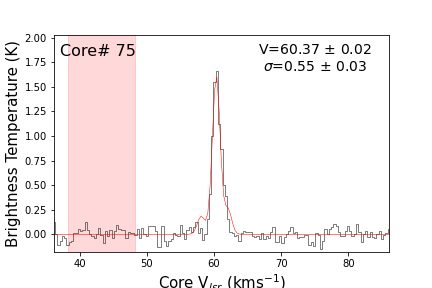}
\includegraphics[width=0.24\textwidth]{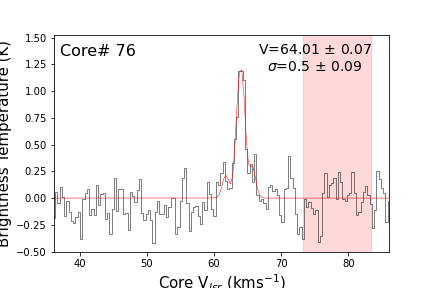}
\includegraphics[width=0.24\textwidth]{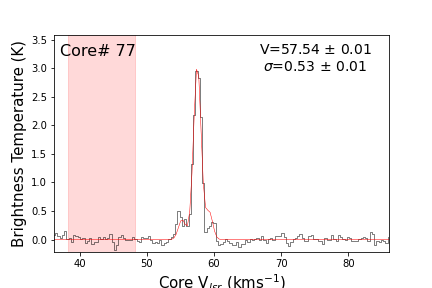}
\includegraphics[width=0.24\textwidth]{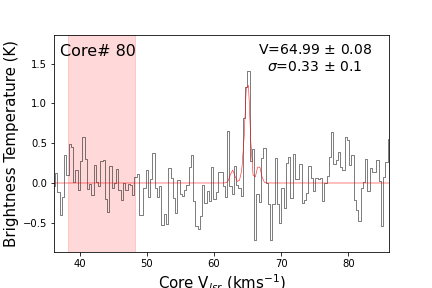}
\includegraphics[width=0.24\textwidth]{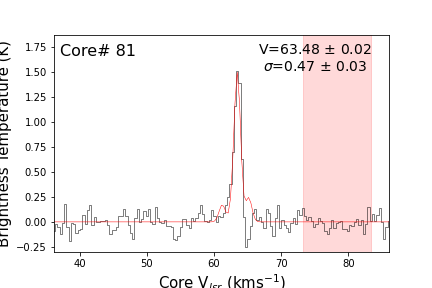}
\includegraphics[width=0.24\textwidth]{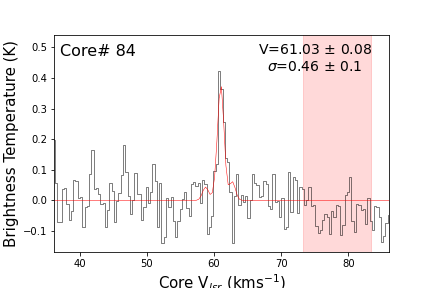}
\includegraphics[width=0.24\textwidth]{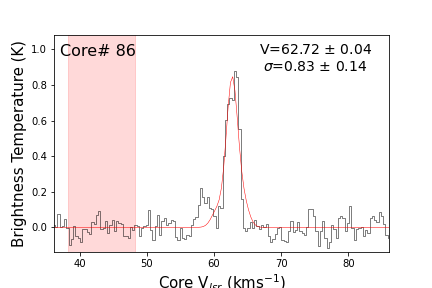}
\includegraphics[width=0.24\textwidth]{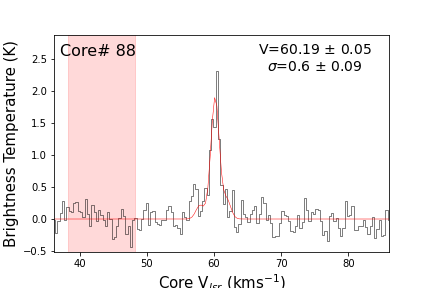}
\includegraphics[width=0.24\textwidth]{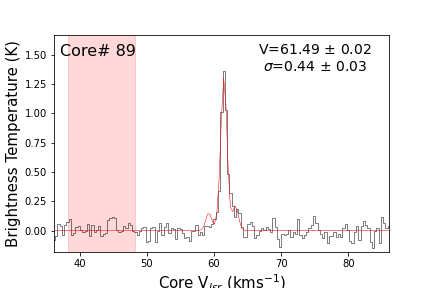}
\includegraphics[width=0.24\textwidth]{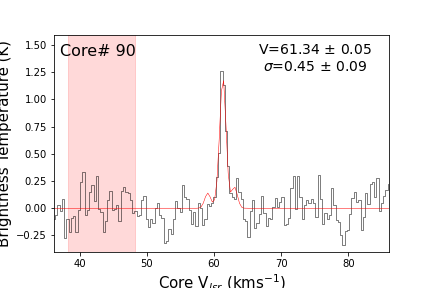}
\includegraphics[width=0.24\textwidth]{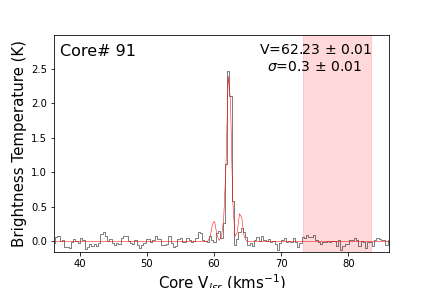}
\includegraphics[width=0.24\textwidth]{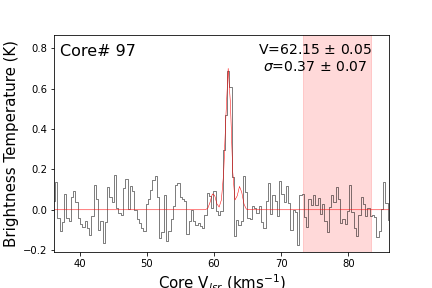}
\includegraphics[width=0.24\textwidth]{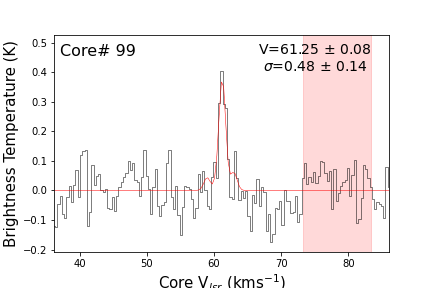}
\includegraphics[width=0.24\textwidth]{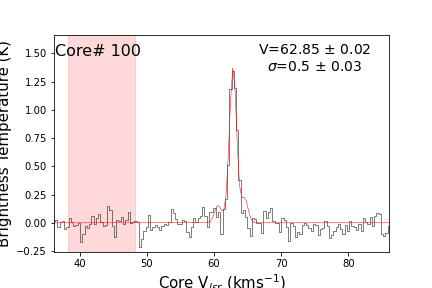}
\includegraphics[width=0.24\textwidth]{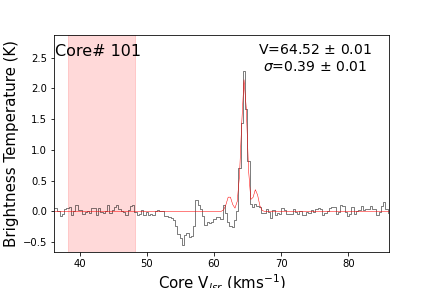}
\includegraphics[width=0.24\textwidth]{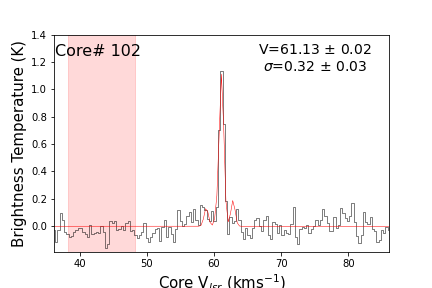}
\includegraphics[width=0.24\textwidth]{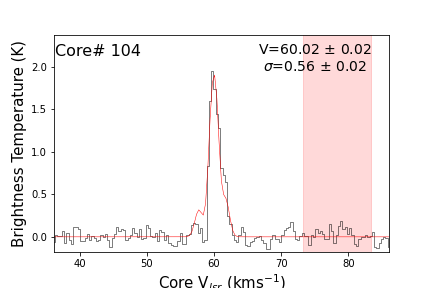}
\includegraphics[width=0.24\textwidth]{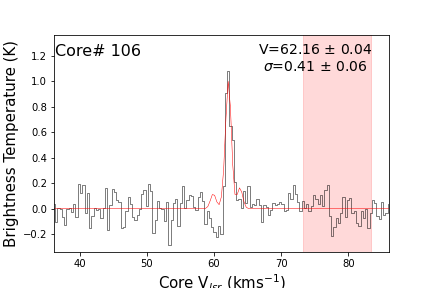}
\includegraphics[width=0.24\textwidth]{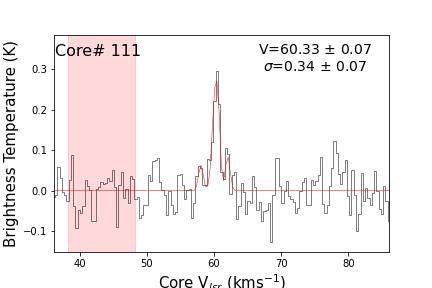}
\includegraphics[width=0.24\textwidth]{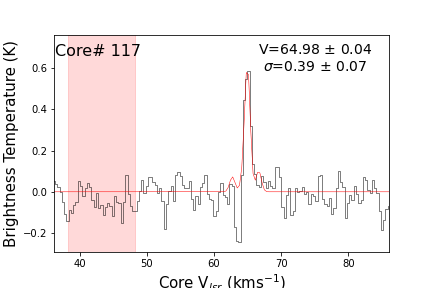}
\includegraphics[width=0.24\textwidth]{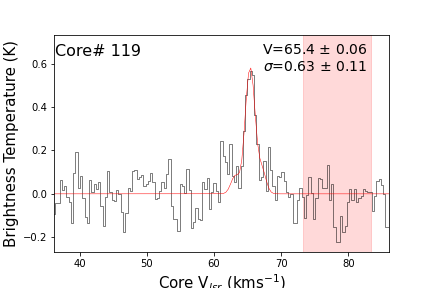}
\caption{Continued: Single-type core-averaged, background-subtracted DCN spectra extracted from the cores in the evolved protocluster W51-IRS2. Readers can refer to \cref{w51irs2_corefits} for the line fit parameters for each core.}
\end{figure*}

\begin{figure*}
    \centering
    
\includegraphics[width=0.24\textwidth]{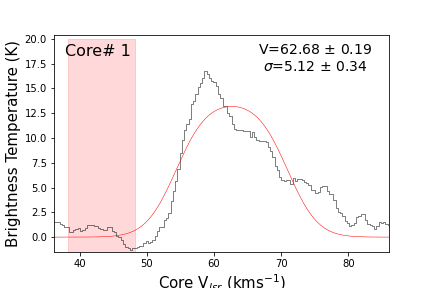}
\includegraphics[width=0.24\textwidth]{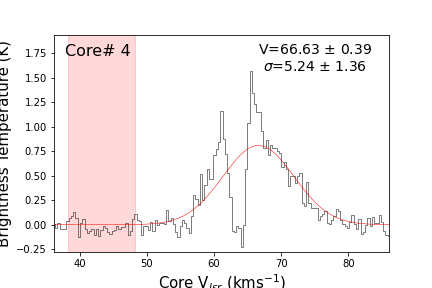}
\includegraphics[width=0.24\textwidth]{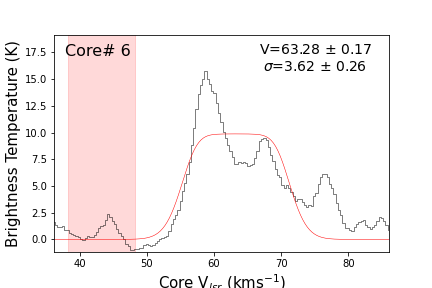}
\includegraphics[width=0.24\textwidth]{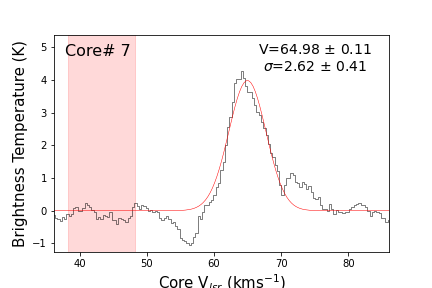}
\includegraphics[width=0.24\textwidth]{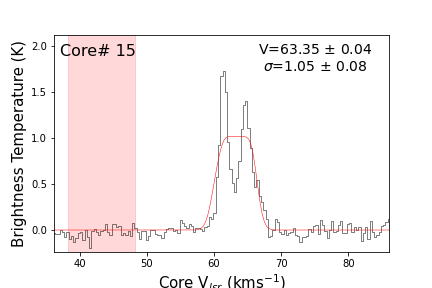}
\includegraphics[width=0.24\textwidth]{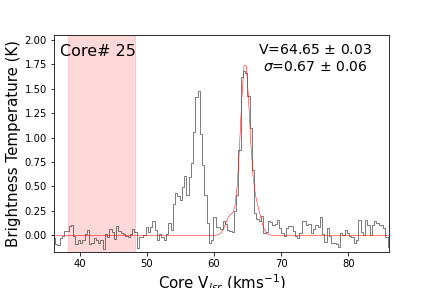}
\includegraphics[width=0.24\textwidth]{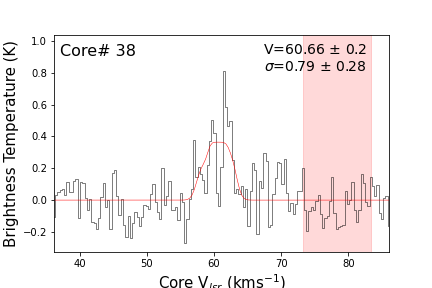}
\includegraphics[width=0.24\textwidth]{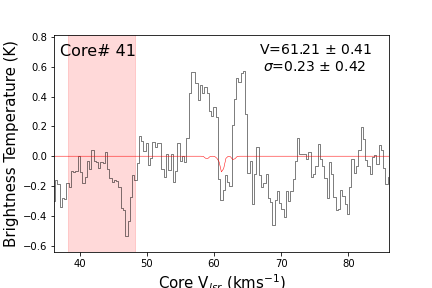}
\includegraphics[width=0.24\textwidth]{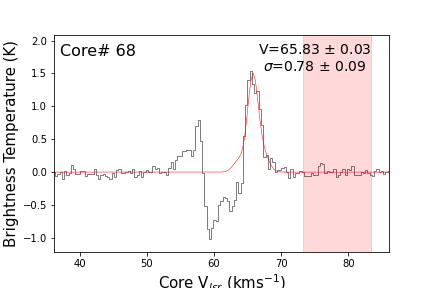}
\includegraphics[width=0.24\textwidth]{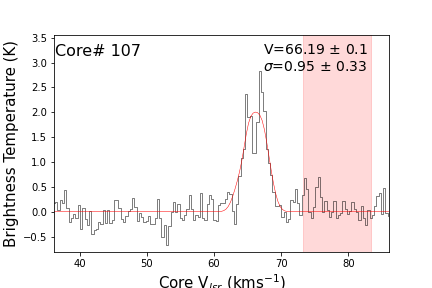}
\includegraphics[width=0.24\textwidth]{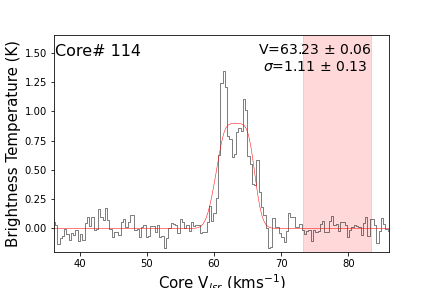}
\includegraphics[width=0.24\textwidth]{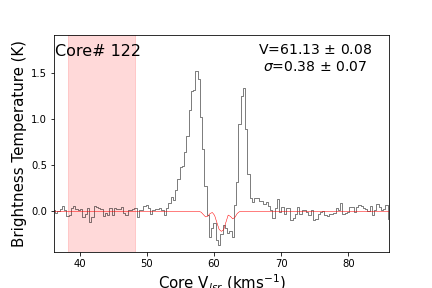}
\caption{Complex-type core-averaged, background-subtracted DCN spectra extracted from the cores in the evolved protocluster W51-IRS2. Readers can refer to \cref{w51irs2_corefits} for the line fit parameters for each core.} 
\end{figure*}
\section{DCN core kinematics}
\label{appendix:dcn_coredisperionplots}
We present the core \vlsr and linewidth of the cores with DCN (3-2) detections for the remaining 13 protoclusters. G338.93 and G333.60 are presented in the main body of the text.
\begin{figure*}
    \centering
    \includegraphics[width=0.97\textwidth]{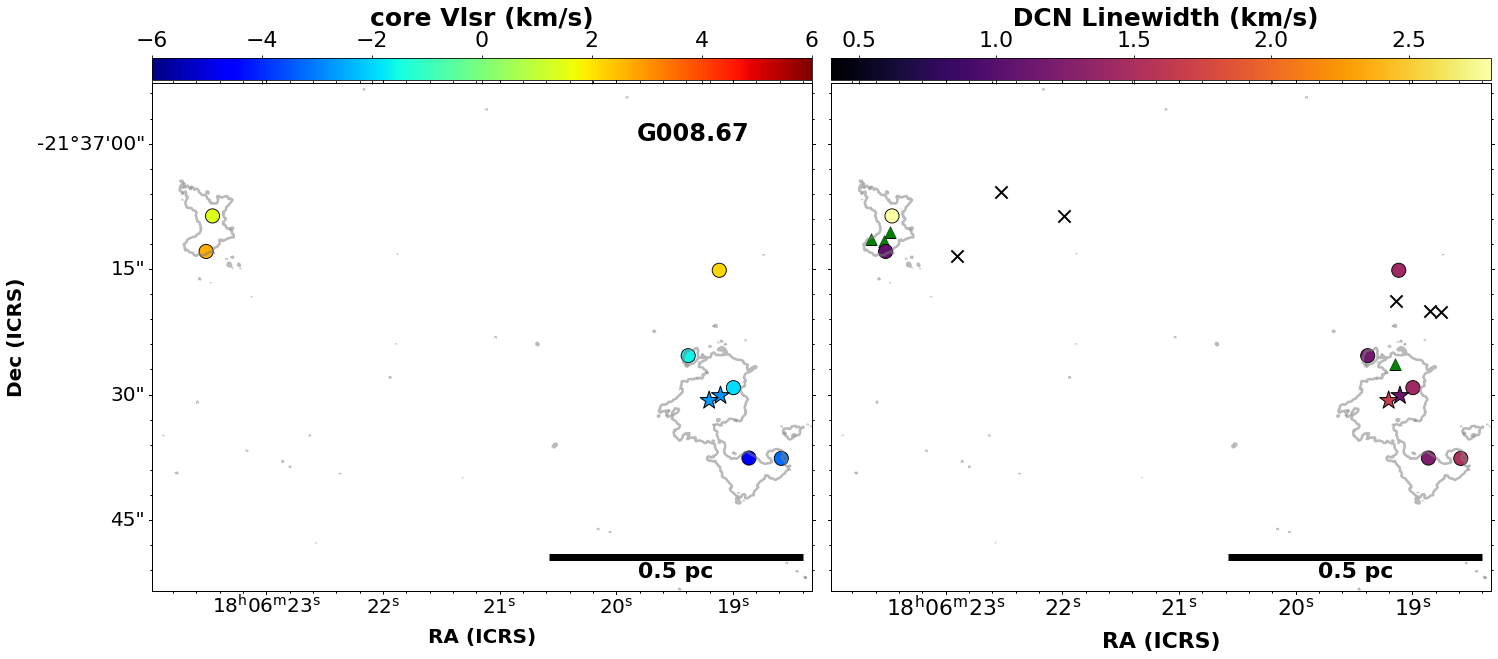}
    \includegraphics[width=0.97\textwidth]{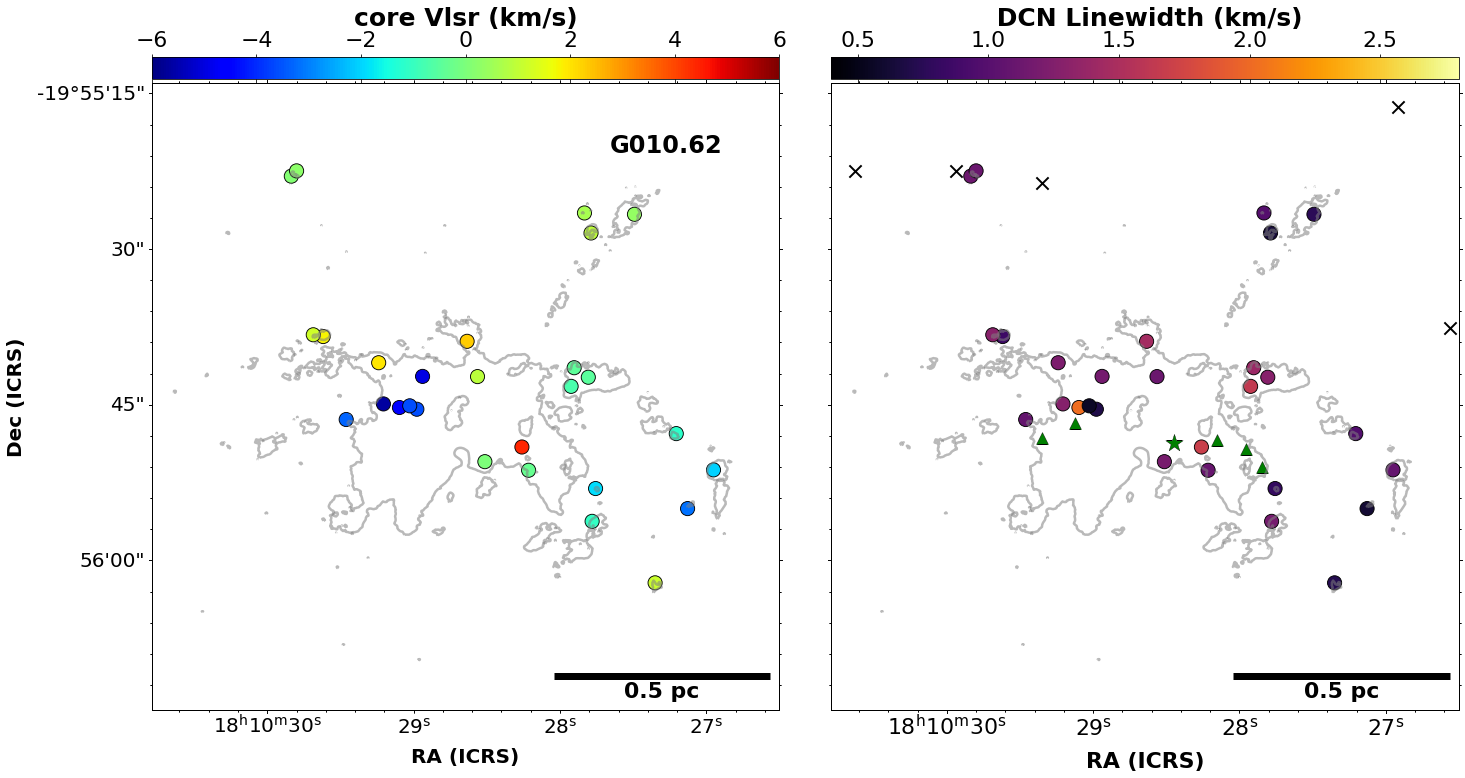}
    \caption{Core \vlsr (left) and DCN (3-2) linewidths (right) estimated from the DCN (3-2) fits to the continuum cores towards the intermediate, and evolved protocluster G008.67 (top) and G010.62 (bottom), respectively. The circles and stars represent the detected cores with mass estimates below and above 8~\msun, respectively, with the colour scale displaying the fitted parameters from the DCN (3-2) fits (left: core \vlsr, right: linewidth). The core \vlsr is the centroid velocity of the DCN (3-2) fit minus the cloud \vlsr (taken as 37~\kms, and -2~\kms for G008.67, and G010.62, respectively). The grey contours are the 4 $\sigma$ level of the DCN (3-2) moment 0 map. In the right panels, the positions of cores without a DCN (3-2) detection are marked with a black cross and black star for cores with a mass estimate below and above 8\msun, respectively, and green triangles and stars represent cores with a Complex-type DCN (3-2) spectra, with a mass estimate below and above 8\msun, respectively.}
\end{figure*}

\begin{figure*}
\ContinuedFloat
    \centering
    \includegraphics[width=0.97\textwidth]{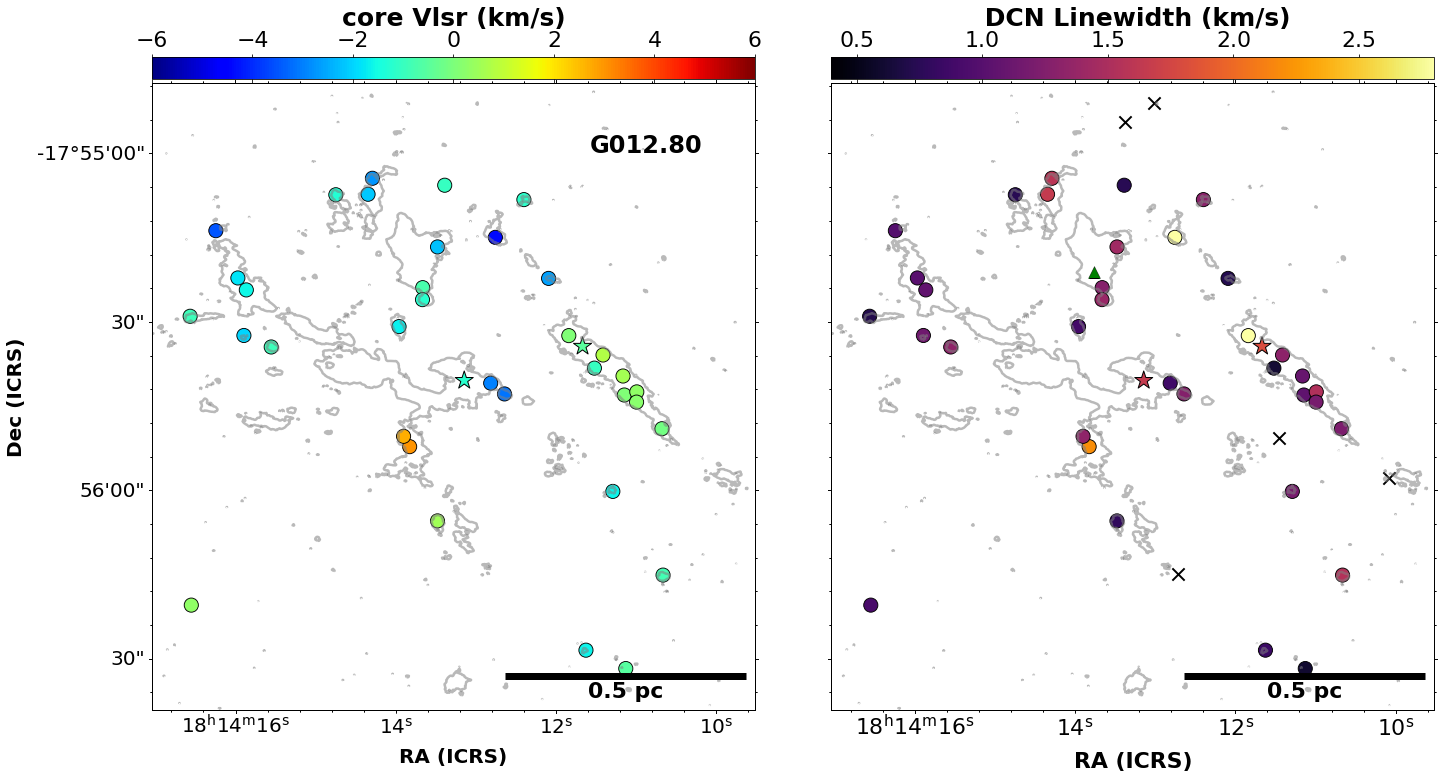}
    \includegraphics[width=0.97\textwidth]{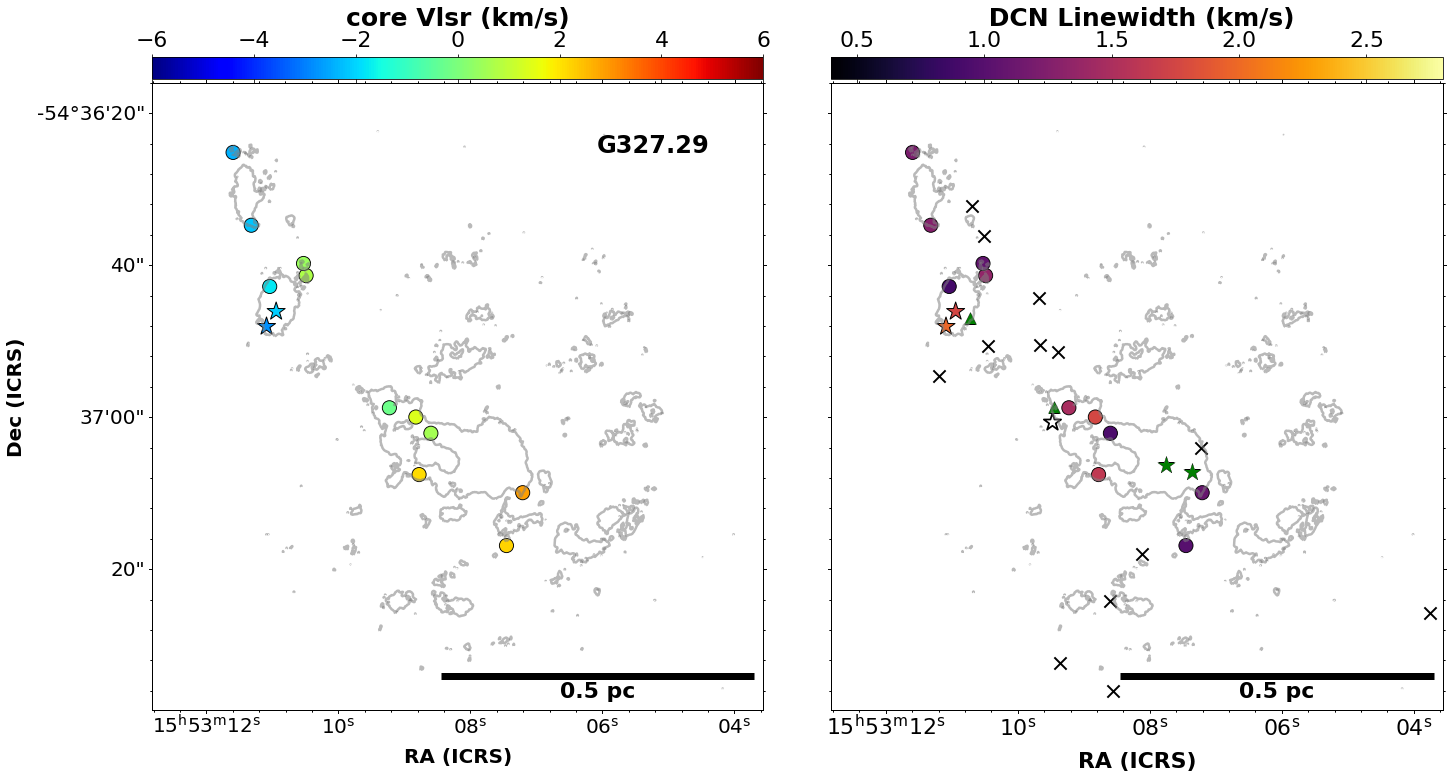}
    \caption{Continued: Core \vlsr (left) and DCN (3-2) linewidths (right) estimated from the DCN (3-2) fits to the continuum cores towards the evolved and young protocluster G012.80 (top) and G327.29 (bottom), respectively. The core \vlsr is the centroid velocity of the DCN (3-2) fit minus the cloud \vlsr (taken as 37~\kms and -45~\kms for G012.80 and G327.29, respectively).}
\end{figure*}

\begin{figure*}
\ContinuedFloat
    \centering
    \includegraphics[width=1\textwidth]{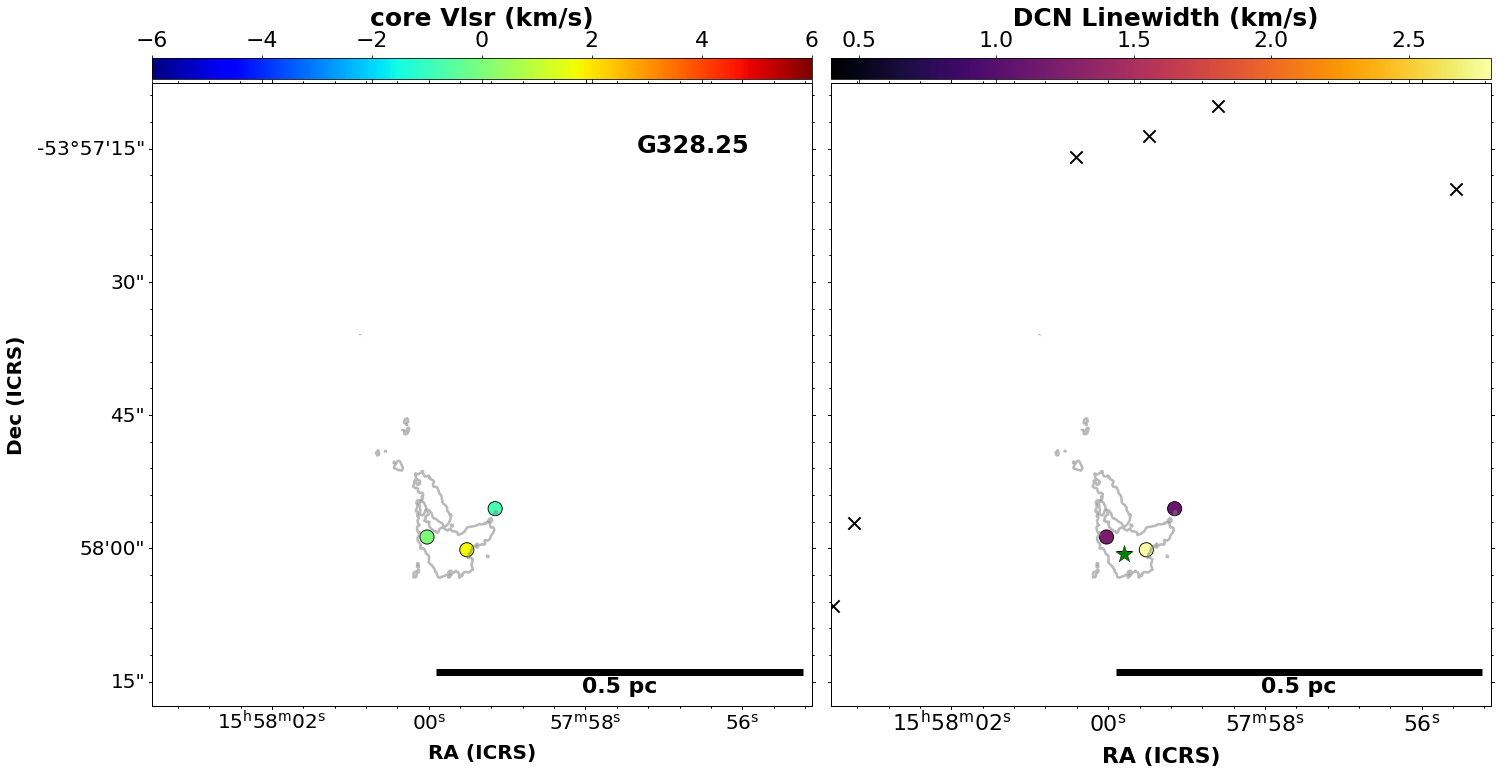}
    \includegraphics[width=1\textwidth]{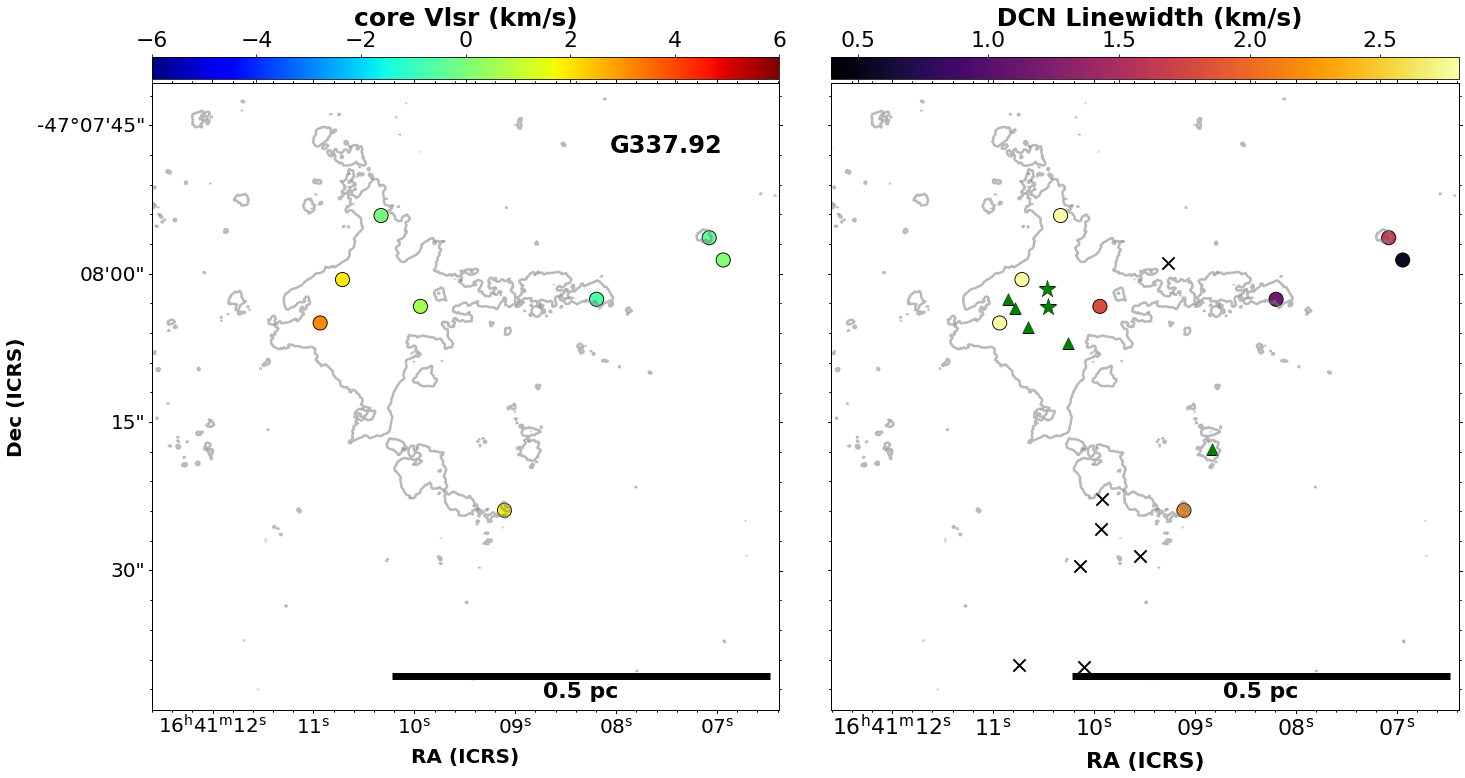}
    \caption{Continued: Core \vlsr (left) and DCN (3-2) linewidths (right) estimated from the DCN (3-2) fits to the continuum cores towards the young protoclusters G328.25 (top) and G337.92 (bottom), respectively. The core \vlsr is the centroid velocity of the DCN (3-2) fit minus the cloud \vlsr (taken as -43~\kms and -40~\kms for G328.25 and G337.92, respectively).}
\end{figure*}

\begin{figure*}
\ContinuedFloat
    \centering
    \includegraphics[width=1\textwidth]{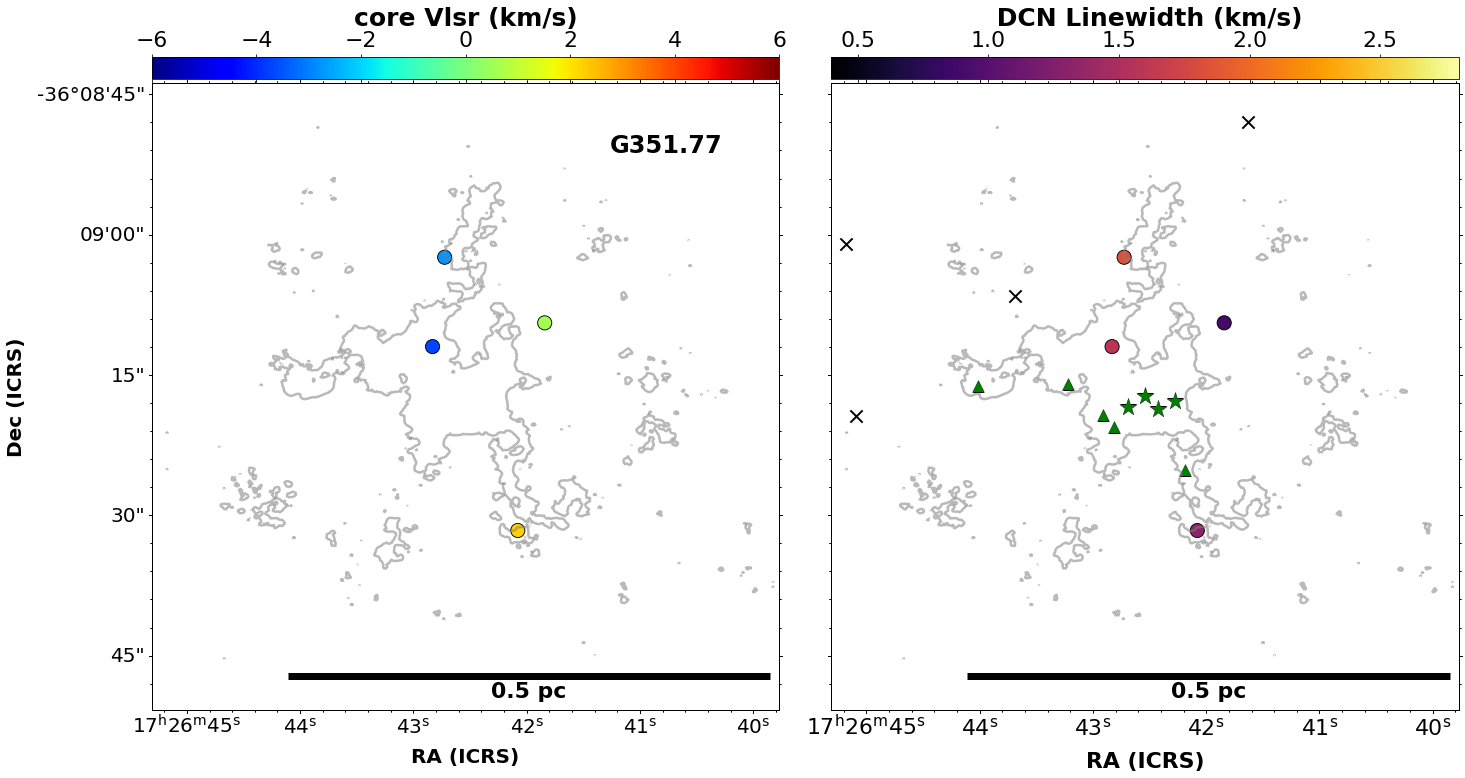}
    \includegraphics[width=1\textwidth]{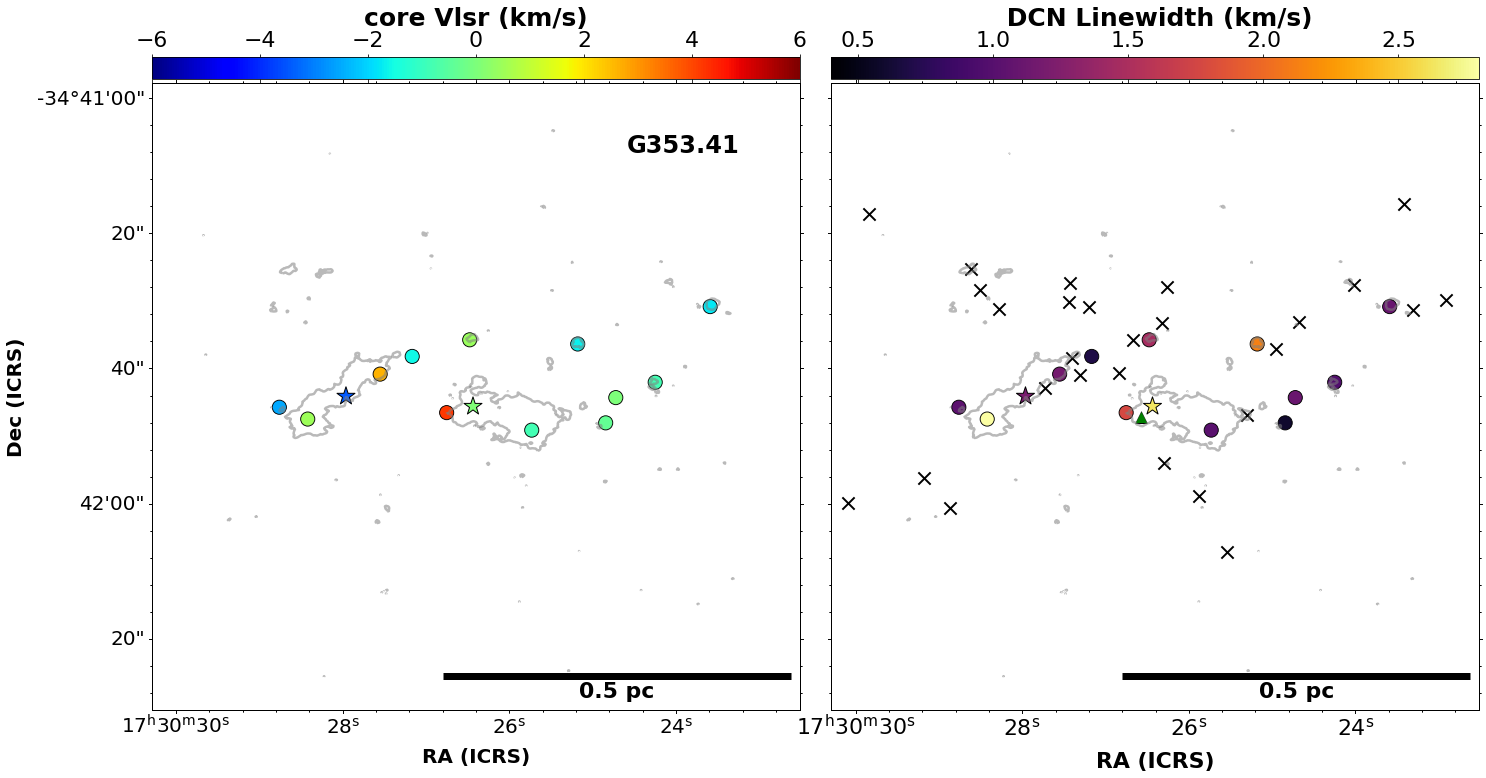}
    \caption{Core \vlsr (left) and DCN (3-2) linewidths (right) estimated from the DCN (3-2) fits to the continuum cores towards the intermediate protoclusters G351.77 (top) and G353.41 (bottom), respectively. The core \vlsr is the centroid velocity of the DCN (3-2) fit minus the cloud \vlsr (taken as -3~\kms, and -17~\kms for G351.77 and G353.41, respectively).}
\end{figure*}

\begin{figure*}
\ContinuedFloat
    \centering
    \includegraphics[width=1\textwidth]{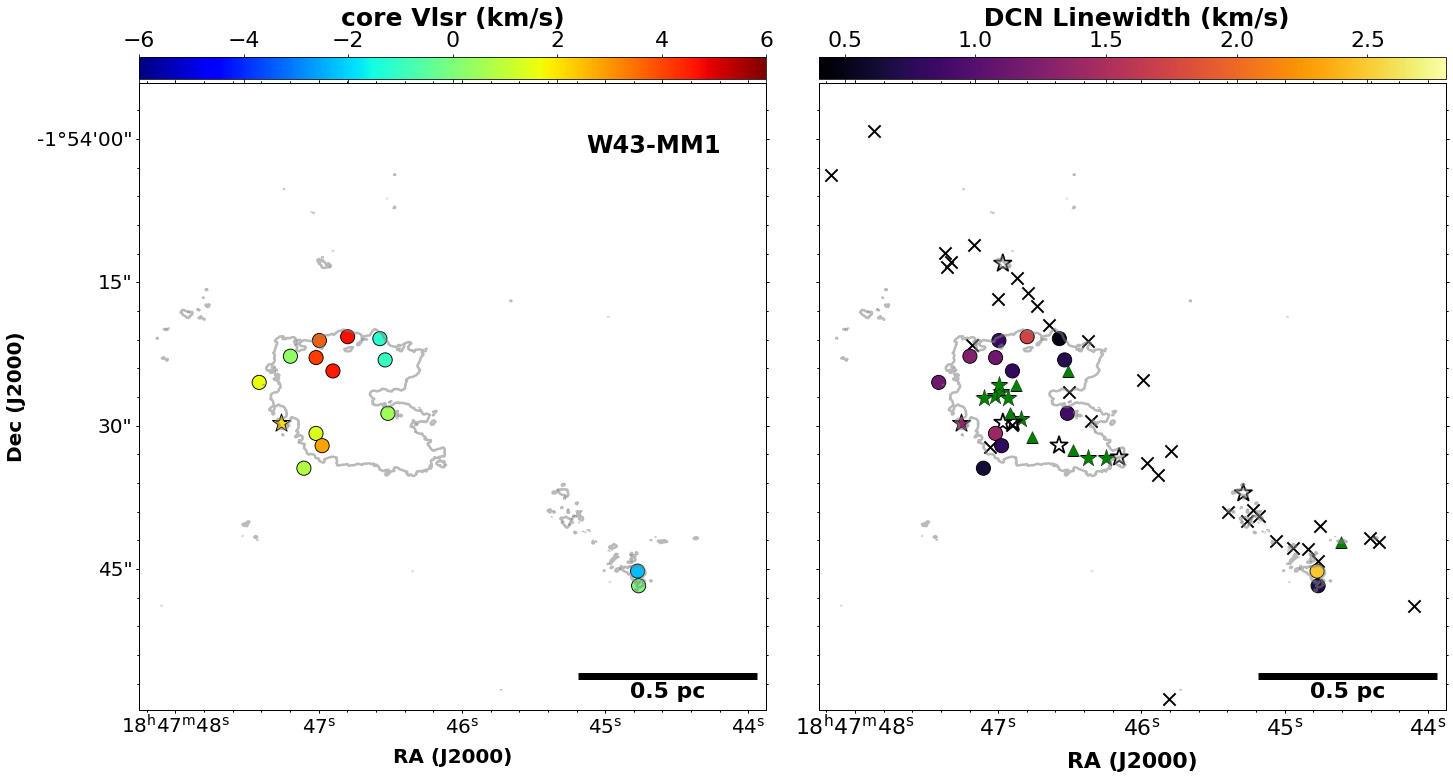}
    \includegraphics[width=1\textwidth]{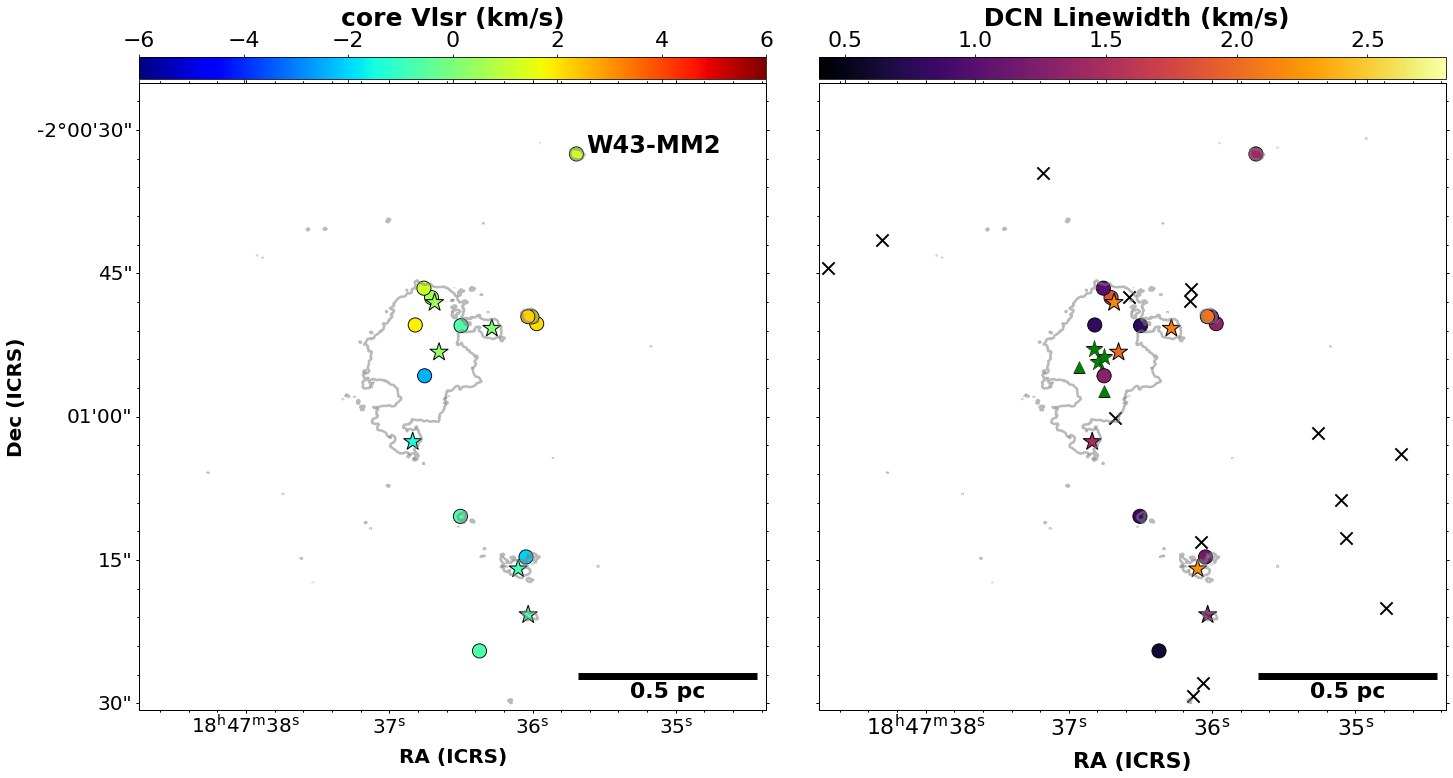}
    \caption{Continued: Core \vlsr (left) and DCN (3-2) linewidths (right) estimated from the DCN (3-2) fits to the continuum cores towards the young protoclusters W43-MM1 (top) and W43-MM2 (bottom), respectively. The core \vlsr is the centroid velocity of the DCN (3-2) fit minus the cloud \vlsr (taken as 97~\kms, and 91~\kms for W43-MM1, and W43-MM2, respectively).}
\end{figure*}

\begin{figure*}
\ContinuedFloat
\centering
\includegraphics[width=1\textwidth]{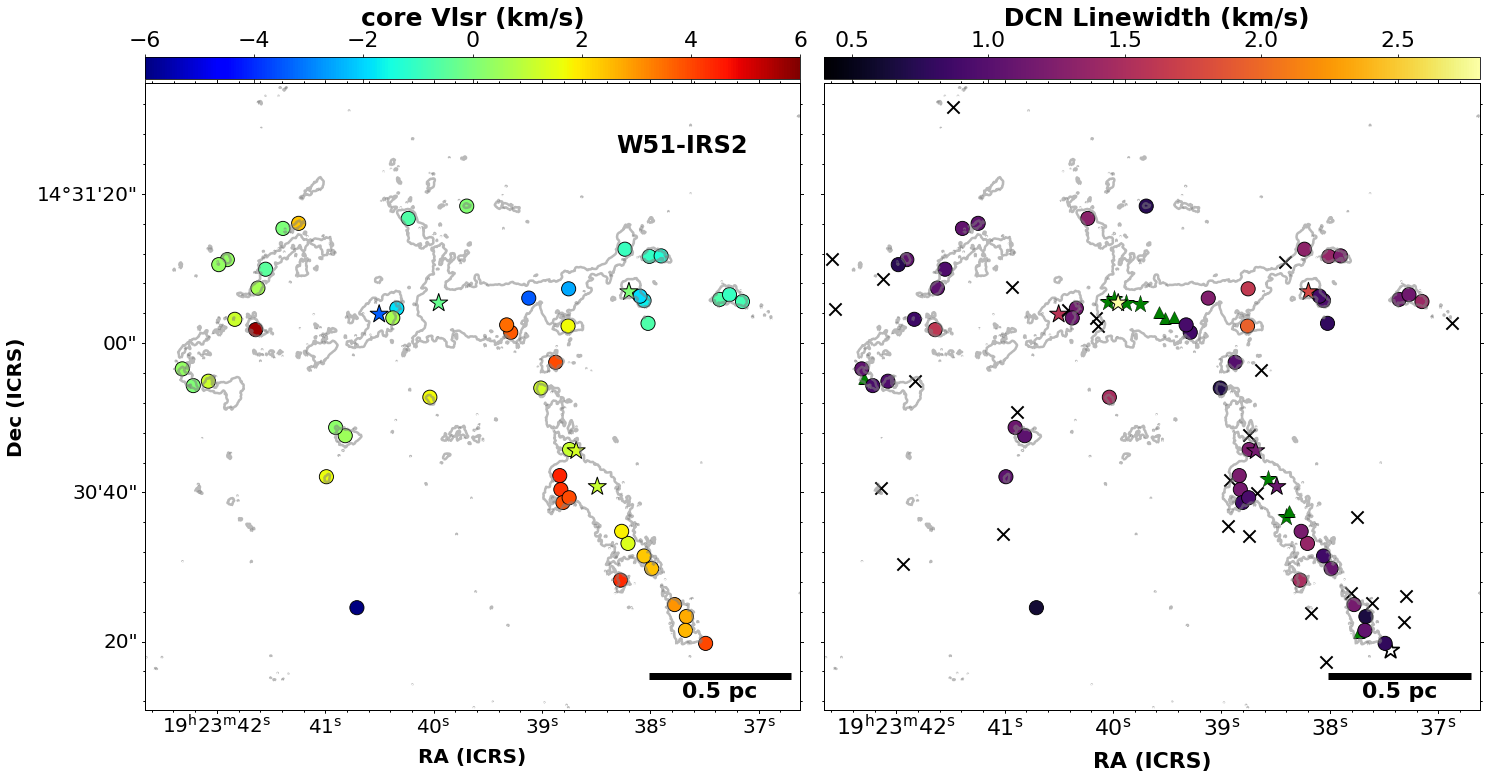}
\caption{Continued: Core \vlsr (left) and DCN (3-2) linewidths (right) estimated from the DCN (3-2) fits to the continuum cores towards the evolved protocluster W51-IRS2. The core \vlsr is the centroid velocity of the DCN (3-2) fit minus the cloud \vlsr (taken as 61~\kms for W51-IRS2).}
\end{figure*}

\begin{figure*}
\ContinuedFloat
\centering
\includegraphics[width=1\textwidth]{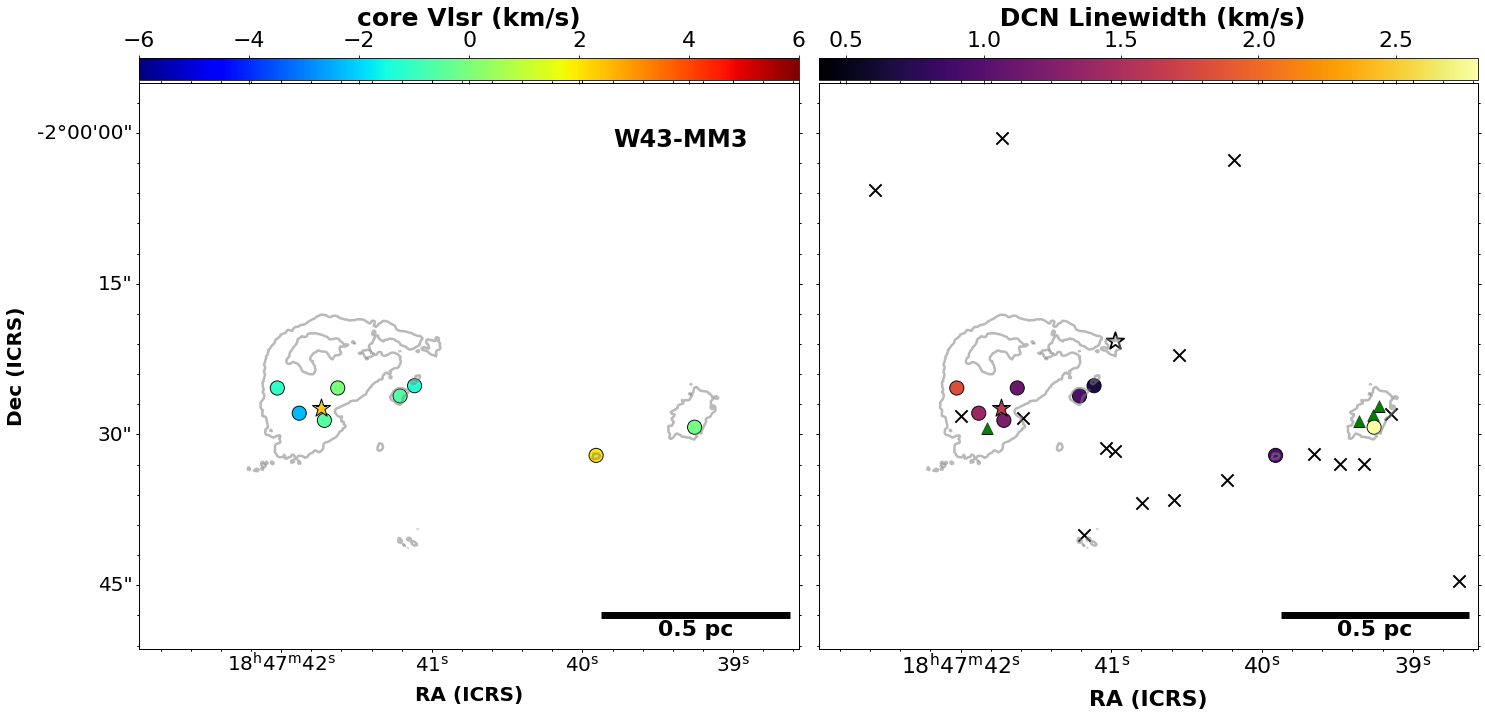}
\includegraphics[width=1\textwidth]{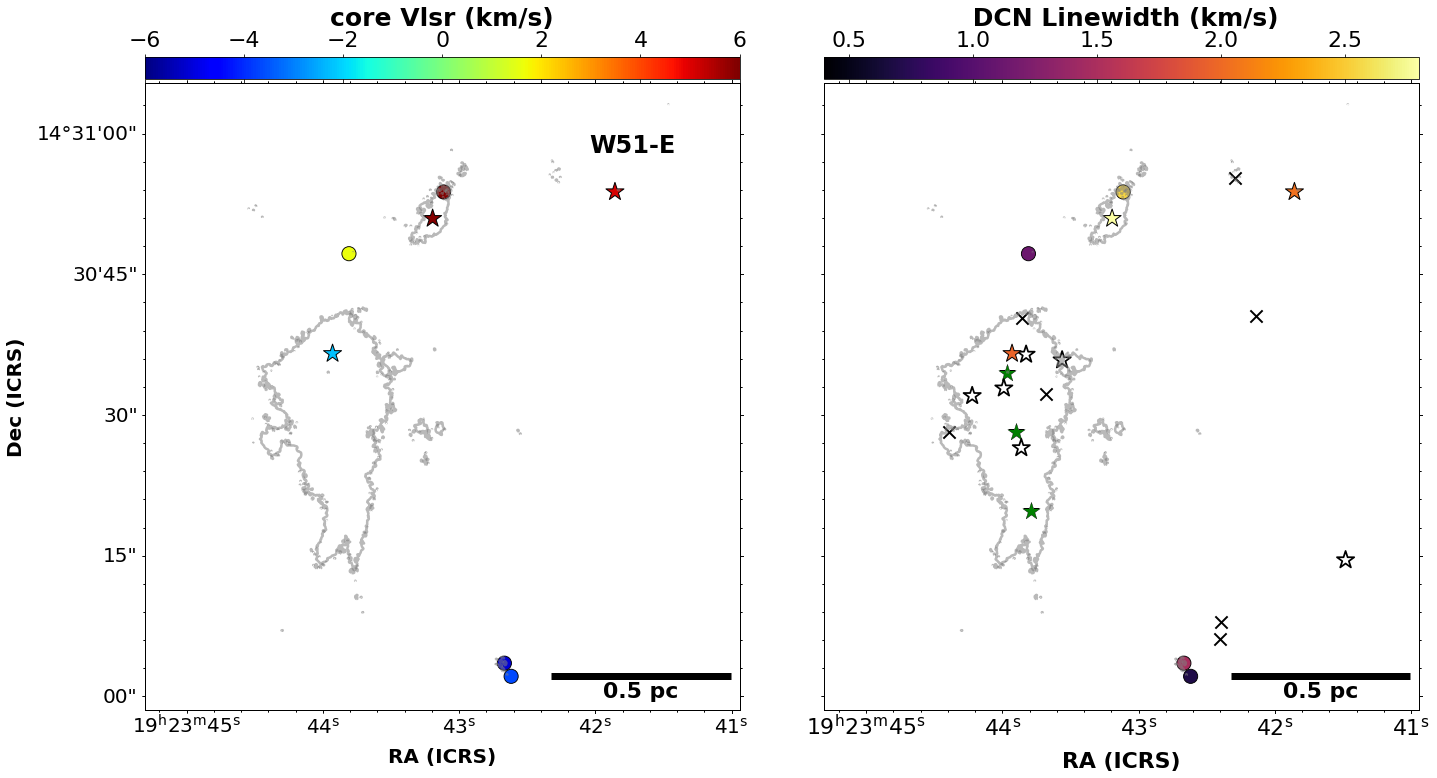}
\caption{Continued: Core \vlsr (left) and DCN (3-2) linewidths (right) estimated from the DCN (3-2) fits to the continuum cores towards the intermediate protoclusters W43-MM3 (top) and W51-E (bottom), respectively. The core \vlsr is the centroid velocity of the DCN (3-2) fit minus the cloud \vlsr (taken as 55~\kms, and 93~\kms for W43-MM3, and W51-E, respectively).}
\end{figure*}
\end{appendix}
\end{document}